\documentclass[preprint,amsmath,amssymb,nofootinbib,floatfix,superscriptaddress]{aastex}
\usepackage{graphics}
\usepackage{graphicx}
\usepackage{dcolumn}
\usepackage{longtable}
\usepackage{bm}
\usepackage{color}
\usepackage{hyperref}
\usepackage{natbib}
\usepackage{tocloft}
\def\prc{ Phys.\ Rev.\ C }
\def\prd{ Phys.\ Rev.\ D }
\begin{document}
\title{STARLIB:  A Next-Generation Reaction-Rate Library for Nuclear Astrophysics} 
\author{A. L. Sallaska\altaffilmark{*}, C. Iliadis, and A. E. Champange}
\affil{University of North Carolina at Chapel Hill, Chapel Hill, NC 27599-3255; and Triangle Universities Nuclear Laboratory, Durham, NC 27708-0308, USA}

\author{S. Goriely}
\affil{Institut d'Astronomie et d'Astrophysique, Universit\'{e} Libre de Bruxelles, C.P. 226, B-1050 Brussels, Belgium}
%

%
%
%
\author{S. Starrfield and F. X. Timmes}
\affil{Arizona State University, Tempe, AZ 85287-1504}

\altaffiltext{*}{Corresponding author:  \url{anne.sallaska@nist.gov}.  Present address:  National Institute of Standards and Technology, Gaithersburg, MD 20899-8462.}

\date{\today}
\begin{abstract}
%


STARLIB is a next-generation, all-purpose nuclear reaction-rate library. For the first time, this library provides the rate probability density at all temperature grid points for convenient implementation in models of stellar phenomena. The recommended rate and its associated uncertainties are also included. Currently, uncertainties are absent from all other rate libraries, and, although estimates have been attempted in previous evaluations and compilations, these are generally not based on rigorous statistical definitions. A common standard for deriving uncertainties is clearly warranted. STARLIB represents a first step in addressing this deficiency by providing a tabular, up-to-date database that supplies not only the rate and its uncertainty but also its distribution. Because a majority of rates are lognormally distributed, this allows the construction of rate probability densities from the columns of STARLIB. This structure is based on a recently suggested Monte Carlo method to calculate reaction rates, where uncertainties are rigorously defined. In STARLIB, experimental rates are supplemented with: (i) theoretical TALYS rates for reactions for which no experimental input is available, and (ii) laboratory and theoretical weak rates. STARLIB includes all types of reactions of astrophysical interest to $Z=83$, such as $(p,\gamma)$, $(p,\alpha)$, $(\alpha,n)$, and corresponding reverse rates. Strong rates account for thermal target excitations. Here, we summarize our Monte Carlo formalism, introduce the library, compare methods of correcting rates for stellar environments, and discuss how to implement our library in Monte Carlo nucleosynthesis studies. We also present a method for accessing STARLIB on the Internet and outline updated Monte Carlo-based rates.

\end{abstract}
%
%
\maketitle
\begin{singlespace}
\end{singlespace}


%
\section{Introduction}

In the mid 20th century, Fowler and collaborators revolutionized stellar evolution simulations by emphasizing standardization of recommended reaction rates for ease of collaboration and comparison in the stellar modeling community (\citet{CF88} and references therein).  The NACRE collaboration followed, providing refined nuclear physics data, and attempted to estimate reaction rate uncertainties never before addressed~\citep{nacre}.  However, the statistical meaning of their ``lower'' and ``upper'' rate limits is unclear, and the need to establish a consistent definition is evident.  The turn of the century brought forth an additional evaluation by~\citet{ilcomp}, which introduced radioactive target nuclei and recommended standard resonance strengths to use as normalizations.  However, no rigorously determined uncertainties were presented.  \citet{des} then presented an evaluation of reaction rates important to Big Bang nucleosynthesis and attempted to quantify uncertainties in a more statistically meaningful manner.  Similar evaluations were published by~\citep{adelberger1,adelberger2} that focused on solar fusion cross sections and their uncertainties.  


Another paradigm shift is currently upon us.  Although stellar reaction rate libraries embrace experimental rates over theoretical ones and cull their experimental data from the above evaluations, {\it none} of the libraries include any estimation of uncertainties (JINA REACLIB: \citet{reaclib}; BRUSLIB: \citet{bruslib}).  This information is essential for realistic simulations of the energy generation and nucleosynthesis that occurs in various stellar phenomena.  In the past, reporting rates without uncertainties was accepted because the additional computing time to allow for uncertainties was exorbitant and because it was unclear how to proceed in a statistically rigorous way.  This, however, can lead to erroneous conclusions drawn from the output of stellar models, as it is not uncommon for the uncertainties in many astrophysically-relevant reaction rates to span orders of magnitude.  To produce realistic nucleosynthesis results and confidence levels, it is imperative for modelers to be able to access a library that includes not only meaningful uncertainties but also probability density functions for each reaction rate at a given temperature.  The advent of readily-available high-powered computing allows not only for applying a Monte Carlo technique to evaluate reaction rates but also for incorporating the reaction rate probability densities into stellar models.  This paper presents a next-generation reaction rate library, STARLIB, built specifically to begin addressing these problems.



STARLIB is a tabular, stellar reaction rate library that includes all types of incident particles, including neutrons, protons, $\alpha$ particles, and $\gamma$ rays.  Target nuclei range between $Z =1-83$.  Experimental data are used when available, and, if not, theory is considered.  The structure of STARLIB rests on a new method to quantitatively define reaction rate uncertainties~\citep{iliadis_1,iliadis_2,iliadis_3,iliadis_4}, although only a fraction of the library as yet contains quantitative uncertainties.  This method uses experimentally determined nuclear physics quantities (resonance strengths and energies, S-factors, partial widths, etc.) as inputs to a Monte-Carlo algorithm.  Each quantity and its associated uncertainty is sampled according to its physically-motivated probability density function.  In 2010, experimental Monte Carlo rates for 62 charged-particle nuclear reactions on $A=14-40$ target nuclei were evaluated~\citep{iliadis_2}.  The library presented here includes updates to 7 Monte-Carlo rates and 1 entirely new rate.  \citet{nicXIIa} is an overview of preliminary work for STARLIB.  In a recently published work \citep{il_al26}, a preliminary version of STARLIB was employed\footnote{In the preliminary 2011 version, the Hauser-Feshbach rates were adopted from \citet{partition}.}.

This work is organized as follows.  We begin with a discussion of the thermonuclear reaction rates in Sec.~\ref{meth}, including analytical evaluation techniques (Sec.~\ref{analmeth}), Monte Carlo methods (Sec.~\ref{MCsec}), theoretical rates based on the TALYS code (Sec.~\ref{theory}), and modification of laboratory rates for stellar environments (Sec.~\ref{sef}).  We present our prescription for building STARLIB in Sec.~\ref{starlib}, and implementation of STARLIB in Monte Carlo nucleosynthesis studies is discussed in Sec.~\ref{use}.  Appendix~\ref{format} details the formatting of STARLIB, and Appendix~\ref{website} specifies how to access the STARLIB library and our Monte Carlo rate calculator on the Internet at~\url{starlib.physics.unc.edu}.  Appendix~\ref{updates} details updated Monte Carlo rates since 2010~\citep{iliadis_3}.  The complete library, with references, is provided on our website and on the Astrophysical Journal website.  



\section{Charged-Particle Reaction Rates}\label{meth}




For a charged particle reaction in a stellar plasma at thermal equilibrium, the reaction rate, $N_A \langle\sigma v\rangle$, for two interacting particles with a Maxwell-Boltzmann distribution of velocities at a temperature $T$ is given by~\citep{iliadisbook,RR}:

\begin{equation}
N_A \langle\sigma v\rangle = N_A \Bigg(\frac{8}{\pi \mu} \Bigg)^{1/2} \frac{1}{(kT)^{3/2}} \int_0^\infty E \sigma(E) e^{-E/kT} \, dE,\label{rxn}
\end{equation}
where $\mu$ is the reduced mass of the particles in the entrance channel, $E$ is the center-of-mass energy of the reaction, and $\sigma$ is the reaction cross section.  The rate may be simplified for resonant (non-interfering and interfering) and non resonant reaction mechanisms.  Details of each are discussed in \citet{iliadis_1} and references therein.


\subsection{Analytical Methodology for Total Reaction Rates}\label{analmeth}


When applying the above formalism to compute the total thermonuclear reaction rate, two main questions arise:  how does one account for uncertainties, and how does one incorporate experimental or theoretical upper limits of partial widths and resonance strengths in a meaningful way?  Only in special circumstances are analytical error propagation methods applicable for reaction rates, for example, when uncertainties in resonance energies are small (i.e., errors less than a few keV) and when upper limits on strengths or partial widths can be disregarded.  In such cases, the uncertainties in both the strength and energy may be propagated using standard techniques of derivatives.  However, in most cases there are many different contributions to the total reaction rate, and analytic methods quickly begin to fail.  

Fowler and collaborators~\citep{CF88} provided no uncertainty information whatsoever.  Although the NACRE collaboration~\citep{nacre} and \citet{ilcomp} did attempt to address this void, their ``upper'' and ``lower'' limits were not established from rigorous statistical methods.  Often what is reported are limits where the authors have guessed the largest sources of uncertainty, and, although this undertaking is a laudable step in the right direction, it has no statistical merit.  There is also the perpetually troublesome problem of how to handle upper limits of nuclear physics input quantities (resonance strengths, partial widths, etc.) in the calculation of the total reaction rate.  The Monte Carlo formalism allows inclusion of upper limits in a natural, statistically precise way.

The interpretation and the meaning of ``upper'' and ``lower'' limits can also be ambiguous.  Because rate probability density functions are usually not reported, these limits have been treated as sharp boundaries in most nucleosynthesis simulations, if uncertainties were taken into account at all.  Several studies attempted to quantify the actual reaction rate probability density.  While \citet{krauss,coc2,izzard} used unphysical, uniform distributions, a few modelers proceeded a step further and assumed Gaussian distributions~\citep{coc04} and even lognormal distributions for rates~\citep{hix,parikh}.  One of the main thrusts of the present work is to begin supplying not only rigorously determined values for recommended rates but also to provide reliable probability densities for a given rate at any temperature.  







\subsection{Monte Carlo Rates}\label{overview}\label{MCsec}\label{matchsec}

To address the persistent problem of quantifying reaction rate uncertainties in a statistically meaningful way, \citet{iliadis_1} and \citet{iliadis_2,iliadis_3,iliadis_4} detailed a Monte Carlo method founded on physically motivated distributions assigned to each nuclear physics input quantity.  Here we present an overview of this method, which motivates the structure of the library presented in this work.

\begin{table}[ht]
\caption{Summary of probability density functions used to estimate reaction rates.   }
\begin{tabular}{ll}
\hline\hline
Parameter & Distribution\\
\hline
Resonance energies & Gaussian \\
Resonance strengths & lognormal\\
Partial widths & lognormal\\
Nonresonant S-factors &lognormal \\
Upper limits ($\Gamma$, $\omega\gamma$) & Porter-Thomas\\
Interference & binary \\
\hline\hline
\end{tabular}
\label{tab}
\end{table}

To calculate the thermonuclear reaction rate, each input parameter must be described by a probability density function.  The distributions employed in the present work are summarized in Table~\ref{tab}.  Once distributions are assigned for each input parameter, a random value can be sampled for each, and the reaction rate is computed from Eq.~\ref{rxn} (or its subsequent simplifications) at a given temperature using standard Monte Carlo sampling techniques.  The sampling continues until the user's desired precision is reached, with a minimum of 5,000 samples in order to achieve reproducibility within a few percent.  All Monte Carlo calculations reported in this work are performed using the code {\tt RatesMC}~\citep{iliadis_3}.  The method allows for a statistically rigorous definition of the low rate, median (recommended) rate, and high rate as the 16th, 50th, and 84th percentiles of the cumulative rate distribution, respectively.  The values chosen here represent a coverage probability of 68\%.

\begin{figure}[ht]
\centering
\includegraphics[scale=0.5]{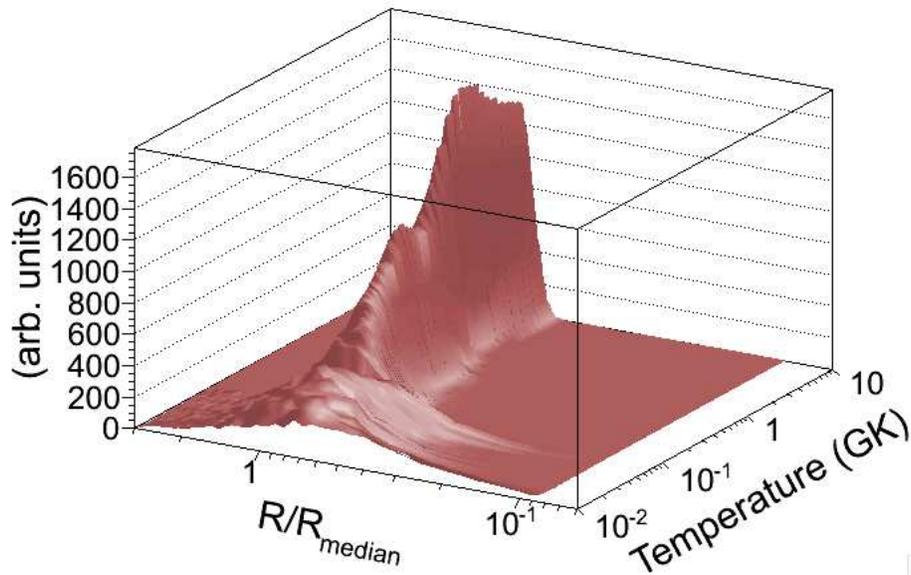}
\caption{(color online) Three-dimensional reaction rate probability density function, normalized to the median (recommended) rate, for $^{22}$Na($p,\gamma)^{23}$Mg (see Appendix~\ref{22naupdate}).  Notice how the distribution becomes narrower with increasing temperature, i.e., the reaction rate uncertainty decreases because the nuclear physics input becomes more reliable at higher energies.    }
\label{3d}
\end{figure}

Although the ``upper limit" and ``lower limit" values of rates presented in the literature are sometimes interpreted by  nuclear astrophysicists as sharp boundaries, it should be clear from the above discussion that sharp reaction rate boundaries have no statistical meaning.  In fact, the values of the bounds simply depend on the rate probability density function and the associated desired coverage probability.  To emphasize this point more clearly, we show in Fig.~\ref{3d} a 3-dimensional reaction rate probability density function, for the $^{22}$Na$(p,\gamma)^{23}$Mg reaction (Appendix~\ref{22naupdate}).  Although we chose (arbitrarily) to associate the 16th and 84th percentiles with the ``low" and ``high" rates, this plot clearly illustrates that the probability density is not zero outside of these bounds.  Notice that the sharp peak at high temperatures indicates a smaller uncertainty band, whereas, at low temperatures, the probability density is much broader because of more uncertain nuclear physics input parameters.

Every experiment performed will be subject to a cutoff at higher energies.  In other words, if the energy window of effective stellar burning at a given temperature is only partially covered by data, then the experimental rate will be smaller than the actual rate.  Two questions then arise:  (i) what is the actual effective stellar energy window? (ii) what is the highest temperature at which experimental reaction rates are still based on data?  These questions are addressed in detail in \citet{newton} and \citet{iliadis_2}.  Methods of extrapolating  rates to high temperatures will be discussed in Sec.~\ref{sef}.

The evaluation of \citet{iliadis_2} contains 62 reaction rates determined with the Monte Carlo method for $A=14$ to 40 target nuclei.  In this work, we present a new Monte Carlo rate for $^{38}$Ar($p,\gamma)^{39}$K and updates for the reactions $^{18}$O$(p,\gamma)^{19}$F, $^{18}$O$(p,\alpha)^{15}$N, $^{22}$Ne$(p,\gamma)^{23}$Na, $^{22}$Ne$(\alpha,\gamma)^{26}$Mg, $^{22}$Ne$(\alpha,n)^{25}$Mg, $^{22}$Na$(p,\gamma)^{23}$Mg, and $^{29}$P$(p,\gamma)^{30}$S.  These results are discussed in Appendix~\ref{updates}.



\subsubsection{Lognormal Reaction Rates}\label{logsec}\label{logdist}

The reaction rate probability density function (the Monte Carlo output) is determined by the interplay of the probability densities for all input quantities.  Depending on which contribution dominates the total rate, the rate probability density may assume many different shapes.  Experience has shown, however, that in the majority of cases (but certainly not all) the rate probability density is well approximated by a lognormal function.  This observation is important for two reasons, as will be discussed in more detail in Sec.~\ref{starlib}:  (i) it allows for a convenient implementation of the Monte Carlo rate probability density in reaction rate libraries, and (ii) it allows us to adopt a lognormal rate probability density for all other reactions for which Monte Carlo rates are not yet available.

The lognormal distribution is a close cousin of the Gaussian distribution and allows for an asymmetric sampling of a manifestly positive quantity.  Consider a continuous random variable $y$ that is Gaussian distributed.  If $y=\rm{ln} (x)$, then $x$ follows a lognormal distribution.  In other words, a lognormal distribution is simply a distribution where the natural logarithm of a continuous random variable follows a Gaussian distribution.  It is parameterized by two variables, $\mu$ and $\sigma$:


\begin{equation}
f(x)  = \frac{1}{\sigma\sqrt{2\pi}} \frac{1}{x} e^{-(\rm{ln}\, x-\mu)^2/(2\sigma^2)}\qquad {\rm for  }\quad 0 < x < \infty.\label{logneq}
\end{equation}  
However, the distribution is no longer symmetric, and the parameters do not represent the mean and square root of the variance, as in the case of the Gaussian distribution.  A corollary to the central limit theorem illustrates the power of this distribution:  the {\it product} of a large number of distributions {\it of any shape} tend toward a lognormal distribution, whereas the sum follows a Gaussian distribution.

The mean, median, and respective uncertainties are given as:

\begin{equation}
{\rm mean} \equiv E[x] = e^{\mu+\sigma^2/2}, \qquad V[x] = e^{2\mu+\sigma^2}(e^{\sigma^2}-1)\label{muinv}
\end{equation}

\begin{equation}
{\rm median}  \equiv  x_{med}= e^\mu, \qquad f.u. \, {\rm (68\% \, coverage)} = e^{\sigma}\label{starmu},
\end{equation}
where $E[x]$ is the expectation value and $V[x]$ is the variance (square of the standard deviation) of a normalized distribution.  $f.u.$ denotes the factor uncertainty.  Given values for the expectation value and variance, Eq.~\ref{muinv} can be inverted to derive the lognormal parameters, $\mu$ and $\sigma$:

\begin{equation}
\mu=\rm{ln}(E[x])-\frac{1}{2}\rm{ln} \rm{\Big( 1+\frac{V[x]}{E[x]^2}\Big)}, \qquad \sigma=\sqrt{\rm{ln}\Big( 1+\frac{V[x]}{E[x]^2}\Big)}\label{mu}
\end{equation}
For a manifestly positive data set $\{ z_i\}$, Eq.~\ref{starmu} is derived from the geometric mean and geometric standard deviation:

\begin{equation}
\mu_g \equiv \sqrt[n]{z_1 \cdot z_2\cdot \cdot \cdot z_n} = e^\mu, \qquad \sigma_g \equiv {\rm exp} [ (1/n \sum_{i=1}^n ({\rm ln} \, z_i/\mu_g)^2] = e^\sigma. \label{eq6}
\end{equation}
If the data are lognormal, the quantity $e^\mu$ is given by the {\it median} value.  Then, the lower and upper bounds are defined as:

\begin{equation}
 x_{low} = \mu_g/\sigma_g = e^{\mu-\sigma}, \qquad x_{high} = \mu_g\sigma_g = e^{\mu+\sigma}, \label{eq7}
 \end{equation}
for a coverage probability of 68\%.  From either relation, one may determine the factor uncertainty such that,

\begin{equation}
f.u. = e^{\mu+\sigma}/e^\mu = e^\mu/e^{\mu-\sigma} = e^\sigma. \label{eq8}
\end{equation}
These quantities are crucially important for our Monte Carlo method (Sec.~\ref{MCsec}) and for using our library to calculate rate probability density functions (Sec.~\ref{starlib}).



\begin{figure}
\centering
\includegraphics[scale=.6]{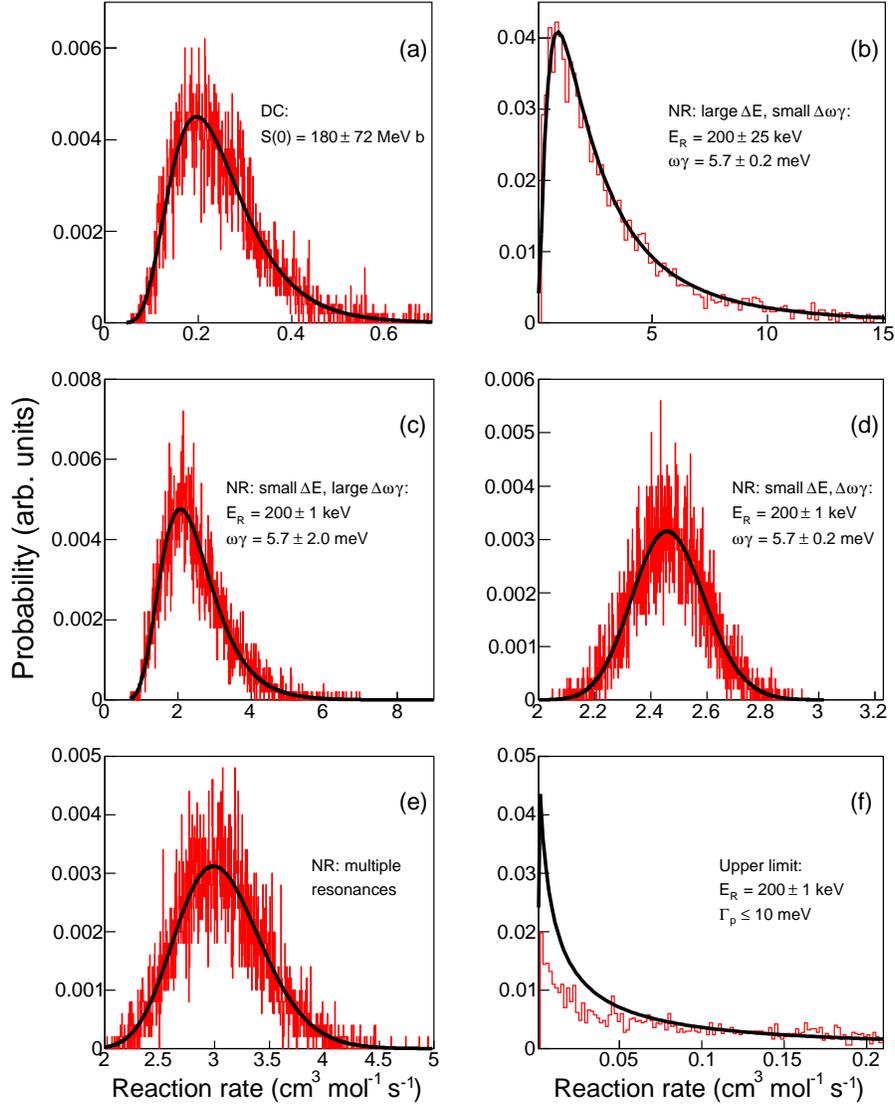}
\caption{(color online) Monte Carlo reaction rate probability density functions (red histograms) for hypothetical input for the $^{22}$Na($p,\gamma)^{23}$Mg reaction. The distributions have been obtained using 5,000 samples at $T = 0.3$ GK.  The solid curves are the lognormal approximations to the rates.   (a) direct capture (DC) only; (b) single narrow resonance (NR) at $E_R=200$ keV with large energy uncertainty; (c) single NR with large resonance strength uncertainty; (d) single NR with both small uncertainties in resonance energy and strength; (e) multiple narrow resonances:  $E_R = 204\pm1$ keV, $274\pm1$ keV, $434\pm1$ keV, $480\pm2$ keV, $583\pm1$ keV with $\omega\gamma = 6\pm1$ meV, $39\pm8$ meV, $170\pm20$ meV, $93\pm36$ meV, $590\pm70$ meV, respectively; and (f) upper limit to a particle partial width at the same resonance energy as (a,b,c).  The rate distribution in (f) deviates from a lognormal distribution since the reduced width input is based on a Porter-Thomas distribution. }
\label{MCillustration}
\end{figure}

We will now discuss the reasons for the observation that most experimental reaction rates follow a lognormal probability density.  This is illustrated in Fig.~\ref{MCillustration} for various cases in one sample reaction.  Panel (a) displays the Monte Carlo reaction rate probability density obtained by assuming that only the (non-resonant) direct capture process contributes to the total rate.  Since we assumed a lognormal distribution for the S-factor input, the reaction rate output distribution is also lognormal.  Shown in panel (b) is the rate probability density for a single narrow resonance with a large uncertainty in the resonance energy and small uncertainty in the resonance strength.  Because we assume that $E_R$ is distributed in a Gaussian fashion, the Boltzmann factor $e^{-E_R/kT}$ that enters into the reaction rate formalism \citep[see][]{iliadis_1,iliadisbook} will be, by definition, lognormal, as is evident in the figure.  Panel (c) illustrates the opposite limit case wherein the uncertainty in the resonance strength is large for a single narrow resonance with a small uncertainty for the resonance energy.  The reaction rate is directly proportional to the lognormally-distributed strength \citep[see][]{iliadis_1,iliadisbook} and will thus also follow a lognormal distribution.  Panel (d) shows the reaction rate probability density function for a single narrow resonance with small uncertainties in both the resonance energy and strength.  In this case, the rate is still lognormally distributed but with a small spread parameter, thus appearing more Gaussian.  This is similar to panel (e), which presents the reaction rate probability density for multiple contributing resonances.  Here, five resonances were chosen within the Gamow window, with relatively small uncertainties.  As per the central limit theorem, the sum of these resonances appears Gaussian, which can be well-approximated by a lognormal distribution.  Panel (f) shows the reaction rate distribution for a single resonance with an upper limit on the particle partial width (and, by extension, strength).  Reduced widths of upper limits are assumed to be Porter-Thomas distributed, not lognormal, hence the deviation from the lognormal approximation.  Other special cases where the rate is not lognormally distributed are discussed in \citet{iliadis_2}.



\subsection{Theoretical Reaction Rates:  TALYS}\label{theory}

In the following sections, we will discuss and employ theoretical estimates of reaction rates.  These are important for a number of issues, including:  (i) rate estimates for reactions where no experimental nuclear physics information exists; (ii) estimation of stellar rates from experimental rates, and extrapolation of stellar rates to temperatures at which no experimental rates exist; (iii) calculation of stellar enhancement factors, stellar rate ground state fractions, and nuclear partition functions (Sec.~\ref{sef}).

In this work, we adopt theoretical estimates for reaction rates based on the statistical (Hauser-Feshbach) model of nuclear reactions, computed using the code TALYS~\citep{Koning04,Koning08a,Goriely08a}.  TALYS is a modern nuclear reaction code, which includes many state-of-the-art nuclear models to cover the main reaction mechanisms.  
Recently TALYS was updated to estimate nuclear reaction rates of particular relevance to astrophysics~\citep{Goriely08a}.  
All available experimental information on nuclear masses \citep[including the latest release of the 2012 atomic mass evaluation by][]{ame2012}, deformations, and low-lying states spectra \citep[from the RIPL3 database,][]{Capote09} are considered.  Otherwise, various local and global models are used to describe the nuclear structure properties, optical potentials, level densities, $\gamma$-ray strengths, and fission properties.  

\begin{figure}[ht]
\centering
\includegraphics[scale=0.30]{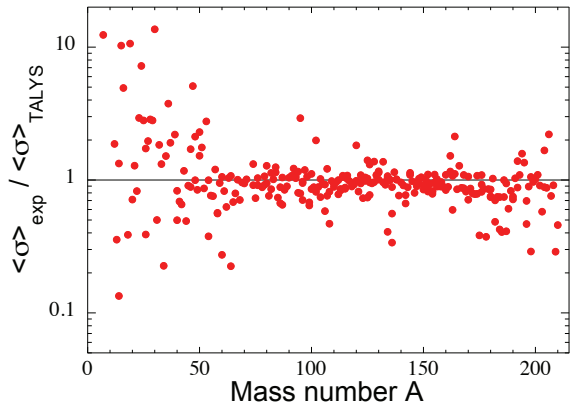}
\caption{(Color online) Comparison between experimental and TALYS Maxwellian-averaged (laboratory) $(n,\gamma)$ cross sections at 30~keV for 281 nuclei between Li and Bi.  Experimental data are taken from the compilation of \citet{Bao00}, as well as from more recent measurements compiled in \citet{bruslib}.  Note that for the heavier target nuclei ($A\ge50$), most experimental and theoretical values agree within a factor of 2.
}
\label{fig_talys}
\end{figure}

In the special case of radiative neutron captures, Maxwellian-averaged cross sections have been measured mostly at 30 keV with relatively high accuracy for about 281 stable or long-lived target nuclei.  See, in particular, the compilation of \citet{Bao00} and references found in \citet{bruslib}.  For those reactions, TALYS (laboratory) Maxwellian-averaged cross sections are normalized using the experimental value at 30 keV to ensure an accurate description in the whole temperature range, the Hauser-Feshbach being known to provide a proper description of the energy dependence of the reaction cross section.  Figure~\ref{fig_talys} displays a comparison of experimental and {\it unnormalized} TALYS Maxwellian-averaged laboratory $(n,\gamma)$ cross sections at 30 keV for 281 target nuclei between Li and Bi.  It is apparent that for medium and heavy target nuclei ($A \ge 50$) most of the radiative capture rates agree within 40\%, although deviations up to a factor of 3 can be found in some cases.  This represents an improvement in comparison with the former calculation based on the MOST code \citep[][see in particular Fig. 39]{arnould07} where globally an 85\% accuracy was obtained.
For light nuclei, larger deviations are found either because the resonance contribution is overestimated, or the direct capture contribution has been disregarded.

For neutron rich and neutron deficient unstable nuclei, the nuclear reaction rates are estimated using microscopic models, in particular nuclear structure properties based on the HFB-21 mass model
~\citep{Goriely10}, the Hartree-Fock-Bogolyuobov (HFB) plus combinatorial nuclear level density model~\citep{Goriely08}, and the HFB plus Quasi-Random Phase Approximation (QRPA) model for the dipole strength~\citep{Goriely04}.  While the particle-induced reaction rates are directly computed with the TALYS code, the reverse photodisintegration rates are derived from the detailed balance relation~\citep[see][]{iliadisbook}.

Finally, we comment briefly on the predictive power of TALYS rates.  The uncertainties involved are mainly of two origins:  (i) the description of the reaction mechanism, i.e., the model of formation and de-excitation of the compound nucleus, including a possible pre-equilibrium and direct capture contribution, and (ii) the evaluation of the nuclear quantities entering the calculation of the transmission coefficients for each entrance or exit channel.  Our estimate of factor uncertainties for the adopted TALYS rates will be discussed in Sec.~\ref{talys}.

TALYS not only provides reaction rate predictions for target nuclei in their ground state (``laboratory" reaction rates), but also reaction rates that include contributions from thermally populated excited target states (``stellar" rates).  These topics, including theoretical estimates of stellar enhancement factors, stellar rate ground state fractions, and partition functions, will be discussed in the next section.

\subsection{Modification of Laboratory Rates for Stellar Environment}\label{sef}\label{part}

Before the reaction rates discussed above can be used in stellar model calculations, they must be corrected for a number of effects.  Almost all measurements necessarily determine the reaction rate with the target in its ground state, $x^L=N_A \langle\sigma v\rangle$, referred to as ``laboratory rate" or ``ground state rate".  However, in stellar environments, excited levels in the target nucleus may be populated that can participate in the stellar burning, giving rise to a ``stellar rate", $x^*=N_A \langle\sigma v\rangle^*$.  This issue is particularly important at elevated stellar temperatures.  Because there is significant confusion in the community regarding how to address this correction, we present a discussion of the different methods.

Consider a reaction $0+1\rightarrow2+3$, with 0 and 3 denoting the target and final (residual) nucleus, respectively.  Species 1 and 2 are assumed to be light particles (p, n or $\alpha$).  The laboratory reaction rate is obtained by summing over all relevant transitions to excited states, $\nu$, in the final nucleus 3:
\begin{equation}  
x^L=N_A \langle\sigma v\rangle=\sum\limits_\nu N_A \langle \sigma v \rangle^{g.s.  \to \nu}
\end{equation}
For a non-degenerate plasma in thermodynamic equilibrium, the ratio of the number density $N_{i\mu}$ of nuclei~$i$ in excited state $\mu$ and the total number density $N_i$ of nuclei~$i$ is given by a Boltzmann distribution:
\begin{equation}   
P_{i\mu} = \frac{N_{i\mu}}{N_i} = \frac{g_{i\mu} e^{-E_{i\mu}/kT}}{\sum_\mu g_{i\mu} e^{-E_{i\mu}/kT}} = \frac{g_{i\mu} e^{-E_{i\mu}/kT}}{G_i}  \label{excitedtototalratio}
\end{equation}
with $g_{i\mu} \equiv (2j_{i\mu} + 1)$, $j_{i\mu}$, and $E_{i\mu}$ the statistical weight, spin, and excitation energy, respectively, of state $\mu$ in nucleus $i$.  $T$ is the plasma temperature.  The sum over $\mu$ in the denominator includes the ground state and is referred to as the \emph{partition function}, $G_i$, of nucleus $i$.  The quantity $G_i^n \equiv G_i/g_{ig.s.}$ is called \emph{normalized partition function}, since $g_{ig.s.}$ refers to the statistical weight of the ground state in nucleus $i$.  When experimental information on excited states is not available, the partition functions can be obtained from nuclear level density estimates.  In this work, we adopt the results computed with the combinatorial model of \citet{Goriely08} as used in the nuclear reaction code TALYS (Sec.~\ref{theory}).  The stellar rate is then obtained by appropriately averaging over excited states, $\mu$, in the target nucleus 0~\citep{iliadisbook}:
\begin{equation}  
x^*=N_A \langle\sigma v\rangle^*= \sum\limits_\mu P_{0\mu} \sum\limits_\nu N_A \langle \sigma v \rangle^{\mu \to \nu} = \frac{\sum_\mu g_{0\mu} e^{-E_{0\mu}/kT} \sum_\nu N_A \langle \sigma v \rangle^{\mu \to \nu}}{G_0}
\end{equation}

From these expression we can derive a number of useful quantities.  The ratio of stellar to laboratory reaction rates, referred to as \emph{stellar enhancement factor}, is defined by:
\begin{equation}
SEF
=\frac{N_A \langle\sigma v\rangle^*}{N_A \langle\sigma v\rangle}
=\frac{\frac{1}{G_0} \sum_\mu g_{0\mu} e^{-E_{0\mu}/kT} \sum_\nu N_A \langle \sigma v \rangle^{\mu \to \nu}}{\sum_\nu N_A \langle \sigma v \rangle^{g.s.  \to \nu}} \label{sefeq}
\end{equation}
Given a theoretical estimate for $SEF$, a stellar rate can be derived from a laboratory rate via $N_A \langle\sigma v\rangle^*=SEF~N_A \langle\sigma v\rangle$.  Numerical examples for deriving stellar rates from laboratory rates are presented in \citet{iliadisbook}.  In the present work, we compute the stellar enhancement factors using the nuclear reaction code TALYS (Sec.~\ref{theory}).  It is frequently overlooked that the numerical value of the stellar ``enhancement" factor can be smaller than unity.  This may happen, for example, if a significant fraction of target nuclei, 0, reside in excited states, and if, at the same time, the reaction rates involving these excited states for some reason (e.g., angular momentum, parity, or isospin selection rules) are much smaller compared to the ground state rate.  For the same reason, one should \emph{not} conclude that a value of $SEF=1$ implies a negligible rate contribution from excited target states: the interplay of significant excited target state population, $P_{0\mu}$, and small reaction rates  from excited target states can indeed give rise to a stellar enhancement factor near unity.

The fractional contribution of the laboratory rate to the stellar rate, referred to as \emph{stellar rate ground state fraction}, can be defined as~\citep{rauscher1}:
\begin{equation} 
GSF
=\frac{P_{0g.s.} N_A \langle \sigma v \rangle}{N_A \langle\sigma v\rangle^*}
= \frac{\frac{g_{0 g.s.}}{G_0}  \sum_\nu N_A \langle \sigma v \rangle^{g.s.  \to \nu}}{\sum\limits_\mu P_{0\mu} \sum\limits_\nu N_A \langle \sigma v \rangle^{\mu \to \nu}} 
=\frac{1}{G_0^n~SEF}
\end{equation}
The range of possible values amounts to $0\le GSF \le1$.  For the limiting value of $GSF=1$, the stellar rate is equal to the laboratory rate, implying $SEF=1$.  It is apparent that the stellar rate ground state fraction contains more information than the stellar enhancement factor.  The interplay of population, $P_{0\mu}$, and reaction rates, $N_A \langle \sigma v \rangle^{\mu \to \nu}$, involving excited target states, $\mu$, may give rise to a significant overall contribution of excited target states to the total rate ($GSF<1$), although the corresponding stellar ``enhancement" factor may not be affected ($SEF=1$).  

\begin{figure}
\centering
\includegraphics[scale=.55]{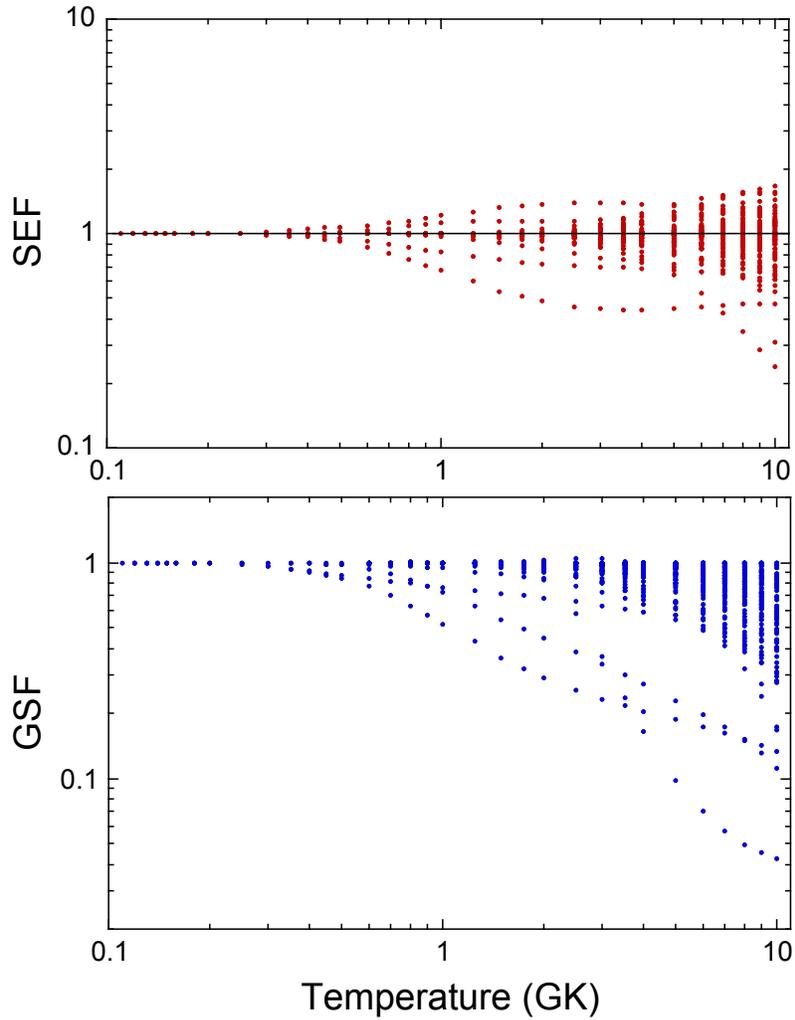}
\caption{Values of (top) $SEF$ and (bottom) $GSF$ versus stellar temperature for $\sim 100$ charged-particle induced reactions on $A\leq40$ targets, including all reactions with experimental rates estimated using the Monte Carlo method.  Only those data points are shown for which either value deviates from unity (20\% of cases).  
\label{sefgsf}}
\end{figure}

Numerical values of $SEF$ and $GSF$ versus temperature for $\sim 100$ charged-particle induced reactions on $A\leq40$ targets, including all reactions with experimental rates estimated using the Monte Carlo method, are shown in Fig.~\ref{sefgsf}.  With few exceptions, values of $SEF$ range from 0.5 to 1.8,  while $GSF$ exceeds a value of 0.3.

As described in Sec.~\ref{matchsec}, an experimental laboratory rate is typically determined by data over a wide range of temperatures.  In some instances, for example, neutron captures, an experimental rate is only available at a single or a few temperature values only.  For the sake of simplicity, we will assume in the following discussion that an experimental laboratory rate, $x_{EX}^L$, has been measured at a single temperature, $T_m$.  Ideally, a theoretical laboratory rate, $x_{TH}^L$, should reproduce the experimental value.  However, this is rarely the case in practice (Sec.~\ref{matchsec}).  At this point we are faced with a number of questions.  Given an experimental value for a laboratory rate at some temperature, $T_m$, what is the stellar rate at that very same temperature? What is the stellar rate at temperatures not covered by experimental data? And what uncertainties should be adopted for the stellar rate? 

\begin{figure}
\centering
\includegraphics[scale=.55,angle=-90]{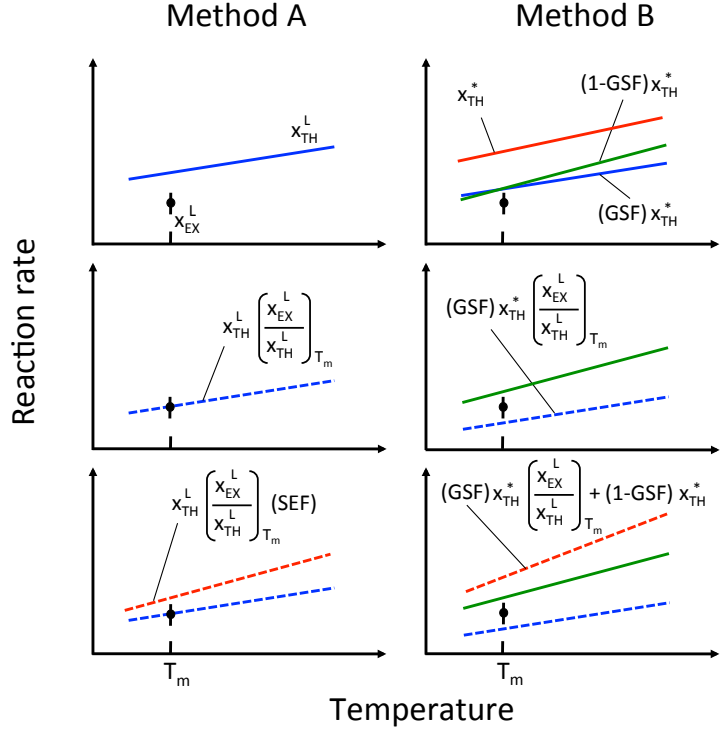}
\caption{Comparison of methods for estimating stellar rates when an experimental laboratory rate is known.  (Left panels) $x_{TH}^L$ denotes a theoretical laboratory (i.e., ground state) rate (blue solid line), while $x_{EX}^L$ is an experimental rate (data point).  The theoretical laboratory rate is first normalized to the experimental rate (dashed blue line in middle left) and then multiplied by the stellar enhancement factor (SEF), to yield the adopted stellar rate (dashed red line in lower left).  (Right panels) $x_{TH}^*$ denotes a theoretical stellar rate (solid red line), and the fractional ground and excited state contributions are displayed as blue and green solid lines, respectively (top right).  The fractional ground state rate is first normalized to the experimental rate, giving the dashed blue line (middle right), and then added to the fractional excited state contribution, to yield the adopted stellar rate (dashed red line in lower right).
\label{sefrates}}
\end{figure}

To explain the different methods applied so far, it is instructive to chose a simple example.  Suppose that only a ground state, $E_{0g.s.}=0$ keV, and one excited state, $E_{01}=50$ keV, can participate in the stellar burning.  For simplicity, we will set the statistical weights, $g_{0\mu}$, equal to unity.  For a temperature of $T_m=1$ GK, we find $kT=86.173$ keV, $G_0=1.56$, $P_{0g.s.}=0.64$, and $P_{01}=0.36$, that is, the population probabilities for ground and excited state amount to 64\% and 36\%, respectively.  Furthermore, assume that the theoretical ground state (laboratory) and excited state rates are $x_{TH}^L=40.0$ cm$^3$~s$^{-1}$~mol$^{-1}$ and $x_{TH}^1=30.0$~cm$^3$~s$^{-1}$~mol$^{-1}$, respectively.  Consequently, the stellar rate becomes $x_{TH}^*=0.64\times40.0+0.36\times30.0=51.4$ cm$^3$~s$^{-1}$~mol$^{-1}$.  This value is entirely based on theory, and the idea is to scale this value, using a suitable method, if experimental information is available for the laboratory rate.  Let us assume that the experimental laboratory rate amounts to $x_{EX}^L=20.0$ cm$^3$~s$^{-1}$~mol$^{-1}$.  Two methods are applied in practice, which are depicted in Fig.~\ref{sefrates}.

\emph{Method A:} In our example, the theoretical and experimental rates are different at $T_m$, as shown in the top left of the figure.  In the first step, the theoretical laboratory rate at all temperatures, $x_{TH}^L$ (solid blue line), is normalized to the experimental rate, $x_{EX}^L$, at $T_m$.  This procedure yields the dashed blue line in the middle left panel.  In the second step, the normalized laboratory rate is multiplied by the stellar enhancement factor, which in our case, for example, at $T_m=1$ GK, amounts to $(SEF)_{T_m}=(x_{TH}^*/x_{TH}^L)_{T_m}=51.4/40.0=1.29$ (see Eq.~\ref{sefeq}).  The adopted stellar rate for this method, $x_{AD}^{*A}$, shown as a dashed red line on the lower left, is given by:
\begin{equation} 
x_{AD}^{*A}
= x_{TH}^L \left(\frac{x_{EX}^L}{x_{TH}^L}\right)_{T_m} (SEF) 
= (x_{EX}^L)_{T_m} \frac{x_{TH}^*}{x_{TH}^L}  \frac{x_{TH}^L}{(x_{TH}^L)_{T_m}} 
\label{methodA}
\end{equation}
For our numerical example at $T_m=1$~GK, we find a value of $(x_{AD}^{*A})_{T_m}=40.0\times(20/40)\times1.29=25.8$ cm$^3$~s$^{-1}$~mol$^{-1}$.  This method is conceptually simple, in the sense that a theoretical stellar enhancement factor is applied to a theoretical laboratory rate that is normalized to an experimental rate.  Its advantage derives from the fact that the quantities $SEF=x_{TH}^*/x_{TH}^L$ and $x_{TH}^L/(x_{TH}^L)_{T_m}$ are ratios of two theoretical model predictions.  For example, if a theoretical laboratory rate deviates from the experimental one, it is assumed that the theoretical stellar rate is affected by a similar deviation, hence lessening the systematic effect on the ratio, $SEF$.  The disadvantage is that no information on the fractional contribution of the laboratory rate to the stellar rate enters in this method.

\emph{Method B:} The starting point is the theoretical stellar rate (red solid line on the upper right) and the fractional contribution of the laboratory (ground state) rate (solid blue line; the solid green line shows the fractional contribution of the excited state).  For our numerical example, we find for the stellar rate ground state fraction at $T_m=1$~GK a value of $(GSF)_{T_m}=(P_{0g.s.}x_{TH}^L/x_{TH}^*)_{T_m}=0.64\times40.0/51.4=0.50$.  That is, the ground and excited state each contribute 50\% ($51.4/2=25.7$ cm$^3$~s$^{-1}$~mol$^{-1}$) to the total stellar rate at this temperature.  In the first step, only the ground state rate fraction, $(GSF) x_{TH}^*$, is normalized to the experimental laboratory rate at $T_m$, yielding the dashed blue line in the middle right panel.  At 1 GK, for example, we find a normalized value of $(GSF)_{T_m}~(x_{TH}^*)_{T_m}~(x_{EX}^L/x_{TH}^L)_{T_m}=0.50\times51.4\times(20/40)=12.9$~cm$^3$~s$^{-1}$~mol$^{-1}$.  In the second step, the normalized rate (dashed blue line) is added to the (unnormalized) excited state fractional contribution to the total rate, $(1-GSF)x_{TH}^*$, shown as green solid line.  This sum represents the adopted rate for this method, $x_{AD}^{*B}$, displayed as dashed red line on the lower right, given by:
\begin{equation} 
x_{AD}^{*B}
= (GSF)~x_{TH}^*~\left(\frac{x_{EX}^L}{x_{TH}^L}\right)_{T_m} + (1-GSF)~x_{TH}^* 
= \frac{1}{G_0^n} \left[ (x_{EX}^L)_{T_m} \frac{x_{TH}^L}{(x_{TH}^L)_{T_m}} - x_{TH}^L \right] + x_{TH}^*
\label{methodB}
\end{equation}
In our numerical example we obtain at 1 GK a value of $(x_{AD}^{*B})_{T_m}=0.50\times51.4\times(20/40)+(1-0.50)\times51.4=38.6$ cm$^3$~s$^{-1}$~mol$^{-1}$.  The advantage of this method is that it accounts for the stellar rate ground state fraction.  A disadvantage is that it normalizes only the theoretical ground state rate to the experimental rate, while the theoretical excited state rate is adopted at face value.  In other words, if experimental and theoretical laboratory rates deviate from each other, necessitating a normalization of the theoretical laboratory rate, then a correlated systematic deviation in the excited state fractional contribution to the total stellar rate remains unaccounted for.  Furthermore, in the extreme case of $GSF\ll1$, the experimental laboratory rate becomes irrelevant and provides no constraint whatsoever on the adopted stellar rate.

Both methods give identical results if laboratory experimental and theoretical rates agree at $T_m$.  The same holds for situations where $GSF=1$, implying $SEF=1$.  In both cases, the adopted and theoretical stellar rates are equal, $x_{AD}^{*}=x_{TH}^*$.  Differences in the two methods arise when $GSF<1$, as demonstrated in the numerical example above.  Method A has been adopted in most previous investigations \citep[e.g.,][]{nacre,ilcomp} and is also used in the present work.  Method B has been advocated recently by Rauscher and collaborators~\citep{rauscher1,rauscher2,rauscher3}.  In this context it should be remembered that the higher the temperature, the more likely are contributions from excited target states, i.e., $GSF<1$.  However, at the same time reverse reactions become more important and, in those cases where rate equilibrium between a pair of forward and reverse reactions is achieved~\citep{iliadisbook}, the actual reaction rates do not impact the nucleosynthesis and the distinction between $SEF$ and $GSF$, or Method A and Method B, is unimportant.

For Method A, the uncertainties in the adopted stellar rate at a temperature of $T_m$, according to Eq.~\ref{methodA}, arise from: (i) the experimental rate, $x_{EX}^L$, and (ii) the theoretical quantity SEF.  If the adopted stellar rate needs to be estimated at temperatures for which no experimental rate is available, then an additional uncertainty arises from (iii) the temperature dependence of the theoretical ratio, $x_{TH}^L/(x_{TH}^L)_{T_m}$.  For Method B (see Eq.~\ref{methodB}), the uncertainties in the adopted stellar rate at $T_m$ arise from: (i) the normalized partition function, $G_0^n$; (ii) the experimental rate, $x_{EX}^L$; (iii) the \emph{absolute} uncertainty of the theoretical laboratory rate, $x_{TH}^L$; and (iv) the \emph{absolute} uncertainty of the theoretical stellar rate, $x_{TH}^*$.  At temperatures for which no experimental rate is available, the uncertainty of the adopted stellar rate has an additional contribution from (v) the temperature dependence of the theoretical ratio, $x_{TH}^L/(x_{TH}^L)_{T_m}$.  

It should be obvious to the reader that all of the uncertainties estimated from theory are difficult to quantify and that  either method makes some questionable assumptions. In our opinion, it would be worthwhile to pursue another procedure in the future, which we call ``Method C". Here, the starting point is the basic assumption that nuclear data provide a constraint for the nuclear reaction model predictions, no matter the actual fraction of the ground state contribution. More precisely, instead of normalizing a theoretical prediction to an experimental \emph{reaction rate}, the nuclear model should be fine-tuned by adjusting nuclear physics ingredients of the Hauser-Feshbach calculation, for example, nuclear level densities and $\gamma$-ray strength functions (see Sec.~\ref{theory}), in order to reproduce both the absolute magnitude and energy dependence of the experimental \emph{cross section}.  Once the nuclear reaction model prediction reproduces the nuclear physics data, i.e., the ingredients for the calculation of the experimental laboratory rate, the stellar rate predicted by that very same model should be more trustworthy.  However, this procedure is time consuming. It requires a detailed look at each reaction, taking also into account other constraints, for example, the s-wave spacing at the neutron binding energy and the Giant Dipole Resonance strength, including the corresponding uncertainties.  For this reason, Method C has rarely been applied for reaction rate libraries, although it is common practice in some other fields of nuclear science~\citep{Capote09,koningrochman}. 

Finally, it must be pointed out that at elevated densities the {\it bare nucleus} reaction rates discussed above have to be corrected for electron screening effects.   For more information on this topic, see \citet{iliadisbook}, and references therein.


For stellar model calculations, it is also necessary to include the corresponding reverse rate for each forward rate.  This rate, $N_A \langle \sigma v \rangle^*_r$ can be calculated from the forward stellar rate, $N_A \langle \sigma v \rangle^*_f$, using the reciprocity theorem~\citep{iliadisbook}.  From the present results for rates and probability density functions of a forward reaction, it is straightforward to calculate, after proper corrections for thermal target excitations have been taken into account, the rates and probability density functions of the corresponding reverse reaction.  We may write the reverse rate as, 

\begin{equation}
N_A \langle \sigma v \rangle_r^* = a(e^{cQ})N_A \langle \sigma v \rangle^*_f,\label{last}
\end{equation}
with $a$ equal to a constant factor \citep[see][]{iliadisbook} and $c = -11.605/T_9$.  The Q-value is expected to follow a Gaussian probability density function, where the value of Q and the uncertainty $\Delta Q$ reported in the literature represent the expectation value and square root of the variance, respectively (Sec.~\ref{MCsec}).  Therefore, $e^Q$ will be represented by a lognormal distribution.  In most cases the forward rate, $N_A \langle \sigma v \rangle^*_f$, is also described by a lognormal distribution, with location and spread parameters of $\mu$ and $\sigma$ that are tabulated here.  Consequently, the reverse reaction rate, $N_A \langle \sigma v \rangle^*_r$, will follow a lognormal distribution as well, with location and spread parameters of $\mu_r = {\rm ln}\, a + \mu + cQ$ and $\sigma_r^2 = \sigma^2 + c^2(\Delta Q)^2$, respectively (see \citet{iliadis_1} or the discussion of combining lognormal distributions in Sec.~\ref{beta}).  The reverse rate factor uncertainties of all Monte Carlo based rates are derived in this work from the lognormal spread parameter, according to $f.u._r = e^{\sigma_r}$.

\section{A New Nuclear Rate Library for Stellar Models: STARLIB}\label{starlib}



The ideas presented above are the foundation of a new kind of stellar reaction rate library, which we call STARLIB.  Other rate libraries commonly used for stellar models, for example, JINA REACLIB~\citep{reaclib} or BRUSLIB~\citep{bruslib}, provide reaction and decay rates versus temperature for all stable and unstable nuclei.  In contrast, STARLIB contains  information not only on rates versus temperature but also on rate uncertainties and probability density functions.  With this additional information, more realistic nucleosynthesis simulations can be performed than previously possible.  In particular, STARLIB begins to enable rigorous quantification of how improvements in determining stellar nuclear reaction rates impact the structure, evolution, and observational consequences of stellar phenomena. 

The new information on rate uncertainties and probability densities can be presented by providing only three quantities: temperature, rate, and factor uncertainty.  To better understand how this can be accomplished, recall the arguments from Sec.~\ref{logsec} that most reaction rates can be described by a lognormal probability density function.  The lognormal function has only two parameters, $\mu$ and $\sigma$. Applying some of its properties to a reaction rate, $x$, we find that:

(i) the median rate, $x_{med}$, is related to the location parameter $\mu$ via (combining Eqs.~\ref{starmu} and~\ref{eq6}):

\begin{equation}
x_{med} = e^\mu = \sqrt{x_{low} x_{high}} \label{aa}
\end{equation}

(ii) the factor uncertainty, $f.u.$, of the rate (for a coverage probability of 68\%) is related to the spread parameter $\sigma$ via (combining Eqs.~\ref{eq6},~\ref{eq7}, and~\ref{eq8}):

\begin{equation}
f.u.= e^\sigma = \sqrt{x_{high}/x_{low}}\label{bb}
\end{equation}

(iii) the high and low rates (for a coverage probability of 68\%) are given by (rewriting Eq.~\ref{eq7} using Eq.~\ref{starmu}):

\begin{equation}
x_{low}=e^\mu/e^\sigma = x_{med}/f.u. \, \qquad x_{high}=e^\mu e^\sigma = x_{med}\, f.u.\label{cc}
\end{equation}
Therefore, by providing temperature, recommended rate ($x_{med}$) and factor uncertainty ($f.u.$), not only can the low rate, $x_{low}$, and high rate, $x_{high}$, be obtained easily but, in addition, the lognormal parameters can be conveniently computed from Eq.~\ref{starmu} or Eqs.~\ref{aa} and~\ref{bb}, and hence the lognormal probability density is determined at any given temperature grid point.  We assume a lognormal distribution for all reaction and decay rates included in STARLIB.


Next, the issue of how to present this information in a data library must be addressed.  For example, the JINA REACLIB~\citep{reaclib} provides rates versus temperature as an analytical fit formula, whereas BRUSLIB~\citep{bruslib} presents temperature and rates in tabular (i.e., two-column) format.  While the factor uncertainty could, in principle, be presented as an analytical formula, we chose instead the simpler solution of a tabular format.  Thus, STARLIB consists of three columns, listing, for each reaction or decay, temperature, recommended rate, and factor uncertainty.

It should be clear that a general-purpose nuclear reaction and decay library must encompass tens of thousands of nuclear interactions.  On the other hand, experimental Monte Carlo rates are available so far for only 63 reactions (i.e., those published in \citet{iliadis_2} and those listed in Appendix~\ref{updates}).  In order for the reader to better grasp what kind of information is provided in STARLIB, we will discuss the step-by-step construction of STARLIB in the following subsections.  Our procedure for treating and evaluating the nuclear physics input is described in detail in Sec. 2 of \citet{iliadis_3}.  Table~\ref{starliblist} presents an overview of the nuclear rate sources used in STARLIB, including the number of rates and the assumption for uncertainties.  STARLIB formatting and reference labels are discussed in Appendix~\ref{format}.

\begin{table}[ht]
\caption{Summary of STARLIB rates and assumption for uncertainties.  Number indicates forward rate only.  See Table~\ref{reftable} for comprehensive list of rate sources.  JINA REACLIB~\citep{reaclib} supplements STARLIB for rates not available from sources listed below.
}
\begin{tabular}{lll}

\hline\hline
Rate Evaluation & Number & Uncertainties \\ 
\hline
Monte Carlo (experimental) &63 &individual \\
 NACRE& 24 &individual\\ 
 Big Bang&9 &  individual \\ 
$\beta$-decays (experimental) &2225  & individual \\ 
$\beta$-decays (theoretical) &5556 & individual \\ 
  TALYS (theory)& 47,721  & factor of 10\\ 
  Neutron capture& 281  & individual\\ 
$^{26}$Al $\gamma$-ray transitions &17  & individual \\
\hline\hline

\end{tabular}
\label{starliblist}
\end{table}

\subsection{Starting point:  Experimental Monte Carlo Based Reaction Rates}\label{starlibMC}


The three-column structure of STARLIB is motivated by the experimental Monte Carlo based thermonuclear reaction rates, as explained above.  These rates consist of those published in \citet{iliadis_2} and those evaluated in the present work (Appendix~\ref{updates}).  A list of reactions for which Monte Carlo based rates are available is given in Table~\ref{master}.  Notice that in this case the recommended rate, listed in column 2 of STARLIB, is directly obtained from the Monte Carlo procedure as the ``median" rate (Sec.~\ref{MCsec}).  In particular, no assumption is made regarding the rate probability density following a lognormal distribution or not, though assumptions have been made on the input parameter distributions.  On the other hand, the factor uncertainty, listed in column 3 of STARLIB, is explicitly derived under the assumption that the rate distribution is lognormal.

\begin{table}
\caption{Reactions with available experimental Monte Carlo based reaction rates.  Most rates and Q values are adopted from \citet{iliadis_2}; those that have been evaluated after 2010 are indicated by an asterisk and bolded; for more details on the latter reactions, see Appendix~\ref{updates}. } 
\begin{tabular}{ll|ll}
\hline\hline

\multicolumn{1}{l}{Reaction\tablenotemark{a}} & \multicolumn{1}{c}{Q\tablenotemark{b}(keV) } &\multicolumn{1}{l}{Reaction\tablenotemark{a}} & \multicolumn{1}{c}{Q\tablenotemark{b} (keV) } \\
\hline

$^{14}$C(p,$\gamma$)$^{15}$N       	&	 10207.42$\pm$0.00		&	$^{25}$Mg(p,$\gamma$)$^{26}$Al$^{m}$	&	 6078.15$\pm$0.05	\\
$^{14}$C($\alpha$,$\gamma$)$^{18}$O      	&	 6226.3$\pm$0.6		&	$^{26}$Mg(p,$\gamma$)$^{27}$Al     	&	 8271.05$\pm$0.12	\\
$^{14}$N($\alpha$,$\gamma$)$^{18}$F      	&	 4414.6$\pm$0.5		&	$^{23}$Al(p,$\gamma$)$^{24}$Si     	&	 3304$\pm$27    	\\
$^{15}$N($\alpha$,$\gamma$)$^{19}$F      	&	 4013.74$\pm$0.07		&	$^{24}$Al(p,$\gamma$)$^{25}$Si     	&	 3408$\pm$10    	\\
$^{15}$O($\alpha$,$\gamma$)$^{19}$Ne      	&	 3529.1$\pm$0.6		&	$^{25}$Al(p,$\gamma$)$^{26}$Si     	&	 5513.7$\pm$0.5	\\
$^{16}$O(p,$\gamma$)$^{17}$F      	&	 600.27$\pm$0.25		&	$^{26}$Al$^{g}$(p,$\gamma$)$^{27}$Si	&	 7462.96$\pm$0.16	\\
$^{16}$O($\alpha$,$\gamma$)$^{20}$Ne      	&	 4729.85$\pm$0.00		&	$^{27}$Al(p,$\gamma$)$^{28}$Si    	&	11585.11$\pm$0.12	\\
$^{17}$O(p,$\gamma$)$^{18}$F      	&	 5606.5$\pm$0.5		&	$^{27}$Al(p,$\alpha$)$^{24}$Mg     	&	 1600.96$\pm$0.12	\\
$^{17}$O(p,$\alpha$)$^{14}$N	&	 1191.82$\pm$0.11		&	$^{26}$Si(p,$\gamma$)$^{27}$P      	&	 861$\pm$27    	\\
{\bf $^{18}$O(p,$\gamma$)$^{19}$F$^*$}     	&	 7994.8$\pm$0.6\tablenotemark{c} 		&	$^{27}$Si(p,$\gamma$)$^{28}$P      	&	 2063$\pm$3	\\
{\bf $^{18}$O(p,$\alpha$)$^{15}$N$^*$} 	 &	 3981.09$\pm$0.62		&	$^{28}$Si(p,$\gamma$)$^{29}$P      	&	 2748.8$\pm$0.6	\\
$^{18}$O($\alpha$,$\gamma$)$^{22}$Ne	&	 9668.1$\pm$0.6   		&	$^{29}$Si(p,$\gamma$)$^{30}$P      	&	 5594.5$\pm$0.3	\\
$^{17}$F(p,$\gamma$)$^{18}$Ne   	&	 3923.5$\pm$0.4		&	$^{30}$Si(p,$\gamma$)$^{31}$P      	&	 7296.93$\pm$0.19	\\
$^{18}$F(p,$\gamma$)$^{19}$Ne   	&	 6411.2$\pm$0.6		&	$^{27}$P(p,$\gamma$)$^{28}$S       	&	 2460$\pm$30 	\\
$^{18}$F(p,$\alpha$)$^{15}$O	&	 2882.15$\pm$0.73		&	{\bf $^{29}$P$(p,\gamma)^{30}$S$^*$}	&	$4398.72\pm 3.06$\tablenotemark{c}	\\
$^{19}$Ne(p,$\gamma$)$^{20}$Na	&	 2193$\pm$7		&	$^{31}$P(p,$\gamma$)$^{32}$S       	&	 8863.78$\pm$0.21	\\
$^{20}$Ne(p,$\gamma$)$^{21}$Na	&	 2431.69$\pm$0.14		&	$^{31}$P(p,$\alpha$)$^{28}$Si      	&	 1915.97$\pm$0.18	\\
$^{20}$Ne($\alpha$,$\gamma$)$^{24}$Mg	&	 9316.55$\pm$0.01		&	$^{30}$S(p,$\gamma$)$^{31}$Cl      	&	 290$\pm$50    	\\
$^{21}$Ne(p,$\gamma$)$^{22}$Na     	&	 6739.6$\pm$0.4		&	$^{31}$S(p,$\gamma$)$^{32}$Cl      	&	 1574$\pm$7	\\
$^{21}$Na(p,$\gamma$)$^{22}$Mg     	&	 5504.18$\pm$0.34		&	$^{32}$S(p,$\gamma$)$^{33}$Cl      	&	 2276.7$\pm$0.4	\\
{\bf $^{22}$Ne$(p,\gamma)^{23}$Na$^*$}	&	$8794.11 \pm 0.02$\tablenotemark{c} 		&	$^{31}$Cl(p,$\gamma$)$^{32}$Ar     	&	 2420$\pm$50    	\\
{\bf $^{22}$Ne$(\alpha,\gamma)^{26}$Mg$^*$}	&	$10614.78 \pm 0.03 $		&	$^{32}$Cl(p,$\gamma$)$^{33}$Ar     	&	 3343$\pm$7    	\\
{\bf $^{22}$Ne$(\alpha,n)^{25}$Mg$^*$}	&	$-478.29 \pm 0.04$		&	$^{35}$Cl(p,$\gamma$)$^{36}$Ar     	&	 8506.97$\pm$0.05	\\
{\bf $^{22}$Na$(p,\gamma)^{23}$Mg$^*$}	&	$7580.717 \pm 0.708$\tablenotemark{c}		&	$^{35}$Cl(p,$\alpha$)$^{32}$S      	&	 1866.21$\pm$0.13	\\
$^{23}$Na(p,$\gamma$)$^{24}$Mg     	&	 11692.68$\pm$0.01		&	$^{34}$Ar(p,$\gamma$)$^{35}$K      	&	 84.5$\pm$0.7     	\\
$^{23}$Na(p,$\alpha$)$^{20}$Ne     	&	 2376.13$\pm$0.00		&	$^{35}$Ar(p,$\gamma$)$^{36}$K      	&	 1668$\pm$8	\\
$^{22}$Mg(p,$\gamma$)$^{23}$Al     	&	 122$\pm$19    		&	$^{36}$Ar(p,$\gamma$)$^{37}$K      	&	1857.63$\pm$0.09	\\
$^{23}$Mg(p,$\gamma$)$^{24}$Al     	&	 1872$\pm$3		&	{\bf $^{38}$Ar(p,$\gamma$)$^{39}$K$^*$}      	&	6381.43$\pm$0.24\tablenotemark{c}	\\
$^{24}$Mg(p,$\gamma$)$^{25}$Al     	&	 2271.6$\pm$0.5		&	$^{35}$K(p,$\gamma$)$^{36}$Ca      	&	 2556$\pm$40    	\\
$^{24}$Mg($\alpha$,$\gamma$)$^{28}$Si	&	 9984.14$\pm$0.01		&	$^{39}$Ca(p,$\gamma$)$^{40}$Sc     	&	 538$\pm$3  	\\
$^{25}$Mg(p,$\gamma$)$^{26}$Al$^{t}$	&	 6306.45$\pm$0.05		&	$^{40}$Ca(p,$\gamma$)$^{41}$Sc     	&	1085.09$\pm$0.08	\\
$^{25}$Mg(p,$\gamma$)$^{26}$Al$^{g}$	&	 6306.45$\pm$0.05		&		&		\\

\hline\hline
\tablenotetext{a}{The ground, isomeric, and total states of $^{26}$Al are denoted by the superscripts $g$, $m$, and $t$, respectively.   }
\tablenotetext{b}{Reaction Q values from \citet{iliadis_2}, unless otherwise noted.  }
\tablenotetext{c}{Reaction Q value from \citet{ame2012}.}
\end{tabular}

 \label{master} 
\end{table}

All experimental ``laboratory" Monte Carlo rates are corrected for thermal target excitations, according to Method A (see Sec.~\ref{sef} and Eq.~\ref{methodA}), before they are introduced into the rate library.  The necessary stellar enhancement factors and the extrapolation to higher temperatures are based on TALYS calculations (Sec.~\ref{theory}).  For the uncertainties, we have to distinguish the temperature ranges below and above the matching temperature, $T_{match}$ \citep[see Sec.~\ref{matchsec} and][]{newton,iliadis_2}.  For $T\le T_{match}$, we adopt the uncertainties of the experimental Monte Carlo rates for the stellar rates.  This disregards any additional uncertainties arising from the calculated stellar enhancement factors, $SEF=x_{TH}^*/x_{TH}^L$ (Sec.~\ref{sef}), and, hence, we may underestimate the actual uncertainty.  However, inspection of the computed $SEF$ values below $T=T_{match}$ (Fig.~\ref{sefgsf}, top) reveal that they are close to unity, and thus the additional uncertainty is likely to be small.  For $T>T_{match}$, we adopt for the stellar rates  the uncertainties of the experimental Monte Carlo rate at $T_{match}$.  This procedure disregards the additional uncertainty from the ratio $x_{TH}^L/(x_{TH}^L)_{T_{match}}$, which is difficult to quantify at present, as explained  in Sec.~\ref{sef}.  From their stellar rates, the rates of the corresponding reverse reactions are computed (see Sec.~\ref{part}), which are also inserted in the new library.

\subsection{Other Experimental Reaction Rates}\label{otherrates}

Experimental Monte Carlo based reaction rates are available for 63 reactions (Table~\ref{master}).  However, for many other reactions for which experimental nuclear physics information exists, Monte Carlo rates have not been computed yet.  Thus, the question arises of how to implement such rates published in the literature.  In the simplest case, when only the experimental reaction rate versus temperature has been published and no estimate of a rate uncertainty is provided, the published rate was introduced as the recommended rate into our new library, and the factor uncertainty was set equal to 10.  This represents our best estimate given the absence of information, and the user is encouraged to change this if a more appropriate value is suspected. 

A number of publications do provide reaction rates including estimates of uncertainties.  For example, in \citet{nacre} and \citet{des}, these uncertainties are expressed in terms of ``lower limits" and ``upper limits".  Since we assume for all rates in STARLIB that the rate probability densities are lognormally distributed, the recommended rate and its factor uncertainty are obtained from the geometric mean and geometric standard deviation of the ``upper limit" and ``lower limit" values, according to Eqs.~\ref{aa} and~\ref{bb}, respectively.  Notice that Eq.~\ref{bb} only applies for a coverage probability of 68\%, whereas the literature rates have been estimated without any reference to probability densities.  Therefore, our adopted geometric mean value may differ from the reported recommended rate\footnote{We emphasize that any such differences between {\it recommended} rates simply reflect our choice of a probability density shape and do not imply any indication of quantitative improvement.}, especially if the literature rates have very large uncertainties and, in addition, are non-symmetric on a logarithmic scale (i.e., are not lognormal; see Sec.~\ref{logsec}).  However, this difference is found to be very small compared to the large reported rate uncertainties.  Each of these experimental rates has been corrected for thermal target excitations, according to Method A, as explained in Sec.~\ref{sef} (see Eq.~\ref{methodA}).  The necessary stellar enhancement factors, $SEF$, are computed using the TALYS code.  The additional uncertainty introduced by $SEF$ has been disregarded (see Sec.~\ref{starlibMC}).

\subsection{Experimental Laboratory $\beta$ Decay Rates}\label{beta}

STARLIB lists laboratory $\beta$-decay constants, $\lambda={\rm ln} \,2/t_{1/2}$, with $t_{1/2}$ as the half-life, in units of $s^{-1}$.  For a given decay, the same decay constant is listed at all 60 temperature grid points, as laboratory decay rates are temperature-independent.  In many cases, a $\beta$-decay competes with a $\beta$-delayed particle decay. This distinction has not always been made in previous rate libraries but is important because these processes represent different links in a reaction network.  Consider, for example, the nucleus $^{29}$S.  Not only does it $\beta$ decay to the ground state of $^{29}$P, it also $\beta$ decays to unbound states of $^{29}$P with comparable probability.  These excited states, in turn, may decay by proton-emission, which results in the final nucleus $^{28}$Si.  Consequently, the $\beta$ decay $^{29}$S $\longrightarrow ^{29}$P and the $\beta$-delayed proton decay $^{29}$S$\longrightarrow ^{28}$Si  compete with each other and must be accounted for as separate links in nucleosynthesis calculations.  

The partial decay constant, $\lambda^p$, for a given link (i.e., either $\beta$ decay or $\beta$-delayed particle decay) is related to the partial half-life, $t^p_{1/2}$, and the branching ratio, $B_\beta$, via:

\begin{equation}
\lambda^p = \frac{{\rm ln} \, 2}{t^p_{1/2}} =\frac{ {\rm ln} \, 2 \, B_\beta}{ t_{1/2}},
\end{equation}
where $t^p_{1/2}$ denotes the total laboratory half-life.  Since $\beta$-decay half-lives are manifestly positive quantities, we assume that they are associated with a lognormal probability density.  It may be argued that half-lives are more properly described by a Poissonian distribution, especially if the uncertainty is determined by counting statistics.  However, almost all of the measured laboratory half-lives have relatively small uncertainties, and thus the Poissonian is well approximated by a Gaussian (for sample sizes $> 15$), which in turn is closely approximated by a lognormal distribution. Therefore, we can derive the lognormal parameters $\mu$ and $\sigma$ for the partial half-life, $t_{1/2}^p$, from the reported recommended value (i.e., the expectation value, $E[x]$) and its uncertainty (i.e., the square-root of the variance, $\sqrt{V[x]}$), according to Eq.~\ref{mu}.

To obtain the probability density for the corresponding partial decay constant, $\lambda_p$, we use a well-known property of lognormal distributions~\citep{iliadis_1,lognormal}:  for two independent, lognormally distributed, random variables $x_1$ and $x_2$, with location and spread parameters of $\mu_1$, $\sigma_1$ and $\mu_2$, $\sigma_2$, the product $\alpha x_1^{\beta_1} x_2^{\beta_2}$ (where $\alpha>0$) is also lognormally distributed, with location and spread parameters of $\mu^\prime ={\rm ln} \, \, \alpha +\beta_1 \mu_1 +\beta_2 \mu_2$ and $\sigma^{\prime 2} = \beta_1^2\sigma_1^2 + \beta_2^2\sigma_2^2$, respectively.  From $\beta_1=0$, $x_1=1$, $\beta_2=-1$, and $x_2=t_{1/2}^p$, we find that the partial decay constant is also lognormally distributed, with the median value and factor uncertainty of:


\begin{equation}
\lambda^p = e^{\mu^\prime}=e^{[{\rm ln} \,({\rm ln} \,2)-\mu]}, \qquad f.u. =e^{\sigma^\prime}= e^\sigma,
\end{equation}
where the lognormal parameters $\mu$ and $\sigma$ for the partial half-life are related to E[x] and V[x] according to Eq.~\ref{mu}.  The latter two quantities are assumed to be equal to the mean value of the reported or calculated partial half-life and the square of the associated uncertainty, respectively.  Therefore, the probability density for the $\beta$-decay constants listed in STARLIB is obtained in exactly the same manner as for reactions. That is, the lognormal location parameter $\mu^\prime$ is equal to the natural logarithm of the listed partial decay constant, while the lognormal spread parameter $\sigma^\prime$ is equal to the natural logarithm of the factor uncertainty.  For example, if the reported partial half-life amounts to $t_{1/2}^p=10\pm1$ s, we obtain from the above expressions $\lambda^p=0.0697$ s$^{-1}$ and $f.u.=1.105$. The lognormal parameters of the partial decay constant are $\mu^\prime=-2.66$ and $\sigma^\prime=0.0998$.  

Measured $\beta$-decay half lives and branching ratios for all nuclides up to $^{121}$Te are included in STARLIB.  Most values have been adopted from \citet{audi,ame2012}, except when newer information was available.  Theoretical $\beta$-decay rates were taken from other sources (see Table~\ref{reftable}).  

For use at elevated temperatures and densities, the $\beta$-decay half lives in STARLIB need to be replaced by stellar $\beta$-decay rates.  A number of possible complications arise in this regard.  First, weak interaction rates depend ,in general, on both temperature {\it and density}.  Therefore, libraries that are available to the community (e.g., REACLIB) only include {\it laboratory} $\beta$-decay rates, and we follow the same strategy in the format of STARLIB.  In actual stellar model calculations, all laboratory weak rates need to be replaced by temperature and density dependent $\beta$-decay rates, which are usually contained in a separate input file \citep[based, e.g., on the results presented in][]{fuller,oda,karl}.  Second, when a stellar modeler uses any of these ``stellar" weak rates from a theoretical model, the question arises of what uncertainty to assign to the estimated values.  At present, there is no simple answer.  The only compilation of $\beta$-decay rates that includes a coherent treatment of the temperature and density dependence concerns the work of \citet{takahashi}.  A specific detailed analysis of the uncertainties related to unknown transition log ft values is given in \citet{goriely99} (see Table 1).  The latter work concerns nuclei close to the valley of $\beta$-stability of relevance for s-process nucleosynthesis.  For nuclei far away from stability, for which no experimental data exist, the uncertainties related to the population of excited states are usually smaller compared to those associated with the adopted theoretical model (as inferred from different model predictions), and more future work on those models is required before reliable uncertainties can be derived.  For other nuclei, it remains difficult to estimate the uncertainties related to the population of excited states if all the transitions and their relative strengths cannot be determined experimentally.  Clearly, more work is needed in the future to estimate reliable uncertainties of ``stellar" weak interaction rates.

\subsection{Theoretical Reaction Rates}\label{talys}

For the vast majority of reaction rates in STARLIB, no experimental cross section information exists, and theoretical estimates using the code TALYS have been adopted (see Sec.~\ref{theory}).  These theoretical rates have been obtained for target nuclei in the region $Z=3-83$. As already mentioned, the latest experimental information on Q-values and level structure is adopted in the TALYS calculations.  Only when no experimental information was available for these input quantities did the calculations resort to nuclear mass and structure models (Sec.~\ref{theory}).

Reliable uncertainties for theoretical reaction rates are difficult to assess.  Various claims have been made in the literature (``on average within a factor of 2"), which may have been too optimistic.  Previously, uncertainties have been systematically evaluated for each target and each reaction channel from the use of different sets of nuclear input models \citep[see, in particular, Fig. 22 of][]{arnould}.  A similar approach could be followed to estimate the uncertainties affecting the TALYS rates.  However, for the present version of STARLIB, we restrict ourselves to recommend a factor of 10 uncertainty for any given reaction rate for which no experimental cross section information exists.  Of course, the theoretical rates are labeled (see Table~\ref{reftable}), and thus it is a simple matter for the user to modify the reaction rate uncertainty value.  More work will be required in the future to refine the estimated uncertainties of purely theoretical reaction rates. 

Since we assume lognormal probability densities for all rates in STARLIB, Eqs.~\ref{aa}-\ref{cc} apply for theoretical rates as well, i.e., the probability densities are easily obtained from the listed recommended rate and its factor uncertainty at each temperature grid point.





\subsection{Experimental Neutron Capture Rates}\label{neutronsec}

Experimental neutron capture rates are compiled in \citet{Bao00} (see also references provided in \citet{bruslib}, and the KADoNiS project at~\url{www.kadonis.org}).  Most of these experimental results provide the Maxwellian-averaged cross sections at only a few energies, or sometimes even a single energy, near $kT=30$ keV, i.e., close to canonical s-process conditions.  For a total of 281 neutron capture reactions involving stable or long-lived target nuclei, the experimental laboratory rates are used to estimate the corresponding stellar rates at all temperatures of astrophysical interest, according to Method A in Sec.~\ref{sef} (see Eq.~\ref{methodA}).  The stellar enhancement factors and the extrapolation to temperatures outside the measured region are based on TALYS calculations (Sec.~\ref{theory}). 


For the uncertainties of the adopted stellar rates we took, at all temperatures, only the uncertainties of the experimental rates into account, as described in Sec.~\ref{starlibMC}, although the radiative neutron capture rates for those 281 nuclei are also affected by the additional uncertainties related to the prediction of the SEF.  Such uncertainties can be estimated by a systematic modification of the nuclear input models, as discussed above \citep[see also][]{arnould}.  They remain however relatively small and are neglected in the present version of STARLIB.  Again, this procedure underestimates the actual uncertainties, and more work is required in the future (see Method C in Sec.~\ref{sef}).



Probability densities for the neutron capture rates listed in STARLIB are obtained, as for all other reactions, using Eqs.~\ref{aa}-\ref{cc}.

\subsection{$^{26}$Al $\beta$-Decays and $\gamma$-Ray Transitions}\label{otheral26}

The nuclide $^{26}$Al has an isomeric state ($E_x$ = 228 keV; $J^\pi = 0^+$) that does not always achieve thermal equilibrium with the ground state ($J^\pi = 5^+$) in a stellar plasma.  Therefore, the ground and isomeric state need to be treated separately in nuclear reaction networks, depending on the temperatures involved.   Stellar modelers generally follow the recommendation of \citet{ward}:  the ground and isomeric states in $^{26}$Al will be in thermal equilibrium at temperatures above $\sim 400$ MK and can be treated as one species in the network.  Below this temperature, the ground and isomeric states are not in thermal equilibrium and should be adopted as distinct species.  

This extreme assumption based on a sharp temperature boundary was questioned by \citet{runkle} and \citet{coc}, who studied the equilibration of ground and isomeric state via $\gamma$-ray excitations involving higher-lying levels of $^{26}$Al.  They showed that it is sufficient to consider just three excited levels for the equilibration:  $E_x$ = 417 keV ($J^\pi = 3^+$), 1058 keV ($J^\pi = 1^+$), and 2070 keV ($J^\pi = 2^+$).  STARLIB includes all levels discussed above, and in a nucleosynthesis simulation the user may select only a single species (i.e., $^{26}$Al in thermal equilibrium, presumably because temperatures are very high), two species \citep[i.e., the ground and isomer, assuming that the equilibration can be disregarded; see, for example,][]{il_al26}, or all five species (i.e., if the equilibration needs to be taken into account explicitly).  The safest method of implementation in a reaction network is to consider all five species such that the equilibration of $^{26}$Al is always explicitly incorporated.  However, processing time will increase significantly.    


The necessary $\gamma$-ray transition and $\beta$-decay constants are listed in Appendix A of \citet{il_al26}, and all of these transition rates are included in STARLIB.  A factor uncertainty of 10 was estimated for those transition rates that have been derived from nuclear theory \citep[see][for details]{il_al26}.  


\subsection{Other Rates}\label{otherreaclib}

For other rates not available from the above sources, STARLIB is supplemented by rates from the JINA REACLIB library~\citep{reaclib}.  Their reaction rates, provided as a seven parameter fit, were converted into a tabular format, and a factor of 10 uncertainty was assumed for each rate, in absence of more reliable information.

\section{Use of STARLIB in Monte Carlo Nucleosynthesis Studies}\label{use}
We will briefly comment on how to randomly sample reaction and decay rates in a Monte Carlo nucleosynthesis calculation, since there seems to be no consensus on a procedure in the literature.  We anticipate that such studies will become increasingly important in the future, allowing for more realistic elemental and isotopic abundance estimates.  We will follow here the ideas of \citet{longlandtobesub}, to which the reader is referred for more details.

Recall that STARLIB provides for each nuclear interaction (nuclear reaction, weak interaction) a lognormal rate probability density function at 60 temperature grid points between 1 MK and 10 GK: the lognormal parameter $\mu$ can be approximated by the natural logarithm of the recommended rate listed in column 2 (Eq.~\ref{aa}), and the lognormal parameter $\sigma$ is obtained from the natural logarithm of the rate factor uncertainty listed in column 3 (Eq.~\ref{bb}).  With these two parameters, the lognormal probability density is precisely defined (see Eq.~\ref{logneq}).  In a Monte Carlo nucleosynthesis simulation, all rates can be simultaneously sampled according to their individual probability densities, or the rates for a subset of links can be sampled, depending on the application.  

For a lognormal probability density, the samples are drawn according to the function~\citep{evansbook}:

\begin{equation}
x_i = e^{\mu + p_i \sigma} = e^\mu \left( e^\sigma \right)^{p_i} = x_{med} (f.u.)^{p_i}\label{above}
\end{equation}
where $x_i$ denotes the sampled rate, and $p_i$ is a standard normal deviate (i.e., a Gaussian deviate with a mean of zero and a standard deviation of unity).  See also Eqs.~\ref{aa} and~\ref{bb}.  For a random sample of $p_i=0$, the recommended rate ($e^\mu$) is obtained.  It is apparent from the right hand side of Eq.~\ref{above} that the first term ($e^\mu$) is listed in column 2 of STARLIB (i.e., the recommended rate), while the base ($e^\sigma$) is listed in column 3 of STARLIB (i.e., the factor uncertainty).  Thus, it becomes obvious how the structure of STARLIB is conveniently tailored for Monte Carlo nucleosynthesis simulations.  

In practical terms, we find it useful to sample the rates according to Eq.~\ref{above} only once at the beginning of each network calculation.  That is, in the beginning of each network run (for example, post-processing studies using constant or varying temperatures and densities, or multi-zone calculations), the probability factor $p_i$ is sampled for each reaction, $i$, in the network, where for a given reaction $p_i$ has the same value at all temperatures.  Notice that this procedure does {\it not} imply that, for a given reaction, the rate variation is constant with temperature.  Rather, the lognormal parameter $\sigma$ depends on the temperature and, therefore, randomly sampling according to Eq.~\ref{above} automatically takes into account the temperature dependence of the rate factor uncertainty.  Furthermore, the parameters $p_i$ could be saved for each sample reaction network run in order to study correlations between rates and abundances.


Note that the procedure described above is different from Monte Carlo variations in previous studies.  Simulations of big bang nucleosynthesis~\citep{krauss90,smith93} employed Gaussian rather than lognormal rate probability densities.  Their assumption is only approximately correct when the rate uncertainties are relatively small (say, less than $20\%$) for all interactions in the network.  Lognormal rate probability densities were used in simulations of classical novae~\citep{smithnova} and Type I x-ray bursts~\citep{nicIX,parikh}.  In these latter works, the assumed enhancement factor multiplying each rate, i.e., the factor $e^{\sigma \,p_i}$, instead of $p_i$ in Eq.~\ref{above}, was sampled randomly.  That procedure is applicable only for the special case that the rate uncertainty is constant with temperature, for example, if rigorous rate uncertainties are not available and global factor uncertainties (e.g., ``factor of 10") need to be used in the simulations instead.  On the other hand, our method described above, of sampling the rate according to Eq.~\ref{above} using the information provided by STARLIB, is more generally applicable, i.e., if the rates of a given reaction have large uncertainties or if they vary with temperature.  For more involved methods of randomly sampling rates in nucleosynthesis simulations, the reader is referred to \citet{lo12}.


\begin{figure}[ht]
\centering
\includegraphics[scale=0.5]{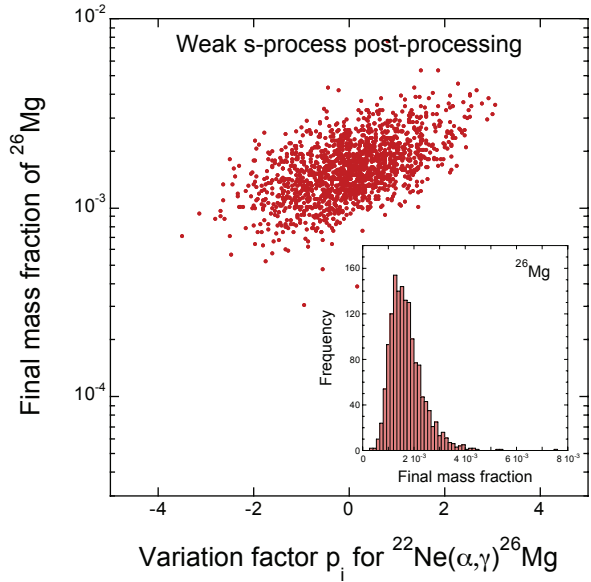}
\caption{(color online) Results of Monte Carlo post-processing nucleosynthesis study for the evolution of a 25 M$_\odot$ star of initial solar metallicity from the end of core hydrogen burning to the end of core helium burning.  Temperature and density evolutions are extracted from the models of \citet{meyer}.  The results are obtained for 1500 reaction network samples.  (Inset) Distribution of $^{26}$Mg abundance at the end of helium burning, where the spread in the values  provides a realistic estimate for the abundance uncertainty;  (Main) final $^{26}$Mg abundance versus random variable $p_i$ for the $^{22}$Ne($\alpha$,$\gamma$)$^{26}$Mg reaction rate (see Eq.~\ref{above}).  The impact of the $^{22}$Ne($\alpha$,$\gamma$)$^{26}$Mg rate on the final $^{26}$Mg abundance is apparent.}
\label{newfigure}
\end{figure}

An example is shown in Fig.~\ref{newfigure}, displaying results of a Monte Carlo post-processing reaction network simulation for the evolution of a 25 M$_\odot$ star with initial solar composition from the end of hydrogen burning to the end of helium burning.  The time evolution of temperature and density in the stellar core is adopted from the models of \citet{meyer}.  All rates are sampled randomly,  simultaneously, and independently according to Eq.~\ref{above}.  As explained above, this is achieved by drawing random variables $p_i$ for each reaction at the beginning of each sample network calculation and by using the lognormal parameters $\mu$ and $\sigma$ provided in STARLIB.  The results are obtained for 1500 reaction network samples.  The main part of the figure displays the $^{26}$Mg abundance versus the random variable $p_i$ for the $^{22}$Ne($\alpha$,$\gamma$)$^{26}$Mg reaction rate.  There is a noticeable slope in the data scatter, indicating that the $^{26}$Mg abundance at the end of helium burning varies as a result of changing the $^{22}$Ne($\alpha$,$\gamma$)$^{26}$Mg reaction rate. The inset shows the resulting distribution of final $^{26}$Mg abundances, which is given by a projection of the scatter plot onto the y-axis. The location and width of this distribution can be used to determine a mean value {\it and an uncertainty} for the final $^{26}$Mg abundance. The scatter plot also indicates that the  $^{22}$Ne($\alpha$,$\gamma$)$^{26}$Mg reaction rate is not the only important factor determining the $^{26}$Mg abundance (otherwise there would be less scatter in the y-direction).


In the example above we assumed that all rates are independent.  However, correlations certainly exist between reaction rates.  For example, the rates of the competing $^{22}$Ne($\alpha$,$\gamma$)$^{26}$Mg and $^{22}$Ne($\alpha$,n)$^{25}$Mg reactions are computed using the same values of particle separation energies and partial widths since the same compound levels are populated.  Therefore, these two rates are indeed correlated.  Such input parameter correlations may indeed be important for the nucleosynthesis, and it would be of interest to study their impact.  As of yet, they are not included in the STARLIB format.  Furthermore, although STARLIB contains probability densities for all reverse rates, these should in general {\it not} be sampled independently because forward and reverse rates are correlated according to detailed balance.  A proper Monte Carlo reaction network study would involve the following steps:  (i) sampling over the rate of the forward reaction, as described above; (ii) sampling over the reaction Q-value, which is described by a Gaussian probability density with a mean of $Q$ and a standard deviation of $\Delta Q$; (iii) calculating the resulting reverse rate from Eqs. 3.44 and 3.45 of \citet{iliadisbook}, and Eq.~\ref{last} for the rate of the forward reaction that was sampled under step (i).  Disregarding the correlation between forward and reverse rate may greatly exaggerate the Monte Carlo derived abundance uncertainties at elevated temperatures, where reverse rates become important.\footnote{For the example shown in Fig. 6, the effects of the correlations discussed here are small.  First, the main uncertainties in the $^{22}$Ne+$\alpha$ rates at $s$-process temperatures are determined by the contributions from unobserved resonances, which are uncorrelated. Uncertainties from correlated quantities, such as excitation energies, properties of observed resonances, etc., are small. Second, $s$-process temperatures are typically $< 300$ MK, which is too low for a significant abundance flow via reverse reactions to occur.}

For the reverse rate sampling method described above, an estimate for the Q-value uncertainty is needed.  Since this information is not provided in the present version of STARLIB, the user needs to supplement STARLIB with other sources.  Future releases will include the Q-value uncertainties.


\section{Conclusions}

In this work, we have presented STARLIB, a next-generation reaction-rate library for nuclear astrophysics.  This library fills a void in the community and urges a paradigm shift for stellar modelers to incorporate not only rate uncertainties but also their probability densities, as no current library provides this information.  We have outlined the Monte Carlo procedure  that enables assignments of rigorous uncertainties for experimental reaction rates.  Monte Carlo rates are supplemented by a variety of other types to facilitate an all-purpose nuclear reaction library, including weak rates and theoretical rates for which no experimental information is known.  Methods to correct laboratory rates for a stellar environment have been discussed, and an application of STARLIB in a Monte Carlo nucleosynthesis study was demonstrated.

\acknowledgments

This work was supported in part by the National Science Foundation under award number AST-1008355 and
by the Department of Energy under grant number DE-FG02-97ER41041.  We would also like to thank Richard Longland, Jordi Jos{\'e}, Alain Coc, and Matthew Buckner.  In addition, we thank Arjan Koning and St{\'e}phane Hilaire for their valuable contribution in writing the TALYS code and making it publicly available.  Our collaborators at ASU acknowledge partial support from NASA.

\appendix

The following appendix details STARLIB's formatting, its website, and new and updated Monte Carlo rates.

\section{STARLIB Format}\label{format}

For each nuclear interaction, STARLIB lists a single-line header, followed by three columns:  temperature (in GK), recommended rate (cm$^3$~s$^{-1}$~mol$^{-1}$ for reactions; s$^{-1}$ for photodisintegration and $\beta$-decay constants), and factor uncertainty of the rate.  The temperature grid consists of 60 values between 1 MK and 10 GK.  STARLIB reference labels are listed in Table~\ref{reftable}.

\setlongtables
\begin{longtable}{ll}
\caption{Labels and references in STARLIB}
 \\
\hline \hline 
 Label &Reference\\
\hline 
\endfirsthead
\multicolumn{2}{c}{{\tablename} \thetable{} -- continued} \\
\hline \hline 
 Label &Reference\\
 \hline 
\endhead
	 \hline \hline
\endfoot
	\hline \hline
\endlastfoot
\multicolumn{2}{l}{REACTION RATES BASED ON EXPERIMENT	}		\\
\hline
{\tt cf88}	&	Caughlan \& Fowler, At. Data Nucl. Data Tab. 40, 283 (1988) [CF88]	\\
{\tt de04	}&	Descouvemont {\it et al.}, At. Data Nucl. Data Tab. 88, 203 (2004)	\\
{\tt mc10}	&	Iliadis {\it et al.}, Nucl. Phys. A 841, 31 (2010) [Monte Carlo rates]	\\
{\tt mc13} 	&	 {\bf Present work [updated Monte Carlo rates]; see Appendix~\ref{updates}	}\\
{\tt 	nacr	}	&	Angulo {\it et al.}, Nucl. Phys. A 656, 3 (1999) [NACRE]	\\
{\tt 	taex	}	&	Bao et al. At. Data Nucl. Data Tab. 76, 70 (2000), and updates; {\bf present work} \\ & {\bf [extrapolated using TALYS]; see Sections~\ref{theory} and~\ref{neutronsec}}\\	
	\hline				
	\multicolumn{2}{l}{REACTION RATES BASED ON HAUSER-FESHBACH THEORY}			\\	
	\hline				
					
{\tt 	taly	}	&	{\bf Present work [calculated using TALYS]; see Sections~\ref{theory} and~\ref{talys}}\\	
	\hline				
					
	\multicolumn{2}{l}{DECAY RATES BASED ON EXPERIMENT		}		\\
	\hline				
{\tt 	au03w	}	&	Audi et al., NPA 729, 3 (2003) \\	
{\tt 	auecw	}	&	Audi et al., NPA 729, 3 (2003)\footnote{Rates need to be multiplied by the electron density, $\rho Y_e$.}\\	
{\tt 	  bet-w	}	&	$\beta^-$-decay \\	
{\tt 	 bet+w	}	&	$\beta^+$-decay \\	
{\tt 	     ~~ecw	}	&	pep reaction; 3He $\rightarrow$ t decay\footnotemark[1]\\	
{\tt 	ru09w	}	&		Rugel et al., Phys. Rev. Lett. 103, 072502 (2009) \\
{\tt 	wc12w	}	&	Tuli, wallet cards, National Nuclear Data Center (2012) \\	
{\tt 	wi11w	}	&	Wietfeldt and Greene, RMP 83, 1173 (2011) \\	
	\hline				
	\multicolumn{2}{l}{DECAY RATES BASED ON THEORY		}		\\
	\hline				
{\tt 	bkmow	}	&	Klapdor, Metzinger \& Oda, At. Data Nucl. Data Tab. 31, 81 (1984) [$\beta^-$-decay]	\\
{\tt 	btykw	}	&	 Takahashi, Yamada \& Kondo, At. Data Nucl. Data Tab. 12, 101 (1973) [$\beta^+$-decay]	\\
{\tt 	il11g	}	&	Iliadis et al., ApJS 193, 16 (2011) [$\gamma$-ray transitions]	\\
					
{\tt 	ka88w	}	&	Kajino {\it et al.}, Nucl. Phys. A 480, 175 (1988) [$\beta$-decay of excited $^{26}$Al levels]	\\
{\tt 	mo92w	}	&	 Moeller {\it et al.} (1992) [$\beta^-$-decay]	\\
{\tt 	mo03w} &		Moeller, Pfeiffer and Kratz, Phys. Rev. C 67, 055802 (2003) \\		
	\hline				
	\multicolumn{2}{l}{INDIVIDUAL RATES, MAINLY BASED ON EXPERIMENT		}		\\
	\hline				
{\tt 	an06	}	&	Ando {\it et al.}, Phys. Rev. C 74, 025809 (2006)	\\
{\tt 	ar12	}	&	 Arnold {\it et al.}, Phys. Rev. C 85, 044605 (2012)	\\
{\tt 	bb92	}		&Rauscher et al., ApJ 429, 499 (1994) \\	
					
{\tt 	be01	}	&	Beaumel {\it et al.}, Phys. Lett. B 514, 226 (2001)	\\
{\tt 	bu12	}	&	 Buckner {\it et al.}, submitted to Phys. Rev. C (2012)	\\
					
{\tt 	cy08	}	&	 Cybert \& Davids, Phys. Rev. C 78, 064614 (2008)	\\
{\tt 	fu90	}	&	 Fukugita \& Kajino, Phys. Rev. D 42, 4251 (1990)	\\
{\tt 	ha10	}	&	 Hammache {\it et al.}, Phys. Rev. C 82, 065803 (2010)	\\
{\tt 	il11	}	&	Iliadis {\it et al.}, Astrophys. J. Suppl. 193, 16 (2011)	\\
{\tt 	im05	}	&	Imbriani {\it et al.}, Eur. Phys. J. A 25, 455 (2005)	\\
{\tt 	ku02	}	&	Kunz {\it et al.}, Astrophys. J. 567, 643 (2002)	\\
{\tt 	lo12	}	&	 Longland {\it et al.}, Phys. Rev. C 85, 065809 (2012)\\	
{\tt 	mafo	}	&	 Malaney \& Fowler, Astrophys. J. 345, L5 (1989)	\\
{\tt 	nk06	}	&	Nagai {\it et al.}, Phys. Rev. C 74, 025804 (2006)	\\
{\tt 	re98	}	&	Rehm {\it et al.}, Phys. Rev. Lett. 80, 676 (1998)	\\
{\tt 	rolf	}	&	 C. Rolfs and collaborators	\\
{\tt 	ta03	}	&	 Tang {\it et al.}, Phys. Rev. C 67, 015804 (2003)	\\
{\tt 	wies	}	&	 M. Wiescher and collaborators	
  \label{reftable}
\end{longtable}

\noindent
The header consists of one line of 64 fields, indicating  (i) the type of interaction, (ii) the interacting nuclei, (iii) a reference label, (iv) an interaction specific label, and (v) the recommended Q-value. 
\\
\\
(i) {\it Type of interaction (left-aligned in fields 1-2):}  these follow the JINA REACLIB chapter numbers, specifically their ``format 2, with chapters 9-11" version \citep[see][]{reaclibformat}.  They are in principle redundant since the type of interaction can always be reconstructed from the interaction label (see below).  However, they have been included in STARLIB as convenient labels for some users.  The chapter numbers are ($e_i$ stand for interacting nuclei):

\begin{table}[ht]
\begin{tabular}{lllll}
&&	1&&	$e_1 \longrightarrow e_2$\\
&&	2&&	$e_1 \longrightarrow e_2 + e_3$\\
&&	3&&	$e_1 \longrightarrow e_2 + e_3 + e_4$\\
&&   4&&	$e_1 + e_2 \longrightarrow e_3 $\\
&&	5&&	$e_1 + e_2 \longrightarrow e_3 + e_4$\\
&&	6&&	$e_1 + e_2 \longrightarrow e_3 + e_4 + e_5$\\
&&	7&&	$e_1 + e_2 \longrightarrow e_3 + e_4 + e_5 + e_6$\\
&&	8&&	$e_1 + e_2 + e_3 \longrightarrow e_4$\\
&&	9&&	$e_1 + e_2 + e_3 \longrightarrow e_4 + e_5$\\
&&	10&&	$e_1 + e_2 + e_3 + e_4 \longrightarrow e_5 + e_6$\\
&&	11&&	$e_1 \longrightarrow e_2 + e_3 + e_4 + e_5$\\
\end{tabular}
\end{table}
\noindent
(ii) {\it Interaction label (fields 6-35):}  all species in the entrance channel are listed first (lightest to heaviest), followed by species in the exit channel; each species in the interaction occupies 5 fields (right-aligned), with the element symbol given first, followed by the mass number; exceptions are neutrons (``n"), protons (``p"), and tritons (``t"); levels in $^{26}$Al (see Sec.~\ref{otheral26}) have special labels: $E_x=0$ keV (ground state; 5$^+$), 228 keV (isomeric state; 0$^+$), 417 keV (3$^+$), 1058 keV (1$^+$) and 2070 keV (2$^+$) are denoted by ``al-6", ``al*6", ``al01", ``al02", and ``al03", respectively; the label ``al26" is used to denote $^{26}$Al with all levels in thermal equilibrium (presumably at sufficiently high temperatures).
\\
\\
(iii) {\it Reference label (right-aligned in fields 44-47):}  these labels, which usually have a length of four characters, provide information on the source of the rates; a complete list of STARLIB reference labels is given in Table~\ref{reftable}.  Note that the characters ``ec" in fields 46-47 indicate electron capture, and, therefore, these decay rates need to be multiplied by $\rho Y_e$, where $\rho$ and $Y_e$ denote the density and electron mole fraction, respectively.
\\
\\(iv) {\it Interaction specific label (field 48):}  indicates a weak interaction (``w"), a $\gamma$-ray transition (``g"), or if the reaction rate has been calculated from a forward rate using the principle of detailed balance (``v"). 
\\
\\
(v) {\it The Q-value (fields 53-64):}  indicating the energy release in the nuclear interaction, in units of MeV; values are obtained from the recommendations of \citet{iliadisbook,ilthomp} or, if no experimental information is available, from nuclear models (see Sec.~\ref{theory}).\\

 A number of examples are listed below:

\begin{table}[ht]
\begin{tabular}{rlrr}
{\tt 1~~~~}  & {\tt ~~o15~~n15 }&{\tt ~~~~~~~~~~~~~~~~~~~~~~~~au03w}&  {\tt ~~~~~~2.75420e+00}\\
{\tt 1~~~~}     & {\tt ~al-6~al01}&{\tt ~~~~~~~~~~~~~~~~~~~~~~~~il11g}& {\tt~~~~~~-4.16800E-01} \\
{\tt 4~~~~}   & {\tt  ~~~~p~~~~p~~~~d}&{\tt ~~~~~~~~~~~~~~~~~~~~~~~~nacr~~}& {\tt ~~~~~~1.44222E+00}\\
{\tt 4~~~~}   &{\tt  ~~~~p~~~~p~~~~d}&{\tt ~~~~~~~~~~~~~~~~~~~~~~~~~~~ecw}& {\tt ~~~~~~1.44206e+00}\\
{\tt 5~~~~}   & {\tt ~~he4~~n14~~~~p~~o17}&{\tt ~~~~~~~~~~~~~~~~~~~~~~~~~mc10v} &{\tt   ~~~~~~-1.19182e+00 }\\
\end{tabular}
\end{table}

The first example denotes the $\beta$-decay $^{15}$O$\rightarrow$$^{15}$N, ``au03" indicates that the decay rate was adopted from \citet{audi}, the ``w" label indicates a weak interaction, and the total energy release (not corrected for neutrino losses) amounts to 2.7542 MeV.  The second example illustrates the $\gamma$-ray transition from the ground state of $^{26}$Al (5$^+$) to the third excited state at E$_x$=417 keV (3$^+$), listing the source of the rate~\citet{il_al26} and the label ``g" for a $\gamma$-ray transitions, and providing the energy for the transition (-0.4168 MeV).  The third example refers to the $p+p \rightarrow d$ reaction, with the source listed as \citet{nacre}, and an energy release of 1.44222 MeV.  The fourth example illustrates the $p+e+p +\bar{\nu} \rightarrow d$ reaction, which depends on the electron density in the plasma and thus is distinct from the previous example.  In this case, the label ``ec" indicates that the reaction rate needs to be multiplied by $\rho Y_e$ in a reaction network calculation, similar to a weak interaction (hence the label ``w"). The fifth and last example denotes the $^{14}$N($\alpha,p$)$^{17}$O reaction.  The source of the reaction rate is the Monte Carlo based evaluation of \citet{iliadis_2}, with the label ``v" (reverse) indicating that this rate has been obtained from the rate of the forward reaction, $^{17}$O($p,\alpha$)$^{14}$N, using detailed balance.  The reaction Q-value amounts to -1.19182 MeV (endothermic reaction).




Below we summarize the meaning of each of the three columns that follow the single-line header: \\

\noindent
(i) {\it Column 1:}  Temperature, in GK ($10^9$ K). 
\\
\\
(ii) {\it Column 2:}  Recommended rate, either the thermonuclear reaction rate in units of cm$^3$ mol$^{-1}$ s$^{-1}$, or the decay constant of a photodisintegration or a $\beta$-decay in units of s$^{-1}$. 

\noindent
All rates in STARLIB are assumed to be lognormally distributed (see Sec.~\ref{logsec}), however, only for the Monte Carlo based rates do the listed rate and factor uncertainty have a precise statistical meaning.  See Secs.~\ref{MCsec} and~\ref{starlib} for a detailed discussion.   
The lognormal probability density is given by Eq.~\ref{logneq}:

\begin{equation}
 f(x)  = \frac{1}{\sigma\sqrt{2\pi}} \frac{1}{x} e^{-(\rm{ln}\, x-\mu)^2/(2\sigma^2)}\qquad {\rm for  }\quad 0 < x < \infty,
 \end{equation}
  where $x$ is the reaction rate.  The lognormal parameters are related to the median, low, and high rates via $x_{low}=e^{\mu-\sigma}$, $x_{med} = e^\mu$, and $x_{high}=e^{\mu+\sigma}$, respectively, for a coverage probability of 68\%.  Thus, the natural logarithm of the recommended rate provides the lognormal parameter $\mu$. 
 \\
\\
(iii) {\it Column 3:}  Factor uncertainty, $f.u$.  One may multiply and divide the recommended rate (Column 2) by the factor uncertainty to calculate the high and low rates.   The factor uncertainty is related to the shape parameter, $\sigma$, by $f.u. = e^\sigma$.  For small values of $\sigma$, below 0.1, the lognormal distribution appears more Gaussian (i.e., symmetric) in shape and deviates from symmetry as $\sigma$ increases in value.  The natural logarithm of the factor uncertainty yields the lognormal parameter $\sigma$.  \\
\\

Once both the lognormal parameters $\mu$ and $\sigma$ have been determined by the natural logarithms of the values listed in Column 2 and 3, respectively, the probability density function may be calculated using Eq.~\ref{logneq}.


\section{STARLIB Website}\label{website}

The STARLIB library is available at:
\noindent
\begin{center}\url{starlib.physics.unc.edu}  
\end{center}
\noindent 
where one may display an individual rate, download the entire library, or download a user-specified subset of the library.  The website allows ease of dissemination into the community and a portal for real-time updating of new rates.  Also included is a reaction rate calculator, where one may use Monte-Carlo simulation code {\tt RatesMC} to calculate an experimental reaction rate.  The following sections outline each major function of the site.  Details on STARLIB, its construction, hyperlinked references, and information on updates may be accessed at~\url{starlib.physics.unc.edu/Details.php}.  Step-by-step instructions for the website are shown in \citet{nicXII}.

\subsection{Library}

The entire STARLIB library can be downloaded from:

\noindent
\begin{center}\url{starlib.physics.unc.edu/RateLib.php}  
\end{center}
\noindent
As we update new versions of STARLIB, archived versions will also be available for comparison purposes.  Because some stellar models may not require the full STARLIB library, one may also download a subset of the library for use in a reduced network, after uploading a user-specified list of nuclides.  The formatting of the list may have the chemical name before or after the mass number (for example, ``na22'' or ``22na'') and is case insensitive.  The site will accept either left or right aligned lists.  

Individual rates may also be displayed with the current or archived versions of the library.  There are three available menus from which to choose.  The left menu requires {\it all} interacting {\it nuclei} to be entered (i.e., electrons or $\gamma$-rays should be omitted), up to three initial and three final species.  This menu also allows a choice of special reactions, for example, those involving four interacting nuclei or the $p+e+p\rightarrow d$ reaction.  

The middle menu allows input in the form of a target nuclide and reaction.  The reaction may be formatted as ``x,y" or ``(x,y)".  A complete list of formatting options is shown in Table~\ref{options}.  

The right menu is for $\beta$-decays and $\gamma$-ray transitions, where parent and daughter nuclei may be entered.  There is also a button that will display examples of the various options for each menu; multiple clicks cycle through examples.  

Upon submission of any of the three menu options, the individual rate will be presented along with its uncertainty, its source (including a link to the source description and Digital Online Identification), Q-value, and the library version.

\begin{table}[ht]
\caption{Formatting options for user input on the STARLIB website.  }
\begin{tabular}{l|lll | ll | l}
\hline\hline

Nuclide& \multicolumn{3}{c|}{Format}&\multicolumn{2}{c|}{$^{26}$Al} & Reactions\\

\hline
proton & p & h1 & 1h & al26 & al*6& x,y \\
deuteron & d&h2&2h&al-6&al01&(x,y)\\
triton& t&h3&3h&al02&al03&\\
$\alpha$-particle & a&he4&4he&&&\\
neutron & n&n1&1n&&&\\
general &- & na22 & 22na &&&\\

\hline\hline
\end{tabular}
\label{options}
\end{table}

\subsection{Calculator for Experimental Monte Carlo Based Reaction Rates} 

In addition to accessing the STARLIB library, the Monte Carlo code {\tt RatesMC} \citep[written by Richard Longland; see][]{iliadis_1} is available to users to run on our website at:

\noindent
\begin{center}\url{starlib.physics.unc.edu/RateCalc.php}  
\end{center}
\noindent
We currently employ an Intel Xeon Processor X5650 server with dual 6-core, hyper-threaded, 2.66 GHz processors (maximum of 24 simultaneous simulations of {\tt RatesMC}).  

The first set of menu items on the Rate Calculator webpage allows access to a repository of current input files for experimental Monte Carlo rates.  The same formatting applies as outlined in the previous subsection, as well as displaying examples via the ``?'' buttons.  For the reaction $^{25}$Mg$(p,\gamma)^{26}$Al, options for choosing the isomeric state, ground state, or both combined are available.  This database is available freely to all users.  However, to run simulations, we require users to register their email addresses.    

We recommend that the user edit an existing input file from our online database.  Once the input file is uploaded or edited, the Monte Carlo simulation may be started by logging in with a registered email address.  Our server employs a PBS queue system, and, after submission, the website displays the job ID, status (``R'' for running, ``Q'' for in the queue), the run time, simulation progress (in percent), and estimated time remaining.  The site employs several shell scripts, one of which determines if any resonances are being integrated.  Depending on the circumstance, the jobs are filtered into different queues to allow jobs without integrated resonances to be completed immediately and instantaneously.  A Monte Carlo simulation requires significantly more CPU time if resonance contributions to the total rate have to be determined by numerical integration.  As discussed in Sec.~\ref{MCsec}, a $(p,\gamma)$ reaction with one integrated resonance will take approximately 45 minutes on our current server.  Simulations may be aborted by supplying the job ID.  

When the simulation is complete, an email alert will be sent to the address provided, including the job ID and a link to access the output files.  Files may also be accessed from:

\noindent
\begin{center}\url{starlib.physics.unc.edu/RateCalc.php#output}

\end{center}
\noindent
or by scrolling to the bottom of the Rate Calculator page to log in.  Files are delineated by the reaction name, the job ID, and the appropriate file extensions (``.in'', ``.out'', ``.hist'', ``.latex'').  For ease of bookkeeping, both the input file (``.in'') and output file (``.out'') are in the data repository.  The ``.hist'' file includes the binned reaction rate probability density function for each of the 60 temperature values, and the ``.latex'' file contains the reaction rate table in Latex format.  Files may be sorted by name, size, or date and may also be deleted by specifying the job ID.

\section{Updated Experimental Monte Carlo Based Rates}\label{updates}


Here we provide updated Monte Carlo based thermonuclear rates for the reactions $^{18}$O$(p,\gamma)^{19}$F, $^{18}$O$(p,\alpha)^{15}$N, $^{22}$Ne$(p,\gamma)^{23}$Na, $^{22}$Ne$(\alpha,\gamma)^{26}$Mg, $^{22}$Ne$(\alpha,n)^{25}$Mg, $^{22}$Na$(p,\gamma)^{23}$Mg, $^{29}$P$(p,\gamma)^{30}$S, and new rate $^{38}$Ar($p,\gamma)^{39}$K.  Similarly, to the procedure in \citet{iliadis_1,iliadis_2,iliadis_3,iliadis_4}, we provide for each reaction (i) a rate table, (ii) the complete input file used to calculate the rate; (iii) the figure with the rate probability densities at six sample temperatures; and (iv) a figure comparing the new Monte Carlo based rate with the latest previously reported rate.  Some of the reaction rates listed here have been published elsewhere~\citep{lo12,bu12}, but the information according to (ii), (iii), and (iv) has not been reported elsewhere.  Details are given below.  Energies are in the center-of-mass system, unless otherwise noted.  

\noindent
(i) {\it Reaction rate table:}



\setlongtables
\begin{longtable}{p{2cm}p{15cm}}
\\

\endfirsthead
\multicolumn{2}{c}{} \\

\endhead
\endfoot
\endlastfoot

T &  Temperature in GK.\\
Low rate & 0.16 quantile of the cumulative reaction rate distribution. \\ 
Median rate & 0.50 quantile of the cumulative reaction rate distribution.\\
High rate & 0.84 quantile of the cumulative reaction rate distribution.\\
&All reaction rates are in units of cm$^3$~s$^{-1}$~mol$^{-1}$.\\
$\mu$&Parameter determining the location of the lognormal reaction rate probability density function.\\
$\sigma$ & Parameter determining the width of the lognormal reaction rate probability density function.\\
 A-D & AndersonÐ-Darling test statistic, t$^*_{AD}$, indicating how well the Monte Carlo reaction rates are approximated by a lognormal distribution \citep[see][]{iliadis_1,iliadis_2,iliadis_3,iliadis_4}.   \\
 () & Values given in parenthesis are  {\it not} obtained from the Monte Carlo method but are found from extrapolation to elevated temperatures (see Sec.~\ref{matchsec}).\\

\label{rrtable}
\end{longtable}

\noindent
(ii) {\it Nuclear data input table:} an example input file and the corresponding detailed explanations are reproduced from Appendix A of \citet{iliadis_3}, and are repeated here for convenience.

\footnotesize
\begin{verbatim}
01 17X(p,a)14Y
02 ***************************************************************************************************************
03 1               ! Zproj
04 8               ! Ztarget
05 2               ! Zexitparticle (=0 when only 2 channels open)
06 1.0078          ! Aproj
07 16.999          ! Atarget
08 4.0026          ! Aexitparticle (=0 when only 2 channels open)
09 0.5             ! Jproj
10 1.5             ! Jtarget
11 0.0             ! Jexitparticle (=0 when only 2 channels open)
12 5261.3          ! projectile separation energy (keV)
13 4015.3          ! exit particle separation energy (=0 when only 2 channels open)
14 1.25            ! Radius parameter R0 (fm)
15 3               ! Gamma-ray channel number (=2 if ejectile is a g-ray; =3 otherwise)
16 ***************************************************************************************************************
17 1.0             ! Minimum energy for numerical integration (keV)
18 5000            ! Number of random samples (>5000 for better statistics)
19 0               ! =0 for rate output at all temperatures; =NT for rate output at selected temperatures
20 *************************************************************************************************************** 
21 Nonresonant Contribution
22 S(keVb)         S'(b)           S''(b/keV)      fracErr         Cutoff Energy (keV)
23 3.1e2           -2.1e-1         4.5e-6          0.4             1200.0                 
24 0.0             0.0             0.0             0.0              0.0
25 *************************************************************************************************************** 
26 Resonant Contribution
27 Note: G1 = entrance channel, G2 = exit channel, G3 = spectator channel !! Ecm, Exf in (keV); wg,Gx in (eV) !!
28 Note: if Er<0,
29 Ecm     DEcm  wg      Dwg     Jr     G1      DG1     L1  G2    DG2   L2  G3      DG3     L3  Exf   Int
30  -5.2   0.6   0       0       1      0.022   0.010   2   30.8  2.6   0   0.77    0.21    1   0.0   1 
31 477.3   1.1   0       0       4      140     32      1   110   12    3   0.13    0.02    1   0.0   1 
32 510.5   2.0   6.0e-1  1.8e-1  0      0       0       0   0     0     0   0       0       0   0.0   0 
33 *************************************************************************************************************** 
34 Upper Limits of Resonances
35 Note: enter partial width upper limit by chosing non-zero value for PT, where PT=<theta^2> for particles and... 
36 Note: ...PT=<B> for g-rays [enter: "upper_limit 0.0"]; for each resonance: # upper limits < # open channels! 
37 Ecm     DEcm   Jr   G1      DG1     L1   PT     G2    DG2    L2   PT    G3     DG3   L3  PT     Exf   Int
38 -3.4    0.7    1    7.9e-3  0.0     1    0.0045 15.0  2.9    1    0.0   0.33   0.07  1   0      0.0   1
39 73.5    0.4    1    4.5e-7  1.8e-7  1    0.0    150.0 0.0    1    0.010 0.21   0.02  1   0      0.0   1
40 *************************************************************************************************************** 
41 Interference between Resonances [numerical integration only]
42 Note: + for positive, - for negative interference; +- if interference sign is unknown
43 Ecm     DEcm   Jr   G1      DG1     L1   PT     G2    DG2    L2   PT    G3     DG3   L3  PT     Exf   
44 +-
45 -2.3    0.6    1    6.2e-3  0.0     1    0.0045 17.0  1.1    1    0.0   0.71   0.07  1   0      0.0  
46 89.0    0.8    1    2.9e-9  0.8e-9  1    0.0    230.0 9.0    1    0.0   0.66   0.03  1   0      0.0  
47 *************************************************************************************************************** 
48 Comments:
49 1. Narrow resonance information is adopted from Peterson et al. 1973.
\end{verbatim}
\normalsize

\setlongtables
\begin{longtable}{p{2cm}p{15cm}}
\\

\endfirsthead
\multicolumn{2}{c}{} \\

\endhead
\endfoot
\endlastfoot

Row 01:  & Reaction label. \\
Row 02:  & Separator.\\
Row 03:  & Projectile charge.\\
Row 04:  & Target charge.\\
Row 05:  & Charge of exit particle; it refers to the channel other than the entrance and the $\gamma$-ray channel; = 0 if only two channels are open and radiative capture is the only possible reaction.\\
Row 06:  & Projectile mass in atomic mass units. \\
Row 07:  & Target mass in atomic mass units.\\
Row 08:  & Mass of exit particle in atomic mass units. \\
Row 09:  & Projectile spin.\\
Row 10:  & Target spin.\\
Row 11:  & Spin of exit particle.\\
Row 12:  & Separation energy of incident particle in keV.\\
Row 13:  & Separation energy of exit particle in keV.\\
Row 14:  & Radius parameter in fm, used for calculating penetration factor.\\
Row 15:  &Label of $\gamma$-ray channel; channel 1 refers to the incident particle,
channel 2 to the emitted quantum, and channel 3 to the spectator quantum. \\
Row 16:  &Separator.\\
Row 17:  &Minimum energy cutoff (in keV) for numerical integration of rates. \\
Row 18:  &Number of random samples.\\
Row 19:  &Flag for temperature output; = 0 outputs results at all temperatures.\\
Row 20:  &Separator.\\
Row 21-22:  &Comments.\\
Row 23:  & Input for nonresonant contribution; S, SÕ, SÓ are the parameters S(0), S$^\prime$(0), S$^{\prime\prime}$(0) of the astrophysical S-factor in units of keV b, b, b/keV, respectively; {\tt fracErr} is the fractional uncertainty, $\sqrt{V [x]/E[x]}$, of the effective S-factor; Cutoff Energy labels the energy $E_{\rm cutoff}$ (in keV) at which the S-factor is cut off at higher energies; it is related to the cutoff temperature by the expression\\
& $T_{9,{\rm cutoff}} = 19.92 E_{\rm cutoff}^{3/2} /\sqrt{Z_0^2 Z_1^2\frac{M_0 M_1}{M_0+M_1}}$ \\
&where $E_{\rm cutoff}$ is in MeV and all other quantities have the same meaning as in Sec.~\ref{meth}.\\ 
Row 24:  &Input for a second nonresonant contribution, if needed.\\
Row 25:  &Separator.\\
Row 26-29:  &Comments.\\
Row 30:  &Input for resonance contribution, one input row for each resonance; {\tt Ecm}, {\tt DEcm}: resonance energy and 1$\sigma$ uncertainty; {\tt wg}, {\tt Dwg}: resonance strength, $\omega\gamma$, and associated uncertainty; {\tt Jr}: resonance spin; {\tt G1,DG1,L1}: incident particle partial width, uncertainty, orbital angular momentum quantum number; for a subthreshold resonance ($E_r < 0$) the dimensionless reduced width is listed instead of the entrance channel partial width; {\tt G2,DG2,L2}: partial width, uncertainty, angular momentum quantum number (or multipolarity for $\gamma$-rays) of emitted quantum; {\tt G3,DG3,L3}: partial width, uncertainty, angular momentum quantum number (or multipolarity for $\gamma$-rays) of spectator quantum; {\tt Exf}: excitation energy of level in residual nucleus that is populated in primary transition; {\tt Int}: = 0 for analytical rate calculation; = 1 for numerical rate calculation; for $E_r < 0$ the rate contribution is always computed numerically; when the resonance strength is entered the rate contribution is always computed analytically, regardless of the flag value; {\tt Ecm,DEcm,Exf} are in units of keV, while resonance strengths and partial widths are in eV.\\
Row 31:  &Input for second resonance; in this example, the rate contribution is calculated from partial widths and the rate is chosen to be integrated numerically.\\
Row 32:  &Input for third resonance; in this example, the rate contribution is calculated from the resonance strength and the rate is necessarily computed analytically.\\
Row 33:  &Separator. \\
Row 34-37:  &Comments.\\
Row 38:  &Input for resonance contribution when only a partial (or reduced) width upper limit is available for at least one reaction channel; the number of upper limit channels must be less than the number of open channels;
the meaning of the input quantities is the same as for row 30, except that (i) resonance strengths are not allowed as input, and (ii) the mean value, PT, for the PorterÐThomas distribution of dimensionless reduced widths is entered for each upper limit channel; upper limit channels are identified by a non-zero value for the partial (or reduced) width and a zero value for the corresponding uncertainty; in this example, an upper limit for the dimensionless reduced proton width is entered (since $E_r < 0$).\\
Row 39:  &Input for second resonance; in this example, an upper limit for the dimensionless reduced $\alpha$-particle width is entered.\\
Row 40:  &Separator.\\
Row 41-43:  &Comments.\\
Row 44:  &Flag for interference between resonances of same spin and parity; +: positive interference; -: negative interference; +-: unknown interference sign (a binary probability density function
is then used for the random sampling.\\
Row 45-46:  & Input for two interfering resonances; the meaning of the input quantities is the same as for row 38, except that the rate contribution is always computed numerically.\\
Row 47:  & Separator.\\
Row 48-49:  &Comments.\\

\label{rrtable}
\end{longtable}

\noindent
(iii) {\it Rate probability density figure:}  This plot consists of six panels showing the Monte Carlo reaction rate probability density function (in red) at sample temperatures of $T= 0.03, 0.06, 0.1, 0.3, 0.6$, and $1.0$ GK.  The lognormal approximation to the rate is shown in black.  Note that it does not represent a least-squares fits to the red histogram, but the lognormal parameters $\mu$ and $\sigma$ are computed directly from the expectation value and the variance of the logarithm of the rate distribution, according to Eq.~\ref{mu}.  For a reaction, each panel displays the temperature, T, and the Anderson\---Darling test statistic, A\---D \citep[see][]{iliadis_1}.  

\noindent
(iv) {\it Comparison to literature figure:}  This plot displays reaction rate ratios as a function of temperature in two panels, using the following notation:

\begin{tabbing}
AAAAAAAAAAAAA\= \kill

$N_A \langle\sigma v \rangle_{c,high}$\>Present high Monte Carlo reaction rate \\
$N_A \langle\sigma v \rangle_{c,med}$\>Present median Monte Carlo reaction rate\\
$N_A \langle\sigma v \rangle_{c,low}$\>Present low Monte Carlo reaction rate\\
$N_A \langle\sigma v \rangle_{p,high}$\>Previous high Monte Carlo reaction rate~\citep{iliadis_2} \\
$N_A \langle\sigma v \rangle_{p,med}$\> Previous median Monte Carlo reaction rate~\citep{iliadis_2} \\
$N_A \langle\sigma v \rangle_{p,low}$\>Previous low Monte Carlo reaction rate~\citep{iliadis_2}\\
\>\\
Top graph: \>\\
 \>\\
 upper solid \>$N_A \langle\sigma v \rangle_{c,high}/N_A \langle\sigma v \rangle_{c,med}$\\
lower solid\>$N_A \langle\sigma v \rangle_{c,low}/N_A \langle\sigma v \rangle_{c,med}$\\
upper dashed line\>$N_A \langle\sigma v \rangle_{p,high}/N_A \langle\sigma v \rangle_{p,med}$\\
 lower dashed line\>$N_A \langle\sigma v \rangle_{p,low}/N_A \langle\sigma v \rangle_{p,med}$\\
 \>\\
Bottom graph:\>\\
 \>\\
upper thin solid line\>$N_A \langle\sigma v \rangle_{c,high}/N_A \langle\sigma v \rangle_{p,med}$\\
thick solid line \>$N_A \langle\sigma v \rangle_{c,med}/N_A \langle\sigma v \rangle_{p,med}$\\
lower thin solid line\>$N_A \langle\sigma v \rangle_{c,low}/N_A \langle\sigma v \rangle_{p,med}$ \\

\label{}
\end{tabbing}

\clearpage
\setlongtables
\begin{longtable}{cccc | ccc}
\caption{Total thermonuclear reaction rates for $^{18}$O(p,$\gamma$)$^{19}$F.}  \label{tab:o18pg} \\
\hline \hline 
	\multicolumn{1}{c}{T (GK)} & \multicolumn{1}{c}{Low rate} & \multicolumn{1}{c}{Median rate} & \multicolumn{1}{c}{High rate}  & \multicolumn{1}{c}{lognormal $\mu$} & \multicolumn{1}{c}{lognormal $\sigma$} & \multicolumn{1}{c}{A-D} \\ \hline 
\endfirsthead
\multicolumn{6}{c}{{\tablename} \thetable{} -- continued} \\
\hline \hline 
	\multicolumn{1}{c}{T (GK)} & \multicolumn{1}{c}{Low rate} & \multicolumn{1}{c}{Median rate} & \multicolumn{1}{c}{High rate} & \multicolumn{1}{c}{lognormal $\mu$} & \multicolumn{1}{c}{lognormal $\sigma$} & \multicolumn{1}{c}{A-D} \\ \hline 
\endhead
	 \hline \hline
\endfoot
	\hline \hline
\endlastfoot
0.010 &  2.92$\times$10$^{-24}$  &  4.97$\times$10$^{-24}$  &
       8.79$\times$10$^{-24}$  &  -5.364$\times$10$^{+01}$  &
       5.49$\times$10$^{-01}$  &  7.33$\times$10$^{+00}$  \\
0.011 &  2.58$\times$10$^{-23}$  &  4.22$\times$10$^{-23}$  &
       7.34$\times$10$^{-23}$  &  -5.149$\times$10$^{+01}$  &
       5.18$\times$10$^{-01}$  &  1.12$\times$10$^{+01}$  \\
0.012 &  1.72$\times$10$^{-22}$  &  2.68$\times$10$^{-22}$  &
       4.47$\times$10$^{-22}$  &  -4.964$\times$10$^{+01}$  &
       4.78$\times$10$^{-01}$  &  1.63$\times$10$^{+01}$  \\
0.013 &  9.30$\times$10$^{-22}$  &  1.37$\times$10$^{-21}$  &
       2.18$\times$10$^{-21}$  &  -4.800$\times$10$^{+01}$  &
       4.28$\times$10$^{-01}$  &  2.33$\times$10$^{+01}$  \\
0.014 &  4.22$\times$10$^{-21}$  &  5.97$\times$10$^{-21}$  &
       8.99$\times$10$^{-21}$  &  -4.653$\times$10$^{+01}$  &
       3.83$\times$10$^{-01}$  &  2.60$\times$10$^{+01}$  \\
0.015 &  1.71$\times$10$^{-20}$  &  2.32$\times$10$^{-20}$  &
       3.31$\times$10$^{-20}$  &  -4.519$\times$10$^{+01}$  &
       3.40$\times$10$^{-01}$  &  2.10$\times$10$^{+01}$  \\
0.016 &  6.18$\times$10$^{-20}$  &  8.15$\times$10$^{-20}$  &
       1.11$\times$10$^{-19}$  &  -4.393$\times$10$^{+01}$  &
       3.01$\times$10$^{-01}$  &  1.47$\times$10$^{+01}$  \\
0.018 &  6.28$\times$10$^{-19}$  &  8.03$\times$10$^{-19}$  &
       1.04$\times$10$^{-18}$  &  -4.166$\times$10$^{+01}$  &
       2.54$\times$10$^{-01}$  &  6.81$\times$10$^{+00}$  \\
0.020 &  4.82$\times$10$^{-18}$  &  6.04$\times$10$^{-18}$  &
       7.72$\times$10$^{-18}$  &  -3.964$\times$10$^{+01}$  &
       2.40$\times$10$^{-01}$  &  6.73$\times$10$^{+00}$  \\
0.025 &  3.23$\times$10$^{-16}$  &  4.04$\times$10$^{-16}$  &
       5.17$\times$10$^{-16}$  &  -3.543$\times$10$^{+01}$  &
       2.44$\times$10$^{-01}$  &  1.47$\times$10$^{+01}$  \\
0.030 &  8.73$\times$10$^{-15}$  &  1.09$\times$10$^{-14}$  &
       1.40$\times$10$^{-14}$  &  -3.212$\times$10$^{+01}$  &
       3.29$\times$10$^{-01}$  &  2.41$\times$10$^{+02}$  \\
0.040 &  1.21$\times$10$^{-12}$  &  1.55$\times$10$^{-12}$  &
       2.34$\times$10$^{-12}$  &  -2.701$\times$10$^{+01}$  &
       7.52$\times$10$^{-01}$  &  1.06$\times$10$^{+03}$  \\
0.050 &  9.35$\times$10$^{-11}$  &  1.23$\times$10$^{-10}$  &
       2.06$\times$10$^{-10}$  &  -2.259$\times$10$^{+01}$  &
       8.58$\times$10$^{-01}$  &  1.07$\times$10$^{+03}$  \\
0.060 &  9.55$\times$10$^{-09}$  &  1.27$\times$10$^{-08}$  &
       1.82$\times$10$^{-08}$  &  -1.809$\times$10$^{+01}$  &
       5.78$\times$10$^{-01}$  &  6.00$\times$10$^{+02}$  \\
0.070 &  3.57$\times$10$^{-07}$  &  4.76$\times$10$^{-07}$  &
       6.50$\times$10$^{-07}$  &  -1.452$\times$10$^{+01}$  &
       4.14$\times$10$^{-01}$  &  2.16$\times$10$^{+02}$  \\
0.080 &  5.50$\times$10$^{-06}$  &  7.30$\times$10$^{-06}$  &
       9.89$\times$10$^{-06}$  &  -1.181$\times$10$^{+01}$  &
       3.46$\times$10$^{-01}$  &  5.98$\times$10$^{+01}$  \\
0.090 &  4.55$\times$10$^{-05}$  &  6.05$\times$10$^{-05}$  &
       8.14$\times$10$^{-05}$  &  -9.706$\times$10$^{+00}$  &
       3.18$\times$10$^{-01}$  &  1.74$\times$10$^{+01}$  \\
0.100 &  2.44$\times$10$^{-04}$  &  3.24$\times$10$^{-04}$  &
       4.35$\times$10$^{-04}$  &  -8.031$\times$10$^{+00}$  &
       3.05$\times$10$^{-01}$  &  6.39$\times$10$^{+00}$  \\
0.110 &  9.50$\times$10$^{-04}$  &  1.26$\times$10$^{-03}$  &
       1.69$\times$10$^{-03}$  &  -6.673$\times$10$^{+00}$  &
       2.99$\times$10$^{-01}$  &  2.81$\times$10$^{+00}$  \\
0.120 &  2.92$\times$10$^{-03}$  &  3.88$\times$10$^{-03}$  &
       5.19$\times$10$^{-03}$  &  -5.551$\times$10$^{+00}$  &
       2.96$\times$10$^{-01}$  &  1.39$\times$10$^{+00}$  \\
0.130 &  7.48$\times$10$^{-03}$  &  9.95$\times$10$^{-03}$  &
       1.33$\times$10$^{-02}$  &  -4.610$\times$10$^{+00}$  &
       2.94$\times$10$^{-01}$  &  8.23$\times$10$^{-01}$  \\
0.140 &  1.66$\times$10$^{-02}$  &  2.21$\times$10$^{-02}$  &
       2.96$\times$10$^{-02}$  &  -3.812$\times$10$^{+00}$  &
       2.93$\times$10$^{-01}$  &  5.51$\times$10$^{-01}$  \\
0.150 &  3.30$\times$10$^{-02}$  &  4.39$\times$10$^{-02}$  &
       5.87$\times$10$^{-02}$  &  -3.127$\times$10$^{+00}$  &
       2.92$\times$10$^{-01}$  &  4.28$\times$10$^{-01}$  \\
0.160 &  5.97$\times$10$^{-02}$  &  7.94$\times$10$^{-02}$  &
       1.06$\times$10$^{-01}$  &  -2.533$\times$10$^{+00}$  &
       2.91$\times$10$^{-01}$  &  3.57$\times$10$^{-01}$  \\
0.180 &  1.58$\times$10$^{-01}$  &  2.10$\times$10$^{-01}$  &
       2.81$\times$10$^{-01}$  &  -1.560$\times$10$^{+00}$  &
       2.90$\times$10$^{-01}$  &  3.07$\times$10$^{-01}$  \\
0.200 &  3.39$\times$10$^{-01}$  &  4.51$\times$10$^{-01}$  &
       6.03$\times$10$^{-01}$  &  -7.970$\times$10$^{-01}$  &
       2.90$\times$10$^{-01}$  &  2.95$\times$10$^{-01}$  \\
0.250 &  1.27$\times$10$^{+00}$  &  1.69$\times$10$^{+00}$  &
       2.27$\times$10$^{+00}$  &  5.265$\times$10$^{-01}$  &
       2.89$\times$10$^{-01}$  &  2.87$\times$10$^{-01}$  \\
0.300 &  2.93$\times$10$^{+00}$  &  3.90$\times$10$^{+00}$  &
       5.21$\times$10$^{+00}$  &  1.360$\times$10$^{+00}$  &
       2.89$\times$10$^{-01}$  &  2.80$\times$10$^{-01}$  \\
0.350 &  5.15$\times$10$^{+00}$  &  6.85$\times$10$^{+00}$  &
       9.13$\times$10$^{+00}$  &  1.923$\times$10$^{+00}$  &
       2.87$\times$10$^{-01}$  &  2.71$\times$10$^{-01}$  \\
0.400 &  7.70$\times$10$^{+00}$  &  1.02$\times$10$^{+01}$  &
       1.36$\times$10$^{+01}$  &  2.322$\times$10$^{+00}$  &
       2.85$\times$10$^{-01}$  &  2.69$\times$10$^{-01}$  \\
0.450 &  1.04$\times$10$^{+01}$  &  1.37$\times$10$^{+01}$  &
       1.82$\times$10$^{+01}$  &  2.616$\times$10$^{+00}$  &
       2.83$\times$10$^{-01}$  &  2.73$\times$10$^{-01}$  \\
0.500 &  1.30$\times$10$^{+01}$  &  1.71$\times$10$^{+01}$  &
       2.27$\times$10$^{+01}$  &  2.842$\times$10$^{+00}$  &
       2.79$\times$10$^{-01}$  &  2.91$\times$10$^{-01}$  \\
0.600 &  1.84$\times$10$^{+01}$  &  2.39$\times$10$^{+01}$  &
       3.12$\times$10$^{+01}$  &  3.175$\times$10$^{+00}$  &
       2.66$\times$10$^{-01}$  &  3.86$\times$10$^{-01}$  \\
0.700 &  2.46$\times$10$^{+01}$  &  3.12$\times$10$^{+01}$  &
       3.99$\times$10$^{+01}$  &  3.445$\times$10$^{+00}$  &
       2.43$\times$10$^{-01}$  &  7.54$\times$10$^{-01}$  \\
0.800 &  3.36$\times$10$^{+01}$  &  4.14$\times$10$^{+01}$  &
       5.13$\times$10$^{+01}$  &  3.727$\times$10$^{+00}$  &
       2.14$\times$10$^{-01}$  &  1.30$\times$10$^{+00}$  \\
0.900 &  4.76$\times$10$^{+01}$  &  5.71$\times$10$^{+01}$  &
       6.94$\times$10$^{+01}$  &  4.050$\times$10$^{+00}$  &
       1.91$\times$10$^{-01}$  &  3.36$\times$10$^{+00}$  \\
1.000 &  6.92$\times$10$^{+01}$  &  8.17$\times$10$^{+01}$  &
       9.82$\times$10$^{+01}$  &  4.412$\times$10$^{+00}$  &
       1.79$\times$10$^{-01}$  &  1.13$\times$10$^{+01}$  \\
1.250 &  1.72$\times$10$^{+02}$  &  2.00$\times$10$^{+02}$  &
       2.38$\times$10$^{+02}$  &  5.312$\times$10$^{+00}$  &
       1.73$\times$10$^{-01}$  &  3.37$\times$10$^{+01}$  \\
1.500 &  3.63$\times$10$^{+02}$  &  4.21$\times$10$^{+02}$  &
       4.98$\times$10$^{+02}$  &  6.053$\times$10$^{+00}$  &
       1.68$\times$10$^{-01}$  &  2.35$\times$10$^{+01}$  \\
1.750 &  6.40$\times$10$^{+02}$  &  7.43$\times$10$^{+02}$  &
       8.73$\times$10$^{+02}$  &  6.619$\times$10$^{+00}$  &
       1.63$\times$10$^{-01}$  &  1.33$\times$10$^{+01}$  \\
2.000 &  9.92$\times$10$^{+02}$  &  1.15$\times$10$^{+03}$  &
       1.34$\times$10$^{+03}$  &  7.053$\times$10$^{+00}$  &
       1.58$\times$10$^{-01}$  &  8.27$\times$10$^{+00}$  \\
2.500 &  1.84$\times$10$^{+03}$  &  2.13$\times$10$^{+03}$  &
       2.49$\times$10$^{+03}$  &  7.668$\times$10$^{+00}$  &
       1.52$\times$10$^{-01}$  &  4.13$\times$10$^{+00}$  \\
3.000 &  2.80$\times$10$^{+03}$  &  3.23$\times$10$^{+03}$  &
       3.77$\times$10$^{+03}$  &  8.086$\times$10$^{+00}$  &
       1.53$\times$10$^{-01}$  &  7.12$\times$10$^{+00}$  \\
3.500 &  3.78$\times$10$^{+03}$  &  4.37$\times$10$^{+03}$  &
       5.13$\times$10$^{+03}$  &  8.389$\times$10$^{+00}$  &
       1.55$\times$10$^{-01}$  &  1.05$\times$10$^{+01}$  \\
4.000 &  4.76$\times$10$^{+03}$  &  5.51$\times$10$^{+03}$  &
       6.51$\times$10$^{+03}$  &  8.625$\times$10$^{+00}$  &
       1.63$\times$10$^{-01}$  &  2.03$\times$10$^{+01}$  \\
5.000 &  6.60$\times$10$^{+03}$  &  7.73$\times$10$^{+03}$  &
       9.35$\times$10$^{+03}$  &  8.968$\times$10$^{+00}$  &
       1.80$\times$10$^{-01}$  &  2.77$\times$10$^{+01}$  \\
6.000 & (8.73$\times$10$^{+03}$) & (1.06$\times$10$^{+04}$) &
      (1.28$\times$10$^{+04}$) & (9.265$\times$10$^{+00}$) &
      (1.90$\times$10$^{-01}$) &  \\
7.000 & (1.17$\times$10$^{+04}$) & (1.41$\times$10$^{+04}$) &
      (1.71$\times$10$^{+04}$) & (9.555$\times$10$^{+00}$) &
      (1.90$\times$10$^{-01}$) &  \\
8.000 & (1.45$\times$10$^{+04}$) & (1.76$\times$10$^{+04}$) &
      (2.13$\times$10$^{+04}$) & (9.774$\times$10$^{+00}$) &
      (1.90$\times$10$^{-01}$) &  \\
9.000 & (1.72$\times$10$^{+04}$) & (2.08$\times$10$^{+04}$) &
      (2.51$\times$10$^{+04}$) & (9.942$\times$10$^{+00}$) &
      (1.90$\times$10$^{-01}$) &  \\
10.000 & (2.03$\times$10$^{+04}$) & (2.46$\times$10$^{+04}$) &
      (2.97$\times$10$^{+04}$) & (1.011$\times$10$^{+01}$) &
      (1.90$\times$10$^{-01}$) &  \\

\end{longtable}

\subsection{$^{18}$O($p,\gamma)^{19}$F}\label{mattsec}

Comments: This rate was adopted from \citet{bu12}, which updated the upper and lower limits on the E$^{\mathrm{lab}}_{\mathrm{R}}=95$ keV resonance strength.  In this work, a new upper limit of $\omega\gamma \leq 7.8 \times 10^{-9}$ eV (90$\%$ CL) was determined.  A lognormal probability density function was constructed from the improved strength constraints, and the expectation value and standard deviation of this function were input into the reaction rate calculation.  A new Q-value ($7993.5994 \pm 0.0011$ keV) was also adopted \citep{ame2012}.  Proton energies were calculated from the excitation energies presented in \citet{tilley} for all resonances except the E$^{\mathrm{lab}}_{\mathrm{R}}=150.8 \pm 0.1$ keV resonance measured directly by \citet{becker}.  The direct capture expansion was adopted from \citet{bu12}, and it was determined using the code $\mathtt{TEDCA}$ \citep{tedca} and parameters presented in \citet{ilw}.   In \citet{bu12}, a zero-scattering potential was used based on the argument presented in \citet{ilw}.  The S-factor fit agreed with the $\mathtt{TEDCA}$ calculations up to an energy of E$^{\mathrm{cm}}_{\mathrm{cutoff}}=2500$ keV.  For temperatures $T > 5$ GK, the rate was extrapolated with the $\mathtt{BRUSLIB}$ Hauser-Feshbach rates \citep{Goriely08a}.



\footnotesize
\begin{verbatim}
18O(p,g)19F
****************************************************************************************************************
1               ! Zproj                                                             
8               ! Ztarget                                                           
2               ! Zexitparticle (=0 when only 2 channels open)                      
1.0078          ! Aproj                                                             
17.999          ! Atarget                                                           
4.0026          ! Aexitparticle (=0 when only 2 channels open)                      
0.5             ! Jproj                                                             
0.0             ! Jtarget                                                           
0.0             ! Jexitparticle (=0 when only 2 channels open)                      
7993.60         ! projectile separation energy (keV)   					                             
4013.80         ! exit particle separation energy (=0 when only 2 channels open)    	
1.25            ! Radius parameter R0 (fm)  						                                        
2               ! Gamma-ray channel number (=2 if ejectile is a g-ray; =3 otherwise)
****************************************************************************************************************
1.0             ! Minimum energy for numerical integration (keV)
10000           ! Number of random samples (>5000 for better statistics)
0               ! =0 for rate output at all temperatures; =NT for rate output at selected temperatures
****************************************************************************************************************
Non-Resonant Contribution
S(keVb)         S'(b)           S''(b/keV)      fracErr         Cutoff Energy (keV)
7.06e0          2.98e-3         -5.20e-7        0.5             2500.0                 
0.0             0.0             0.0             0.0             0.0
****************************************************************************************************************
Resonant Contribution
Note: G1 = entrance channel, G2 = exit channel, G3 = spectator channel !! Ecm, Exf in (keV); wg, Gx in (eV) !!
Note: if Er<0, theta^2=C2S*theta_sp^2 must be entered instead of entrance channel partial width
Ecm     DEcm  wg      Dwg     Jr     G1      DG1     L1  G2    DG2   L2  G3      DG3     L3  Exf   Int
  20.4  0.7   0       0       2.5    2.3e-19 0.5e-19 2   2.3   1.0   1   2.5e3   1.0e3   3   0.0   1 
  90.4  3.0   5.3e-8  9.0e-6  0      0       0       0   0     0     0   0       0       0   0.0   0 
 142.8  0.1   0       0       0.5    1.67e-1 0.12e-1 0   0.72  0.15  1   1.23e2  0.24e2  1   0.0   1 
 205.4  1.0   5.0e-6  1.0e-6  0      0       0       0   0     0     0   0       0       0   0.0   0 
 260.7  2.6   3.7e-5  0.5e-5  0      0       0       0   0     0     0   0       0       0   0.0   0 
 316.4  1.3   0       0       2.5    1.9e-2  0.3e-2  2   0.78  0.34  1   47.0    19.0    3   0.0   1 
 589.9  1.7   1.0e-2  0.2e-2  0      0       0       0   0     0     0   0       0       0   0.0   0 
 598.3  1.2   0       0       1.5    1.4e2   0.7e2   1   0.71  0.39  1   2.0e3   0.1e3   2   0.0   1 
 799.6  1.6   0       0       0.5    24.6e3  1.4e3   0   2.5   0.4   1   20.e3   1.0e3   1   0.0   1 
 933.1  2.8   0       0       1.5    76.0    7.0     1   0.34  0.06  1   3.5e3   0.3e3   2   0.0   1 
1106.1  4.0   0.29    0.03    0      0       0       0   0     0     0   0       0       0   0.0   0 
1173.4  1.5   0       0       0.5    0.38e3  0.03e3  0   1.4   1.0   1   5.4e3   0.38e3  1   0.0   1 
1324.4  2.1   0.08    0.01    0      0       0       0   0     0     0   0       0       0   0.0   0 
1327.4  1.2   0       0       0.5    0.22e3  0.02e3  0   3.4   1.7   1   4.7e3   0.4e3   1   0.0   1 
1542.8  2.1   0.025   0.005   0      0       0       0   0     0     0   0       0       0   0.0   0 
1572.4  3.0   0.041   0.010   0      0       0       0   0     0     0   0       0       0   0.0   0
1581.4  4.0   0.06    0.01    0      0       0       0   0     0     0   0       0       0   0.0   0 
1592.4  3.0   0.025   0.004   0      0       0       0   0     0     0   0       0       0   0.0   0 
1673.9  1.6   0       0       1.5    2.0e3   0.6e3   2   1.0   0.4   1   1.4e3   0.4e3   1   0.0   1 
1826.4  1.2   2.8     0.7     0      0       0       0   0     0     0   0       0       0   0.0   0 
1880.4  1.9   0.13    0.04    0      0       0       0   0     0     0   0       0       0   0.0   0 
1893.4  3.0   0       0       0.5    11.e3   3.0e3   0   0.36  0.20  1   18.0e3  5.4e3   1   0.0   1 
****************************************************************************************************************
Upper Limits of Resonances
Note: enter partial width upper limit by chosing non-zero value for PT, where PT=<theta^2> for particles and...
Note: ...PT=<B> for g-rays [enter: "upper_limit 0.0"]; for each resonance: # upper limits < # open channels!  
Ecm     DEcm   Jr   G1      DG1     L1   PT   G2    DG2    L2   PT   G3     DG3   L3  PT     Exf   Int
! 0.0   0.0    0.0  0.0     0.0     0    0    0.0   0.0    0    0    0.0    0.0   0   0      0.0   0
****************************************************************************************************************
Interference between Resonances [numerical integration only]
Note: + for positive, - for negative interference; +- if interference sign is unknown
Ecm	DEcm	Jr    G1    DG1   L1   PT    G2    DG2    L2  PT   G3    DG3    L3  PT   Exf  
!+- 
0.0     0.0     0.0   0.0   0.0   0    0     0.0   0.0    0   0    0.0   0.0    0   0    0.0  
0.0     0.0     0.0   0.0   0.0   0    0     0.0   0.0    0   0    0.0   0.0    0   0    0.0  
****************************************************************************************************************
Reaction Rate and PDF at NT selected temperatures only
Note: default values are used for reaction rate range if Min=Max=0.0
T9	Min	Max
0.01	0.0	0.0
0.1	0.0	0.0
****************************************************************************************************************
Comments: 
1. Almost the same input data as in Angulo et al. 1999 (NACRE) are adopted here; however, we obtain the resonant
   rate contributions by numerical integration, whenever possible.
2. Proton width for Er=20.4 keV from Champagne and Pitt 1986 and La Cognata et al. 2008, while total and 
   radiative widths are from a private communication quoted in Wiescher et al. 1980. 
3. Er=90.4 keV wg expectation value and variance from Buckner et al. 2012.
4. Direct capture S-factor adopted from Buckner et al. 2012.
5. Above T=5 GK the rate is extrapolated using the MOST Hauser-Feshbach rate.






\end{verbatim}
\normalsize
\vspace{5mm}

\begin{figure}[ht]
\centering
\includegraphics[scale=0.5]{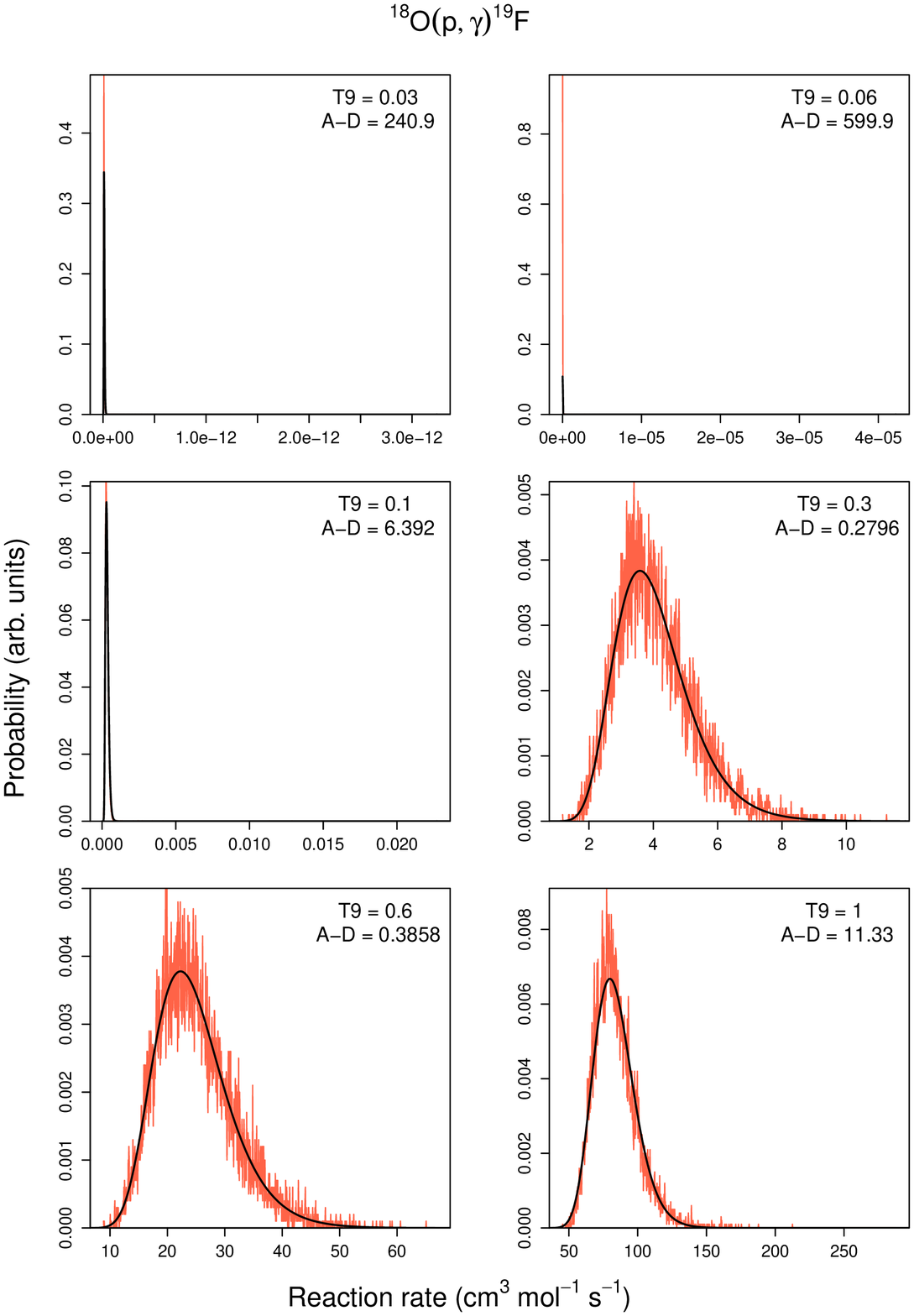}
\end{figure}

\begin{figure}[ht]
\centering
\includegraphics[scale=0.5]{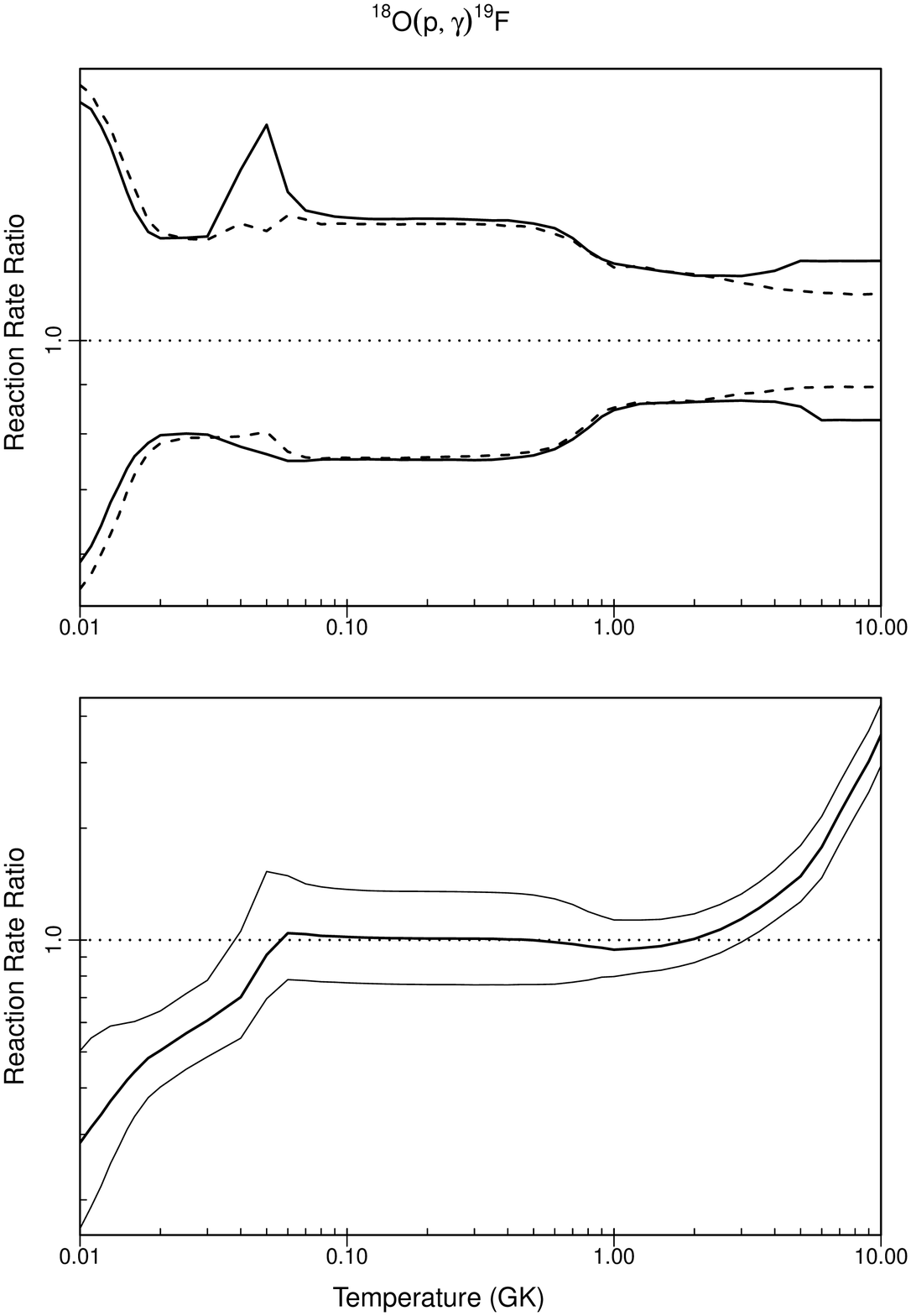}
\end{figure}
\clearpage

\clearpage
\setlongtables
\begin{longtable}{cccc | ccc}
\caption{Total thermonuclear reaction rates for $^{18}$O(p,$\alpha$)$^{15}$N.}  \label{tab:o18pa} \\
\hline \hline 
	\multicolumn{1}{c}{T (GK)} & \multicolumn{1}{c}{Low rate} & \multicolumn{1}{c}{Median rate} & \multicolumn{1}{c}{High rate}  & \multicolumn{1}{c}{lognormal $\mu$} & \multicolumn{1}{c}{lognormal $\sigma$} & \multicolumn{1}{c}{A-D} \\ \hline 
\endfirsthead
\multicolumn{6}{c}{{\tablename} \thetable{} -- continued} \\
\hline \hline 
	\multicolumn{1}{c}{T (GK)} & \multicolumn{1}{c}{Low rate} & \multicolumn{1}{c}{Median rate} & \multicolumn{1}{c}{High rate} & \multicolumn{1}{c}{lognormal $\mu$} & \multicolumn{1}{c}{lognormal $\sigma$} & \multicolumn{1}{c}{A-D} \\ \hline 
\endhead
	 \hline \hline
\endfoot
	\hline \hline
\endlastfoot
0.010 & 7.18$\times$10$^{-21}$ & 8.66$\times$10$^{-21}$ &
      1.04$\times$10$^{-20}$ &  -4.620$\times$10$^{+01}$ &
       1.87$\times$10$^{-01}$  & 3.07$\times$10$^{-01}$ \\ 
0.011 & 6.56$\times$10$^{-20}$ & 8.00$\times$10$^{-20}$ &
      9.76$\times$10$^{-20}$ &  -4.397$\times$10$^{+01}$ &
       1.97$\times$10$^{-01}$  & 3.15$\times$10$^{-01}$ \\ 
0.012 & 4.66$\times$10$^{-19}$ & 5.71$\times$10$^{-19}$ &
      6.97$\times$10$^{-19}$ &  -4.201$\times$10$^{+01}$ &
       2.03$\times$10$^{-01}$  & 5.47$\times$10$^{-01}$ \\ 
0.013 & 2.66$\times$10$^{-18}$ & 3.32$\times$10$^{-18}$ &
      4.10$\times$10$^{-18}$ &  -4.025$\times$10$^{+01}$ &
       2.16$\times$10$^{-01}$  & 1.80$\times$10$^{-01}$ \\ 
0.014 & 1.31$\times$10$^{-17}$ & 1.63$\times$10$^{-17}$ &
      2.04$\times$10$^{-17}$ &  -3.865$\times$10$^{+01}$ &
       2.26$\times$10$^{-01}$  & 1.09$\times$10$^{+00}$ \\ 
0.015 & 5.60$\times$10$^{-17}$ & 6.98$\times$10$^{-17}$ &
      8.89$\times$10$^{-17}$ &  -3.719$\times$10$^{+01}$ &
       2.38$\times$10$^{-01}$  & 2.59$\times$10$^{+00}$ \\ 
0.016 & 2.13$\times$10$^{-16}$ & 2.68$\times$10$^{-16}$ &
      3.44$\times$10$^{-16}$ &  -3.585$\times$10$^{+01}$ &
       2.44$\times$10$^{-01}$  & 1.76$\times$10$^{+00}$ \\ 
0.018 & 2.37$\times$10$^{-15}$ & 3.08$\times$10$^{-15}$ &
      3.95$\times$10$^{-15}$ &  -3.342$\times$10$^{+01}$ &
       2.55$\times$10$^{-01}$  & 4.88$\times$10$^{-01}$ \\ 
0.020 & 1.96$\times$10$^{-14}$ & 2.52$\times$10$^{-14}$ &
      3.34$\times$10$^{-14}$ &  -3.130$\times$10$^{+01}$ &
       2.70$\times$10$^{-01}$  & 1.51$\times$10$^{+00}$ \\ 
0.025 & 1.39$\times$10$^{-12}$ & 1.83$\times$10$^{-12}$ &
      2.41$\times$10$^{-12}$ &  -2.702$\times$10$^{+01}$ &
       2.76$\times$10$^{-01}$  & 9.46$\times$10$^{-01}$ \\ 
0.030 & 3.69$\times$10$^{-11}$ & 4.84$\times$10$^{-11}$ &
      6.40$\times$10$^{-11}$ &  -2.375$\times$10$^{+01}$ &
       2.76$\times$10$^{-01}$  & 8.57$\times$10$^{-01}$ \\ 
0.040 & 4.51$\times$10$^{-09}$ & 5.86$\times$10$^{-09}$ &
      7.67$\times$10$^{-09}$ &  -1.895$\times$10$^{+01}$ &
       2.67$\times$10$^{-01}$  & 9.51$\times$10$^{-01}$ \\ 
0.050 & 1.52$\times$10$^{-07}$ & 1.92$\times$10$^{-07}$ &
      2.47$\times$10$^{-07}$ &  -1.546$\times$10$^{+01}$ &
       2.47$\times$10$^{-01}$  & 1.03$\times$10$^{+00}$ \\ 
0.060 & 3.91$\times$10$^{-06}$ & 4.48$\times$10$^{-06}$ &
      5.30$\times$10$^{-06}$ &  -1.230$\times$10$^{+01}$ &
       1.54$\times$10$^{-01}$  & 1.16$\times$10$^{+01}$ \\ 
0.070 & 9.27$\times$10$^{-05}$ & 1.00$\times$10$^{-04}$ &
      1.08$\times$10$^{-04}$ &  -9.208$\times$10$^{+00}$ &
       8.06$\times$10$^{-02}$  & 2.33$\times$10$^{+00}$ \\ 
0.080 & 1.25$\times$10$^{-03}$ & 1.34$\times$10$^{-03}$ &
      1.44$\times$10$^{-03}$ &  -6.612$\times$10$^{+00}$ &
       7.07$\times$10$^{-02}$  & 5.24$\times$10$^{-01}$ \\ 
0.090 & 1.00$\times$10$^{-02}$ & 1.07$\times$10$^{-02}$ &
      1.15$\times$10$^{-02}$ &  -4.535$\times$10$^{+00}$ &
       7.01$\times$10$^{-02}$  & 2.53$\times$10$^{-01}$ \\ 
0.100 & 5.29$\times$10$^{-02}$ & 5.67$\times$10$^{-02}$ &
      6.09$\times$10$^{-02}$ &  -2.870$\times$10$^{+00}$ &
       7.09$\times$10$^{-02}$  & 2.24$\times$10$^{-01}$ \\ 
0.110 & 2.05$\times$10$^{-01}$ & 2.20$\times$10$^{-01}$ &
      2.36$\times$10$^{-01}$ &  -1.515$\times$10$^{+00}$ &
       7.11$\times$10$^{-02}$  & 3.97$\times$10$^{-01}$ \\ 
0.120 & 6.29$\times$10$^{-01}$ & 6.73$\times$10$^{-01}$ &
      7.24$\times$10$^{-01}$ &  -3.944$\times$10$^{-01}$ &
       7.12$\times$10$^{-02}$  & 4.06$\times$10$^{-01}$ \\ 
0.130 & 1.61$\times$10$^{+00}$ & 1.73$\times$10$^{+00}$ &
      1.86$\times$10$^{+00}$ &  5.462$\times$10$^{-01}$ &
       7.13$\times$10$^{-02}$  & 2.71$\times$10$^{-01}$ \\ 
0.140 & 3.58$\times$10$^{+00}$ & 3.84$\times$10$^{+00}$ &
      4.13$\times$10$^{+00}$ &  1.345$\times$10$^{+00}$ &
       7.11$\times$10$^{-02}$  & 3.75$\times$10$^{-01}$ \\ 
0.150 & 7.11$\times$10$^{+00}$ & 7.61$\times$10$^{+00}$ &
      8.19$\times$10$^{+00}$ &  2.031$\times$10$^{+00}$ &
       7.12$\times$10$^{-02}$  & 3.61$\times$10$^{-01}$ \\ 
0.160 & 1.29$\times$10$^{+01}$ & 1.38$\times$10$^{+01}$ &
      1.48$\times$10$^{+01}$ &  2.626$\times$10$^{+00}$ &
       7.12$\times$10$^{-02}$  & 3.26$\times$10$^{-01}$ \\ 
0.180 & 3.43$\times$10$^{+01}$ & 3.67$\times$10$^{+01}$ &
      3.95$\times$10$^{+01}$ &  3.605$\times$10$^{+00}$ &
       7.10$\times$10$^{-02}$  & 4.34$\times$10$^{-01}$ \\ 
0.200 & 7.42$\times$10$^{+01}$ & 7.94$\times$10$^{+01}$ &
      8.53$\times$10$^{+01}$ &  4.375$\times$10$^{+00}$ &
       7.05$\times$10$^{-02}$  & 3.73$\times$10$^{-01}$ \\ 
0.250 & 2.88$\times$10$^{+02}$ & 3.08$\times$10$^{+02}$ &
      3.31$\times$10$^{+02}$ &  5.732$\times$10$^{+00}$ &
       6.99$\times$10$^{-02}$  & 5.87$\times$10$^{-01}$ \\ 
0.300 & 7.09$\times$10$^{+02}$ & 7.62$\times$10$^{+02}$ &
      8.19$\times$10$^{+02}$ &  6.636$\times$10$^{+00}$ &
       7.24$\times$10$^{-02}$  & 2.06$\times$10$^{-01}$ \\ 
0.350 & 1.41$\times$10$^{+03}$ & 1.53$\times$10$^{+03}$ &
      1.67$\times$10$^{+03}$ &  7.336$\times$10$^{+00}$ &
       8.48$\times$10$^{-02}$  & 1.05$\times$10$^{+00}$ \\ 
0.400 & 2.58$\times$10$^{+03}$ & 2.86$\times$10$^{+03}$ &
      3.23$\times$10$^{+03}$ &  7.968$\times$10$^{+00}$ &
       1.13$\times$10$^{-01}$  & 7.92$\times$10$^{+00}$ \\ 
0.450 & 4.73$\times$10$^{+03}$ & 5.47$\times$10$^{+03}$ &
      6.36$\times$10$^{+03}$ &  8.612$\times$10$^{+00}$ &
       1.48$\times$10$^{-01}$  & 4.23$\times$10$^{+00}$ \\ 
0.500 & 8.94$\times$10$^{+03}$ & 1.06$\times$10$^{+04}$ &
      1.27$\times$10$^{+04}$ &  9.272$\times$10$^{+00}$ &
       1.77$\times$10$^{-01}$  & 4.04$\times$10$^{+00}$ \\ 
0.600 & 3.08$\times$10$^{+04}$ & 3.73$\times$10$^{+04}$ &
      4.61$\times$10$^{+04}$ &  1.054$\times$10$^{+01}$ &
       2.06$\times$10$^{-01}$  & 1.94$\times$10$^{+00}$ \\ 
0.700 & 9.01$\times$10$^{+04}$ & 1.11$\times$10$^{+05}$ &
      1.37$\times$10$^{+05}$ &  1.162$\times$10$^{+01}$ &
       2.14$\times$10$^{-01}$  & 1.29$\times$10$^{+00}$ \\ 
0.800 & 2.17$\times$10$^{+05}$ & 2.71$\times$10$^{+05}$ &
      3.36$\times$10$^{+05}$ &  1.251$\times$10$^{+01}$ &
       2.17$\times$10$^{-01}$  & 5.57$\times$10$^{-01}$ \\ 
0.900 & 4.53$\times$10$^{+05}$ & 5.56$\times$10$^{+05}$ &
      6.91$\times$10$^{+05}$ &  1.323$\times$10$^{+01}$ &
       2.10$\times$10$^{-01}$  & 1.95$\times$10$^{+00}$ \\ 
1.000 & 8.27$\times$10$^{+05}$ & 1.02$\times$10$^{+06}$ &
      1.25$\times$10$^{+06}$ &  1.383$\times$10$^{+01}$ &
       2.04$\times$10$^{-01}$  & 7.73$\times$10$^{-01}$ \\ 
1.250 & 2.56$\times$10$^{+06}$ & 3.08$\times$10$^{+06}$ &
      3.71$\times$10$^{+06}$ &  1.494$\times$10$^{+01}$ &
       1.89$\times$10$^{-01}$  & 2.33$\times$10$^{+00}$ \\ 
1.500 & 5.49$\times$10$^{+06}$ & 6.50$\times$10$^{+06}$ &
      7.81$\times$10$^{+06}$ &  1.569$\times$10$^{+01}$ &
       1.77$\times$10$^{-01}$  & 2.42$\times$10$^{+00}$ \\ 
1.750 & 9.44$\times$10$^{+06}$ & 1.11$\times$10$^{+07}$ &
      1.31$\times$10$^{+07}$ &  1.623$\times$10$^{+01}$ &
       1.65$\times$10$^{-01}$  & 2.24$\times$10$^{+00}$ \\ 
2.000 & 1.40$\times$10$^{+07}$ & 1.63$\times$10$^{+07}$ &
      1.92$\times$10$^{+07}$ &  1.661$\times$10$^{+01}$ &
       1.61$\times$10$^{-01}$  & 1.33$\times$10$^{+00}$ \\ 
2.500 & 2.38$\times$10$^{+07}$ & 2.75$\times$10$^{+07}$ &
      3.23$\times$10$^{+07}$ &  1.714$\times$10$^{+01}$ &
       1.55$\times$10$^{-01}$  & 3.67$\times$10$^{+00}$ \\ 
3.000 & 3.30$\times$10$^{+07}$ & 3.79$\times$10$^{+07}$ &
      4.42$\times$10$^{+07}$ &  1.746$\times$10$^{+01}$ &
       1.48$\times$10$^{-01}$  & 2.20$\times$10$^{+00}$ \\ 
3.500 & 4.09$\times$10$^{+07}$ & 4.70$\times$10$^{+07}$ &
      5.44$\times$10$^{+07}$ &  1.767$\times$10$^{+01}$ &
       1.45$\times$10$^{-01}$  & 1.46$\times$10$^{+00}$ \\ 
4.000 & 4.72$\times$10$^{+07}$ & 5.43$\times$10$^{+07}$ &
      6.28$\times$10$^{+07}$ &  1.781$\times$10$^{+01}$ &
       1.44$\times$10$^{-01}$  & 9.12$\times$10$^{-01}$ \\ 
5.000 & 5.68$\times$10$^{+07}$ & 6.49$\times$10$^{+07}$ &
      7.47$\times$10$^{+07}$ &  1.799$\times$10$^{+01}$ &
       1.37$\times$10$^{-01}$  & 9.16$\times$10$^{-01}$ \\ 
6.000 & 6.40$\times$10$^{+07}$ & 7.26$\times$10$^{+07}$ &
      8.30$\times$10$^{+07}$ &  1.810$\times$10$^{+01}$ &
       1.32$\times$10$^{-01}$  & 1.34$\times$10$^{+00}$ \\ 
7.000 & 6.98$\times$10$^{+07}$ & 7.86$\times$10$^{+07}$ &
      8.97$\times$10$^{+07}$ &  1.819$\times$10$^{+01}$ &
       1.25$\times$10$^{-01}$  & 3.57$\times$10$^{+00}$ \\ 
8.000 & 7.62$\times$10$^{+07}$ & 8.52$\times$10$^{+07}$ &
      9.56$\times$10$^{+07}$ &  1.826$\times$10$^{+01}$ &
       1.15$\times$10$^{-01}$  & 2.65$\times$10$^{+00}$ \\ 
9.000 & 8.29$\times$10$^{+07}$ & 9.17$\times$10$^{+07}$ &
      1.02$\times$10$^{+08}$ &  1.834$\times$10$^{+01}$ &
       1.06$\times$10$^{-01}$  & 3.06$\times$10$^{+00}$ \\ 
10.000 & 8.98$\times$10$^{+07}$ & 9.89$\times$10$^{+07}$ &
      1.09$\times$10$^{+08}$ &  1.841$\times$10$^{+01}$ &
       9.83$\times$10$^{-02}$  & 2.30$\times$10$^{+00}$ \\

\end{longtable}

\subsection{$^{18}$O($p,\alpha)^{15}$N}

Comments:  In the previous input file for $^{18}$O$(p,\alpha)^{15}$N \citep[see][]{iliadis_3}, only two interfering resonances were taken into account, and the interference sign was assumed to be unknown.  Here, we take interference between three $1/2^+$ states into account at $E_R = 143, 609$, and 807 keV.  The interference sign between the 609 and 807 keV resonances is known to be (-) (i.e., destructive outside the resonances and constructive between).  This was determined by calculating the cross section for each sign and comparing with the data from \citet{LW}.  Energies and partial widths were adopted from the weighted average of \citet{yagi,mak,lacogb}, using the method discussed in \citet{neutron} to expand uncertainties by adding an unknown systematic uncertainty in quadrature.  For the 143 and 609 keV resonances, the (-) interference sign is known according to the yield curve shown in \cite{becker,LW} \citep[strength from][]{becker}.   Because the {\tt RatesMC} code can only include interference from two resonances, we ran the code twice with two input files.  Input file \#1:  the 143 and 609 keV resonances interfering, with the 807 resonance in the resonant contribution section (shown in the input file below).  Input file \#2: the 609 and 807 keV resonances interfering, with the 143 keV resonance in the resonant contribution section.  Between temperatures of 0.07 and 0.16 GK, the rates deviate from one another by less than a percent.  Therefore, the rate tabulated here is from input file \#1 for $T\le 0.06$ GK, while above this threshold, the rate is from input file \#2.  Only one input file is shown for clarity (input file \#1).  To construct input file \#2, move the "!" comment from the 143 keV resonance to the 807 keV resonance in the resonant contribution section and from the 807 keV resonance to the 143 keV resonance in the interference section.

Additional changes from the previous input file include moving the 90-keV resonance from the upper limit section to the resonant contribution section with a strength of $1.6\pm0.5$ eV directly measured by \citet{LW}.  In addition, resonance energies less than 2 MeV were adopted from the $^{18}$O$(p,\gamma)^{19}$F reaction input file (see \ref{mattsec}).  \citet{sellin,orihara,almanza,murillo} provided values for the higher resonance energies and partial widths, and Q-values and masses are from \citet{ame2012}.  Our adopted strengths for the 20 and 90 keV resonances agree with \citet{lacoga}.

\footnotesize
\begin{verbatim}

18O(p,a)15N
****************************************************************************************************************
1               ! Zproj                                                             
8               ! Ztarget                                                           
2               ! Zexitparticle (=0 when only 2 channels open)                      
1.0078          ! Aproj                                                             
17.999          ! Atarget                                                           
4.0026          ! Aexitparticle (=0 when only 2 channels open)                      
0.5             ! Jproj                                                             
0.0             ! Jtarget                                                           
0.0             ! Jexitparticle (=0 when only 2 channels open)                      
7993.6          ! projectile separation energy (keV)                                
4013.8          ! exit particle separation energy (=0 when only 2 channels open)    
1.25            ! Radius parameter R0 (fm)                                          
3               ! Gamma-ray channel number (=2 if ejectile is a g-ray; =3 otherwise)
****************************************************************************************************************
1.0             ! Minimum energy for numerical integration (keV)
5000            ! Number of random samples (>5000 for better statistics)
0               ! =0 for rate output at all temperatures; =NT for rate output at selected temperatures
****************************************************************************************************************
Non-Resonant Contribution
S(keVb)        S'(b)        S''(b/keV)    fracErr        Cutoff Energy (keV)
 0.0           0.0          0.0           0.0            0.0
 0.0           0.0          0.0           0.0            0.0
****************************************************************************************************************
Resonant Contribution
Note: G1 = entrance channel, G2 = exit channel, G3 = spectator channel !! Ecm, Exf in (keV); wg, Gx in (eV) !!
Note: if Er<0, theta^2=C2S*theta_sp^2 must be entered instead of entrance channel partial width
Ecm       DEcm    wg    Dwg   Jr     G1       DG1      L1   G2        DG2      L2   G3     DG3   L3   Exf    Int
   20.4   0.7     0     0     2.5    2.3e-19  0.5e-19  2    2.5e3     1.0e3    3    2.3    1.0   1    0.0    1 
   90.4   3.0  1.6e-7 0.5e-7  0      0        0        0    0         0        0    0      0     0    0.0    0 
! 142.8   0.1     0     0     0.5    1.67e-1  0.12e-1  0    1.23e2    0.24e2   1    0.72   0.15  1    0.0    1   
  315.2   1.3     0     0     2.5    1.9e-2   0.3e-2   2    4.7e1     1.9e1    3    0.78   0.34  1    0.0    1 
  597.1   1.2     0     0     1.5    1.4e2    0.7e2    1    2.0e3     0.1e3    2    0.71   0.39  1    0.0    1 
  807.0   3.0     0     0     0.5    2.46e4   0.14e4   0    2.2e4     4.0e3    1    2.5    0.4   1    0.0    1  
  931.9   2.8     0     0     1.5    7.6e1    0.7e1    1    3.5e3     0.3e3    2    0.34   0.06  1    0.0    1  
 1106.2   4.0     0     0     3.5    4.7      0.6      4    5.6e2     0.76e2   3    0      0     0    0.0    1  
 1172.2   1.5     0     0     0.5    3.8e2    0.3e2    0    5.4e3     0.38e3   1    1.4    1.0   1    0.0    1  
 1326.2   1.2     0     0     0.5    2.2e2    0.2e2    0    4.7e3     0.4e3    1    3.4    1.7   1    0.0    1 
 1672.7   1.6     0     0     1.5    2.0e3    0.6e3    2    1.4e3     0.4e3    1    1.0    0.4   1    0.0    1 
 1825.2   1.2     0     0     2.5    9.0e1    3.0e1    3    7.0e1     2.0e1    2    0      0     0    0.0    1 
 1892.0   3.0     0     0     0.5    1.1e4    0.3e4    0    1.8e4     0.54e4   1    0.36   0.20  1    0.0    1 
 2167.0   3.0     0     0     0.5    2.2e3    0.7e3    0    0.9e3     0.3e3    1    0      0     0    0.0    1 
 2237.0   3.0     0     0     0.5    2.7e3    0.8e3    0    1.6e3     0.5e3    1    0      0     0    0.0    1 
 2259.0   3.0     0     0     0.5    1.0e4    3.0e3    0    12.0e3    4.0e3    1    0      0     0    0.0    1 
 2313.0   4.0     0     0     1.5    4.9e3    1.5e3    2    4.3e3     1.3e3    1    0      0     0    0.0    1 
 2501.5   1.4     0     0     1.5    2.3e3    0.7e3    2    0.95e3    0.3e3    1    0      0     0    0.0    1 
 2619.5   1.7     0     0     2.5    0.66e3   0.20e3   2    1.0e3     0.3e3    3    0      0     0    0.0    1 
 2768.5   2.6     0     0     0.5    4.3e3    1.3e3    1    1.1e3     0.3e3    0    0      0     0    0.0    1 
 2864.9   2.0     0     0     2.5    12.3e3   3.7e3    2    5.4e3     1.6e3    3    0      0     0    0.0    1 
 2980.2   2.6     0     0     1.5    4.7e3    1.4e3    2    4.3e3     1.3e3    1    0      0     0    0.0    1 
 3291.0   7.0     0     0     2.5    4.07e3   0.95e3   2    7.7e3     4.8e3    3    0      0     0    0.0    1 
 3355.0   25.0    0     0     0.5    228.3e3  1.9e3    0    43.0e3    31.0e3   1    0      0     0    0.0    1 
 3455.0   3.5     0     0     0.5    16.1e3   2.8e3    1    22.0e3    7.0e3    0    0      0     0    0.0    1 
 3507.0   5.0     0     0     1.5    11.4e3   1.9e3    1    16.0e3    6.0e3    2    0      0     0    0.0    1 
 3545.0   7.0     0     0     2.5    3.5e3    1.0e3    2    18.3e3    4.8e3    3    0      0     0    0.0    1 
 3608.0   12.0    0     0     1.5    26.0e3   8.0e3    1    43.0e3    16.0e3   2    0      0     0    0.0    1 
 3658.0   4.0     0     0     1.5    11.2e3   1.8e3    2    19.0e3    8.0e3    1    0      0     0    0.0    1 
 4045.0   20.0    0     0     0.5    70.0e3   60.0e3   1    64.0e3    16.0e3   0    0      0     0    0.0    1 
 4141.0   8.0     0     0     1.5    61.0e3   15.0e3   1    51.0e3    9.0e3    2    0      0     0    0.0    1 
 4227.0   12.     0     0     1.5    39.0e3   10.0e3   2    36.0e3    9.0e3    1    0      0     0    0.0    1 
 4527.0   7.0     0     0     0.5    2.6e3    0.9e3    1    13.4e3    4.4e3    0    0      0     0    0.0    1 
 4582.0   10.0    0     0     2.5    4.3e3    1.6e3    2    44.4e3    7.8e3    3    0      0     0    0.0    1 
 4585.0   25.0    0     0     0.5    112.0e3  28.0e3   1    226.0e3   33.0e3   0    0      0     0    0.0    1 
 4785.0   10.0    0     0     2.5    12.3e3   6.2e3    2    82.0e3    33.0e3   3    0      0     0    0.0    1 
 4865.0   30.0    0     0     1.5    118.0e3  25.0e3   2    161.0e3   24.0e3   1    0      0     0    0.0    1 
 4945.0   25.0    0     0     2.5    11.0e3   8.0e3    2    76.0e3    14.0e3   3    0      0     0    0.0    1 
 4985.0   50.0    0     0     0.5    18.0e3   10.0e3   1    105.0e3   33.0e3   0    0      0     0    0.0    1 
 5095.0   75.0    0     0     1.5    71.0e3   27.0e3   1    213.0e3   56.0e3   2    0      0     0    0.0    1 
 5322.0   8.0     0     0     3.5    9.1e3    2.1e3    3    22.0e3    10.0e3   4    0      0     0    0.0    1 
 5365.0   25.0    0     0     1.5    1.9e3    1.2e3    1    36.0e3    18.0e3   2    0      0     0    0.0    1 
 5737.0   11.0    0     0     3.5    11.6e3   2.2e3    3    43.0e3    9.0e3    4    0      0     0    0.0    1 
 6045.0   20.0    0     0     2.5    11.4e3   2.8e3    2    129.0e3   29.0e3   3    0      0     0    0.0    1  
 6105.0   21.0    0     0     1.5    7.6e3    2.8e3    1    76.0e3    29.0e3   2    0      0     0    0.0    1 
 6335.0   20.0    0     0     1.5    8.5e3    2.8e3    1    67.0e3    29.0e3   2    0      0     0    0.0    1 
 6705.0   20.0    0     0     1.5    19.9e3   4.7e3    1    103.0e3   38.0e3   2    0      0     0    0.0    1 
 6745.0   50.0    0     0     0.5    95.0e3   24.0e3   0    265.0e3   70.0e3   1    0      0     0    0.0    1 
 6785.0   20.0    0     0     2.5    29.9e3   5.7e3    2    179.0e3   48.0e3   3    0      0     0    0.0    1 
 6925.0   30.0    0     0     3.5    19.0e3   3.8e3    3    178.0e3   29.0e3   4    0      0     0    0.0    1 
 7365.0   20.0    0     0     0.5    5.7e3    1.9e3    1    61.0e3    10.0e3   0    0      0     0    0.0    1 
 7405.0   30.0    0     0     2.5    6.6e3    1.9e3    2    73.0e3    24.0e3   3    0      0     0    0.0    1 
 7775.0   21.0    0     0     1.5    7.6e3    2.8e3    1    89.0e3    24.0e3   2    0      0     0    0.0    1 
 8205.0   40.0    0     0     1.5    15.2e3   3.8e3    2    155.0e3   29.0e3   1    0      0     0    0.0    1 
 8235.0   30.0    0     0     3.5    12.3e3   3.8e3    3    209.0e3   38.0e3   4    0      0     0    0.0    1 
 8285.0   20.0    0     0     1.5    12.3e3   3.8e3    1    154.0e3   29.0e3   2    0      0     0    0.0    1 
 9055.0   40.0    0     0     1.5    37.0e3   7.6e3    1    293.0e3   67.0e3   2    0      0     0    0.0    1 
 9165.0   40.0    0     0     3.5    28.4e3   7.6e3    3    294.0e3   67.0e3   4    0      0     0    0.0    1 
 9455.0   30.0    0     0     1.5    2.8e3    1.9e3    1    29.0e3    19.0e3   2    0      0     0    0.0    1 
 9655.0   60.0    0     0     3.5    4.7e3    2.8e3    3    90.0e3    57.0e3   4    0      0     0    0.0    1 
 9935.0   40.0    0     0     1.5    21.8e3   4.7e3    1    232.0e3   57.0e3   2    0      0     0    0.0    1 
10035.0   40.0    0     0     3.5    30.3e3   6.6e3    3    333.0e3   57.0e3   4    0      0     0    0.0    1 
11075.0   60.0    0     0     1.5    20.8e3   6.6e3    1    532.0e3   142.0e3  2    0      0     0    0.0    1 
11835.0   150.0   0     0     2.5    12.3e3   5.7e3    3    355.0e3   57.0e3   2    0      0     0    0.0    1 
11895.0   30.0    0     0     1.5    37.0e3   7.6e3    1    435.0e3   57.0e3   2    0      0     0    0.0    1 
12815.0   50.0    0     0     0.5    30.3e3   4.7e3    1    381.0e3   57.0e3   0    0      0     0    0.0    1 
12935.0   50.0    0     0     1.5    11.4e3   3.8e3    1    305.0e3   48.0e3   2    0      0     0    0.0    1 
13055.0   50.0    0     0     3.5    23.7e3   4.7e3    3    423.0e3   29.0e3   4    0      0     0    0.0    1 
****************************************************************************************************************
Upper Limits of Resonances
Note: enter partial width upper limit by chosing non-zero value for PT, where PT=<theta^2> for particles and...
Note: ...PT=<B> for g-rays [enter: "upper_limit 0.0"]; for each resonance: # upper limits < # open channels!  
Ecm      DEcm  Jr     G1       DG1       L1   PT     G2      DG2   L2   PT      G3   DG3   L3   PT    Exf   Int
204.2    1.0   2.5    7.7e-4   2.0e-4    2    0.0    0.8e3   0.0   3    0.010   0.0  0.0   0    0.0   0.0   1  
****************************************************************************************************************
Interference between Resonances [numerical integration only]
Note: + for positive, - for negative interference; +- if interference sign is unknown
Ecm    DEcm   Jr     G1       DG1      L1   PT      G2      DG2     L2   PT     G3     DG3    L3  PT    Exf  
-
142.8  0.1    0.5    1.67e-1  0.12e-1  0    0.0     1.23e2  0.24e2  1    0.0    0.72   0.15   1   0.0   0.0 
609.0  2.0    0.5    7.0e3    2.0e3    0    0.0     1.5e5   0.3e5   1    0.0    0.1    0.001  1   0.0   0.0 
!807.0  3.0    0.5    2.46e4   0.14e4   0    0.0     2.2e4   4.0e3   1   0.0    2.5    0.4    1   0.0   0.0  
****************************************************************************************************************
Reaction Rate and PDF at NT selected temperatures only
Note: default values are used for reaction rate range if Min=Max=0.0
T9     Min    Max
0.01   0.0    0.0
0.1    0.0    0.0
****************************************************************************************************************
Comments: 
1. Up to Er=2 MeV, input data are taken from the same source as for the 18O(p,g)19F reaction (see previous 
   table), except for comments below; for higher energies the partial widths are adopted from Sellin et al. 1969, 
   Orihara et al. 1973, Almanza et al. 1975 and Murillo et al. 1979.
2. Q-values and masses from Audi and Meng 2011.
3. For 90 keV resonance, directly measured strength adopted from Lorenz-Wirzba et al. (1979).
4. Resonance energy for 143 keV from Becker et al. 1995.
5. For 20 and 90 keV resonances, strengths agree with La Cognata et al. (2010b).
6. For energies and partial widths of broad 609 keV and 807 keV resonances, we adopted the weighted average from
   Yagi 1962, Mak et al. 1978, and La Cognata et al. 2010a, using the "additive variable method"; see Sec. 8.2.;
   note that the results of Lorenz-Wirzba et al. 1979 are explicitly included in the values reported by La Cognata
   et al. 2010a.
7. Interferences between 143, 609, and 807 keV resonances: we assumed that the two-level interference is 
   destructive (constructive) outside (between) (i) 143-609 keV (see Becker et al. 1995 and Lorenz-Wirzba et al. 
   1979); (ii) 906-807 keV (see Lorenz-Wirzba et al. 1979 and La Cognata et al. 2010a). 




\end{verbatim}
\normalsize
\vspace{5mm}

\begin{figure}[ht]
\centering
\includegraphics[scale=0.5]{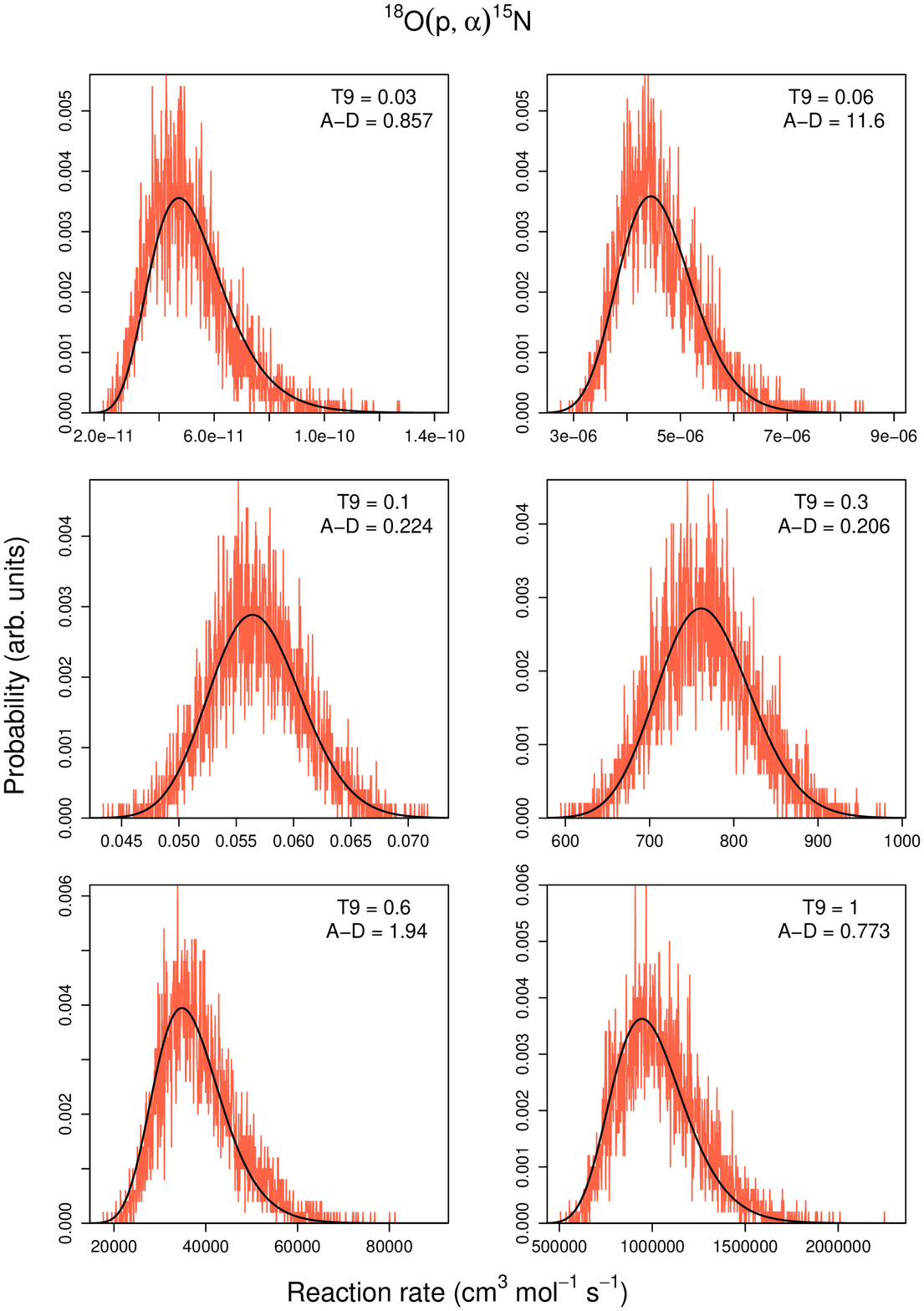}
\end{figure}

\begin{figure}[ht]
\centering
\includegraphics[scale=0.5]{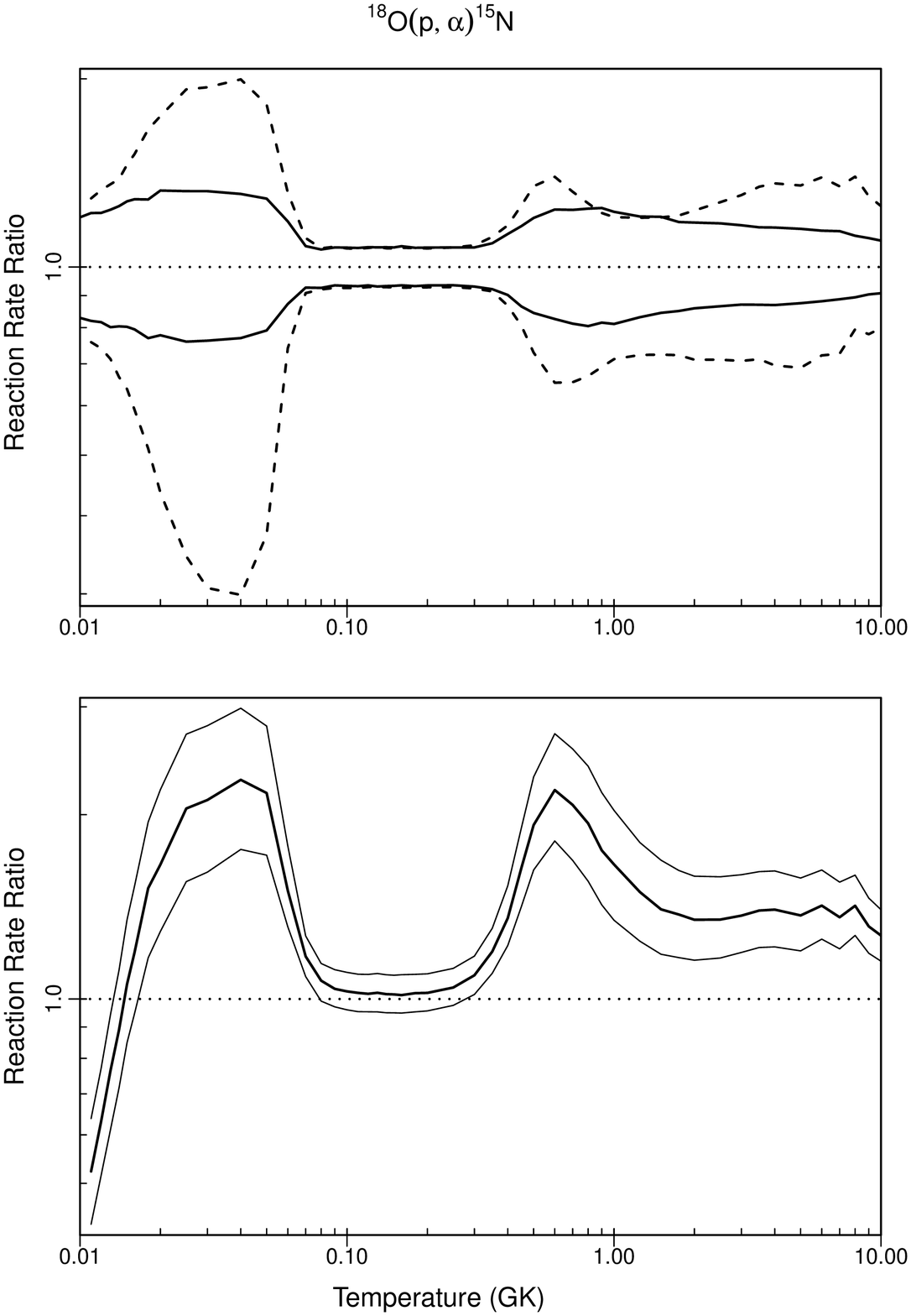}
\end{figure}
\clearpage

\clearpage
\setlongtables
\begin{longtable}{cccc | ccc}
\caption{Total thermonuclear reaction rates for $^{22}$Ne(p,$\gamma$)$^{23}$Na.}  \label{tab:ne22pg} \\
\hline \hline 
	\multicolumn{1}{c}{T (GK)} & \multicolumn{1}{c}{Low rate} & \multicolumn{1}{c}{Median rate} & \multicolumn{1}{c}{High rate}  & \multicolumn{1}{c}{lognormal $\mu$} & \multicolumn{1}{c}{lognormal $\sigma$} & \multicolumn{1}{c}{A-D} \\ \hline 
\endfirsthead
\multicolumn{6}{c}{{\tablename} \thetable{} -- continued} \\
\hline \hline 
	\multicolumn{1}{c}{T (GK)} & \multicolumn{1}{c}{Low rate} & \multicolumn{1}{c}{Median rate} & \multicolumn{1}{c}{High rate} & \multicolumn{1}{c}{lognormal $\mu$} & \multicolumn{1}{c}{lognormal $\sigma$} & \multicolumn{1}{c}{A-D} \\ \hline 
\endhead
	 \hline \hline
\endfoot
	\hline \hline
\endlastfoot
0.010 &  4.21$\times$10$^{-25}$  &  6.76$\times$10$^{-25}$  &
       1.07$\times$10$^{-24}$  &  -5.566$\times$10$^{+01}$  &
       4.68$\times$10$^{-01}$  &  4.60$\times$10$^{-01}$  \\
0.011 &  1.60$\times$10$^{-23}$  &  2.46$\times$10$^{-23}$  &
       3.71$\times$10$^{-23}$  &  -5.206$\times$10$^{+01}$  &
       4.27$\times$10$^{-01}$  &  4.88$\times$10$^{-01}$  \\
0.012 &  3.25$\times$10$^{-22}$  &  4.85$\times$10$^{-22}$  &
       7.12$\times$10$^{-22}$  &  -4.908$\times$10$^{+01}$  &
       4.00$\times$10$^{-01}$  &  4.80$\times$10$^{-01}$  \\
0.013 &  4.07$\times$10$^{-21}$  &  6.00$\times$10$^{-21}$  &
       8.64$\times$10$^{-21}$  &  -4.657$\times$10$^{+01}$  &
       3.83$\times$10$^{-01}$  &  3.83$\times$10$^{-01}$  \\
0.014 &  3.52$\times$10$^{-20}$  &  5.10$\times$10$^{-20}$  &
       7.35$\times$10$^{-20}$  &  -4.442$\times$10$^{+01}$  &
       3.73$\times$10$^{-01}$  &  2.28$\times$10$^{-01}$  \\
0.015 &  2.25$\times$10$^{-19}$  &  3.24$\times$10$^{-19}$  &
       4.70$\times$10$^{-19}$  &  -4.257$\times$10$^{+01}$  &
       3.70$\times$10$^{-01}$  &  2.06$\times$10$^{-01}$  \\
0.016 &  1.13$\times$10$^{-18}$  &  1.63$\times$10$^{-18}$  &
       2.37$\times$10$^{-18}$  &  -4.096$\times$10$^{+01}$  &
       3.69$\times$10$^{-01}$  &  2.80$\times$10$^{-01}$  \\
0.018 &  1.64$\times$10$^{-17}$  &  2.37$\times$10$^{-17}$  &
       3.44$\times$10$^{-17}$  &  -3.828$\times$10$^{+01}$  &
       3.76$\times$10$^{-01}$  &  3.33$\times$10$^{-01}$  \\
0.020 &  1.35$\times$10$^{-16}$  &  1.98$\times$10$^{-16}$  &
       2.91$\times$10$^{-16}$  &  -3.616$\times$10$^{+01}$  &
       3.87$\times$10$^{-01}$  &  4.12$\times$10$^{-01}$  \\
0.025 &  5.71$\times$10$^{-15}$  &  8.68$\times$10$^{-15}$  &
       1.30$\times$10$^{-14}$  &  -3.239$\times$10$^{+01}$  &
       4.19$\times$10$^{-01}$  &  5.82$\times$10$^{-01}$  \\
0.030 &  6.55$\times$10$^{-14}$  &  1.02$\times$10$^{-13}$  &
       1.57$\times$10$^{-13}$  &  -2.992$\times$10$^{+01}$  &
       4.47$\times$10$^{-01}$  &  6.44$\times$10$^{-01}$  \\
0.040 &  1.25$\times$10$^{-12}$  &  2.04$\times$10$^{-12}$  &
       3.27$\times$10$^{-12}$  &  -2.693$\times$10$^{+01}$  &
       4.87$\times$10$^{-01}$  &  6.52$\times$10$^{-01}$  \\
0.050 &  6.93$\times$10$^{-12}$  &  1.16$\times$10$^{-11}$  &
       1.89$\times$10$^{-11}$  &  -2.519$\times$10$^{+01}$  &
       5.08$\times$10$^{-01}$  &  5.88$\times$10$^{-01}$  \\
0.060 &  2.27$\times$10$^{-11}$  &  3.74$\times$10$^{-11}$  &
       6.08$\times$10$^{-11}$  &  -2.401$\times$10$^{+01}$  &
       4.97$\times$10$^{-01}$  &  3.37$\times$10$^{-01}$  \\
0.070 &  7.06$\times$10$^{-11}$  &  1.05$\times$10$^{-10}$  &
       1.59$\times$10$^{-10}$  &  -2.297$\times$10$^{+01}$  &
       4.13$\times$10$^{-01}$  &  1.83$\times$10$^{+00}$  \\
0.080 &  2.85$\times$10$^{-10}$  &  3.84$\times$10$^{-10}$  &
       5.18$\times$10$^{-10}$  &  -2.168$\times$10$^{+01}$  &
       3.00$\times$10$^{-01}$  &  2.97$\times$10$^{-01}$  \\
0.090 &  1.32$\times$10$^{-09}$  &  1.89$\times$10$^{-09}$  &
       2.67$\times$10$^{-09}$  &  -2.010$\times$10$^{+01}$  &
       3.35$\times$10$^{-01}$  &  9.47$\times$10$^{+00}$  \\
0.100 &  5.52$\times$10$^{-09}$  &  9.22$\times$10$^{-09}$  &
       1.56$\times$10$^{-08}$  &  -1.850$\times$10$^{+01}$  &
       4.66$\times$10$^{-01}$  &  3.13$\times$10$^{+01}$  \\
0.110 &  2.08$\times$10$^{-08}$  &  4.02$\times$10$^{-08}$  &
       7.67$\times$10$^{-08}$  &  -1.704$\times$10$^{+01}$  &
       5.73$\times$10$^{-01}$  &  4.61$\times$10$^{+01}$  \\
0.120 &  6.51$\times$10$^{-08}$  &  1.45$\times$10$^{-07}$  &
       3.00$\times$10$^{-07}$  &  -1.578$\times$10$^{+01}$  &
       6.62$\times$10$^{-01}$  &  5.75$\times$10$^{+01}$  \\
0.130 &  1.80$\times$10$^{-07}$  &  4.48$\times$10$^{-07}$  &
       9.60$\times$10$^{-07}$  &  -1.468$\times$10$^{+01}$  &
       7.23$\times$10$^{-01}$  &  6.86$\times$10$^{+01}$  \\
0.140 &  4.42$\times$10$^{-07}$  &  1.18$\times$10$^{-06}$  &
       2.61$\times$10$^{-06}$  &  -1.373$\times$10$^{+01}$  &
       7.65$\times$10$^{-01}$  &  6.94$\times$10$^{+01}$  \\
0.150 &  9.89$\times$10$^{-07}$  &  2.76$\times$10$^{-06}$  &
       6.27$\times$10$^{-06}$  &  -1.288$\times$10$^{+01}$  &
       7.92$\times$10$^{-01}$  &  7.35$\times$10$^{+01}$  \\
0.160 &  2.01$\times$10$^{-06}$  &  5.90$\times$10$^{-06}$  &
       1.33$\times$10$^{-05}$  &  -1.214$\times$10$^{+01}$  &
       8.08$\times$10$^{-01}$  &  7.61$\times$10$^{+01}$  \\
0.180 &  7.34$\times$10$^{-06}$  &  2.09$\times$10$^{-05}$  &
       4.73$\times$10$^{-05}$  &  -1.086$\times$10$^{+01}$  &
       7.91$\times$10$^{-01}$  &  7.67$\times$10$^{+01}$  \\
0.200 &  2.68$\times$10$^{-05}$  &  6.30$\times$10$^{-05}$  &
       1.33$\times$10$^{-04}$  &  -9.715$\times$10$^{+00}$  &
       6.81$\times$10$^{-01}$  &  7.40$\times$10$^{+01}$  \\
0.250 &  9.43$\times$10$^{-04}$  &  1.18$\times$10$^{-03}$  &
       1.57$\times$10$^{-03}$  &  -6.722$\times$10$^{+00}$  &
       2.30$\times$10$^{-01}$  &  2.81$\times$10$^{+01}$  \\
0.300 &  1.77$\times$10$^{-02}$  &  1.98$\times$10$^{-02}$  &
       2.23$\times$10$^{-02}$  &  -3.919$\times$10$^{+00}$  &
       1.17$\times$10$^{-01}$  &  9.06$\times$10$^{-01}$  \\
0.350 &  1.56$\times$10$^{-01}$  &  1.73$\times$10$^{-01}$  &
       1.92$\times$10$^{-01}$  &  -1.751$\times$10$^{+00}$  &
       1.06$\times$10$^{-01}$  &  1.06$\times$10$^{+00}$  \\
0.400 &  8.19$\times$10$^{-01}$  &  9.03$\times$10$^{-01}$  &
       9.97$\times$10$^{-01}$  &  -1.006$\times$10$^{-01}$  &
       9.93$\times$10$^{-02}$  &  7.22$\times$10$^{-01}$  \\
0.450 &  3.00$\times$10$^{+00}$  &  3.28$\times$10$^{+00}$  &
       3.60$\times$10$^{+00}$  &  1.190$\times$10$^{+00}$  &
       9.33$\times$10$^{-02}$  &  4.60$\times$10$^{-01}$  \\
0.500 &  8.50$\times$10$^{+00}$  &  9.27$\times$10$^{+00}$  &
       1.01$\times$10$^{+01}$  &  2.228$\times$10$^{+00}$  &
       8.81$\times$10$^{-02}$  &  3.56$\times$10$^{-01}$  \\
0.600 &  4.13$\times$10$^{+01}$  &  4.47$\times$10$^{+01}$  &
       4.83$\times$10$^{+01}$  &  3.800$\times$10$^{+00}$  &
       8.04$\times$10$^{-02}$  &  2.68$\times$10$^{-01}$  \\
0.700 &  1.30$\times$10$^{+02}$  &  1.41$\times$10$^{+02}$  &
       1.51$\times$10$^{+02}$  &  4.945$\times$10$^{+00}$  &
       7.61$\times$10$^{-02}$  &  2.87$\times$10$^{-01}$  \\
0.800 &  3.14$\times$10$^{+02}$  &  3.38$\times$10$^{+02}$  &
       3.64$\times$10$^{+02}$  &  5.824$\times$10$^{+00}$  &
       7.46$\times$10$^{-02}$  &  4.03$\times$10$^{-01}$  \\
0.900 &  6.32$\times$10$^{+02}$  &  6.81$\times$10$^{+02}$  &
       7.34$\times$10$^{+02}$  &  6.524$\times$10$^{+00}$  &
       7.51$\times$10$^{-02}$  &  5.80$\times$10$^{-01}$  \\
1.000 &  1.12$\times$10$^{+03}$  &  1.21$\times$10$^{+03}$  &
       1.30$\times$10$^{+03}$  &  7.100$\times$10$^{+00}$  &
       7.68$\times$10$^{-02}$  &  7.91$\times$10$^{-01}$  \\
1.250 &  3.28$\times$10$^{+03}$  &  3.54$\times$10$^{+03}$  &
       3.86$\times$10$^{+03}$  &  8.176$\times$10$^{+00}$  &
       8.35$\times$10$^{-02}$  &  2.51$\times$10$^{+00}$  \\
1.500 &  6.90$\times$10$^{+03}$  &  7.49$\times$10$^{+03}$  &
       8.22$\times$10$^{+03}$  &  8.926$\times$10$^{+00}$  &
       9.01$\times$10$^{-02}$  &  4.82$\times$10$^{+00}$  \\
1.750 &  1.20$\times$10$^{+04}$  &  1.30$\times$10$^{+04}$  &
       1.44$\times$10$^{+04}$  &  9.479$\times$10$^{+00}$  &
       9.45$\times$10$^{-02}$  &  6.75$\times$10$^{+00}$  \\
2.000 &  1.82$\times$10$^{+04}$  &  1.99$\times$10$^{+04}$  &
       2.20$\times$10$^{+04}$  &  9.901$\times$10$^{+00}$  &
       9.68$\times$10$^{-02}$  &  7.77$\times$10$^{+00}$  \\
2.500 &  3.31$\times$10$^{+04}$  &  3.60$\times$10$^{+04}$  &
       3.99$\times$10$^{+04}$  &  1.050$\times$10$^{+01}$  &
       9.67$\times$10$^{-02}$  &  8.73$\times$10$^{+00}$  \\
3.000 &  4.93$\times$10$^{+04}$  &  5.35$\times$10$^{+04}$  &
       5.90$\times$10$^{+04}$  &  1.089$\times$10$^{+01}$  &
       9.37$\times$10$^{-02}$  &  8.78$\times$10$^{+00}$  \\
3.500 &  6.51$\times$10$^{+04}$  &  7.05$\times$10$^{+04}$  &
       7.74$\times$10$^{+04}$  &  1.117$\times$10$^{+01}$  &
       8.98$\times$10$^{-02}$  &  8.67$\times$10$^{+00}$  \\
4.000 &  7.96$\times$10$^{+04}$  &  8.59$\times$10$^{+04}$  &
       9.39$\times$10$^{+04}$  &  1.137$\times$10$^{+01}$  &
       8.59$\times$10$^{-02}$  &  8.44$\times$10$^{+00}$  \\
5.000 & (1.07$\times$10$^{+05}$) & (1.16$\times$10$^{+05}$) &
      (1.27$\times$10$^{+05}$) & (1.166$\times$10$^{+01}$) &
      (8.56$\times$10$^{-02}$) &  \\
6.000 & (1.31$\times$10$^{+05}$) & (1.43$\times$10$^{+05}$) &
      (1.56$\times$10$^{+05}$) & (1.187$\times$10$^{+01}$) &
      (8.56$\times$10$^{-02}$) &  \\
7.000 & (1.51$\times$10$^{+05}$) & (1.65$\times$10$^{+05}$) &
      (1.79$\times$10$^{+05}$) & (1.201$\times$10$^{+01}$) &
      (8.56$\times$10$^{-02}$) &  \\
8.000 & (1.68$\times$10$^{+05}$) & (1.83$\times$10$^{+05}$) &
      (2.00$\times$10$^{+05}$) & (1.212$\times$10$^{+01}$) &
      (8.56$\times$10$^{-02}$) &  \\
9.000 & (1.81$\times$10$^{+05}$) & (1.98$\times$10$^{+05}$) &
      (2.15$\times$10$^{+05}$) & (1.219$\times$10$^{+01}$) &
      (8.56$\times$10$^{-02}$) &  \\
10.000 & (1.96$\times$10$^{+05}$) & (2.13$\times$10$^{+05}$) &
      (2.32$\times$10$^{+05}$) & (1.227$\times$10$^{+01}$) &
      (8.56$\times$10$^{-02}$) &  \\

\end{longtable}

\subsection{$^{22}$Ne($p,\gamma)^{23}$Na}
Comments:  The input file from \citet{iliadis_3} has been updated with a new normalization of all previously measured resonance strengths, based on an absolute measurement of the $E_p = 458$ keV resonance ($\omega\gamma = 0.524 \pm 0.051$ eV)~\citep{2010PhRvC..81e5804L}.  Uncertainties from the previous strengths and the uncertainty of the new measurement were added in quadrature.  Particle partial widths for the two subthreshold resonances at $E_p = 35$ and 151 keV have not been renormalized.  Values for upper limits remain unchanged.  For $T>4.0$ GK, rates were extrapolated using Hauser-Feshbach calculations.

\footnotesize
\begin{verbatim}
22Ne(p,g)23Na
****************************************************************************************************************
1               ! Zproj
10              ! Ztarget
0               ! Zexitparticle (=0 when only 2 channels open)
1.0078          ! Aproj		
21.991          ! Atarget
0               ! Aexitparticle (=0 when only 2 channels open)
0.5             ! Jproj
0.0             ! Jtarget
0.0             ! Jexitparticle (=0 when only 2 channels open)
8794.11         ! projectile separation energy (keV)
0.0             ! exit particle separation energy (=0 when only 2 channels open)
1.25            ! Radius parameter R0 (fm)
2               ! Gamma-ray channel number (=2 if ejectile is a g-ray; =3 otherwise)
****************************************************************************************************************
1.0             ! Minimum energy for numerical integration (keV)
5000            ! Number of random samples (>5000 for better statistics)
0               ! =0 for rate output at all temperatures; =NT for rate output at selected temperatures
****************************************************************************************************************
Non-resonant contribution
S(keVb)	S'(b)	S''(b/keV)	fracErr	Cutoff Energy (keV)
6.2e1    0.0     0.0          0.4       1500.0
0.0      0.0     0.0          0.0       0.0
****************************************************************************************************************
Resonant Contribution
Note: G1 = entrance channel, G2 = exit channel, G3 = spectator channel !! Ecm, Exf in (keV); wg, Gx in (eV) !!
Note: if Er<0, theta^2=C2S*theta_sp^2 must be entered instead of entrance channel partial width
Ecm	 DEcm	wg	 Dwg	  Jr	G1	DG1	L1	G2      DG2	L2   G3  DG3  L3    Exf	  Int
  35.4   0.7    0        0        0.5   3.1e-15 1.2e-15 0       2.2     1.0     1    0   0    0     0.0   1
 150.9   2.0    0        0        3.5   2.3e-9  9.2e-10 3       0.02    0.01    1    0   0    0     0.0   1
 417.0   0.8    0.076    0.019    0     0       0       0	      0	      0      	0    0   0    0     0.0   0
 458.2   0.8    0.524    0.051    0     0       0       0	      0	      0      	0    0   0    0     0.0   0
 601.9   0.3    0.035    0.012    0     0       0       0	      0	      0      	0    0   0    0     0.0   0
 610.3   0.3    3.3      0.5      0     0       0       0	      0	      0      	0    0   0    0     0.0   0
 631.4   0.4    0.41     0.12     0     0       0       0	      0	      0      	0    0   0    0     0.0   0
 693.3   0.7    0.15     0.05     0     0       0       0	      0	      0      	0    0   0    0     0.0   0
 813.7   0.2    8.2      3.0      0     0       0       0	      0	      0      	0    0   0    0     0.0   0
 857.3   0.5    2.1      1.1      0     0       0       0	      0	      0      	0    0   0    0     0.0   0
 861.1   1.0    1.22     0.59     0     0       0       0	      0	      0      	0    0   0    0     0.0   0
 879.5   1.0    0.93     0.31     0     0       0       0	      0	      0      	0    0   0    0     0.0   0
 888.2   0.3    0.41     0.12     0     0       0       0	      0	      0      	0    0   0    0     0.0   0
 906.3   1.0    7.0      2.4      0     0       0       0	      0	      0      	0    0   0    0     0.0   0
 937.91  0.07   0.47     0.13     0     0       0       0	      0	      0      	0    0   0    0     0.0   0
 960.9   0.5    2.8      0.9      0     0       0       0	      0	      0      	0    0   0    0     0.0   0
1021.0   0.4    1.0      0.3      0     0       0       0	      0	      0      	0    0   0    0     0.0   0
1040.7   1.0    2.50     0.69     0     0       0       0	      0	      0      	0    0   0    0     0.0   0
1055.2   0.5    2.50     0.69     0     0       0       0	      0	      0      	0    0   0    0     0.0   0
1096.1   0.6    1.7      0.5      0     0       0       0	      0	      0      	0    0   0    0     0.0   0
1122.2   0.6    0.70     0.19     0     0       0       0	      0	      0      	0    0   0    0     0.0   0
1208.5   0.6    1.28     0.37     0     0       0       0	      0	      0      	0    0   0    0     0.0   0
1221.9   0.4    12.2     1.7      0     0       0       0	      0	      0      	0    0   0    0     0.0   0
1254.3   0.6    0.23     0.06     0     0       0       0	      0	      0      	0    0   0    0     0.0   0
1275.9   0.6    3.20     0.87     0     0       0       0	      0	      0      	0    0   0    0     0.0   0
1281.1   0.5    1.40     0.38     0     0       0       0	      0	      0      	0    0   0    0     0.0   0
1290.3   0.2    0.93     0.25     0     0       0       0	      0	      0      	0    0   0    0     0.0   0
1319.9   0.5    0.76     0.19     0     0       0       0	      0	      0      	0    0   0    0     0.0   0
1331.0   0.5    1.63     0.44     0     0       0       0	      0	      0      	0    0   0    0     0.0   0
1374.7   0.2    3.1      0.9      0     0       0       0	      0	      0      	0    0   0    0     0.0   0
1436.7   0.3    5.2      1.3      0     0       0       0	      0	      0      	0    0   0    0     0.0   0
1448.8   1.4    1.3      0.4      0     0       0       0	      0	      0      	0    0   0    0     0.0   0
1486.6   0.6    1.63     0.44     0     0       0       0	      0	      0      	0    0   0    0     0.0   0
1523.1   0.6    5.8      1.8      0     0       0       0	      0	      0      	0    0   0    0     0.0   0
1543.7   0.7    1.22     0.31     0     0       0       0	      0	      0      	0    0   0    0     0.0   0
1551.1   0.7    5.2      1.3      0     0       0       0	      0	      0      	0    0   0    0     0.0   0
1558.9   0.7    3.49     0.94     0     0       0       0	      0	      0      	0    0   0    0     0.0   0
1645.6   1.0    7.6      1.9      0     0       0       0	      0	      0      	0    0   0    0     0.0   0
1653.7   1.2    2.04     0.56     0     0       0       0	      0	      0      	0    0   0    0     0.0   0
1683.8   0.7    2.97     0.81     0     0       0       0	      0	      0      	0    0   0    0     0.0   0
1706.8   0.7    3.84     1.06     0     0       0       0	      0	      0      	0    0   0    0     0.0   0
1712.8   0.7    0.58     0.18     0     0       0       0	      0	      0      	0    0   0    0     0.0   0
1724.1   0.7    2.56     0.69     0     0       0       0	      0	      0      	0    0   0    0     0.0   0
1739.1   0.7    0.87     0.25     0     0       0       0	      0	      0      	0    0   0    0     0.0   0
1754.2   0.9    6.4      1.9      0     0       0       0	      0	      0      	0    0   0    0     0.0   0
1779.5   0.8    1.34     0.37     0     0       0       0	      0	      0      	0    0   0    0     0.0   0
1821.8   0.8    4.37     1.19     0     0       0       0	      0	      0      	0    0   0    0     0.0   0
****************************************************************************************************************
Upper Limits of Resonances
Note: enter partial width upper limit by chosing non-zero value for PT, where PT=<theta^2> for particles and...
Note: ...PT=<B> for g-rays [enter: "upper_limit 0.0"]; for each resonance: # upper limits < # open channels!  
Ecm	DEcm	Jr    G1      DG1   L1   PT     G2    DG2    L2  PT   G3    DG3    L3  PT   Exf  Int
 27.9   3.0     4.5   5.2e-26 0.0   5    0.0045 0.1   0.01   1   0    0     0      0   0    0.0  0   
177.9   2.0     0.5   2.6e-6  0.0   0    0.0045 0.1   0.01   1   0    0     0      0   0    0.0  0
247.9   1.0     3.5   3.3e-8  0.0   4    0.0045 0.04  0.02   1   0    0     0      0   0    0.0  1
277.9   3.0     0.5   2.2e-6  0.0   0    0.0045 0.1   0.01   1   0    0     0      0   0    0.0  0
308.9   3.0     0.5   2.2e-6  0.0   0    0.0045 0.1   0.01   1   0    0     0      0   0    0.0  0 
318.9   3.0     0.5   3.0e-6  0.0   0    0.0045 0.1   0.01   1   0    0     0      0   0    0.0  0
352.9   5.0     0.5   6.0e-4  0.0   0    0.0045 0.1   0.01   1   0    0     0      0   0    0.0  0
376.9   3.0     0.5   6.0e-4  0.0   0    0.0045 0.1   0.01   1   0    0     0      0   0    0.0  0
****************************************************************************************************************
Interference between Resonances [numerical integration only]
Note: + for positive, - for negative interference; +- if interference sign is unknown
Ecm	DEcm	Jr    G1      DG1   L1   PT     G2    DG2    L2  PT   G3    DG3    L3  PT   Exf  
!+- 
0.0     0.0     0.0   0.0     0.0   0    0      0.0   0.0    0   0    0.0   0.0    0   0    0.0  
0.0     0.0     0.0   0.0     0.0   0    0      0.0   0.0    0   0    0.0   0.0    0   0    0.0  
****************************************************************************************************************
Reaction Rate and PDF at NT selected temperatures only
Note: default values are used for reaction rate range if Min=Max=0.0
T9	Min	Max
0.01	0.0	0.0
0.1	0.0	0.0
****************************************************************************************************************
Comments:
1. Information for Er>400 keV from Endt 1990 (strengths normalized relative to Er=458 keV). 
2. Er=178, 278-377 keV: s-wave resonances (Jp=1/2+) assumed for upper limit; value of Gg=0.1+-0.01 eV is a guess
   (inconsequential since Gp<<Gg); Gp upper limit values calculated from strength upper limits of Goerres et al.
   1982.
3. Er=28 keV: h-wave resonance (Jp=9/2-) assumed for upper limit; Gg=0.1+-0.01 eV is a guess (inconsequential).
4. Er=151 keV: contrary to Hale et al. 2001, we adopt C2S=0.0011 (see Hale's Ph.D. thesis).
5. Er=248 keV: g-wave resonance (Jp=7/2+) assumed for upper limit.
6. Direct capture S-factor adopted from Goerres et al. 1983, with uncertainty estimate from Hale et al. 2001.
7. Levels at Ex=8862, 8894 and 9000 keV (Powers et al. 1971) have been disregarded.
8. All strengths have been renormalized to the resonance at 458 keV (Longland 2010a), with the exception of 
   upper limits and resonances at 35 and 151 keV.   
   
   
      
   
\end{verbatim}
\normalsize
\vspace{5mm}

\begin{figure}[ht]
\centering
\includegraphics[scale=0.5]{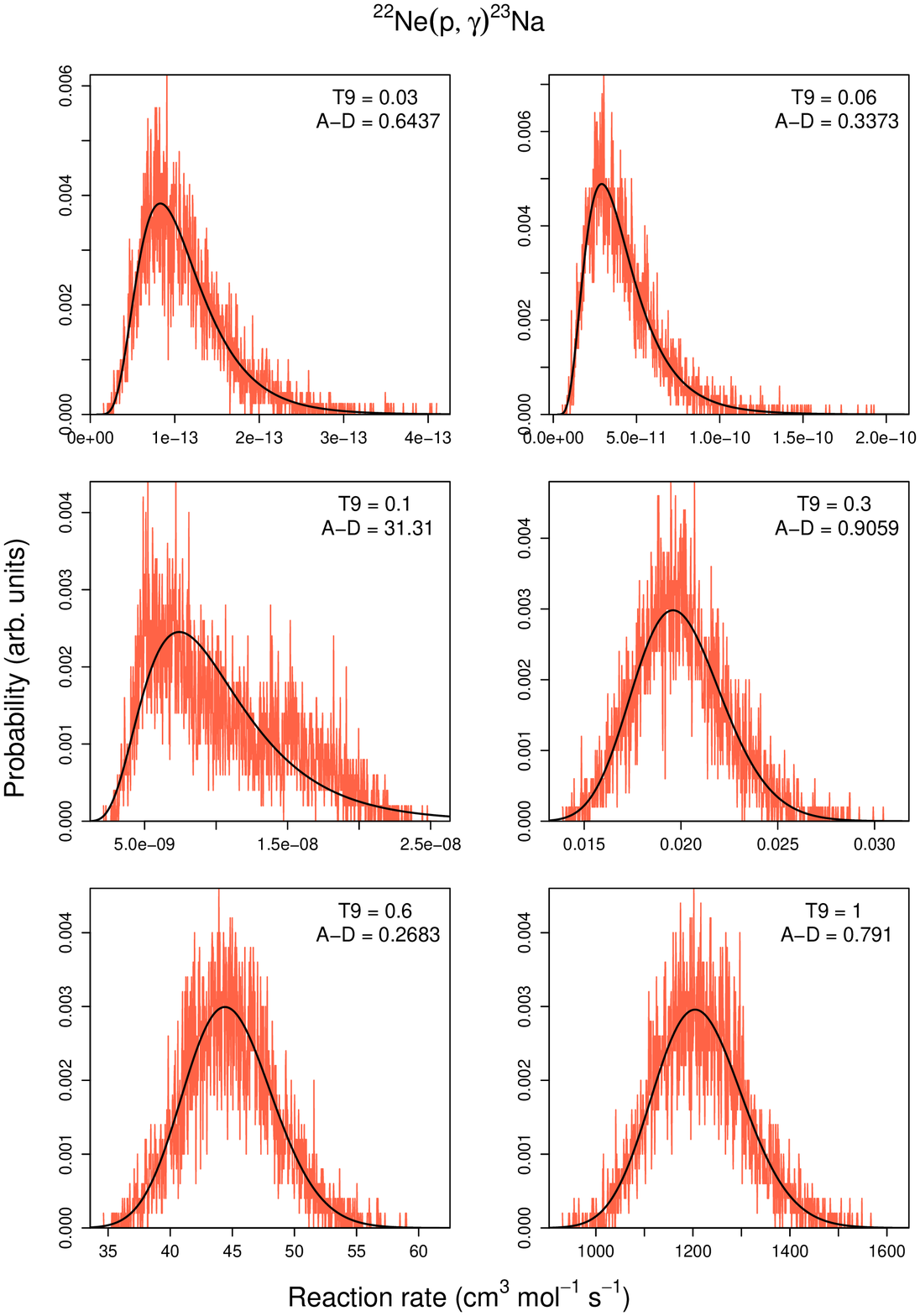}
\label{22ne_ag}
\end{figure}

\begin{figure}[ht]
\centering
\includegraphics[scale=0.5]{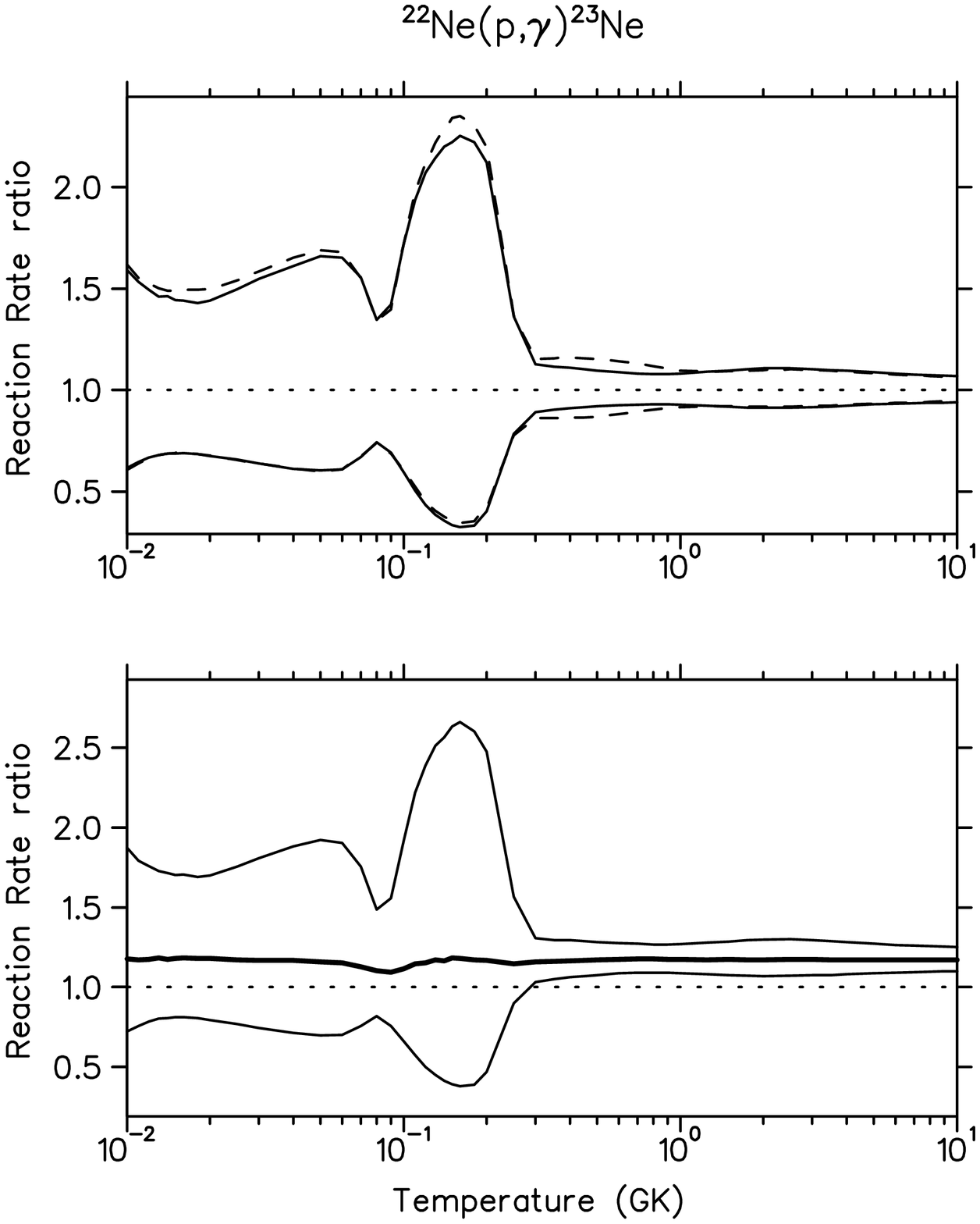}
\label{22ne_ag}
\end{figure}
\clearpage

\setlongtables
\begin{longtable}{cccc | ccc}
\caption{Total thermonuclear reaction rates for $^{22}$Ne($\alpha,\gamma$)$^{26}$Mg.}  \label{tab:ne22ag} \\
\hline \hline 
	\multicolumn{1}{c}{T (GK)} & \multicolumn{1}{c}{Low rate} & \multicolumn{1}{c}{Median rate} & \multicolumn{1}{c}{High rate}  & \multicolumn{1}{c}{lognormal $\mu$} & \multicolumn{1}{c}{lognormal $\sigma$} & \multicolumn{1}{c}{A-D} \\ \hline 
\endfirsthead
\multicolumn{6}{c}{{\tablename} \thetable{} -- continued} \\
\hline \hline 
	\multicolumn{1}{c}{T (GK)} & \multicolumn{1}{c}{Low rate} & \multicolumn{1}{c}{Median rate} & \multicolumn{1}{c}{High rate} & \multicolumn{1}{c}{lognormal $\mu$} & \multicolumn{1}{c}{lognormal $\sigma$} & \multicolumn{1}{c}{A-D} \\ \hline 
\endhead
	 \hline \hline
\endfoot
	\hline \hline
\endlastfoot
0.010 &  1.05$\times$10$^{-77}$  &  2.14$\times$10$^{-77}$  &
       4.52$\times$10$^{-77}$  &  -1.765$\times$10$^{+02}$  &
       7.42$\times$10$^{-01}$  &  3.20$\times$10$^{-01}$  \\
0.011 &  3.99$\times$10$^{-74}$  &  7.28$\times$10$^{-74}$  &
       1.34$\times$10$^{-73}$  &  -1.684$\times$10$^{+02}$  &
       6.15$\times$10$^{-01}$  &  8.94$\times$10$^{-02}$  \\
0.012 &  3.69$\times$10$^{-71}$  &  6.34$\times$10$^{-71}$  &
       1.07$\times$10$^{-70}$  &  -1.617$\times$10$^{+02}$  &
       5.34$\times$10$^{-01}$  &  1.60$\times$10$^{-01}$  \\
0.013 &  1.15$\times$10$^{-68}$  &  1.90$\times$10$^{-68}$  &
       3.09$\times$10$^{-68}$  &  -1.559$\times$10$^{+02}$  &
       4.92$\times$10$^{-01}$  &  1.99$\times$10$^{-01}$  \\
0.014 &  1.55$\times$10$^{-66}$  &  2.52$\times$10$^{-66}$  &
       4.04$\times$10$^{-66}$  &  -1.511$\times$10$^{+02}$  &
       4.80$\times$10$^{-01}$  &  1.92$\times$10$^{-01}$  \\
0.015 &  1.06$\times$10$^{-64}$  &  1.73$\times$10$^{-64}$  &
       2.79$\times$10$^{-64}$  &  -1.468$\times$10$^{+02}$  &
       4.90$\times$10$^{-01}$  &  2.81$\times$10$^{-01}$  \\
0.016 &  4.11$\times$10$^{-63}$  &  6.96$\times$10$^{-63}$  &
       1.14$\times$10$^{-62}$  &  -1.431$\times$10$^{+02}$  &
       5.13$\times$10$^{-01}$  &  4.12$\times$10$^{-01}$  \\
0.018 &  1.80$\times$10$^{-60}$  &  3.26$\times$10$^{-60}$  &
       5.63$\times$10$^{-60}$  &  -1.370$\times$10$^{+02}$  &
       5.75$\times$10$^{-01}$  &  7.20$\times$10$^{-01}$  \\
0.020 &  2.24$\times$10$^{-58}$  &  4.34$\times$10$^{-58}$  &
       8.04$\times$10$^{-58}$  &  -1.321$\times$10$^{+02}$  &
       6.43$\times$10$^{-01}$  &  9.51$\times$10$^{-01}$  \\
0.025 &  1.54$\times$10$^{-54}$  &  3.14$\times$10$^{-54}$  &
       6.30$\times$10$^{-54}$  &  -1.232$\times$10$^{+02}$  &
       7.13$\times$10$^{-01}$  &  6.38$\times$10$^{-01}$  \\
0.030 &  2.82$\times$10$^{-50}$  &  3.35$\times$10$^{-49}$  &
       1.30$\times$10$^{-48}$  &  -1.121$\times$10$^{+02}$  &
       1.87$\times$10$^{+00}$  &  9.43$\times$10$^{+01}$  \\
0.040 &  1.81$\times$10$^{-42}$  &  2.31$\times$10$^{-41}$  &
       8.91$\times$10$^{-41}$  &  -9.413$\times$10$^{+01}$  &
       2.14$\times$10$^{+00}$  &  1.31$\times$10$^{+02}$  \\
0.050 &  8.51$\times$10$^{-38}$  &  1.08$\times$10$^{-36}$  &
       4.17$\times$10$^{-36}$  &  -8.338$\times$10$^{+01}$  &
       2.15$\times$10$^{+00}$  &  1.32$\times$10$^{+02}$  \\
0.060 &  1.05$\times$10$^{-34}$  &  1.34$\times$10$^{-33}$  &
       5.14$\times$10$^{-33}$  &  -7.624$\times$10$^{+01}$  &
       2.08$\times$10$^{+00}$  &  1.25$\times$10$^{+02}$  \\
0.070 &  1.95$\times$10$^{-32}$  &  2.12$\times$10$^{-31}$  &
       8.04$\times$10$^{-31}$  &  -7.104$\times$10$^{+01}$  &
       1.79$\times$10$^{+00}$  &  7.79$\times$10$^{+01}$  \\
0.080 &  2.76$\times$10$^{-30}$  &  1.14$\times$10$^{-29}$  &
       3.67$\times$10$^{-29}$  &  -6.679$\times$10$^{+01}$  &
       1.33$\times$10$^{+00}$  &  3.60$\times$10$^{+01}$  \\
0.090 &  1.76$\times$10$^{-28}$  &  6.30$\times$10$^{-28}$  &
       1.35$\times$10$^{-27}$  &  -6.289$\times$10$^{+01}$  &
       1.15$\times$10$^{+00}$  &  1.26$\times$10$^{+02}$  \\
0.100 &  4.79$\times$10$^{-27}$  &  2.28$\times$10$^{-26}$  &
       6.55$\times$10$^{-26}$  &  -5.931$\times$10$^{+01}$  &
       1.35$\times$10$^{+00}$  &  9.20$\times$10$^{+01}$  \\
0.110 &  8.17$\times$10$^{-26}$  &  5.95$\times$10$^{-25}$  &
       1.86$\times$10$^{-24}$  &  -5.616$\times$10$^{+01}$  &
       1.55$\times$10$^{+00}$  &  1.14$\times$10$^{+02}$  \\
0.120 &  1.11$\times$10$^{-24}$  &  9.63$\times$10$^{-24}$  &
       3.07$\times$10$^{-23}$  &  -5.343$\times$10$^{+01}$  &
       1.64$\times$10$^{+00}$  &  1.34$\times$10$^{+02}$  \\
0.130 &  1.23$\times$10$^{-23}$  &  1.03$\times$10$^{-22}$  &
       3.28$\times$10$^{-22}$  &  -5.102$\times$10$^{+01}$  &
       1.57$\times$10$^{+00}$  &  1.22$\times$10$^{+02}$  \\
0.140 &  1.38$\times$10$^{-22}$  &  8.23$\times$10$^{-22}$  &
       2.50$\times$10$^{-21}$  &  -4.883$\times$10$^{+01}$  &
       1.36$\times$10$^{+00}$  &  8.62$\times$10$^{+01}$  \\
0.150 &  1.53$\times$10$^{-21}$  &  5.57$\times$10$^{-21}$  &
       1.51$\times$10$^{-20}$  &  -4.679$\times$10$^{+01}$  &
       1.10$\times$10$^{+00}$  &  5.66$\times$10$^{+01}$  \\
0.160 &  1.41$\times$10$^{-20}$  &  3.79$\times$10$^{-20}$  &
       8.10$\times$10$^{-20}$  &  -4.484$\times$10$^{+01}$  &
       8.63$\times$10$^{-01}$  &  4.85$\times$10$^{+01}$  \\
0.180 &  8.05$\times$10$^{-19}$  &  1.54$\times$10$^{-18}$  &
       2.84$\times$10$^{-18}$  &  -4.102$\times$10$^{+01}$  &
       6.29$\times$10$^{-01}$  &  8.06$\times$10$^{-01}$  \\
0.200 &  3.41$\times$10$^{-17}$  &  5.43$\times$10$^{-17}$  &
       9.60$\times$10$^{-17}$  &  -3.740$\times$10$^{+01}$  &
       5.19$\times$10$^{-01}$  &  2.31$\times$10$^{+01}$  \\
0.250 &  5.88$\times$10$^{-14}$  &  7.56$\times$10$^{-14}$  &
       1.00$\times$10$^{-13}$  &  -3.019$\times$10$^{+01}$  &
       2.78$\times$10$^{-01}$  &  1.22$\times$10$^{+01}$  \\
0.300 &  9.32$\times$10$^{-12}$  &  1.13$\times$10$^{-11}$  &
       1.38$\times$10$^{-11}$  &  -2.520$\times$10$^{+01}$  &
       1.96$\times$10$^{-01}$  &  1.18$\times$10$^{+00}$  \\
0.350 &  3.46$\times$10$^{-10}$  &  4.08$\times$10$^{-10}$  &
       4.86$\times$10$^{-10}$  &  -2.162$\times$10$^{+01}$  &
       1.69$\times$10$^{-01}$  &  5.63$\times$10$^{-01}$  \\
0.400 &  5.11$\times$10$^{-09}$  &  5.95$\times$10$^{-09}$  &
       6.98$\times$10$^{-09}$  &  -1.894$\times$10$^{+01}$  &
       1.56$\times$10$^{-01}$  &  6.32$\times$10$^{-01}$  \\
0.450 &  4.09$\times$10$^{-08}$  &  4.72$\times$10$^{-08}$  &
       5.50$\times$10$^{-08}$  &  -1.686$\times$10$^{+01}$  &
       1.47$\times$10$^{-01}$  &  6.44$\times$10$^{-01}$  \\
0.500 &  2.13$\times$10$^{-07}$  &  2.44$\times$10$^{-07}$  &
       2.82$\times$10$^{-07}$  &  -1.522$\times$10$^{+01}$  &
       1.41$\times$10$^{-01}$  &  6.51$\times$10$^{-01}$  \\
0.600 &  2.47$\times$10$^{-06}$  &  2.79$\times$10$^{-06}$  &
       3.20$\times$10$^{-06}$  &  -1.278$\times$10$^{+01}$  &
       1.32$\times$10$^{-01}$  &  6.04$\times$10$^{-01}$  \\
0.700 &  1.39$\times$10$^{-05}$  &  1.57$\times$10$^{-05}$  &
       1.78$\times$10$^{-05}$  &  -1.106$\times$10$^{+01}$  &
       1.25$\times$10$^{-01}$  &  5.52$\times$10$^{-01}$  \\
0.800 &  5.15$\times$10$^{-05}$  &  5.77$\times$10$^{-05}$  &
       6.51$\times$10$^{-05}$  &  -9.758$\times$10$^{+00}$  &
       1.18$\times$10$^{-01}$  &  4.75$\times$10$^{-01}$  \\
0.900 &  1.48$\times$10$^{-04}$  &  1.66$\times$10$^{-04}$  &
       1.88$\times$10$^{-04}$  &  -8.701$\times$10$^{+00}$  &
       1.19$\times$10$^{-01}$  &  1.95$\times$10$^{+00}$  \\
1.000 &  3.65$\times$10$^{-04}$  &  4.11$\times$10$^{-04}$  &
       4.73$\times$10$^{-04}$  &  -7.788$\times$10$^{+00}$  &
       1.35$\times$10$^{-01}$  &  1.02$\times$10$^{+01}$  \\
1.250 &  2.33$\times$10$^{-03}$  &  2.77$\times$10$^{-03}$  &
       3.43$\times$10$^{-03}$  &  -5.867$\times$10$^{+00}$  &
       2.02$\times$10$^{-01}$  &  2.19$\times$10$^{+01}$  \\
1.500 & (1.45$\times$10$^{-02}$) & (1.79$\times$10$^{-02}$) &
      (2.21$\times$10$^{-02}$) & (-4.024$\times$10$^{+00}$) &
      (2.12$\times$10$^{-01}$) &  \\
1.750 & (7.64$\times$10$^{-02}$) & (9.45$\times$10$^{-02}$) &
      (1.17$\times$10$^{-01}$) & (-2.360$\times$10$^{+00}$) &
      (2.12$\times$10$^{-01}$) &  \\
2.000 & (3.00$\times$10$^{-01}$) & (3.70$\times$10$^{-01}$) &
      (4.58$\times$10$^{-01}$) & (-9.932$\times$10$^{-01}$) &
      (2.12$\times$10$^{-01}$) &  \\
2.500 & (2.55$\times$10$^{+00}$) & (3.15$\times$10$^{+00}$) &
      (3.89$\times$10$^{+00}$) & (1.147$\times$10$^{+00}$) &
      (2.12$\times$10$^{-01}$) &  \\
3.000 & (1.24$\times$10$^{+01}$) & (1.53$\times$10$^{+01}$) &
      (1.89$\times$10$^{+01}$) & (2.729$\times$10$^{+00}$) &
      (2.12$\times$10$^{-01}$) &  \\
3.500 & (4.18$\times$10$^{+01}$) & (5.17$\times$10$^{+01}$) &
      (6.39$\times$10$^{+01}$) & (3.945$\times$10$^{+00}$) &
      (2.12$\times$10$^{-01}$) &  \\
4.000 & (1.10$\times$10$^{+02}$) & (1.36$\times$10$^{+02}$) &
      (1.68$\times$10$^{+02}$) & (4.913$\times$10$^{+00}$) &
      (2.12$\times$10$^{-01}$) &  \\
5.000 & (4.71$\times$10$^{+02}$) & (5.82$\times$10$^{+02}$) &
      (7.19$\times$10$^{+02}$) & (6.366$\times$10$^{+00}$) &
      (2.12$\times$10$^{-01}$) &  \\
6.000 & (1.33$\times$10$^{+03}$) & (1.64$\times$10$^{+03}$) &
      (2.03$\times$10$^{+03}$) & (7.405$\times$10$^{+00}$) &
      (2.12$\times$10$^{-01}$) &  \\
7.000 & (2.91$\times$10$^{+03}$) & (3.59$\times$10$^{+03}$) &
      (4.44$\times$10$^{+03}$) & (8.186$\times$10$^{+00}$) &
      (2.12$\times$10$^{-01}$) &  \\
8.000 & (5.35$\times$10$^{+03}$) & (6.62$\times$10$^{+03}$) &
      (8.18$\times$10$^{+03}$) & (8.798$\times$10$^{+00}$) &
      (2.12$\times$10$^{-01}$) &  \\
9.000 & (8.68$\times$10$^{+03}$) & (1.07$\times$10$^{+04}$) &
      (1.33$\times$10$^{+04}$) & (9.281$\times$10$^{+00}$) &
      (2.12$\times$10$^{-01}$) &  \\
10.000 & (1.30$\times$10$^{+04}$) & (1.60$\times$10$^{+04}$) &
      (1.98$\times$10$^{+04}$) & (9.681$\times$10$^{+00}$) &
      (2.12$\times$10$^{-01}$) &  \\

\end{longtable}

\subsection{$^{22}$Ne($\alpha,\gamma)^{26}$Mg}\label{22neag}

Comments:  the $^{22}$Ne+$\alpha$ reaction rates have previously been published in \citet{lo12}.  Here, we also include the update input file and additional plots.  These rates differ from those presented in \citet{iliadis_2} in the following ways (see \citet{lo12} for more details):  (i) the inflated weighted average method has been employed to account for the ambiguities in experimental data sets for resonance strengths and energies; (ii) normalization of the relative spectroscopic factors reported in \citet{giesen} is now performed with respect to the E$_r^{lab}$ = 1434 keV resonance rather than the E$_r^{lab}$ = 831 keV resonance used in previous work; and (iii) a previous ($^{6}$Li,d) transfer measurement~\citep{claudio} suggested a spectroscopic factor for the state corresponding to a resonance at E$_r^{lab}$ = 588 keV; in \citet{iliadis_2}, this preliminary value is treated as an upper limit; in the new evaluation, this upper limit has been removed and replaced by C$^2$S=1.  For $T>1.25$ GK, rates were extrapolated using Hauser-Feshbach calculations.

\footnotesize
\begin{verbatim}
22Ne(a,g)26Mg
****************************************************************************************************************
2               ! Zproj
10              ! Ztarget
0               ! Zexitparticle (=0 when only 2 channels open)
4.003           ! Aproj         
21.991          ! Atarget
1.009           ! Aexitparticle (=0 when only 2 channels open)
0.0             ! Jproj
0.0             ! Jtarget
0.5             ! Jexitparticle (=0 when only 2 channels open)
10614.78        ! projectile separation energy (keV)
11093.08        ! exit particle separation energy (=0 when only 2 channels open)
1.25            ! Radius parameter R0 (fm)
2               ! Gamma-ray channel number (=2 if ejectile is a g-ray; =3 otherwise)
****************************************************************************************************************
1.0             ! Minimum energy for numerical integration (keV)
5000            ! Number of random samples (>5000 for better statistics)
0               ! =0 for rate output at all temperatures; =NT for rate output at selected temperatures
****************************************************************************************************************
Nonresonant Contribution
S(keVb)  S'(b)   S''(b/keV)   fracErr   Cutoff Energy (keV)
0.0      0.0     0.0          0.0       0.0
0.0      0.0     0.0          0.0       0.0
****************************************************************************************************************
Resonant Contribution
Note: G1 = entrance channel, G2 = exit channel, G3 = spectator channel !! Ecm, Exf in (keV); wg, Gx in (eV) !!
Note: if Er<0, theta^2=C2S*theta_sp^2 must be entered instead of entrance channel partial width
Ecm      DEcm   wg      Dwg     J   G1        DG1      L1   G2   DG2   L2   G3       DG3     L3  Exf   Int
  78.37  1.7    0       0       4   1.5e-46   1.2e-46  4    3.0  1.5   1    0        0       0   0.0   0
 703.78  2.11   0       0       2   7.2e-6    4.4e-7   2    3.0  1.5   1    2.5e2    1.7e2   1   0.0   1
 826.04  0.19   0       0       4   3.78e-6   4.44e-7  4    3.0  1.5   1    1.47e3   8.0e1   2   0.0   1
 850.44  0.21   0       0       5   4.36e-6   9.09e-7  5    3.0  1.5   1    6.55e3   9.0e1   3   0.0   1
 893.31  0.90   0       0       1   1.17e-4   2.0e-5   1    3.0  1.5   1    1.27e4   2.5e3   1   0.0   1
 911.16  1.69   0       0       1   2.77e-4   2.33e-5  1    3.0  1.5   1    1.80e3   9.0e2   1   0.0   1
1015.22  1.69   0       0       1   2.83e-3   3.33e-4  1    3.0  1.5   1    1.35e4   1.7e3   1   0.0   1 
1133.66  8.46   0       0       1   2.0e-2    3.0e-3   1    3.0  1.5   1    6.35e4   8.5e3   1   0.0   1 
1171.74  3.38   0       0       1   1.67e-2   2.33e-3  1    3.0  1.5   1    2.45e4   2.4e3   1   0.0   1
1213.0   2.0    0       0       2   1.84e-1   1.03e-1  2    3.0  1.5   1    1.10e3   2.5e2   0   0.0   1
1280.0   4.0    2.0e-3  2.0e-4  1   0         0        0    0    0     0    0        0       1   0.0   0
1297.0   3.0    0       0       1   1.89      7.88e-1  1    3.0  1.5   1    5.0e3    2.0e3   1   0.0   1
1338.0   3.0    0       0       3   6.48e-1   3.33e-1  3    3.0  1.5   1    4.0e3    2.0e3   0   0.0   1
1437.0   3.0    0       0       3   8.58e-1   5.81e-1  3    3.0  1.5   1    3.0e3    2.0e3   0   0.0   1
1525.0   3.0    0       0       1   1.67      4.01e-1  1    3.0  1.5   1    1.5e4    2.0e3   1   0.0   1
1569.0   7.0    0       0       0   1.21e1    2.86     0    3.0  1.5   1    3.3e4    5.0e3   2   0.0   1
1658.0   7.0    0       0       0   1.63e2    3.49e1   0    3.0  1.5   1    5.5e4    1.0e4   2   0.0   1
1728.0   4.0    0       0       0   6.30e2    1.22e2   0    3.0  1.5   1    3.5e4    5.0e3   2   0.0   1
****************************************************************************************************************
Upper Limits of Resonances
Note: enter partial width upper limit by chosing non-zero value for PT, where PT=<theta^2> for particles and...
Note: ...PT=<B> for g-rays [enter: "upper_limit 0.0"]; for each resonance: # upper limits < # open channels!   
Ecm      DEcm   Jr  G1        DG1  L1  PT     G2      DG2     L2  PT  G3       DG3     L3  PT   Exf   Int
191.08   0.15   1   1.25e-23  0    1   0.01   3.0     1.5     1   0   0        0       0   0    0.0   0
334.31   0.1    1   1.20e-9   0    1   0.01   3.0     1.5     1   0   0        0       0   0    0.0   0
328.21   2.0    7   3.70e-23  0    7   0.01   3.0     1.5     1   0   0        0       0   0    0.0   0
497.38   0.08   2   9.28e-12  0    2   0.01   1.73    3.1e-2  1   0   2.58e3   2.40e1  0   0    0.0   1
548.16   0.10   2   8.74e-8   0    2   0.01   4.56    2.9e-1  1   0   4.64e3   1.00e2  1   0    0.0   1
556.28   0.16   2   1.25e-7   0    2   0.01   3.0     1.5     1   0   1.44     1.6e-1  2   0    0.0   0
568.27   0.19   1   2.08e-7   0    1   0.01   3.0     1.5     1   0   5.4e-1   8.8e-2  1   0    0.0   0
628.43   0.10   2   9.46e-7   0    2   0.01   7.42    6.0e-1  1   0   4.51e3   1.07e2  1   0    0.0   1
659.32   0.12   2   9.97e-7   0    2   0.01   3.24    3.5e-1  1   0   5.4e2    5.4e1   0   0    0.0   1
665.11   0.11   4   9.16e-8   0    4   0.01   5.9e-1  2.4e-1  1   0   1.51e3   3.4e1   1   0    0.0   1
670.81   0.13   1   1.69e-6   0    1   0.01   7.9e-1  4.6e-1  1   0   1.26e3   1.0e2   1   0    0.0   1
671.59   0.12   2   1.02e-6   0    2   0.01   4.26    6.0e-1  1   0   1.28e1   6.0     2   0    0.0   1
674.36   0.25   2   1.02e-6   0    2   0.01   3.0     1.5     1   0   1.54     4.6e-1  1   0    0.0   0
681.21   0.13   3   7.37e-7   0    3   0.01   3.31    7.3e-1  1   0   8.06e3   1.2e2   1   0    0.0   1
695.95   0.35   1   1.76e-6   0    1   0.01   3.0     1.5     1   0   1.12     4.0e-1  1   0    0.0   0
711.34   0.54   1   1.80e-6   0    1   0.01   3.0     1.5     1   0   6.0e-1   3.2e-1  1   0    0.0   0
713.40   0.14   1   1.81e-6   0    1   0.01   3.63    4.7e-1  1   0   4.2e2    8.6e1   1   0    0.0   1
714.34   0.55   1   1.81e-6   0    1   0.01   3.0     1.5     1   0   2.8      1.0     1   0    0.0   0
722.12   0.56   1   1.83e-6   0    1   0.01   3.0     1.5     1   0   1.42     5.6e-1  1   0    0.0   0
729.15   0.15   2   1.11e-6   0    2   0.01   1.18    2.7e-1  1   0   1.53e2   4.2e1   1   0    0.0   1
730.03   0.16   4   6.16e-7   0    4   0.01   1.82    3.8e-1  1   0   4.13e3   1.9e2   3   0    0.0   1
777.78   0.16   5   1.51e-7   0    5   0.01   3.0     1.5     1   0   2.9e2    1.9e1   2   0    0.0   1
****************************************************************************************************************
Interference between Resonances [numerical integration only]
Note: + for positive, - for negative interference; +- if interference sign is unknown
Ecm    DEcm   Jr    G1      DG1   L1   PT      G2     DG2     L2  PT   G3    DG3    L3  PT   Exf  Int
!+- 
0.0    0.0    0.0   0.0     0.0   0    0       0.0    0.0     0   0    0.0   0.0    0   0    0.0  0
0.0    0.0    0.0   0.0     0.0   0    0       0.0    0.0     0   0    0.0   0.0    0   0    0.0  0
****************************************************************************************************************
Reaction Rate and PDF at NT selected temperatures only
Note: default values are used for reaction rate range if Min=Max=0.0
T9    Min   Max
0.01  0.0   0.0
0.1   0.0   0.0
****************************************************************************************************************
Comments:
1. Reaction Rates using recent FEL results
2. The doublet state at ~330 keV has been included twice as upper limits for the spectroscopic factor
3. The 703 keV resonance is treated as the same as seen in 22Ne(a,n)

\end{verbatim}
\normalsize
\vspace{5mm}

\begin{figure}[ht]
\centering
\includegraphics[scale=0.5]{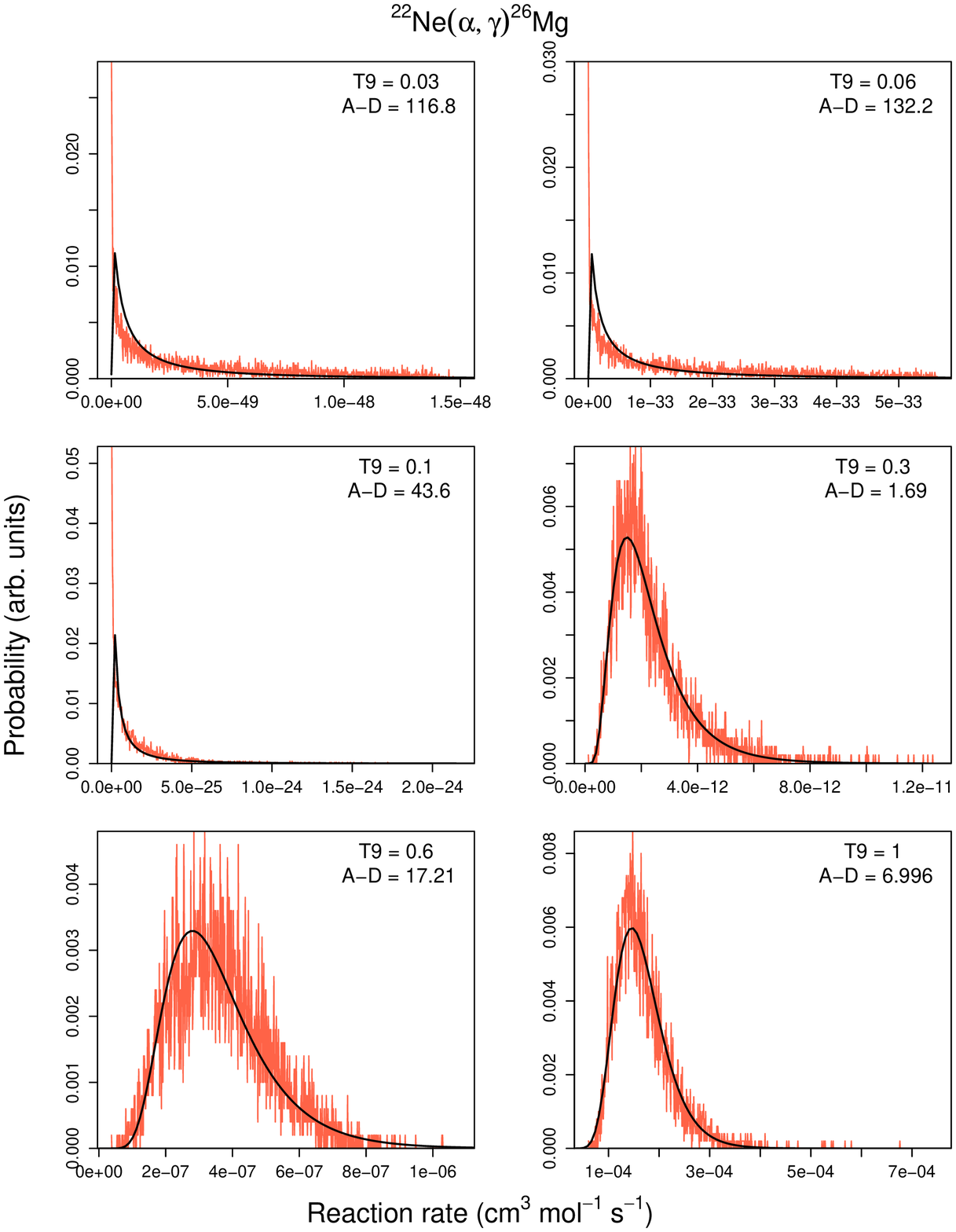}
\label{22ne_ag}
\end{figure}
\begin{figure}[ht]
\centering
\includegraphics[scale=0.5]{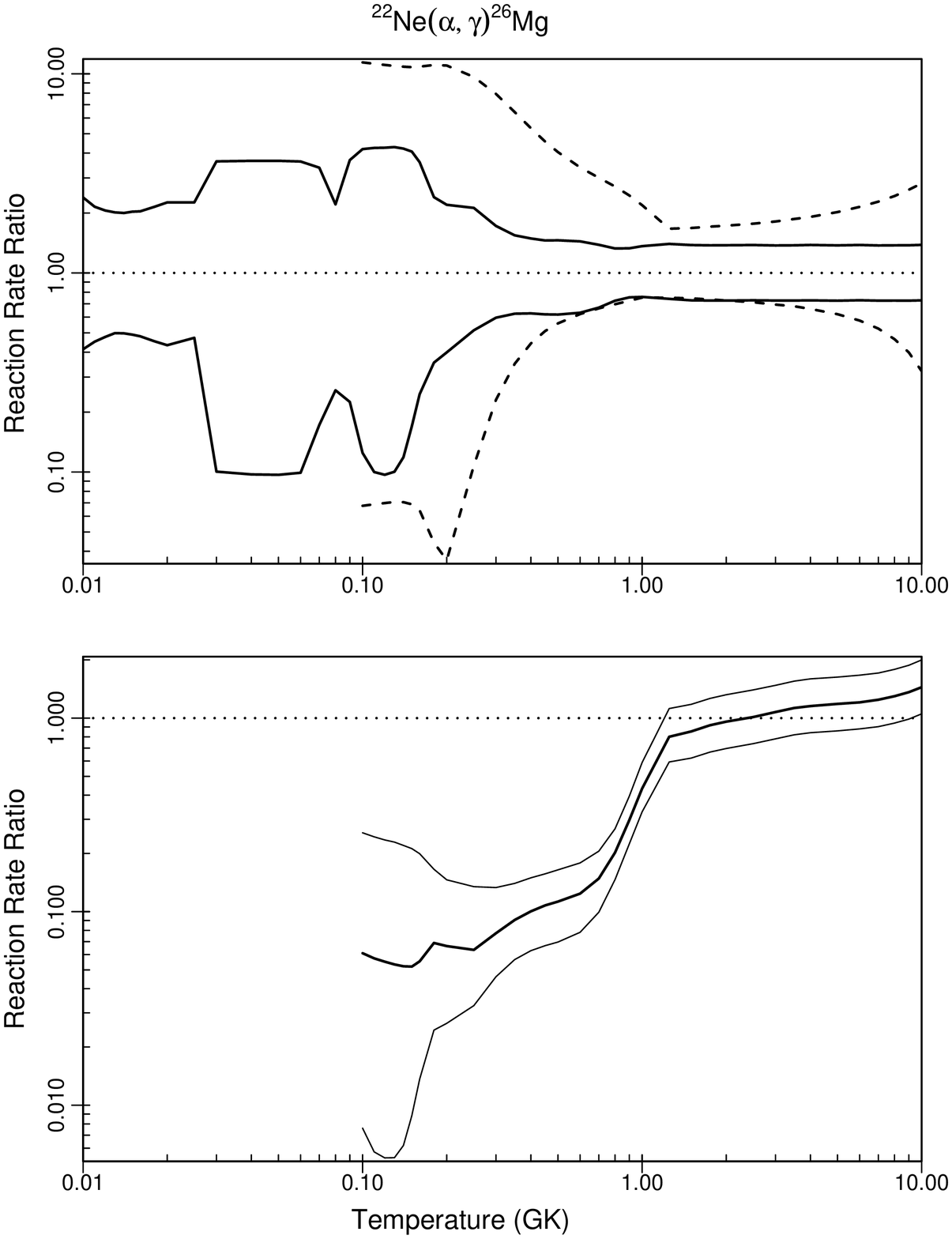}
\label{22ne_ag}
\end{figure}
\clearpage

\setlongtables
\begin{longtable}{cccc | ccc}
\caption{Total thermonuclear reaction rates for $^{22}$Ne($\alpha,n$)$^{25}$Mg.}  \label{tab:ne22an} \\
\hline \hline 
	\multicolumn{1}{c}{T (GK)} & \multicolumn{1}{c}{Low rate} & \multicolumn{1}{c}{Median rate} & \multicolumn{1}{c}{High rate}  & \multicolumn{1}{c}{lognormal $\mu$} & \multicolumn{1}{c}{lognormal $\sigma$} & \multicolumn{1}{c}{A-D} \\ \hline 
\endfirsthead
\multicolumn{6}{c}{{\tablename} \thetable{} -- continued} \\
\hline \hline 
	\multicolumn{1}{c}{T (GK)} & \multicolumn{1}{c}{Low rate} & \multicolumn{1}{c}{Median rate} & \multicolumn{1}{c}{High rate} & \multicolumn{1}{c}{lognormal $\mu$} & \multicolumn{1}{c}{lognormal $\sigma$} & \multicolumn{1}{c}{A-D} \\ \hline 
\endhead
	 \hline \hline
\endfoot
	\hline \hline
\endlastfoot
0.010 &  0.00$\times$10$^{+00}$  &  0.00$\times$10$^{+00}$  &
       0.00$\times$10$^{+00}$  &  -5.708$\times$10$^{+02}$  &
       1.86$\times$10$^{+00}$  &  6.14$\times$10$^{+01}$  \\
0.011 &   0.00$\times$10$^{+00}$  &  0.00$\times$10$^{+00}$  &
       0.00$\times$10$^{+00}$  &  -5.202$\times$10$^{+02}$  &
       1.86$\times$10$^{+00}$  &  6.10$\times$10$^{+01}$  \\
0.012 &   0.00$\times$10$^{+00}$  &  0.00$\times$10$^{+00}$  &
       0.00$\times$10$^{+00}$  &  -4.781$\times$10$^{+02}$  &
       1.85$\times$10$^{+00}$  &  6.06$\times$10$^{+01}$  \\
0.013 &   0.00$\times$10$^{+00}$  &  0.00$\times$10$^{+00}$  &
       0.00$\times$10$^{+00}$ &  -4.424$\times$10$^{+02}$  &
       1.85$\times$10$^{+00}$  &  6.06$\times$10$^{+01}$  \\
0.014 &   0.00$\times$10$^{+00}$  &  0.00$\times$10$^{+00}$  &
       0.00$\times$10$^{+00}$  &  -4.118$\times$10$^{+02}$  &
       1.85$\times$10$^{+00}$  &  6.04$\times$10$^{+01}$  \\
0.015 &   0.00$\times$10$^{+00}$  &  0.00$\times$10$^{+00}$  &
       0.00$\times$10$^{+00}$  &  -3.853$\times$10$^{+02}$  &
       1.85$\times$10$^{+00}$  &  6.02$\times$10$^{+01}$  \\
0.016 &   0.00$\times$10$^{+00}$  &  0.00$\times$10$^{+00}$  &
       0.00$\times$10$^{+00}$  &  -3.621$\times$10$^{+02}$  &
       1.85$\times$10$^{+00}$  &  6.00$\times$10$^{+01}$  \\
0.018 &   0.00$\times$10$^{+00}$  &  0.00$\times$10$^{+00}$  &
       0.00$\times$10$^{+00}$  &  -3.235$\times$10$^{+02}$  &
       1.85$\times$10$^{+00}$  &  6.00$\times$10$^{+01}$  \\
0.020 &   0.00$\times$10$^{+00}$  &  0.00$\times$10$^{+00}$  &
       0.00$\times$10$^{+00}$  &  -2.925$\times$10$^{+02}$  &
       1.86$\times$10$^{+00}$  &  6.03$\times$10$^{+01}$  \\
0.025 &   0.00$\times$10$^{+00}$  &  0.00$\times$10$^{+00}$  &
       0.00$\times$10$^{+00}$  &  -2.365$\times$10$^{+02}$  &
       1.87$\times$10$^{+00}$  &  6.28$\times$10$^{+01}$  \\
0.030 &  5.12$\times$10$^{-88}$  &  5.08$\times$10$^{-87}$  &
       2.25$\times$10$^{-86}$  &  -1.991$\times$10$^{+02}$  &
       1.90$\times$10$^{+00}$  &  6.65$\times$10$^{+01}$  \\
0.040 &  1.46$\times$10$^{-67}$  &  1.49$\times$10$^{-66}$  &
       6.64$\times$10$^{-66}$  &  -1.519$\times$10$^{+02}$  &
       1.94$\times$10$^{+00}$  &  7.20$\times$10$^{+01}$  \\
0.050 &  2.99$\times$10$^{-55}$  &  3.05$\times$10$^{-54}$  &
       1.36$\times$10$^{-53}$  &  -1.236$\times$10$^{+02}$  &
       1.95$\times$10$^{+00}$  &  7.29$\times$10$^{+01}$  \\
0.060 &  4.92$\times$10$^{-47}$  &  4.87$\times$10$^{-46}$  &
       2.17$\times$10$^{-45}$  &  -1.047$\times$10$^{+02}$  &
       1.92$\times$10$^{+00}$  &  6.81$\times$10$^{+01}$  \\
0.070 &  3.70$\times$10$^{-41}$  &  3.48$\times$10$^{-40}$  &
       1.55$\times$10$^{-39}$  &  -9.117$\times$10$^{+01}$  &
       1.84$\times$10$^{+00}$  &  5.72$\times$10$^{+01}$  \\
0.080 &  1.03$\times$10$^{-36}$  &  8.44$\times$10$^{-36}$  &
       3.73$\times$10$^{-35}$  &  -8.101$\times$10$^{+01}$  &
       1.74$\times$10$^{+00}$  &  4.36$\times$10$^{+01}$  \\
0.090 &  3.23$\times$10$^{-33}$  &  2.19$\times$10$^{-32}$  &
       9.43$\times$10$^{-32}$  &  -7.309$\times$10$^{+01}$  &
       1.62$\times$10$^{+00}$  &  3.11$\times$10$^{+01}$  \\
0.100 &  2.17$\times$10$^{-30}$  &  1.20$\times$10$^{-29}$  &
       4.92$\times$10$^{-29}$  &  -6.673$\times$10$^{+01}$  &
       1.50$\times$10$^{+00}$  &  2.12$\times$10$^{+01}$  \\
0.110 &  4.65$\times$10$^{-28}$  &  2.12$\times$10$^{-27}$  &
       8.22$\times$10$^{-27}$  &  -6.151$\times$10$^{+01}$  &
       1.39$\times$10$^{+00}$  &  1.43$\times$10$^{+01}$  \\
0.120 &  4.24$\times$10$^{-26}$  &  1.62$\times$10$^{-25}$  &
       5.82$\times$10$^{-25}$  &  -5.714$\times$10$^{+01}$  &
       1.29$\times$10$^{+00}$  &  9.86$\times$10$^{+00}$  \\
0.130 &  1.94$\times$10$^{-24}$  &  6.61$\times$10$^{-24}$  &
       2.14$\times$10$^{-23}$  &  -5.342$\times$10$^{+01}$  &
       1.19$\times$10$^{+00}$  &  6.93$\times$10$^{+00}$  \\
0.140 &  5.27$\times$10$^{-23}$  &  1.64$\times$10$^{-22}$  &
       4.81$\times$10$^{-22}$  &  -5.020$\times$10$^{+01}$  &
       1.08$\times$10$^{+00}$  &  4.64$\times$10$^{+00}$  \\
0.150 &  9.94$\times$10$^{-22}$  &  2.74$\times$10$^{-21}$  &
       7.18$\times$10$^{-21}$  &  -4.737$\times$10$^{+01}$  &
       9.62$\times$10$^{-01}$  &  2.96$\times$10$^{+00}$  \\
0.160 &  1.43$\times$10$^{-20}$  &  3.39$\times$10$^{-20}$  &
       7.89$\times$10$^{-20}$  &  -4.484$\times$10$^{+01}$  &
       8.29$\times$10$^{-01}$  &  3.27$\times$10$^{+00}$  \\
0.180 &  1.61$\times$10$^{-18}$  &  2.74$\times$10$^{-18}$  &
       5.01$\times$10$^{-18}$  &  -4.040$\times$10$^{+01}$  &
       5.53$\times$10$^{-01}$  &  1.26$\times$10$^{+01}$  \\
0.200 &  9.14$\times$10$^{-17}$  &  1.24$\times$10$^{-16}$  &
       1.79$\times$10$^{-16}$  &  -3.660$\times$10$^{+01}$  &
       3.43$\times$10$^{-01}$  &  1.12$\times$10$^{+01}$  \\
0.250 &  1.68$\times$10$^{-13}$  &  2.06$\times$10$^{-13}$  &
       2.53$\times$10$^{-13}$  &  -2.921$\times$10$^{+01}$  &
       2.06$\times$10$^{-01}$  &  3.49$\times$10$^{-01}$  \\
0.300 &  2.74$\times$10$^{-11}$  &  3.36$\times$10$^{-11}$  &
       4.15$\times$10$^{-11}$  &  -2.411$\times$10$^{+01}$  &
       2.06$\times$10$^{-01}$  &  4.10$\times$10$^{-01}$  \\
0.350 &  1.05$\times$10$^{-09}$  &  1.29$\times$10$^{-09}$  &
       1.59$\times$10$^{-09}$  &  -2.046$\times$10$^{+01}$  &
       2.05$\times$10$^{-01}$  &  4.68$\times$10$^{-01}$  \\
0.400 &  1.64$\times$10$^{-08}$  &  2.00$\times$10$^{-08}$  &
       2.45$\times$10$^{-08}$  &  -1.773$\times$10$^{+01}$  &
       1.99$\times$10$^{-01}$  &  5.25$\times$10$^{-01}$  \\
0.450 &  1.42$\times$10$^{-07}$  &  1.71$\times$10$^{-07}$  &
       2.07$\times$10$^{-07}$  &  -1.558$\times$10$^{+01}$  &
       1.88$\times$10$^{-01}$  &  6.99$\times$10$^{-01}$  \\
0.500 &  8.51$\times$10$^{-07}$  &  1.00$\times$10$^{-06}$  &
       1.19$\times$10$^{-06}$  &  -1.381$\times$10$^{+01}$  &
       1.68$\times$10$^{-01}$  &  1.13$\times$10$^{+00}$  \\
0.600 &  1.74$\times$10$^{-05}$  &  1.92$\times$10$^{-05}$  &
       2.15$\times$10$^{-05}$  &  -1.085$\times$10$^{+01}$  &
       1.07$\times$10$^{-01}$  &  2.15$\times$10$^{+00}$  \\
0.700 &  2.36$\times$10$^{-04}$  &  2.51$\times$10$^{-04}$  &
       2.69$\times$10$^{-04}$  &  -8.287$\times$10$^{+00}$  &
       6.70$\times$10$^{-02}$  &  1.13$\times$10$^{+00}$  \\
0.800 &  2.15$\times$10$^{-03}$  &  2.27$\times$10$^{-03}$  &
       2.42$\times$10$^{-03}$  &  -6.084$\times$10$^{+00}$  &
       5.79$\times$10$^{-02}$  &  3.45$\times$10$^{+00}$  \\
0.900 &  1.36$\times$10$^{-02}$  &  1.43$\times$10$^{-02}$  &
       1.51$\times$10$^{-02}$  &  -4.246$\times$10$^{+00}$  &
       5.33$\times$10$^{-02}$  &  4.07$\times$10$^{+00}$  \\
1.000 &  6.34$\times$10$^{-02}$  &  6.64$\times$10$^{-02}$  &
       6.98$\times$10$^{-02}$  &  -2.711$\times$10$^{+00}$  &
       4.82$\times$10$^{-02}$  &  3.59$\times$10$^{+00}$  \\
1.250 &  1.18$\times$10$^{+00}$  &  1.22$\times$10$^{+00}$  &
       1.27$\times$10$^{+00}$  &  1.998$\times$10$^{-01}$  &
       3.88$\times$10$^{-02}$  &  1.63$\times$10$^{+00}$  \\
1.500 & (1.09$\times$10$^{+01}$) & (1.14$\times$10$^{+01}$) &
      (1.18$\times$10$^{+01}$) & (2.431$\times$10$^{+00}$) &
      (3.89$\times$10$^{-02}$) &  \\
1.750 & (6.79$\times$10$^{+01}$) & (7.06$\times$10$^{+01}$) &
      (7.34$\times$10$^{+01}$) & (4.257$\times$10$^{+00}$) &
      (3.89$\times$10$^{-02}$) &  \\
2.000 & (2.92$\times$10$^{+02}$) & (3.04$\times$10$^{+02}$) &
      (3.16$\times$10$^{+02}$) & (5.717$\times$10$^{+00}$) &
      (3.89$\times$10$^{-02}$) &  \\
2.500 & (2.74$\times$10$^{+03}$) & (2.85$\times$10$^{+03}$) &
      (2.96$\times$10$^{+03}$) & (7.953$\times$10$^{+00}$) &
      (3.89$\times$10$^{-02}$) &  \\
3.000 & (1.41$\times$10$^{+04}$) & (1.46$\times$10$^{+04}$) &
      (1.52$\times$10$^{+04}$) & (9.590$\times$10$^{+00}$) &
      (3.89$\times$10$^{-02}$) &  \\
3.500 & (4.96$\times$10$^{+04}$) & (5.16$\times$10$^{+04}$) &
      (5.37$\times$10$^{+04}$) & (1.085$\times$10$^{+01}$) &
      (3.89$\times$10$^{-02}$) &  \\
4.000 & (1.36$\times$10$^{+05}$) & (1.41$\times$10$^{+05}$) &
      (1.47$\times$10$^{+05}$) & (1.186$\times$10$^{+01}$) &
      (3.89$\times$10$^{-02}$) &  \\
5.000 & (6.10$\times$10$^{+05}$) & (6.34$\times$10$^{+05}$) &
      (6.59$\times$10$^{+05}$) & (1.336$\times$10$^{+01}$) &
      (3.89$\times$10$^{-02}$) &  \\
6.000 & (1.80$\times$10$^{+06}$) & (1.88$\times$10$^{+06}$) &
      (1.95$\times$10$^{+06}$) & (1.444$\times$10$^{+01}$) &
      (3.89$\times$10$^{-02}$) &  \\
7.000 & (4.07$\times$10$^{+06}$) & (4.23$\times$10$^{+06}$) &
      (4.40$\times$10$^{+06}$) & (1.526$\times$10$^{+01}$) &
      (3.89$\times$10$^{-02}$) &  \\
8.000 & (7.70$\times$10$^{+06}$) & (8.01$\times$10$^{+06}$) &
      (8.32$\times$10$^{+06}$) & (1.590$\times$10$^{+01}$) &
      (3.89$\times$10$^{-02}$) &  \\
9.000 & (1.28$\times$10$^{+07}$) & (1.33$\times$10$^{+07}$) &
      (1.39$\times$10$^{+07}$) & (1.640$\times$10$^{+01}$) &
      (3.89$\times$10$^{-02}$) &  \\
10.000 & (1.97$\times$10$^{+07}$) & (2.04$\times$10$^{+07}$) &
      (2.12$\times$10$^{+07}$) & (1.683$\times$10$^{+01}$) &
      (3.89$\times$10$^{-02}$) &  \\

\end{longtable}
\subsection{$^{22}$Ne($\alpha,n)^{25}$Mg}
Comments:  see Appendix~\ref{22neag} for details on the new nuclear data for this reaction.
\footnotesize
\begin{verbatim}
22Ne(a,n)25Mg
****************************************************************************************************************
2               ! Zproj                                                             
10              ! Ztarget                                                           
0               ! Zexitparticle (=0 when only 2 channels open)                      
4.003           ! Aproj                                                             
21.991          ! Atarget                                                           
1.009           ! Aexitparticle (=0 when only 2 channels open)                      
0.0             ! Jproj                                                             
0.0             ! Jtarget                                                           
0.5             ! Jexitparticle (=0 when only 2 channels open)                      
10614.78        ! projectile separation energy (keV)                                
11093.08        ! exit particle separation energy (=0 when only 2 channels open)    
1.25            ! Radius parameter R0 (fm)                                          
3               ! Gamma-ray channel number (=2 if ejectile is a g-ray; =3 otherwise)
****************************************************************************************************************
1.0             ! Minimum energy for numerical integration (keV)
5000            ! Number of random samples (>5000 for better statistics)
0               ! =0 for rate output at all temperatures; =NT for rate output at selected temperatures
****************************************************************************************************************
Nonresonant Contribution
S(keVb)  S'(b)   S''(b/keV)   fracErr   Cutoff Energy (keV)
0.0      0.0     0.0          0.0       0.0
0.0      0.0     0.0          0.0       0.0
****************************************************************************************************************
Resonant Contribution
Note: G1 = entrance channel, G2 = exit channel, G3 = spectator channel !! Ecm, Exf in (keV); wg, Gx in (eV) !!
Note: if Er<0, theta^2=C2S*theta_sp^2 must be entered instead of entrance channel partial width
Ecm      DEcm   wg      Dwg      J   G1        DG1      L1  G2        DG2       L2  G3  DG3   L3  Exf   Int
703.78   2.11   0       0        2   2.36e-5   2.2e-6   2   2.5e2     1.7e2     2   3   1.5   1   0.0   1
826.04   0.19   0       0        4   3.78e-6   4.4e-7   4   1.47e3    8.0e1     4   3   1.5   1   0.0   1
850.44   0.21   0       0        5   4.36e-6   9.1e-7   5   6.55e3    9.0e1     5   3   1.5   1   0.0   1
893.31   0.90   0       0        1   1.17e-4   2.0e-5   1   1.27e4    2.5e3     1   3   1.5   1   0.0   1
911.16   1.69   0       0        1   2.77e-4   2.3e-5   1   1.80e3    9.0e2     1   3   1.5   1   0.0   1
1015.22  1.69   0       0        1   2.83e-3   3.3e-4   1   1.35e4    1.7e3     1   3   1.5   1   0.0   1
1133.66  8.46   0       0        1   2.0e-2    3.0e-3   1   6.35e4    8.5e3     1   3   1.5   1   0.0   1
1171.74  3.38   0       0        1   1.67e-2   2.3e-3   1   2.45e4    2.4e3     1   3   1.5   1   0.0   1
1213.19  2.34   0       0        2   2.13e-1   8.4e-3   2   1.10e3    2.5e2     2   3   1.5   1   0.0   1
1247.88  2.54   0       0        1   1.5e-2    1.0e-2   1   2.45e4    3.4e3     1   3   1.5   1   0.0   1
1264.80  2.54   3.9e-1  5.7e-2   1   0         0        1   0         0         1   0   0     1   0.0   0
1275.80  2.54   5.6e-1  6.0e-2   1   0         0        1   0         0         1   0   0     1   0.0   0
1295.25  2.54   1.5     1.6e-1   1   0         0        1   0         0         1   0   0     1   0.0   0
1336.71  2.54   2.9     3.0e-1   1   0         0        1   0         0         1   0   0     1   0.0   0
1437.38  2.54   6.0     7.7e-1   1   0         0        1   0         0         1   0   0     1   0.0   0
1499.99  4.23   1.0     2.4e-1   1   0         0        1   0         0         1   0   0     1   0.0   0
1526.22  2.54   3.0     3.4e-1   1   0         0        1   0         0         1   0   0     1   0.0   0
1569.36  6.77   9.0e-1  2.1e-1   1   0         0        1   0         0         1   0   0     1   0.0   0
1649.74  8.46   3.1e+1  8.5      1   0         0        1   0         0         1   0   0     1   0.0   0
1730.95  6.77   2.0e+2  3.3e+1   1   0         0        1   0         0         1   0   0     1   0.0   0
1820.63  8.46   2.8e+1  7.0      1   0         0        1   0         0         1   0   0     1   0.0   0
1936.54  12.69  1.2e+2  4.5e+1   1   0         0        1   0         0         1   0   0     1   0.0   0
****************************************************************************************************************
Upper Limits of Resonances
Note: enter partial width upper limit by chosing non-zero value for PT, where PT=<theta^2> for particles and...
Note: ...PT=<B> for g-rays [enter: "upper_limit 0.0"]; for each resonance: # upper limits < # open channels!   
Ecm      DEcm   Jr  G1        DG1 L1  PT     G2       DG2      L2   PT   G3       DG3       L3  PT  Exf   Int
497.38   0.08   2   9.28e-12  0   2   0.01   2.58e3   2.4e1    0    0    1.73     3e-2      1   0   0.0   1
548.16   0.10   2   3.80e-8   0   2   0.01   4.64e3   1.0e2    1    0    4.56     0.29      1   0   0.0   1
556.28   0.16   2   1.50e-8   0   2   0.01   1.44     1.6e-1   2    0    3.0      1.5       1   0   0.0   0
568.27   0.19   1   2.08e-7   0   1   0.01   5.40e-1  8.8e-2   1    0    3.0      1.5       1   0   0.0   0
628.43   0.10   2   2.40e-8   0   2   0.01   4.51e3   1.1e2    1    0    7.42     0.60      1   0   0.0   1
659.32   0.12   2   2.20e-8   0   2   0.01   5.40e2   5.4e1    0    0    3.24     0.35      1   0   0.0   1
665.11   0.11   4   1.44e-8   0   4   0.01   1.51e3   3.4e1    1    0    5.9e-1   2.4e-1    1   0   0.0   1
670.81   0.13   1   2.57e-8   0   1   0.01   1.26e3   1.0e2    1    0    7.9e-1   4.6e-1    1   0   0.0   1
671.59   0.12   2   1.54e-8   0   2   0.01   1.28e1   6.0      2    0    4.26     0.60      1   0   0.0   1
674.36   0.25   2   1.54e-8   0   2   0.01   1.54     0.46     1    0    3.0      1.5       1   0   0.0   0
681.21   0.13   3   1.43e-8   0   3   0.01   8.06e3   1.2e2    1    0    3.31     0.73      1   0   0.0   1
695.95   0.35   1   5.34e-9   0   1   0.01   1.12     0.40     1    0    3.0      1.5       1   0   0.0   0
711.34   0.54   1   4.11e-8   0   1   0.01   6.0e-1   3.2e-1   1    0    3.0      1.5       1   0   0.0   0
713.40   0.14   1   1.67e-7   0   1   0.01   4.24e2   8.6e1    1    0    3.63     0.47      1   0   0.0   1
714.34   0.55   1   4.12e-8   0   1   0.01   2.8      1.0      1    0    3.0      1.5       1   0   0.0   0
722.12   0.56   1   4.17e-8   0   1   0.01   1.42     0.56     1    0    3.0      1.5       1   0   0.0   0
729.15   0.15   2   4.00e-8   0   2   0.01   1.53e2   4.2e1    1    0    1.18     0.27      1   0   0.0   1
730.03   0.16   4   4.68e-9   0   4   0.01   4.13e3   1.9e2    3    0    1.82     0.38      1   0   0.0   1
777.78   0.16   5   3.34e-9   0   5   0.01   2.90e2   1.9e1    2    0    3.0      1.5       1   0   0.0   1
****************************************************************************************************************
Interference between Resonances [numerical integration only]
Note: + for positive, - for negative interference; +- if interference sign is unknown
Ecm    DEcm   Jr    G1      DG1   L1   PT      G2     DG2     L2  PT   G3    DG3    L3  PT   Exf  Int
!+- 
0.0    0.0    0.0   0.0     0.0   0    0       0.0    0.0     0   0    0.0   0.0    0   0    0.0  0
0.0    0.0    0.0   0.0     0.0   0    0       0.0    0.0     0   0    0.0   0.0    0   0    0.0  0
****************************************************************************************************************
Reaction Rate and PDF at NT selected temperatures only
Note: default values are used for reaction rate range if Min=Max=0.0
T9    Min   Max
0.01  0.0   0.0
0.1   0.0   0.0
****************************************************************************************************************
Comments:
1. a,n measurements including results from recent FEL run
2. The 703 keV resonance is treated as the same as seen in 22Ne(a,g)

\end{verbatim}
\normalsize
\vspace{5mm}

\begin{figure}[ht]
\centering
\includegraphics[scale=0.5]{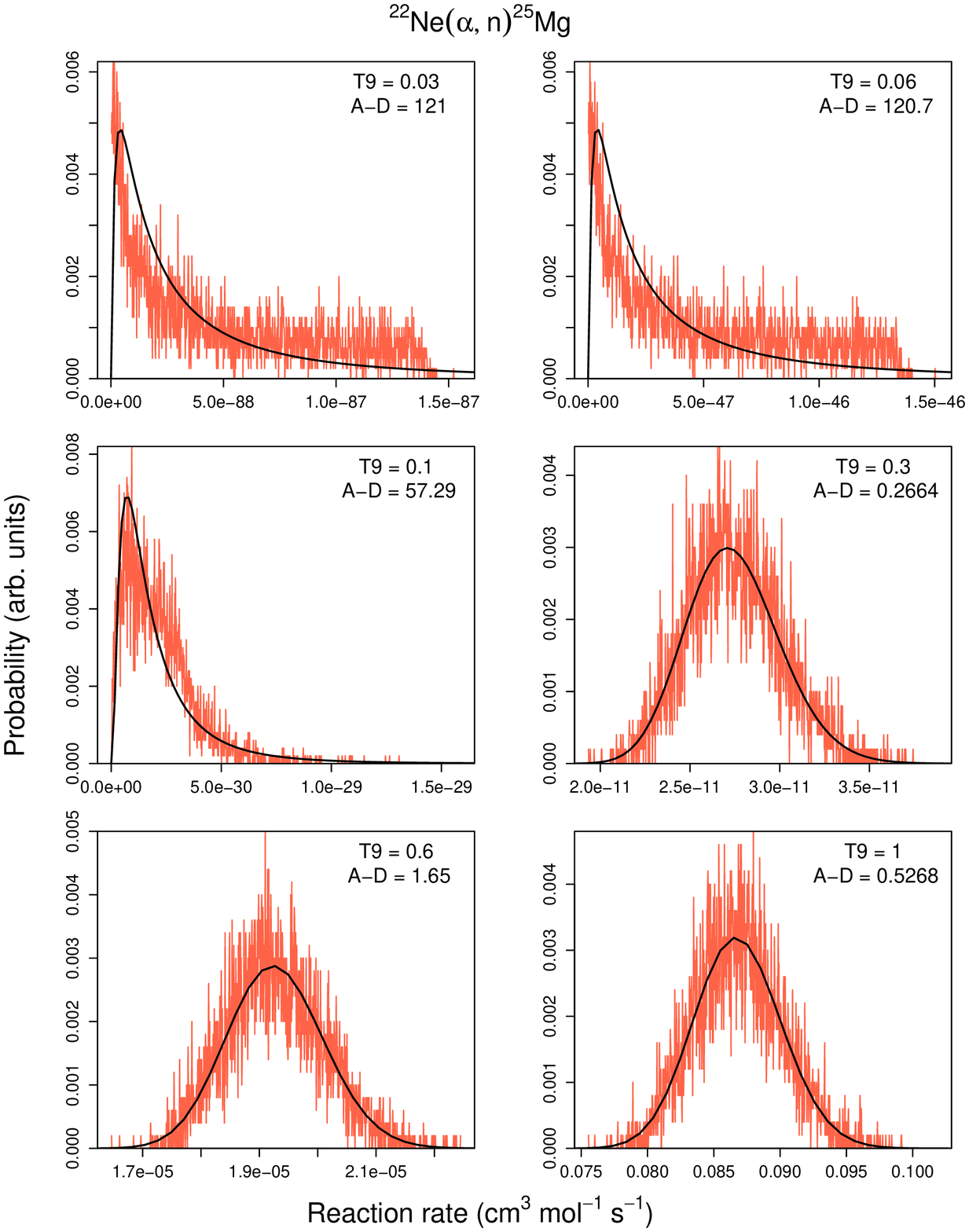}
\label{22ne_an}
\end{figure}
\begin{figure}[ht]
\centering
\includegraphics[scale=0.5]{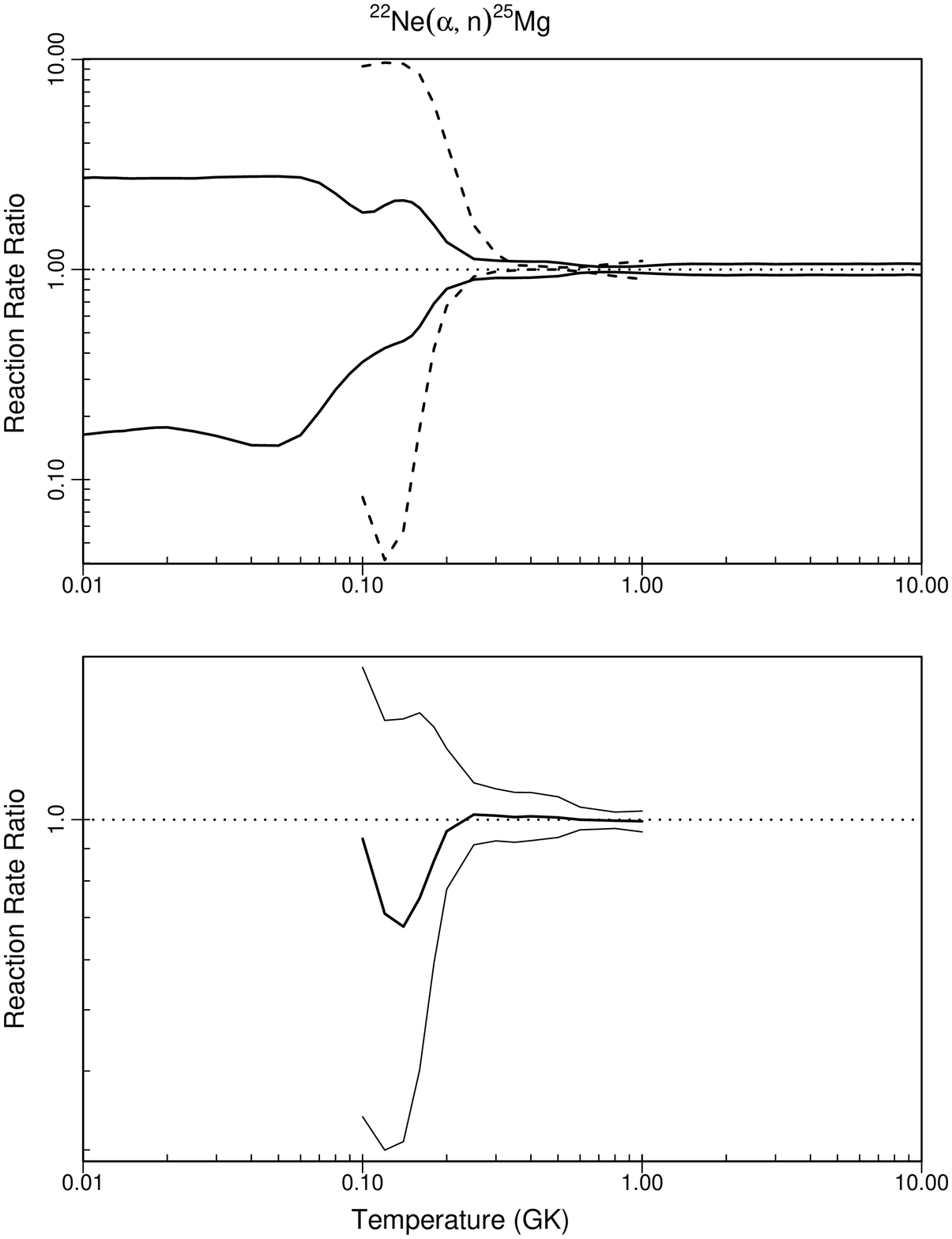}
\label{22ne_ag}
\end{figure}

\clearpage

\clearpage

\setlongtables
\begin{longtable}{cccc | ccc}
\caption{Total thermonuclear reaction rates for $^{22}$Na($p,\gamma$)$^{23}$Mg.}  \label{tab:na22pg} \\
\hline \hline 
	\multicolumn{1}{c}{T (GK)} & \multicolumn{1}{c}{Low rate} & \multicolumn{1}{c}{Median rate} & \multicolumn{1}{c}{High rate}  & \multicolumn{1}{c}{lognormal $\mu$} & \multicolumn{1}{c}{lognormal $\sigma$} & \multicolumn{1}{c}{A-D} \\ \hline 
\endfirsthead
\multicolumn{6}{c}{{\tablename} \thetable{} -- continued} \\
\hline \hline 
	\multicolumn{1}{c}{T (GK)} & \multicolumn{1}{c}{Low rate} & \multicolumn{1}{c}{Median rate} & \multicolumn{1}{c}{High rate} & \multicolumn{1}{c}{lognormal $\mu$} & \multicolumn{1}{c}{lognormal $\sigma$} & \multicolumn{1}{c}{A-D} \\ \hline 
\endhead
	 \hline \hline
\endfoot
	\hline \hline
\endlastfoot
0.010 &  1.22$\times$10$^{-30}$  &  2.52$\times$10$^{-30}$  &
       5.16$\times$10$^{-30}$  &  -6.816$\times$10$^{+01}$  &
       7.17$\times$10$^{-01}$  &  6.24$\times$10$^{-01}$  \\
0.011 &  1.05$\times$10$^{-28}$  &  1.96$\times$10$^{-28}$  &
       3.66$\times$10$^{-28}$  &  -6.380$\times$10$^{+01}$  &
       6.29$\times$10$^{-01}$  &  6.85$\times$10$^{-01}$  \\
0.012 &  4.20$\times$10$^{-27}$  &  7.33$\times$10$^{-27}$  &
       1.27$\times$10$^{-26}$  &  -6.018$\times$10$^{+01}$  &
       5.59$\times$10$^{-01}$  &  6.84$\times$10$^{-01}$  \\
0.013 &  9.35$\times$10$^{-26}$  &  1.57$\times$10$^{-25}$  &
       2.56$\times$10$^{-25}$  &  -5.712$\times$10$^{+01}$  &
       5.07$\times$10$^{-01}$  &  5.62$\times$10$^{-01}$  \\
0.014 &  1.33$\times$10$^{-24}$  &  2.13$\times$10$^{-24}$  &
       3.36$\times$10$^{-24}$  &  -5.451$\times$10$^{+01}$  &
       4.68$\times$10$^{-01}$  &  4.65$\times$10$^{-01}$  \\
0.015 &  1.30$\times$10$^{-23}$  &  2.03$\times$10$^{-23}$  &
       3.11$\times$10$^{-23}$  &  -5.226$\times$10$^{+01}$  &
       4.41$\times$10$^{-01}$  &  3.96$\times$10$^{-01}$  \\
0.016 &  9.55$\times$10$^{-23}$  &  1.44$\times$10$^{-22}$  &
       2.19$\times$10$^{-22}$  &  -5.029$\times$10$^{+01}$  &
       4.22$\times$10$^{-01}$  &  2.66$\times$10$^{-01}$  \\
0.018 &  2.55$\times$10$^{-21}$  &  3.76$\times$10$^{-21}$  &
       5.64$\times$10$^{-21}$  &  -4.702$\times$10$^{+01}$  &
       4.03$\times$10$^{-01}$  &  3.99$\times$10$^{-01}$  \\
0.020 &  3.47$\times$10$^{-20}$  &  5.10$\times$10$^{-20}$  &
       7.61$\times$10$^{-20}$  &  -4.442$\times$10$^{+01}$  &
       3.98$\times$10$^{-01}$  &  3.48$\times$10$^{-01}$  \\
0.025 &  4.19$\times$10$^{-18}$  &  6.06$\times$10$^{-18}$  &
       8.92$\times$10$^{-18}$  &  -3.965$\times$10$^{+01}$  &
       3.80$\times$10$^{-01}$  &  4.33$\times$10$^{-01}$  \\
0.030 &  1.54$\times$10$^{-16}$  &  2.15$\times$10$^{-16}$  &
       2.97$\times$10$^{-16}$  &  -3.608$\times$10$^{+01}$  &
       3.31$\times$10$^{-01}$  &  3.12$\times$10$^{-01}$  \\
0.040 &  3.03$\times$10$^{-14}$  &  4.34$\times$10$^{-14}$  &
       6.24$\times$10$^{-14}$  &  -3.077$\times$10$^{+01}$  &
       3.66$\times$10$^{-01}$  &  2.86$\times$10$^{-01}$  \\
0.050 &  8.55$\times$10$^{-13}$  &  1.33$\times$10$^{-12}$  &
       2.06$\times$10$^{-12}$  &  -2.735$\times$10$^{+01}$  &
       4.45$\times$10$^{-01}$  &  8.19$\times$10$^{-01}$  \\
0.060 &  8.33$\times$10$^{-12}$  &  1.37$\times$10$^{-11}$  &
       2.20$\times$10$^{-11}$  &  -2.502$\times$10$^{+01}$  &
       4.84$\times$10$^{-01}$  &  5.18$\times$10$^{-01}$  \\
0.070 &  1.47$\times$10$^{-10}$  &  1.95$\times$10$^{-10}$  &
       2.60$\times$10$^{-10}$  &  -2.236$\times$10$^{+01}$  &
       2.79$\times$10$^{-01}$  &  1.18$\times$10$^{+00}$  \\
0.080 &  5.29$\times$10$^{-09}$  &  6.79$\times$10$^{-09}$  &
       8.88$\times$10$^{-09}$  &  -1.880$\times$10$^{+01}$  &
       2.50$\times$10$^{-01}$  &  3.54$\times$10$^{+00}$  \\
0.090 &  1.14$\times$10$^{-07}$  &  1.43$\times$10$^{-07}$  &
       1.82$\times$10$^{-07}$  &  -1.575$\times$10$^{+01}$  &
       2.30$\times$10$^{-01}$  &  1.62$\times$10$^{+00}$  \\
0.100 &  1.35$\times$10$^{-06}$  &  1.66$\times$10$^{-06}$  &
       2.06$\times$10$^{-06}$  &  -1.330$\times$10$^{+01}$  &
       2.12$\times$10$^{-01}$  &  7.75$\times$10$^{-01}$  \\
0.110 &  1.01$\times$10$^{-05}$  &  1.23$\times$10$^{-05}$  &
       1.50$\times$10$^{-05}$  &  -1.131$\times$10$^{+01}$  &
       2.00$\times$10$^{-01}$  &  4.32$\times$10$^{-01}$  \\
0.120 &  5.34$\times$10$^{-05}$  &  6.45$\times$10$^{-05}$  &
       7.76$\times$10$^{-05}$  &  -9.650$\times$10$^{+00}$  &
       1.90$\times$10$^{-01}$  &  3.05$\times$10$^{-01}$  \\
0.130 & 2.17$\times$10$^{-04}$ & 2.61$\times$10$^{-04}$  &
       3.11$\times$10$^{-04}$  &  -8.254$\times$10$^{+00}$  &
       1.83$\times$10$^{-01}$  &  2.74$\times$10$^{-01}$  \\
0.140 &  7.21$\times$10$^{-04}$  &  8.59$\times$10$^{-04}$  &
       1.02$\times$10$^{-03}$  &  -7.062$\times$10$^{+00}$  &
       1.76$\times$10$^{-01}$  &  2.79$\times$10$^{-01}$  \\
0.150 &  2.03$\times$10$^{-03}$  &  2.40$\times$10$^{-03}$  &
       2.83$\times$10$^{-03}$  &  -6.032$\times$10$^{+00}$  &
       1.71$\times$10$^{-01}$  &  3.29$\times$10$^{-01}$  \\
0.160 &  5.01$\times$10$^{-03}$  &  5.90$\times$10$^{-03}$  &
       6.94$\times$10$^{-03}$  &  -5.133$\times$10$^{+00}$  &
       1.66$\times$10$^{-01}$  &  3.93$\times$10$^{-01}$  \\
0.180 &  2.25$\times$10$^{-02}$  &  2.63$\times$10$^{-02}$  &
       3.07$\times$10$^{-02}$  &  -3.639$\times$10$^{+00}$  &
       1.57$\times$10$^{-01}$  &  5.19$\times$10$^{-01}$  \\
0.200 &  7.49$\times$10$^{-02}$  &  8.67$\times$10$^{-02}$  &
       1.00$\times$10$^{-01}$  &  -2.445$\times$10$^{+00}$  &
       1.50$\times$10$^{-01}$  &  6.04$\times$10$^{-01}$  \\
0.250 &  6.56$\times$10$^{-01}$  &  7.46$\times$10$^{-01}$  &
       8.53$\times$10$^{-01}$  &  -2.911$\times$10$^{-01}$  &
       1.35$\times$10$^{-01}$  &  6.39$\times$10$^{-01}$  \\
0.300 &  2.81$\times$10$^{+00}$  &  3.16$\times$10$^{+00}$  &
       3.59$\times$10$^{+00}$  &  1.153$\times$10$^{+00}$  &
       1.26$\times$10$^{-01}$  &  6.20$\times$10$^{-01}$  \\
0.350 &  7.93$\times$10$^{+00}$  &  8.94$\times$10$^{+00}$  &
       1.01$\times$10$^{+01}$  &  2.191$\times$10$^{+00}$  &
       1.24$\times$10$^{-01}$  &  4.25$\times$10$^{-01}$  \\
0.400 &  1.73$\times$10$^{+01}$  &  1.96$\times$10$^{+01}$  &
       2.21$\times$10$^{+01}$  &  2.974$\times$10$^{+00}$  &
       1.24$\times$10$^{-01}$  &  4.83$\times$10$^{-01}$  \\
0.450 &  3.19$\times$10$^{+01}$  &  3.62$\times$10$^{+01}$  &
       4.08$\times$10$^{+01}$  &  3.587$\times$10$^{+00}$  &
       1.24$\times$10$^{-01}$  &  6.08$\times$10$^{-01}$  \\
0.500 &  5.25$\times$10$^{+01}$  &  5.94$\times$10$^{+01}$  &
       6.69$\times$10$^{+01}$  &  4.084$\times$10$^{+00}$  &
       1.24$\times$10$^{-01}$  &  6.73$\times$10$^{-01}$  \\
0.600 &  1.14$\times$10$^{+02}$  &  1.28$\times$10$^{+02}$  &
       1.44$\times$10$^{+02}$  &  4.852$\times$10$^{+00}$  &
       1.20$\times$10$^{-01}$  &  8.94$\times$10$^{-01}$  \\
0.700 &  2.05$\times$10$^{+02}$  &  2.28$\times$10$^{+02}$  &
       2.55$\times$10$^{+02}$  &  5.434$\times$10$^{+00}$  &
       1.11$\times$10$^{-01}$  &  1.24$\times$10$^{+00}$  \\
0.800 &  3.30$\times$10$^{+02}$  &  3.65$\times$10$^{+02}$  &
       4.04$\times$10$^{+02}$  &  5.902$\times$10$^{+00}$  &
       1.02$\times$10$^{-01}$  &  1.50$\times$10$^{+00}$  \\
0.900 &  4.92$\times$10$^{+02}$  &  5.39$\times$10$^{+02}$  &
       5.92$\times$10$^{+02}$  &  6.292$\times$10$^{+00}$  &
       9.32$\times$10$^{-02}$  &  1.57$\times$10$^{+00}$  \\
1.000 &  6.91$\times$10$^{+02}$  &  7.52$\times$10$^{+02}$  &
       8.19$\times$10$^{+02}$  &  6.624$\times$10$^{+00}$  &
       8.59$\times$10$^{-02}$  &  1.51$\times$10$^{+00}$  \\
1.250 & (1.40$\times$10$^{+03}$) & (1.52$\times$10$^{+03}$) &
      (1.65$\times$10$^{+03}$) & (7.327$\times$10$^{+00}$) &
      (8.01$\times$10$^{-02}$) &  \\
1.500 & (2.41$\times$10$^{+03}$) & (2.61$\times$10$^{+03}$) &
      (2.82$\times$10$^{+03}$) & (7.865$\times$10$^{+00}$) &
      (8.01$\times$10$^{-02}$) &  \\
1.750 & (3.57$\times$10$^{+03}$) & (3.87$\times$10$^{+03}$) &
      (4.20$\times$10$^{+03}$) & (8.262$\times$10$^{+00}$) &
      (8.01$\times$10$^{-02}$) &  \\
2.000 & (4.85$\times$10$^{+03}$) & (5.25$\times$10$^{+03}$) &
      (5.69$\times$10$^{+03}$) & (8.566$\times$10$^{+00}$) &
      (8.01$\times$10$^{-02}$) &  \\
2.500 & (7.53$\times$10$^{+03}$) & (8.15$\times$10$^{+03}$) &
      (8.83$\times$10$^{+03}$) & (9.006$\times$10$^{+00}$) &
      (8.01$\times$10$^{-02}$) &  \\
3.000 & (1.01$\times$10$^{+04}$) & (1.09$\times$10$^{+04}$) &
      (1.18$\times$10$^{+04}$) & (9.297$\times$10$^{+00}$) &
      (8.01$\times$10$^{-02}$) &  \\
3.500 & (1.24$\times$10$^{+04}$) & (1.35$\times$10$^{+04}$) &
      (1.46$\times$10$^{+04}$) & (9.507$\times$10$^{+00}$) &
      (8.01$\times$10$^{-02}$) &  \\
4.000 & (1.45$\times$10$^{+04}$) & (1.58$\times$10$^{+04}$) &
      (1.71$\times$10$^{+04}$) & (9.665$\times$10$^{+00}$) &
      (8.01$\times$10$^{-02}$) &  \\
5.000 & (1.81$\times$10$^{+04}$) & (1.97$\times$10$^{+04}$) &
      (2.13$\times$10$^{+04}$) & (9.886$\times$10$^{+00}$) &
      (8.01$\times$10$^{-02}$) &  \\
6.000 & (2.10$\times$10$^{+04}$) & (2.28$\times$10$^{+04}$) &
      (2.47$\times$10$^{+04}$) & (1.003$\times$10$^{+01}$) &
      (8.01$\times$10$^{-02}$) &  \\
7.000 & (2.33$\times$10$^{+04}$) & (2.52$\times$10$^{+04}$) &
      (2.73$\times$10$^{+04}$) & (1.013$\times$10$^{+01}$) &
      (8.01$\times$10$^{-02}$) &  \\
8.000 & (2.51$\times$10$^{+04}$) & (2.72$\times$10$^{+04}$) &
      (2.95$\times$10$^{+04}$) & (1.021$\times$10$^{+01}$) &
      (8.01$\times$10$^{-02}$) &  \\
9.000 & (2.66$\times$10$^{+04}$) & (2.88$\times$10$^{+04}$) &
      (3.12$\times$10$^{+04}$) & (1.027$\times$10$^{+01}$) &
      (8.01$\times$10$^{-02}$) &  \\
10.000 & (2.82$\times$10$^{+04}$) & (3.05$\times$10$^{+04}$) &
      (3.30$\times$10$^{+04}$) & (1.033$\times$10$^{+01}$) &
      (8.01$\times$10$^{-02}$) &  \\

\end{longtable}

\subsection{$^{22}$Na($p,\gamma)^{23}$Mg}\label{22naupdate}

Comments:  The input file from \citet{iliadis_3} has been updated with newly measured values for resonance strengths~\citep{2011PhRvC..83c4611S,2010PhRvL.105o2501S} and energies~\citep{2011PhRvC..83c4611S,2010PhRvL.105o2501S,2011PhRvC..83d5808S}.  A new Q value (7580.72 $\pm$ 0.71 keV) has also been adopted~\citep{ame2012}.  This Q-value was used to calculate proton energies from the excitation energies of \citet{jenkins} for resonances at 43, 66, and 189 keV, of \citet{endt98} for resonances at 480, 494, 612, 706, and 760 keV, and of \citet{per} and \citet{iacob} for the resonance at 221 keV (weighted average).  The energy analysis outlined in \citep{2011PhRvC..83c4611S,sallaska} was performed with the new Q value to extract proton energies for resonances at 204, 274, 434, and 583 keV, and the results changed by 0.1 keV or less.  These values were combined in a weighted average with proton energies calculated via excitation energies from \citet{2011PhRvC..83d5808S,jenkins}.  The value calculated from \citet{iacob} was also included for the resonance at 204 keV.  

All resonance strengths and proton partial widths, including upper limits, have been taken from the direct measurement of \citet{2011PhRvC..83c4611S,2010PhRvL.105o2501S}, with the exception of resonances at 43 and 66 keV.  For two resonances measured by Sallaska {\it et al.} (204 and 583 keV), the reported strengths have an inflated upper uncertainty to account for possible unobserved contributions to the rate.  To calculate the reaction rate with the present method, we simply take the symmetric, measured result.  Because of the slight energy change for these resonances, the penetration factors were recalculated, resulting in slightly different particle partial widths from \citet{iliadis_3}.  The resonance strengths not directly measured by Sallaska {\it et al.} but measured by \citet{seuthe} or \citet{stegmuller} have been scaled by a factor of 2.5, as discussed in \citet{2011PhRvC..83c4611S}.  This includes resonances at 480, 706, and 760 keV.  Although not quoted explicitly in \citet{2011PhRvC..83c4611S}, this methodology has also be applied to potential resonances at 493 and 612 keV, originally investigated in \citet{seuthe}.  For $T>1.0$ GK, rates were extrapolated using Hauser-Feshbach calculations.

The direct capture component has roughly been estimated in \citet{seuthe}.  As a test, S(0) was increased and decreased by a factor of 5, which changed the total rate by no more than 3\%.  The conclusion is the direct capture process is insignificant.  

\footnotesize
\begin{verbatim}
22Na(p,g)23Mg
****************************************************************************************************************
1               ! Zproj
11              ! Ztarget
0               ! Zexitparticle (=0 when only 2 channels open)
1.0078          ! Aproj		
21.9944         ! Atarget
0               ! Aexitparticle (=0 when only 2 channels open)
0.5             ! Jproj
3.0             ! Jtarget
0.0             ! Jexitparticle (=0 when only 2 channels open)
7580.72         ! projectile separation energy
0.0             ! exit particle separation energy (=0 when only 2 channels open)
1.25            ! Radius parameter R0 (fm)
2               ! Gamma-ray channel number (=2 if ejectile is a g-ray; =3 otherwise)
****************************************************************************************************************
1.0             ! Minimum energy for numerical integration (keV)
5000            ! Number of random samples (>5000 for better statistics)
0               ! =0 for rate output at all temperatures; =NT for rate output at selected temperatures
****************************************************************************************************************
Non-resonant contribution
S(keVb)  S'(b)   S''(b/keV)   fracErr   Cutoff Energy (keV)
1.8e1    0.0     0.0          0.4       1500.0
0.0      0.0     0.0          0.0       0.0
****************************************************************************************************************
Resonant Contribution
Note: G1 = entrance channel, G2 = exit channel, G3 = spectator channel !! Ecm, Exf in (keV); wg, Gx in (eV) !!
Note: if Er<0, theta^2=C2S*theta_sp^2 must be entered instead of entrance channel partial width
Ecm	DEcm	wg	Dwg	Jr	G1	DG1	L1    	G2      DG2      L2   G3  DG3  L3 Exf	Int
  42.7  1.1     0       0       4.5     7.8e-17 3.2e-17 2	     1.6e-1  0.8e-1   1    0   0    0  0.0   1
  66.2  2.7     0       0       1.5     1.6e-12 6.6e-13 2	     2.0e-2  1.0e-2   1    0   0    0  0.0   0
 204.4  0.9     5.7e-3  0.9e-3  0       0       0       0      0       0        0    0   0    0  0.0   0
 273.8  0.9     3.9e-2  0.8e-2  0       0       0       0      0       0        0    0   0    0  0.0   0
 434.4  0.7     1.7e-1  0.2e-1  0       0       0       0      0       0        0    0   0    0  0.0   0
 480.3  2.1     9.3e-2  3.6e-2  0       0       0       0      0       0        0    0   0    0  0.0   0
 582.5  0.7     5.9e-1  0.7e-1  0       0       0       0      0       0        0    0   0    0  0.0   0
 706.3  2.1     9.1e-1  1.7e-1  0       0       0       0      0       0        0    0   0    0  0.0   0
 760.3  2.1     2.4e-1  0.8e-1  0       0       0       0      0       0        0    0   0    0  0.0   0
****************************************************************************************************************
Upper Limits of Resonances
Note: enter partial width upper limit by chosing non-zero value for PT, where PT=<theta^2> for particles and...
Note: ...PT=<B> for g-rays [enter: "upper_limit 0.0"]; for each resonance: # upper limits < # open channels!  
Ecm	DEcm	Jr    G1     DG1  L1  PT     G2    DG2    L2  PT   G3    DG3  L3  PT  Exf     Int
 188.5  1.2     2.5   0.0012 0    0   0.0045 0.33  0.16   1   0    0     0    0   0   2715.0  0 
 221.3  1.6     2.5   0.0016 0    0   0.0045 0.2   0.1    1   0    0     0    0   0   0.0     0
 493.3  6.0     2.5   0.044  0    0   0.0045 0.2   0.1    1   0    0     0    0   0   0.0     0
 612.3  8.0     2.5   0.052  0    0   0.0045 0.2   0.1    1   0    0     0    0   0   0.0     0
****************************************************************************************************************
Interference between Resonances [numerical integration only]
Note: + for positive, - for negative interference; +- if interference sign is unknown
Ecm	DEcm	Jr    G1     DG1  L1   PT    G2    DG2    L2  PT   G3    DG3  L3  PT   Exf    
!+- 
0.0     0.0     0.0   0.0    0.0  0    0     0.0   0.0    0   0    0.0   0.0  0   0    0.0    
0.0     0.0     0.0   0.0    0.0  0    0     0.0   0.0    0   0    0.0   0.0  0   0    0.0    
****************************************************************************************************************
Reaction Rate and PDF at NT selected temperatures only
Note: default values are used for reaction rate range if Min=Max=0.0
T9	Min	Max
0.01	0.0	0.0
0.1	0.0	0.0
****************************************************************************************************************
Comments:
1. Gg=2.0e-2 eV for Er=66.6 keV resonance is a guess (not important since Gp<<Gg). 
2. Gg=2.0e-1 eV for Er=222, 494 and 614 keV resonances is a guess; we assume for these undetected resonances
   that Gp<<Gg (assumption most likely inconsequential for total rates). 
3. Spin and parity of Er=494 and 614 keV resonances unknown; for upper limit contributions we assume s-waves
   (Jp=5/2+).
4. Strengths taken directly from Sallaska et al. (2010, 2011), except at Er=43 and 66 keV; strengths at 480, 
   493, 612, 706, and 760 have been scaled up from Seuthe et al. (1990) by 2.5, per Sallaska et al. (2011).
5. Energies are from Sallaska et al. (2010, 2011), Perajarvi et al. (2000), Iacob et al. (2006), 
   Saastamoinen et al. (2011), Jenkins (2004), and Endt (1998), with a Q-value from Audi & Meng (2011). 

\end{verbatim}
\normalsize
\vspace{5mm}

\begin{figure}[ht]
\centering
\includegraphics[scale=0.5]{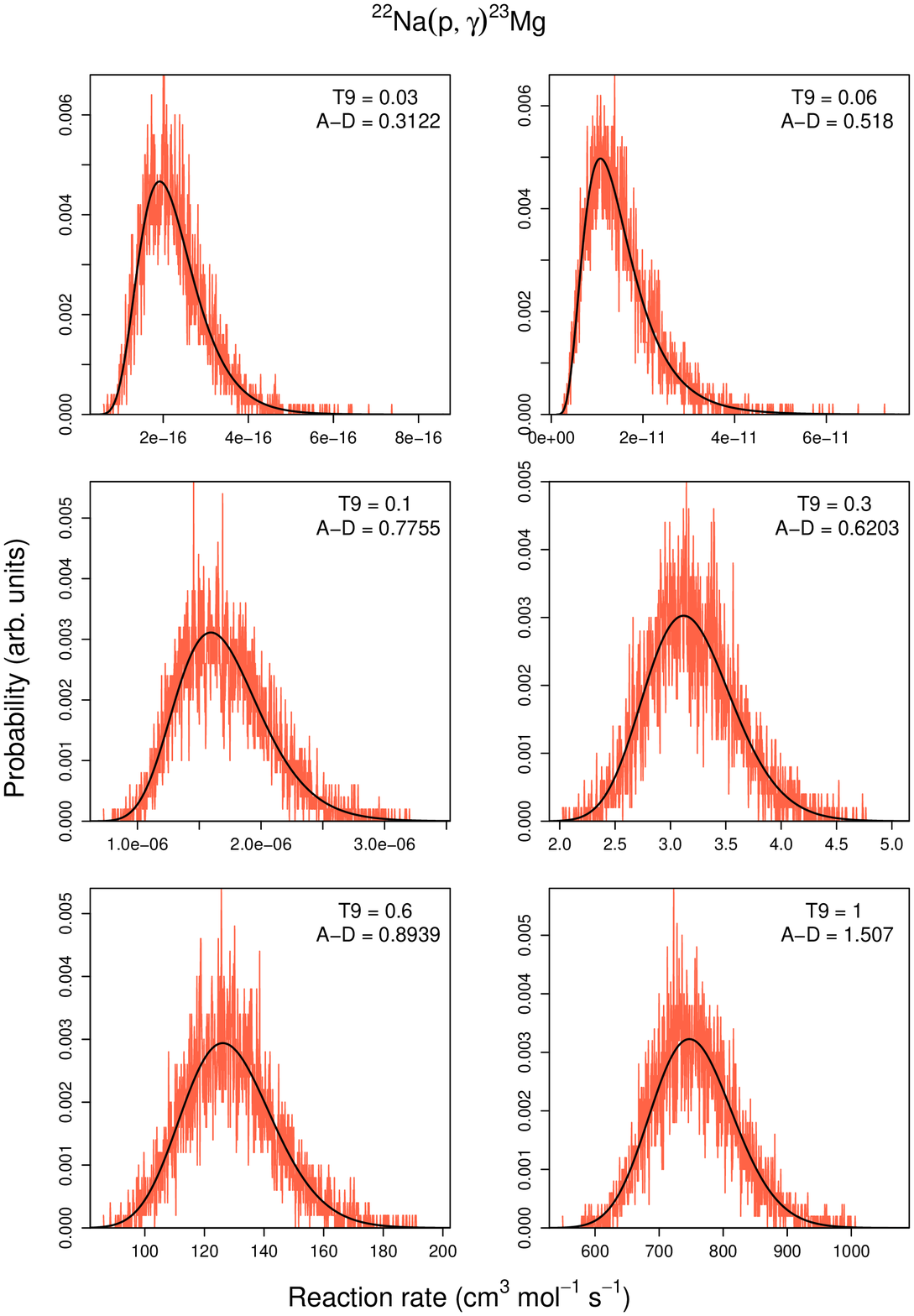}
\label{22ne_ag}
\end{figure}

\begin{figure}[ht]
\centering
\includegraphics[scale=0.5]{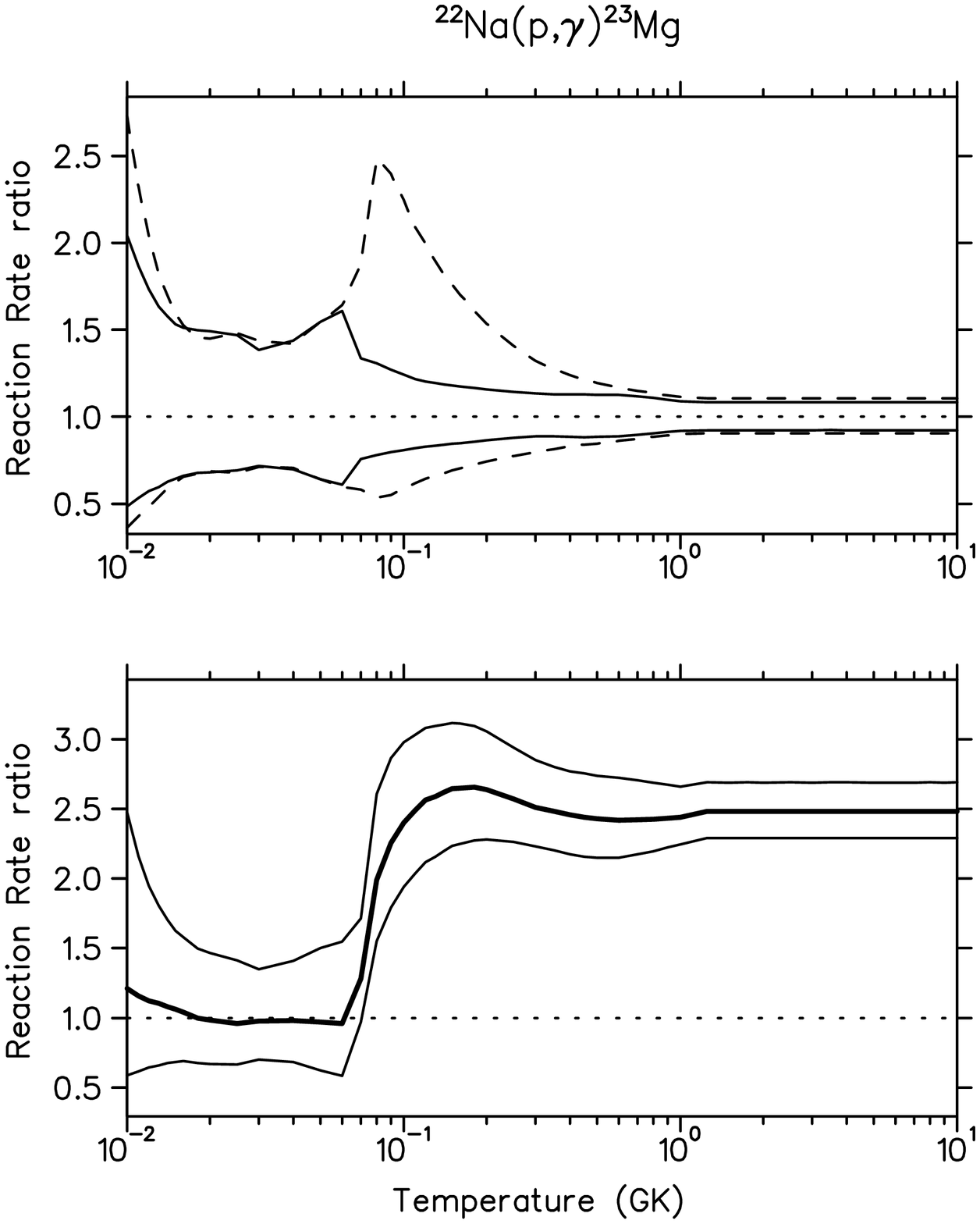}
\label{22ne_ag}
\end{figure}





\clearpage
\setlongtables
\begin{longtable}{cccc | ccc}
\caption{Total thermonuclear reaction rates for $^{29}$P(p,$\gamma$)$^{30}$S.}  \label{tab:p29pg} \\
\hline \hline 
	\multicolumn{1}{c}{T (GK)} & \multicolumn{1}{c}{Low rate} & \multicolumn{1}{c}{Median rate} & \multicolumn{1}{c}{High rate}  & \multicolumn{1}{c}{lognormal $\mu$} & \multicolumn{1}{c}{lognormal $\sigma$} & \multicolumn{1}{c}{A-D} \\ \hline 
\endfirsthead
\multicolumn{6}{c}{{\tablename} \thetable{} -- continued} \\
\hline \hline 
	\multicolumn{1}{c}{T (GK)} & \multicolumn{1}{c}{Low rate} & \multicolumn{1}{c}{Median rate} & \multicolumn{1}{c}{High rate} & \multicolumn{1}{c}{lognormal $\mu$} & \multicolumn{1}{c}{lognormal $\sigma$} & \multicolumn{1}{c}{A-D} \\ \hline 
\endhead
	 \hline \hline
\endfoot
	\hline \hline
\endlastfoot
0.010 &  4.57$\times$10$^{-42}$  &  6.75$\times$10$^{-42}$  &
       9.90$\times$10$^{-42}$  &  -9.480$\times$10$^{+01}$  &
       3.87$\times$10$^{-01}$  &  2.82$\times$10$^{-01}$  \\
0.011 &  1.79$\times$10$^{-40}$  &  2.60$\times$10$^{-40}$  &
       3.85$\times$10$^{-40}$  &  -9.114$\times$10$^{+01}$  &
       3.80$\times$10$^{-01}$  &  4.17$\times$10$^{-01}$  \\
0.012 &  4.50$\times$10$^{-39}$  &  6.70$\times$10$^{-39}$  &
       9.77$\times$10$^{-39}$  &  -8.790$\times$10$^{+01}$  &
       3.93$\times$10$^{-01}$  &  3.56$\times$10$^{-01}$  \\
0.013 &  8.19$\times$10$^{-38}$  &  1.21$\times$10$^{-37}$  &
       1.74$\times$10$^{-37}$  &  -8.501$\times$10$^{+01}$  &
       3.82$\times$10$^{-01}$  &  7.03$\times$10$^{-01}$  \\
0.014 &  1.09$\times$10$^{-36}$  &  1.64$\times$10$^{-36}$  &
       2.39$\times$10$^{-36}$  &  -8.241$\times$10$^{+01}$  &
       3.89$\times$10$^{-01}$  &  4.17$\times$10$^{-01}$  \\
0.015 &  1.19$\times$10$^{-35}$  &  1.73$\times$10$^{-35}$  &
       2.54$\times$10$^{-35}$  &  -8.004$\times$10$^{+01}$  &
       3.84$\times$10$^{-01}$  &  5.09$\times$10$^{-01}$  \\
0.016 &  1.03$\times$10$^{-34}$  &  1.52$\times$10$^{-34}$  &
       2.22$\times$10$^{-34}$  &  -7.787$\times$10$^{+01}$  &
       3.86$\times$10$^{-01}$  &  2.10$\times$10$^{-01}$  \\
0.018 &  4.79$\times$10$^{-33}$  &  7.01$\times$10$^{-33}$  &
       1.03$\times$10$^{-32}$  &  -7.404$\times$10$^{+01}$  &
       3.83$\times$10$^{-01}$  &  3.63$\times$10$^{-01}$  \\
0.020 &  1.30$\times$10$^{-31}$  &  1.90$\times$10$^{-31}$  &
       2.79$\times$10$^{-31}$  &  -7.074$\times$10$^{+01}$  &
       3.86$\times$10$^{-01}$  &  1.88$\times$10$^{-01}$  \\
0.025 &  9.70$\times$10$^{-29}$  &  1.44$\times$10$^{-28}$  &
       2.10$\times$10$^{-28}$  &  -6.412$\times$10$^{+01}$  &
       3.87$\times$10$^{-01}$  &  4.47$\times$10$^{-01}$  \\
0.030 &  1.51$\times$10$^{-26}$  &  2.22$\times$10$^{-26}$  &
       3.25$\times$10$^{-26}$  &  -5.907$\times$10$^{+01}$  &
       3.82$\times$10$^{-01}$  &  1.81$\times$10$^{-01}$  \\
0.040 &  2.34$\times$10$^{-23}$  &  3.39$\times$10$^{-23}$  &
       4.98$\times$10$^{-23}$  &  -5.173$\times$10$^{+01}$  &
       3.83$\times$10$^{-01}$  &  2.73$\times$10$^{-01}$  \\
0.050 &  4.28$\times$10$^{-21}$  &  6.25$\times$10$^{-21}$  &
       9.16$\times$10$^{-21}$  &  -4.652$\times$10$^{+01}$  &
       3.87$\times$10$^{-01}$  &  7.71$\times$10$^{-01}$  \\
0.060 &  2.22$\times$10$^{-19}$  &  3.29$\times$10$^{-19}$  &
       4.82$\times$10$^{-19}$  &  -4.256$\times$10$^{+01}$  &
       3.84$\times$10$^{-01}$  &  4.39$\times$10$^{-01}$  \\
0.070 &  5.70$\times$10$^{-18}$  &  8.23$\times$10$^{-18}$  &
       1.18$\times$10$^{-17}$  &  -3.934$\times$10$^{+01}$  &
       3.72$\times$10$^{-01}$  &  2.75$\times$10$^{-01}$  \\
0.080 &  1.50$\times$10$^{-16}$  &  2.04$\times$10$^{-16}$  &
       2.78$\times$10$^{-16}$  &  -3.613$\times$10$^{+01}$  &
       3.13$\times$10$^{-01}$  &  6.52$\times$10$^{-01}$  \\
0.090 &  5.96$\times$10$^{-15}$  &  8.77$\times$10$^{-15}$  &
       1.33$\times$10$^{-14}$  &  -3.235$\times$10$^{+01}$  &
       4.08$\times$10$^{-01}$  &  1.93$\times$10$^{+00}$  \\
0.100 &  1.87$\times$10$^{-13}$  &  2.82$\times$10$^{-13}$  &
       4.35$\times$10$^{-13}$  &  -2.889$\times$10$^{+01}$  &
       4.30$\times$10$^{-01}$  &  6.69$\times$10$^{-01}$  \\
0.110 &  3.39$\times$10$^{-12}$  &  5.09$\times$10$^{-12}$  &
       7.79$\times$10$^{-12}$  &  -2.600$\times$10$^{+01}$  &
       4.23$\times$10$^{-01}$  &  5.10$\times$10$^{-01}$  \\
0.120 &  3.80$\times$10$^{-11}$  &  5.65$\times$10$^{-11}$  &
       8.59$\times$10$^{-11}$  &  -2.359$\times$10$^{+01}$  &
       4.15$\times$10$^{-01}$  &  4.87$\times$10$^{-01}$  \\
0.130 & 2.91$\times$10$^{-10}$ & 4.32$\times$10$^{-10}$  &
       6.50$\times$10$^{-10}$  &  -2.156$\times$10$^{+01}$  &
       4.07$\times$10$^{-01}$  &  5.06$\times$10$^{-01}$  \\
0.140 &  1.66$\times$10$^{-09}$  &  2.45$\times$10$^{-09}$  &
       3.67$\times$10$^{-09}$  &  -1.982$\times$10$^{+01}$  &
       4.01$\times$10$^{-01}$  &  5.28$\times$10$^{-01}$  \\
0.150 &  7.49$\times$10$^{-09}$  &  1.10$\times$10$^{-08}$  &
       1.64$\times$10$^{-08}$  &  -1.832$\times$10$^{+01}$  &
       3.95$\times$10$^{-01}$  &  5.77$\times$10$^{-01}$  \\
0.160 &  2.78$\times$10$^{-08}$  &  4.08$\times$10$^{-08}$  &
       6.04$\times$10$^{-08}$  &  -1.701$\times$10$^{+01}$  &
       3.90$\times$10$^{-01}$  &  6.68$\times$10$^{-01}$  \\
0.180 &  2.49$\times$10$^{-07}$  &  3.58$\times$10$^{-07}$  &
       5.29$\times$10$^{-07}$  &  -1.483$\times$10$^{+01}$  &
       3.77$\times$10$^{-01}$  &  1.08$\times$10$^{+00}$  \\
0.200 &  1.46$\times$10$^{-06}$  &  2.07$\times$10$^{-06}$  &
       3.00$\times$10$^{-06}$  &  -1.308$\times$10$^{+01}$  &
       3.59$\times$10$^{-01}$  &  1.80$\times$10$^{+00}$  \\
0.250 &  3.98$\times$10$^{-05}$  &  5.31$\times$10$^{-05}$  &
       7.22$\times$10$^{-05}$  &  -9.833$\times$10$^{+00}$  &
       3.01$\times$10$^{-01}$  &  3.74$\times$10$^{+00}$  \\
0.300 &  4.16$\times$10$^{-04}$  &  5.33$\times$10$^{-04}$  &
       6.88$\times$10$^{-04}$  &  -7.532$\times$10$^{+00}$  &
       2.57$\times$10$^{-01}$  &  1.82$\times$10$^{+00}$  \\
0.350 &  2.39$\times$10$^{-03}$  &  3.03$\times$10$^{-03}$  &
       3.86$\times$10$^{-03}$  &  -5.798$\times$10$^{+00}$  &
       2.44$\times$10$^{-01}$  &  4.95$\times$10$^{-01}$  \\
0.400 &  9.07$\times$10$^{-03}$  &  1.16$\times$10$^{-02}$  &
       1.48$\times$10$^{-02}$  &  -4.458$\times$10$^{+00}$  &
       2.48$\times$10$^{-01}$  &  4.46$\times$10$^{-01}$  \\
0.450 &  2.60$\times$10$^{-02}$  &  3.31$\times$10$^{-02}$  &
       4.30$\times$10$^{-02}$  &  -3.401$\times$10$^{+00}$  &
       2.56$\times$10$^{-01}$  &  6.09$\times$10$^{-01}$  \\
0.500 &  6.03$\times$10$^{-02}$  &  7.73$\times$10$^{-02}$  &
       1.02$\times$10$^{-01}$  &  -2.553$\times$10$^{+00}$  &
       2.63$\times$10$^{-01}$  &  6.65$\times$10$^{-01}$  \\
0.600 &  2.11$\times$10$^{-01}$  &  2.75$\times$10$^{-01}$  &
       3.62$\times$10$^{-01}$  &  -1.287$\times$10$^{+00}$  &
       2.71$\times$10$^{-01}$  &  5.70$\times$10$^{-01}$  \\
0.700 &  5.17$\times$10$^{-01}$  &  6.73$\times$10$^{-01}$  &
       8.88$\times$10$^{-01}$  &  -3.918$\times$10$^{-01}$  &
       2.73$\times$10$^{-01}$  &  4.91$\times$10$^{-01}$  \\
0.800 &  1.01$\times$10$^{+00}$  &  1.31$\times$10$^{+00}$  &
       1.73$\times$10$^{+00}$  &  2.775$\times$10$^{-01}$  &
       2.69$\times$10$^{-01}$  &  4.67$\times$10$^{-01}$  \\
0.900 &  1.73$\times$10$^{+00}$  &  2.22$\times$10$^{+00}$  &
       2.90$\times$10$^{+00}$  &  8.021$\times$10$^{-01}$  &
       2.61$\times$10$^{-01}$  &  4.17$\times$10$^{-01}$  \\
1.000 &  2.66$\times$10$^{+00}$  &  3.41$\times$10$^{+00}$  &
       4.41$\times$10$^{+00}$  &  1.230$\times$10$^{+00}$  &
       2.52$\times$10$^{-01}$  &  3.43$\times$10$^{-01}$  \\
1.250 &  6.09$\times$10$^{+00}$  &  7.65$\times$10$^{+00}$  &
       9.62$\times$10$^{+00}$  &  2.037$\times$10$^{+00}$  &
       2.32$\times$10$^{-01}$  &  2.60$\times$10$^{-01}$  \\
1.500 &  1.10$\times$10$^{+01}$  &  1.36$\times$10$^{+01}$  &
       1.70$\times$10$^{+01}$  &  2.616$\times$10$^{+00}$  &
       2.20$\times$10$^{-01}$  &  3.78$\times$10$^{-01}$  \\
1.750 & (1.70$\times$10$^{+01}$) & (2.11$\times$10$^{+01}$) &
      (2.62$\times$10$^{+01}$) & (3.048$\times$10$^{+00}$) &
      (2.17$\times$10$^{-01}$) &  \\
2.000 & (2.48$\times$10$^{+01}$) & (3.08$\times$10$^{+01}$) &
      (3.82$\times$10$^{+01}$) & (3.426$\times$10$^{+00}$) &
      (2.17$\times$10$^{-01}$) &  \\
2.500 & (4.29$\times$10$^{+01}$) & (5.33$\times$10$^{+01}$) &
      (6.62$\times$10$^{+01}$) & (3.976$\times$10$^{+00}$) &
      (2.17$\times$10$^{-01}$) &  \\
3.000 & (6.33$\times$10$^{+01}$) & (7.86$\times$10$^{+01}$) &
      (9.75$\times$10$^{+01}$) & (4.364$\times$10$^{+00}$) &
      (2.17$\times$10$^{-01}$) &  \\
3.500 & (8.38$\times$10$^{+01}$) & (1.04$\times$10$^{+02}$) &
      (1.29$\times$10$^{+02}$) & (4.645$\times$10$^{+00}$) &
      (2.17$\times$10$^{-01}$) &  \\
4.000 & (1.05$\times$10$^{+02}$) & (1.31$\times$10$^{+02}$) &
      (1.63$\times$10$^{+02}$) & (4.875$\times$10$^{+00}$) &
      (2.17$\times$10$^{-01}$) &  \\
5.000 & (1.47$\times$10$^{+02}$) & (1.83$\times$10$^{+02}$) &
      (2.27$\times$10$^{+02}$) & (5.208$\times$10$^{+00}$) &
      (2.17$\times$10$^{-01}$) &  \\
6.000 & (1.88$\times$10$^{+02}$) & (2.34$\times$10$^{+02}$) &
      (2.90$\times$10$^{+02}$) & (5.454$\times$10$^{+00}$) &
      (2.17$\times$10$^{-01}$) &  \\
7.000 & (2.27$\times$10$^{+02}$) & (2.81$\times$10$^{+02}$) &
      (3.50$\times$10$^{+02}$) & (5.640$\times$10$^{+00}$) &
      (2.17$\times$10$^{-01}$) &  \\
8.000 & (2.63$\times$10$^{+02}$) & (3.27$\times$10$^{+02}$) &
      (4.06$\times$10$^{+02}$) & (5.789$\times$10$^{+00}$) &
      (2.17$\times$10$^{-01}$) &  \\
9.000 & (2.97$\times$10$^{+02}$) & (3.69$\times$10$^{+02}$) &
      (4.58$\times$10$^{+02}$) & (5.911$\times$10$^{+00}$) &
      (2.17$\times$10$^{-01}$) &  \\
10.000 & (3.36$\times$10$^{+02}$) & (4.17$\times$10$^{+02}$) &
      (5.18$\times$10$^{+02}$) & (6.034$\times$10$^{+00}$) &
      (2.17$\times$10$^{-01}$) &  \\

\end{longtable}

\subsection{$^{29}$P($p,\gamma)^{30}$S}
Comments:  The input file from \citet{iliadis_3} has been updated with new information on the energies of astrophysically important resonances~\citet{kiana2010,kiana2011}.  \citet{kiana2010} reaffirms the existence of a level predicted to be $3^+$ ($E_x = 4693(5)$ keV), previously observed in \citet{bard}, and measured the energy of the potential $2^+$ level ($E_x = 4814(3)$ keV) for the first time, decreasing the uncertainty in the energy estimated using the IMME from \citet{iliadis_3} by a factor of 10.  In our previous work~\citep{iliadis_3}, the $E_x=5288$ keV $(3^-)$ level of \citet{yokota} had been associated with the $E_x=5217$ keV level of \citet{fyn}.  The new measurements of \citet{kiana2010} indicate that these are separate levels, with energies of 5226(3) and 5318(4) keV.  Because of this development, it has been suggested~\citep{kiana2011} that the doublet level ($0^+$ and $4^+$) of the $E_x=5168$ keV of \citet{bard} is two separate levels with different energies.  The proposed assignments in \citet{kiana2011} leave a $3^+$ level missing; we therefore chose to associate the $0^+$ level with $E_x = 5127$ keV, leaving $E_x = 5168$ keV with a $4^+$ assignment (each with $C^2S \le 0.01$).  This allows the level at $E_x= 5318$ keV to be designated as $3^-$ (with $C^2S = 0.36$).  Other values for spins remain unchanged.  Spectroscopic factors are taken from \citet{mackh}.

Along with a new Q-value~\citep[4398.72 $\pm$ 3.06 keV from][]{ame2012}, all resonance energies were reevaluated.  Weighted averages of excitation energies were calculated for the resonances at 289 keV~\citep{kiana2010,bard,kiana2011}, 412 keV~\citep{kiana2010,kiana2011}, 734 keV~\citep{yokota,kuhl,kiana2011}, 819 keV~\citep{kiana2010,fyn,paddock}, 915 keV~\citep{yokota, kiana2010,paddock}, 993 keV~\citep{fyn,kiana2010,yokota,paddock}.  This was unnecessary for resonances at 769 keV~\citep{bard} and 1443 keV~\citep{fyn}.  All penetration factors were recalculated with the new resonance energies, and subsequent particle partial widths changed slightly from \citet{iliadis_3}.  Gamma-ray partial widths are also unchanged, although which level each is associated with changed for consistency with the above spin assignment changes.  For $T>1.5$ GK, rates were extrapolated using Hauser-Feshbach calculations.

The significance of the direct capture component was estimated through simulation.  Above 0.1 GK, the effect of increasing or decreasing S(0) by a factor of 5 is negligible; however, between 0.01 and 0.1 GK, the rate changes linearly with this factor.

\footnotesize
\begin{verbatim}
29P(p,g)30S
****************************************************************************************************************
1               ! Zproj
15              ! Ztarget
0               ! Zexitparticle (=0 when only 2 channels open)
1.0078          ! Aproj		
28.982          ! Atarget
0               ! Aexitparticle (=0 when only 2 channels open)
0.5             ! Jproj
0.5             ! Jtarget
0.0             ! Jexitparticle (=0 when only 2 channels open)
4398.72         ! projectile separation energy (keV)
0.0             ! exit particle separation energy (=0 when only 2 channels open)
1.25            ! Radius parameter R0 (fm)
2               ! Gamma-ray channel number (=2 if ejectile is a g-ray; =3 otherwise)
****************************************************************************************************************
1.0             ! Minimum energy for numerical integration (keV)
5000            ! Number of random samples (>5000 for better statistics)
0               ! =0 for rate output at all temperatures; =NT for rate output at selected temperatures
****************************************************************************************************************
Non-resonant contribution
S(keVb)	S'(b)		S''(b/keV)	fracErr	Cutoff Energy (keV)
7.3e1    -1.25e-2     0.0          0.4       1000.0
0.0      0.0          0.0          0.0       0.0
****************************************************************************************************************
Resonant Contribution
Note: G1 = entrance channel, G2 = exit channel, G3 = spectator channel !! Ecm, Exf in (keV); wg, Gx in (eV) !!
Note: if Er<0, theta^2=C2S*theta_sp^2 must be entered instead of entrance channel partial width
Ecm	 DEcm	 wg	 Dwg	J     G1	DG1	 L1    G2      DG2      L2   G3  DG3  L3 Exf	Int
 289.4   3.1     0       0      3     1.3e-5    5.3e-6   2     4.9e-3  2.5e-3   1    0   0    0  0.0    1
 411.8   3.1     0       0      2     3.3e-3    1.3e-3   2     4.2e-3  2.1e-3   1    0   0    0  0.0    1
 734.1   3.1     0       0      3     2.2e-1    8.8e-2   2     8.2e-3  4.1e-3   1    0   0    0  0.0    1
 915.0   4.8     0       0      3     1.1e0     4.4e-1   3     9.4e-3  4.7e-3   1    0   0    0  0.0    1
 992.5   3.5     0       0      2     6.4e0     2.6e0    2     0.01    5.0e-3   1    0   0    0  0.0    0
****************************************************************************************************************
Upper Limits of Resonances
Note: enter partial width upper limit by chosing non-zero value for PT, where PT=<theta^2> for particles and...
Note: ...PT=<B> for g-rays [enter: "upper_limit 0.0"]; for each resonance: # upper limits < # open channels!  
Ecm	DEcm	Jr   G1      DG1   L1   PT      G2      DG2     L2  PT   G3    DG3    L3  PT   Exf  Int
 769.3   6.7    4    3.5e-4  0.0   4    0.0045  5.5e-3  2.8e-3  1   0    0     0      0   0    0.0  1
 819.1   3.1    0    1.8e1   0.0   0    0.0045  7.7e-3  3.9e-3  1   0    0     0      0   0    0.0  1
1443.3   5.0    4    1.0e-1  0.0   4    0.0045  2.7e-2  1.4e-2  1   0    0     0      0   0    0.0  1
****************************************************************************************************************
Interference between Resonances [numerical integration only]
Note: + for positive, - for negative interference; +- if interference sign is unknown
Ecm	DEcm	Jr    G1    DG1   L1   PT    G2    DG2    L2  PT   G3    DG3    L3  PT   Exf  
!+- 
0.0     0.0     0.0   0.0   0.0   0    0     0.0   0.0    0   0    0.0   0.0    0   0    0.0  
0.0     0.0     0.0   0.0   0.0   0    0     0.0   0.0    0   0    0.0   0.0    0   0    0.0  
****************************************************************************************************************
Reaction Rate and PDF at NT selected temperatures only
Note: default values are used for reaction rate range if Min=Max=0.0
T9	Min	Max
0.01	0.0	0.0
0.1	0.0	0.0
****************************************************************************************************************
Comments:
1. Resonance energies calculated from excitation energies (Endt 1990, Tab. I of Bardayan et al. 2007,
   Setoodehnia et al. 2010, and Setoodehnia et al., 2011) and Q-value, except for Er=489 keV, which is based on
   the IMME (see Iliadis et al. 2001).
2. The spin-parity assignments are not unambiguous; they are based on experimental Jp restrictions (Endt 1990,
   Bardayan et al. 2007) and application of the IMME (Iliadis et al. 2001). 
3. Proton partial widths are computed using C2S of mirror states from (d,p) transfer experiment (Mackh et al. 
   1973).
4. Gamma-ray partial widths are computed from measured lifetimes of 30Si mirror states, except for Er=990 keV
   for which Gg=0.012 eV is a rough estimate (application of RULs to the known g-ray branchings of the mirror 
   state yields Gg<0.1 eV for this level).


   
\end{verbatim}
\normalsize
\vspace{5mm}

\begin{figure}[ht]
\centering
\includegraphics[scale=0.5]{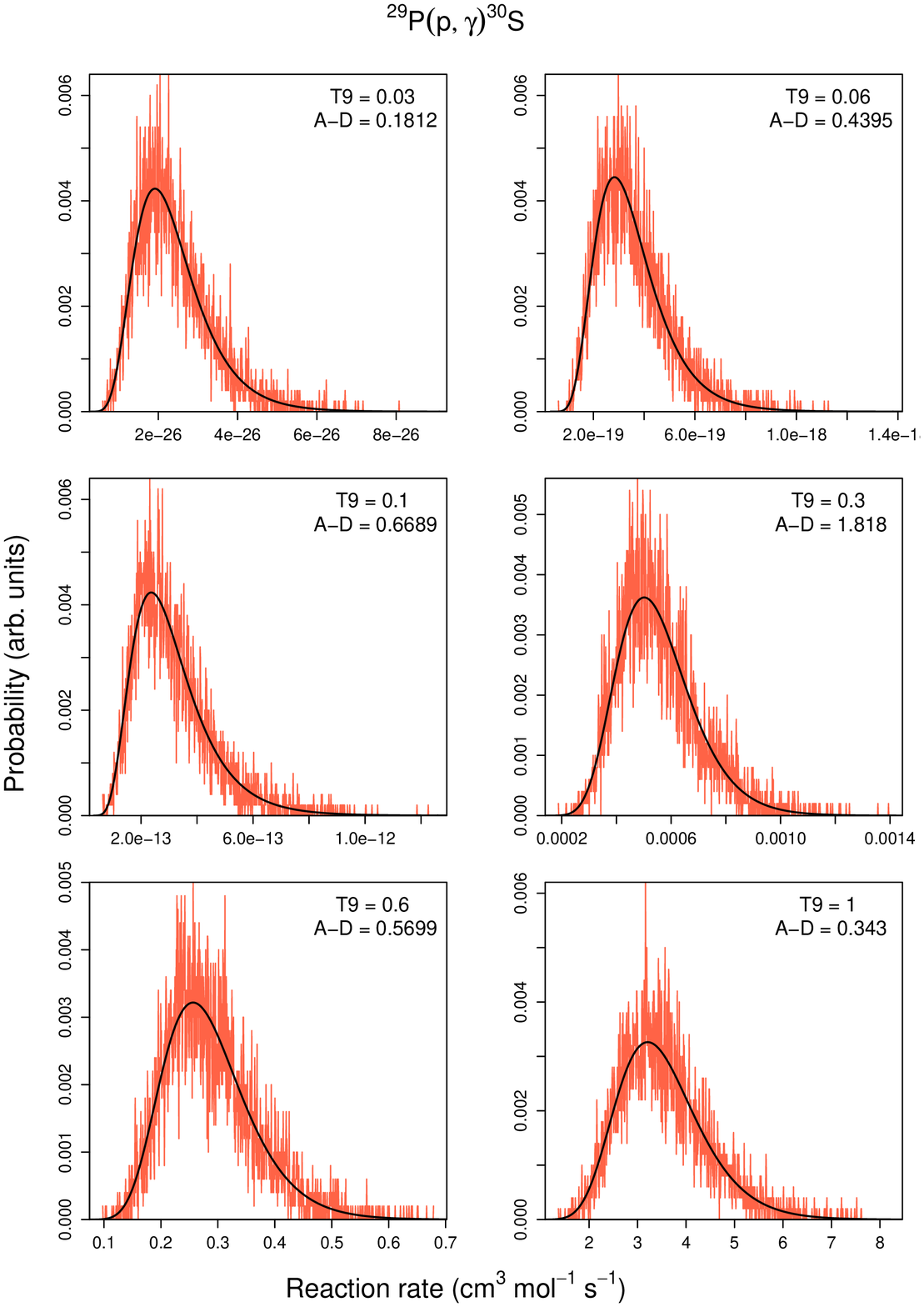}
\label{29p_ag}
\end{figure}
\begin{figure}[ht]
\centering
\includegraphics[scale=0.5]{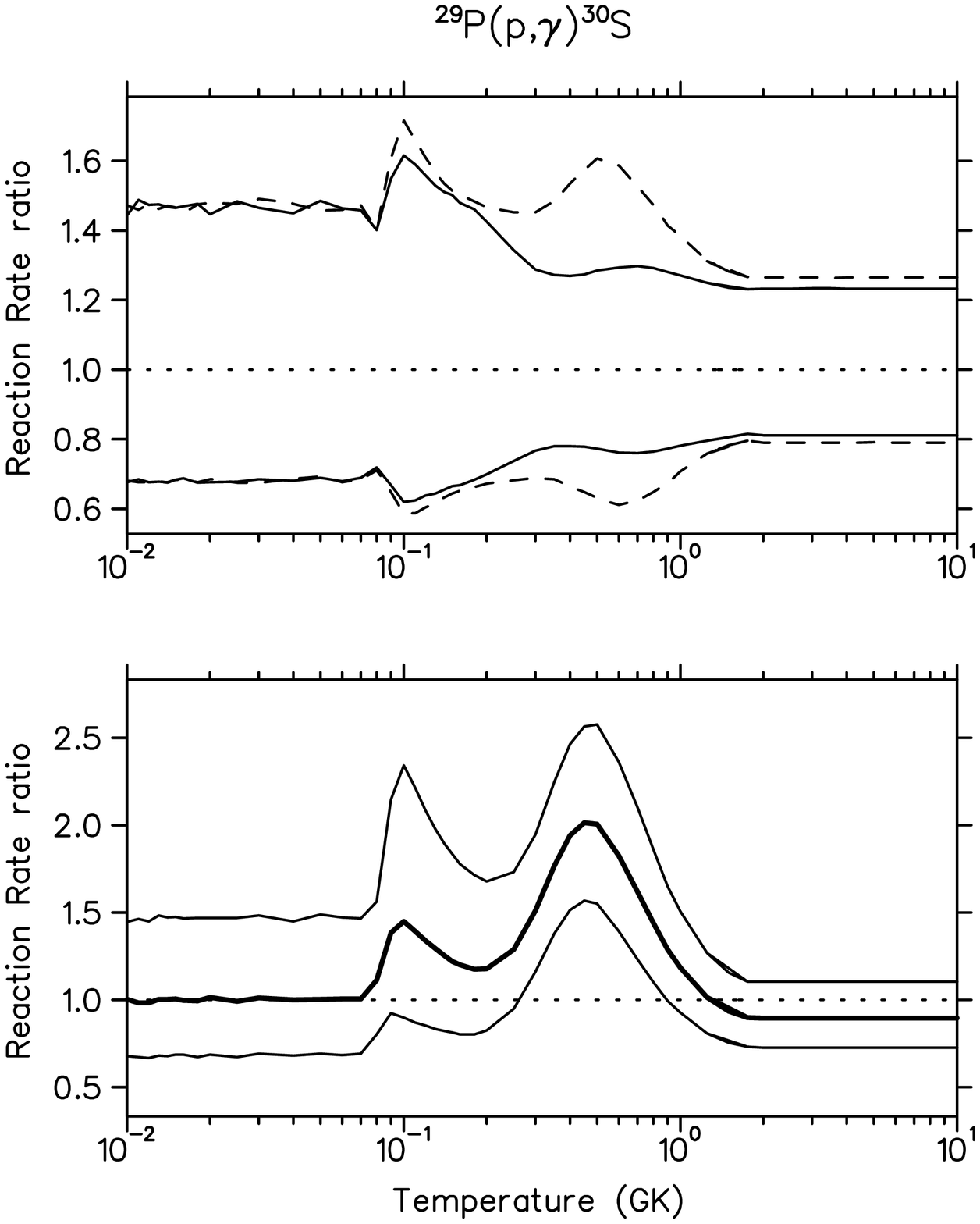}
\label{22ne_ag}
\end{figure}

\clearpage

\setlongtables
\begin{longtable}{cccc | ccc}
\caption{Total thermonuclear reaction rates for $^{38}$Ar(p,$\gamma$)$^{39}$K.}  \label{tab:ar38pg} \\
\hline \hline 
	\multicolumn{1}{c}{T (GK)} & \multicolumn{1}{c}{Low rate} & \multicolumn{1}{c}{Median rate} & \multicolumn{1}{c}{High rate}  & \multicolumn{1}{c}{lognormal $\mu$} & \multicolumn{1}{c}{lognormal $\sigma$} & \multicolumn{1}{c}{A-D} \\ \hline 
\endfirsthead
\multicolumn{6}{c}{{\tablename} \thetable{} -- continued} \\
\hline \hline 
	\multicolumn{1}{c}{T (GK)} & \multicolumn{1}{c}{Low rate} & \multicolumn{1}{c}{Median rate} & \multicolumn{1}{c}{High rate} & \multicolumn{1}{c}{lognormal $\mu$} & \multicolumn{1}{c}{lognormal $\sigma$} & \multicolumn{1}{c}{A-D} \\ \hline 
\endhead
	 \hline \hline
\endfoot
	\hline \hline
\endlastfoot

0.010 & 1.89$\times$10$^{-46}$  &  3.18$\times$10$^{-45}$  &
       2.20$\times$10$^{-44}$  &  -1.028$\times$10$^{+02}$  &
       2.28$\times$10$^{+00}$  &  3.11$\times$10$^{+01}$  \\
0.011 &  3.97$\times$10$^{-45}$  &  5.68$\times$10$^{-44}$  &
       4.43$\times$10$^{-43}$  &  -9.975$\times$10$^{+01}$  &
       2.16$\times$10$^{+00}$  &  1.64$\times$10$^{+01}$  \\
0.012 &  1.53$\times$10$^{-43}$  &  9.00$\times$10$^{-43}$  &
       5.92$\times$10$^{-42}$  &  -9.676$\times$10$^{+01}$  &
       1.76$\times$10$^{+00}$  &  5.80$\times$10$^{+00}$  \\
0.013 &  7.87$\times$10$^{-42}$  &  3.61$\times$10$^{-41}$  &
       1.53$\times$10$^{-40}$  &  -9.316$\times$10$^{+01}$  &
       1.46$\times$10$^{+00}$  &  2.19$\times$10$^{+00}$  \\
0.014 &  4.53$\times$10$^{-40}$  &  2.37$\times$10$^{-39}$  &
       1.02$\times$10$^{-38}$  &  -8.903$\times$10$^{+01}$  &
       1.55$\times$10$^{+00}$  &  3.75$\times$10$^{+00}$  \\
0.015 &  2.42$\times$10$^{-38}$  &  1.39$\times$10$^{-37}$  &
       6.04$\times$10$^{-37}$  &  -8.499$\times$10$^{+01}$  &
       1.61$\times$10$^{+00}$  &  9.28$\times$10$^{+00}$  \\
0.016 &  9.80$\times$10$^{-37}$  &  5.50$\times$10$^{-36}$  &
       2.21$\times$10$^{-35}$  &  -8.135$\times$10$^{+01}$  &
       1.58$\times$10$^{+00}$  &  1.55$\times$10$^{+01}$  \\
0.018 &  5.20$\times$10$^{-34}$  &  2.56$\times$10$^{-33}$  &
       8.77$\times$10$^{-33}$  &  -7.522$\times$10$^{+01}$  &
       1.48$\times$10$^{+00}$  &  2.43$\times$10$^{+01}$  \\
0.020 &  7.76$\times$10$^{-32}$  &  3.56$\times$10$^{-31}$  &
       1.11$\times$10$^{-30}$  &  -7.030$\times$10$^{+01}$  &
       1.40$\times$10$^{+00}$  &  3.31$\times$10$^{+01}$  \\
0.025 &  6.33$\times$10$^{-28}$  &  2.53$\times$10$^{-27}$  &
       7.08$\times$10$^{-27}$  &  -6.144$\times$10$^{+01}$  &
       1.30$\times$10$^{+00}$  &  5.12$\times$10$^{+01}$  \\
0.030 &  2.40$\times$10$^{-25}$  &  9.39$\times$10$^{-25}$  &
       2.56$\times$10$^{-24}$  &  -5.552$\times$10$^{+01}$  &
       1.27$\times$10$^{+00}$  &  5.48$\times$10$^{+01}$  \\
0.040 &  4.52$\times$10$^{-22}$  &  1.53$\times$10$^{-21}$  &
       4.28$\times$10$^{-21}$  &  -4.803$\times$10$^{+01}$  &
       1.11$\times$10$^{+00}$  &  1.77$\times$10$^{+01}$  \\
0.050 &  1.13$\times$10$^{-19}$  &  2.99$\times$10$^{-19}$  &
       6.92$\times$10$^{-19}$  &  -4.272$\times$10$^{+01}$  &
       9.03$\times$10$^{-01}$  &  1.13$\times$10$^{+01}$  \\
0.060 &  7.26$\times$10$^{-18}$  &  2.46$\times$10$^{-17}$  &
       8.73$\times$10$^{-17}$  &  -3.825$\times$10$^{+01}$  &
       1.19$\times$10$^{+00}$  &  5.51$\times$10$^{+00}$  \\
0.070 &  1.95$\times$10$^{-16}$  &  9.25$\times$10$^{-16}$  &
       3.80$\times$10$^{-15}$  &  -3.467$\times$10$^{+01}$  &
       1.36$\times$10$^{+00}$  &  1.60$\times$10$^{+01}$  \\
0.080 &  2.97$\times$10$^{-15}$  &  1.54$\times$10$^{-14}$  &
       6.45$\times$10$^{-14}$  &  -3.188$\times$10$^{+01}$  &
       1.40$\times$10$^{+00}$  &  2.02$\times$10$^{+01}$  \\
0.090 &  2.77$\times$10$^{-14}$  &  1.38$\times$10$^{-13}$  &
       5.78$\times$10$^{-13}$  &  -2.966$\times$10$^{+01}$  &
       1.37$\times$10$^{+00}$  &  2.09$\times$10$^{+01}$  \\
0.100 &  1.92$\times$10$^{-13}$  &  8.13$\times$10$^{-13}$  &
       3.33$\times$10$^{-12}$  &  -2.784$\times$10$^{+01}$  &
       1.28$\times$10$^{+00}$  &  2.04$\times$10$^{+01}$  \\
0.110 &  1.19$\times$10$^{-12}$  &  3.88$\times$10$^{-12}$  &
       1.41$\times$10$^{-11}$  &  -2.624$\times$10$^{+01}$  &
       1.14$\times$10$^{+00}$  &  1.21$\times$10$^{+01}$  \\
0.120 &  6.46$\times$10$^{-12}$  &  1.85$\times$10$^{-11}$  &
       5.18$\times$10$^{-11}$  &  -2.473$\times$10$^{+01}$  &
       1.00$\times$10$^{+00}$  &  4.60$\times$10$^{+00}$  \\
0.130 &  2.87$\times$10$^{-11}$  &  8.26$\times$10$^{-11}$  &
       2.02$\times$10$^{-10}$  &  -2.329$\times$10$^{+01}$  &
       9.56$\times$10$^{-01}$  &  1.30$\times$10$^{+01}$  \\
0.140 &  1.10$\times$10$^{-10}$  &  3.31$\times$10$^{-10}$  &
       8.42$\times$10$^{-10}$  &  -2.190$\times$10$^{+01}$  &
       1.00$\times$10$^{+00}$  &  1.07$\times$10$^{+01}$  \\
0.150 &  3.71$\times$10$^{-10}$  &  1.18$\times$10$^{-09}$  &
       3.39$\times$10$^{-09}$  &  -2.060$\times$10$^{+01}$  &
       1.09$\times$10$^{+00}$  &  5.29$\times$10$^{+00}$  \\
0.160 &  1.16$\times$10$^{-09}$  &  3.92$\times$10$^{-09}$  &
       1.27$\times$10$^{-08}$  &  -1.939$\times$10$^{+01}$  &
       1.16$\times$10$^{+00}$  &  5.70$\times$10$^{+00}$  \\
0.180 &  9.13$\times$10$^{-09}$  &  3.41$\times$10$^{-08}$  &
       1.20$\times$10$^{-07}$  &  -1.722$\times$10$^{+01}$  &
       1.22$\times$10$^{+00}$  &  1.02$\times$10$^{+01}$  \\
0.200 &  6.11$\times$10$^{-08}$  &  2.15$\times$10$^{-07}$  &
       7.38$\times$10$^{-07}$  &  -1.537$\times$10$^{+01}$  &
       1.20$\times$10$^{+00}$  &  7.91$\times$10$^{+00}$  \\
0.250 &  2.58$\times$10$^{-06}$  &  8.16$\times$10$^{-06}$  &
       2.17$\times$10$^{-05}$  &  -1.179$\times$10$^{+01}$  &
       1.04$\times$10$^{+00}$  &  9.50$\times$10$^{+00}$  \\
0.300 &  3.67$\times$10$^{-05}$  &  1.10$\times$10$^{-04}$  &
       2.62$\times$10$^{-04}$  &  -9.208$\times$10$^{+00}$  &
       9.56$\times$10$^{-01}$  &  1.67$\times$10$^{+01}$  \\
0.350 &  2.79$\times$10$^{-04}$  &  7.60$\times$10$^{-04}$  &
       1.83$\times$10$^{-03}$  &  -7.232$\times$10$^{+00}$  &
       9.01$\times$10$^{-01}$  &  7.88$\times$10$^{+00}$  \\
0.400 &  1.49$\times$10$^{-03}$  &  3.52$\times$10$^{-03}$  &
       8.57$\times$10$^{-03}$  &  -5.636$\times$10$^{+00}$  &
       8.37$\times$10$^{-01}$  &  4.05$\times$10$^{+00}$  \\
0.450 &  6.28$\times$10$^{-03}$  &  1.30$\times$10$^{-02}$  &
       3.03$\times$10$^{-02}$  &  -4.299$\times$10$^{+00}$  &
       7.62$\times$10$^{-01}$  &  7.17$\times$10$^{+00}$  \\
0.500 &  2.18$\times$10$^{-02}$  &  4.04$\times$10$^{-02}$  &
       8.68$\times$10$^{-02}$  &  -3.156$\times$10$^{+00}$  &
       6.85$\times$10$^{-01}$  &  7.68$\times$10$^{+00}$  \\
0.600 &  1.59$\times$10$^{-01}$  &  2.69$\times$10$^{-01}$  &
       4.67$\times$10$^{-01}$  &  -1.309$\times$10$^{+00}$  &
       5.55$\times$10$^{-01}$  &  2.12$\times$10$^{+00}$  \\
0.700 &  7.13$\times$10$^{-01}$  &  1.13$\times$10$^{+00}$  &
       1.75$\times$10$^{+00}$  &  1.060$\times$10$^{-01}$  &
       4.67$\times$10$^{-01}$  &  2.52$\times$10$^{+00}$  \\
0.800 &  2.26$\times$10$^{+00}$  &  3.47$\times$10$^{+00}$  &
       5.04$\times$10$^{+00}$  &  1.218$\times$10$^{+00}$  &
       4.09$\times$10$^{-01}$  &  6.04$\times$10$^{+00}$  \\
0.900 &  5.76$\times$10$^{+00}$  &  8.46$\times$10$^{+00}$  &
       1.19$\times$10$^{+01}$  &  2.112$\times$10$^{+00}$  &
       3.70$\times$10$^{-01}$  &  7.80$\times$10$^{+00}$  \\
1.000 &  1.24$\times$10$^{+01}$  &  1.76$\times$10$^{+01}$  &
       2.41$\times$10$^{+01}$  &  2.845$\times$10$^{+00}$  &
       3.40$\times$10$^{-01}$  &  7.73$\times$10$^{+00}$  \\
1.250 &  5.10$\times$10$^{+01}$  &  6.86$\times$10$^{+01}$  &
       8.91$\times$10$^{+01}$  &  4.211$\times$10$^{+00}$  &
       2.84$\times$10$^{-01}$  &  5.01$\times$10$^{+00}$  \\
1.500 &  1.38$\times$10$^{+02}$  &  1.77$\times$10$^{+02}$  &
       2.22$\times$10$^{+02}$  &  5.164$\times$10$^{+00}$  &
       2.39$\times$10$^{-01}$  &  2.78$\times$10$^{+00}$  \\
1.750 &  2.91$\times$10$^{+02}$  &  3.59$\times$10$^{+02}$  &
       4.36$\times$10$^{+02}$  &  5.876$\times$10$^{+00}$  &
       2.03$\times$10$^{-01}$  &  1.47$\times$10$^{+00}$  \\
2.000 &  5.22$\times$10$^{+02}$  &  6.24$\times$10$^{+02}$  &
       7.37$\times$10$^{+02}$  &  6.433$\times$10$^{+00}$  &
       1.72$\times$10$^{-01}$  &  7.61$\times$10$^{-01}$  \\
2.500 &  1.24$\times$10$^{+03}$  &  1.42$\times$10$^{+03}$  &
       1.61$\times$10$^{+03}$  &  7.258$\times$10$^{+00}$  &
       1.29$\times$10$^{-01}$  &  2.78$\times$10$^{-01}$  \\
3.000 &  2.30$\times$10$^{+03}$  &  2.55$\times$10$^{+03}$  &
       2.81$\times$10$^{+03}$  &  7.842$\times$10$^{+00}$  &
       1.01$\times$10$^{-01}$  &  1.89$\times$10$^{-01}$  \\
3.500 & (3.59$\times$10$^{+03}$) & (3.91$\times$10$^{+03}$) &
      (4.26$\times$10$^{+03}$) & (8.271$\times$10$^{+00}$) &
      (8.56$\times$10$^{-02}$) &  \\
4.000 & (4.98$\times$10$^{+03}$) & (5.42$\times$10$^{+03}$) &
      (5.91$\times$10$^{+03}$) & (8.598$\times$10$^{+00}$) &
      (8.56$\times$10$^{-02}$) &  \\
5.000 & (8.18$\times$10$^{+03}$) & (8.92$\times$10$^{+03}$) &
      (9.71$\times$10$^{+03}$) & (9.096$\times$10$^{+00}$) &
      (8.56$\times$10$^{-02}$) &  \\
6.000 & (1.17$\times$10$^{+04}$) & (1.28$\times$10$^{+04}$) &
      (1.39$\times$10$^{+04}$) & (9.455$\times$10$^{+00}$) &
      (8.56$\times$10$^{-02}$) &  \\
7.000 & (1.50$\times$10$^{+04}$) & (1.64$\times$10$^{+04}$) &
      (1.78$\times$10$^{+04}$) & (9.704$\times$10$^{+00}$) &
      (8.56$\times$10$^{-02}$) &  \\
8.000 & (1.70$\times$10$^{+04}$) & (1.85$\times$10$^{+04}$) &
      (2.02$\times$10$^{+04}$) & (9.827$\times$10$^{+00}$) &
      (8.56$\times$10$^{-02}$) &  \\
9.000 & (1.66$\times$10$^{+04}$) & (1.80$\times$10$^{+04}$) &
      (1.97$\times$10$^{+04}$) & (9.801$\times$10$^{+00}$) &
      (8.56$\times$10$^{-02}$) &  \\
10.000 & (1.61$\times$10$^{+04}$) & (1.76$\times$10$^{+04}$) &
      (1.91$\times$10$^{+04}$) & (9.774$\times$10$^{+00}$) &
      (8.56$\times$10$^{-02}$) &  \\

\end{longtable}

\subsection{$^{38}$Ar($p,\gamma)^{39}$K}
Comments:  In total, 118 resonances are taken into account. For 99 resonances in the energy region of E$_r=873-2338$ keV, the measured energies and strengths are adopted from \citet{endt90} and \citet{hanninen}, respectively.  In the latter work, all the resonance strengths were normalized to E$_r=1358$ keV, $\omega\gamma$=1.30$\pm$0.25 eV. This value is in agreement with the earlier measurement of \citet{maripuu}.  For the resonance at E$_r=165$ keV (J$^\pi;T=7/2^-;3/2$), the strength is estimated using the proton spectroscopic factor, $C^2S=0.25$, measured in the $^{38}$Ar($^3$He,d)$^{39}$K transfer study of \citet{knopfle}.  For 18 resonances between the lowest-lying directly measured resonance (E$_r=873$ keV) and the proton threshold, spins, parities and mean lifetimes are adopted from \citet{endt90} or the Evaluated Nuclear Structure Data File (ENSDF).  For these resonances, the dimensionless reduced proton widths are sampled randomly using a Porter-Thomas distribution with a cutoff at $C^2S=1$. The direct capture rate contribution is estimated by weighing the partial S-factors, obtained from a single-particle potential model, with the experimental proton spectroscopic factors of \citet{knopfle}.  At 3.4 GK, the rates need to be matched to Hauser-Feshbach results, for extrapolation of the laboratory rates to higher temperatures.  In the bottom panel of the comparison plot of the new Monte Carlo-based rates to the previous rates of \citet{hanninen}, it can be seen that our results are much higher at temperatures between 0.5 GK and 2.5 GK.  The reason is that the work of \citet{hanninen} disregarded any rate contributions from resonances below E$_r=873$ keV.

\footnotesize
\begin{verbatim}
38Ar(p,g)39K
****************************************************************************************************************
1               ! Zproj
18              ! Ztarget
0               ! Zexitparticle (=0 when only 2 channels open)
1.0078          ! Aproj		
37.962732       ! Atarget
0               ! Aexitparticle (=0 when only 2 channels open)
0.5             ! Jproj
0.0             ! Jtarget
0.0             ! Jexitparticle (=0 when only 2 channels open)
6381.43         ! projectile separation energy (keV)
0.0             ! exit particle separation energy (=0 when only 2 channels open)
1.25            ! Radius parameter R0 (fm)
2               ! Gamma-ray channel number (=2 if ejectile is a g-ray; =3 otherwise)
****************************************************************************************************************
1.0             ! Minimum energy for numerical integration (keV)
5000            ! Number of random samples (>5000 for better statistics)
0               ! =0 for rate output at all temperatures; =NT for rate output at selected temperatures
****************************************************************************************************************
Non-resonant contribution
S(keVb)	   S'(b)	    S''(b/keV)	fracErr	    Cutoff Energy (keV)
741.0	    -0.2598 	  1.208e-4	 	0.5	        1600.0
0.0	       0.0	      0.0		       0.0	        0.0
****************************************************************************************************************
Resonant Contribution
Note: G1 = entrance channel, G2 = exit channel, G3 = spectator channel !! Ecm, Exf in (keV); wg, Gx in (eV) !!
Note: if Er<0, theta^2=C2S*theta_sp^2 must be entered instead of entrance channel partial width
Ecm	     DEcm    wg	      Dwg	   Jr   G1       DG1      L1   G2   DG2     L2   G3     DG3     L3   Exf  Int
 164.6   2.0     0.0      0.0      3.5  7.5e-13  3.8e-13  3    0.1  0.001   1    0      0       0    0     0
 873.2   0.8     0.085    0.025    0    0        0        0    0      0     0    0      0       0    0     0
 897.3   0.6     0.165    0.05     0    0        0        0    0      0     0    0      0       0    0     0
 905.2   0.8     0.02     0.01     0    0        0        0    0      0     0    0      0       0    0     0
 955.0   0.8     0.03     0.01     0    0        0        0    0      0     0    0      0       0    0     0
1000.0   0.6     0.05     0.02     0    0        0        0    0      0     0    0      0       0    0     0
1057.6   0.7     0.175    0.05     0    0        0        0    0      0     0    0      0       0    0     0
1067.0   0.8     0.03     0.01     0    0        0        0    0      0     0    0      0       0    0     0
1080.0   0.7     0.37     0.095    0    0        0        0    0      0     0    0      0       0    0     0
1100.4   0.7     0.115    0.03     0    0        0        0    0      0     0    0      0       0    0     0
1154.3   0.8     0.5      0.2      0    0        0        0    0      0     0    0      0       0    0     0
1159.1   0.8     0.06     0.02     0    0        0        0    0      0     0    0      0       0    0     0
1170.4   0.8     0.04     0.02     0    0        0        0    0      0     0    0      0       0    0     0
1178.9   0.8     0.02     0.01     0    0        0        0    0      0     0    0      0       0    0     0
1220.9   0.8     0.06     0.02     0    0        0        0    0      0     0    0      0       0    0     0
1223.5   0.8     0.07     0.025    0    0        0        0    0      0     0    0      0       0    0     0
1251.8   0.6     0.27     0.085    0    0        0        0    0      0     0    0      0       0    0     0
1318.2   0.8     0.06     0.02     0    0        0        0    0      0     0    0      0       0    0     0
1332.9   0.8     0.145    0.05     0    0        0        0    0      0     0    0      0       0    0     0
1358.0   0.6     1.3      0.25     0    0        0        0    0      0     0    0      0       0    0     0
1374.3   0.8     0.135    0.03     0    0        0        0    0      0     0    0      0       0    0     0
1384.6   0.8     0.05     0.02     0    0        0        0    0      0     0    0      0       0    0     0
1391.5   0.8     0.145    0.04     0    0        0        0    0      0     0    0      0       0    0     0
1403.2   0.8     0.08     0.03     0    0        0        0    0      0     0    0      0       0    0     0
1415.9   0.8     0.435    0.115    0    0        0        0    0      0     0    0      0       0    0     0
1420.5   0.8     0.215    0.05     0    0        0        0    0      0     0    0      0       0    0     0
1423.3   0.8     0.1      0.03     0    0        0        0    0      0     0    0      0       0    0     0
1439.2   0.8     0.205    0.05     0    0        0        0    0      0     0    0      0       0    0     0
1465.3   1.2     0.3      0.095    0    0        0        0    0      0     0    0      0       0    0     0
1486.6   1.0     0.38     0.10     0    0        0        0    0      0     0    0      0       0    0     0
1576.3   0.8     0.39     0.10     0    0        0        0    0      0     0    0      0       0    0     0
1597.1   0.8     0.25     0.06     0    0        0        0    0      0     0    0      0       0    0     0
1600.1   1.0     0.07     0.03     0    0        0        0    0      0     0    0      0       0    0     0
1601.8   0.8     0.5      0.2      0    0        0        0    0      0     0    0      0       0    0     0
1604.8   0.8     0.06     0.02     0    0        0        0    0      0     0    0      0       0    0     0
1610.8   0.8     0.465    0.105    0    0        0        0    0      0     0    0      0       0    0     0
1613.9   0.8     0.07     0.02     0    0        0        0    0      0     0    0      0       0    0     0
1616.9   0.8     0.06     0.02     0    0        0        0    0      0     0    0      0       0    0     0
1649.5   0.8     0.25     0.06     0    0        0        0    0      0     0    0      0       0    0     0
1652.4   1.0     0.105    0.03     0    0        0        0    0      0     0    0      0       0    0     0
1656.9   0.8     0.65     0.20     0    0        0        0    0      0     0    0      0       0    0     0
1698.1   1.2     0.175    0.070    0    0        0        0    0      0     0    0      0       0    0     0
1699.9   1.2     0.260    0.125    0    0        0        0    0      0     0    0      0       0    0     0
1705.5   1.0     0.03     0.01     0    0        0        0    0      0     0    0      0       0    0     0
1711.7   1.0     0.27     0.07     0    0        0        0    0      0     0    0      0       0    0     0
1717.6   1.0     0.175    0.06     0    0        0        0    0      0     0    0      0       0    0     0
1726.2   1.0     0.135    0.070    0    0        0        0    0      0     0    0      0       0    0     0
1736.4   0.8     0.435    0.105    0    0        0        0    0      0     0    0      0       0    0     0
1747.1   1.0     0.09     0.03     0    0        0        0    0      0     0    0      0       0    0     0
1756.8   1.0     0.095    0.030    0    0        0        0    0      0     0    0      0       0    0     0
1788.7   1.0     0.155    0.060    0    0        0        0    0      0     0    0      0       0    0     0
1803.4   1.0     0.280    0.095    0    0        0        0    0      0     0    0      0       0    0     0
1807.6   1.0     0.260    0.085    0    0        0        0    0      0     0    0      0       0    0     0
1809.7   1.0     0.280    0.095    0    0        0        0    0      0     0    0      0       0    0     0
1816.7   1.0     0.6      0.2      0    0        0        0    0      0     0    0      0       0    0     0
1821.6   1.2     0.10     0.03     0    0        0        0    0      0     0    0      0       0    0     0
1871.2   0.8     1.35     0.45     0    0        0        0    0      0     0    0      0       0    0     0
1880.8   1.0     0.70     0.25     0    0        0        0    0      0     0    0      0       0    0     0
1889.4   1.0     0.125    0.04     0    0        0        0    0      0     0    0      0       0    0     0
1898.1   1.0     0.185    0.060    0    0        0        0    0      0     0    0      0       0    0     0
1903.6   1.0     0.095    0.040    0    0        0        0    0      0     0    0      0       0    0     0
1912.6   1.0     0.235    0.080    0    0        0        0    0      0     0    0      0       0    0     0
1923.4   1.0     0.465    0.155    0    0        0        0    0      0     0    0      0       0    0     0
1933.1   1.5     0.145    0.050    0    0        0        0    0      0     0    0      0       0    0     0
1944.4   1.0     0.07     0.03     0    0        0        0    0      0     0    0      0       0    0     0
1959.3   1.0     0.310    0.105    0    0        0        0    0      0     0    0      0       0    0     0
1965.4   1.0     0.185    0.060    0    0        0        0    0      0     0    0      0       0    0     0
1998.2   1.0     0.195    0.060    0    0        0        0    0      0     0    0      0       0    0     0
2004.8   1.0     0.55     0.20     0    0        0        0    0      0     0    0      0       0    0     0
2013.7   1.5     0.260    0.085    0    0        0        0    0      0     0    0      0       0    0     0
2032.4   1.2     0.6      0.2      0    0        0        0    0      0     0    0      0       0    0     0
2045.4   1.2     0.185    0.060    0    0        0        0    0      0     0    0      0       0    0     0
2048.1   1.2     0.475    0.155    0    0        0        0    0      0     0    0      0       0    0     0
2083.9   1.2     0.235    0.085    0    0        0        0    0      0     0    0      0       0    0     0
2094.6   1.5     0.270    0.095    0    0        0        0    0      0     0    0      0       0    0     0
2102.7   1.2     0.50     0.20     0    0        0        0    0      0     0    0      0       0    0     0
2127.8   1.2     0.135    0.040    0    0        0        0    0      0     0    0      0       0    0     0
2132.2   1.2     0.185    0.060    0    0        0        0    0      0     0    0      0       0    0     0
2142.8   1.2     0.195    0.060    0    0        0        0    0      0     0    0      0       0    0     0
2149.0   1.2     0.55     0.20     0    0        0        0    0      0     0    0      0       0    0     0
2166.1   1.3     0.330    0.115    0    0        0        0    0      0     0    0      0       0    0     0
2177.9   1.5     0.105    0.030    0    0        0        0    0      0     0    0      0       0    0     0
2185.7   1.3     0.135    0.040    0    0        0        0    0      0     0    0      0       0    0     0
2202.2   1.3     0.80     0.25     0    0        0        0    0      0     0    0      0       0    0     0
2210.8   1.5     0.205    0.070    0    0        0        0    0      0     0    0      0       0    0     0
2216.6   1.5     0.105    0.050    0    0        0        0    0      0     0    0      0       0    0     0
2243.4   1.5     0.310    0.105    0    0        0        0    0      0     0    0      0       0    0     0
2245.5   1.5     0.85     0.30     0    0        0        0    0      0     0    0      0       0    0     0
2256.9   1.5     0.29     0.10     0    0        0        0    0      0     0    0      0       0    0     0
2273.5   1.5     0.205    0.075    0    0        0        0    0      0     0    0      0       0    0     0
2285.3   1.5     0.125    0.040    0    0        0        0    0      0     0    0      0       0    0     0
2290.8   1.5     0.185    0.060    0    0        0        0    0      0     0    0      0       0    0     0
2292.8   1.5     0.155    0.050    0    0        0        0    0      0     0    0      0       0    0     0
2301.9   1.5     0.415    0.135    0    0        0        0    0      0     0    0      0       0    0     0
2306.4   1.5     0.415    0.135    0    0        0        0    0      0     0    0      0       0    0     0
2311.7   1.5     0.260    0.085    0    0        0        0    0      0     0    0      0       0    0     0
2322.4   1.5     0.310    0.105    0    0        0        0    0      0     0    0      0       0    0     0
2327.6   1.5     0.6      0.2      0    0        0        0    0      0     0    0      0       0    0     0
2332.7   1.5     0.310    0.105    0    0        0        0    0      0     0    0      0       0    0     0
2338.2   1.5     0.290    0.105    0    0        0        0    0      0     0    0      0       0    0     0
****************************************************************************************************************
Upper Limits of Resonances
Note: enter partial width upper limit by chosing non-zero value for PT, where PT=<theta^2> for particles and...
Note: ...PT=<B> for g-rays [enter: "upper_limit 0.0"]; for each resonance: # upper limits < # open channels!  
Ecm	    DEcm	Jr    G1       DG1   L1   PT        G2      DG2     L2  PT    G3    DG3    L3  PT    Exf   Int
  14.6  2.0     0.5   3.6e-53  0.0   0    0.0045    0.1     0.001   1   0     0     0      0   0     0     0
  28.6  2.0     1.5   3.9e-37  0.0   2    0.0045    0.1     0.001   1   0     0     0      0   0     0     0
  76.6  2.0     0.5   1.9e-17  0.0   0    0.0045    0.1     0.001   1   0     0     0      0   0     0     0
  83.6  2.0     0.5   2.9e-16  0.0   0    0.0045    0.1     0.001   1   0     0     0      0   0     0     0
 119.6  2.0     1.5   5.9e-14  0.0   2    0.0045    7.7e-3  3.8e-3  1   0     0     0      0   0     0     1
 146.6  2.0     0.5   8.6e-10  0.0   0    0.0045    5.0e-3  2.5e-3  1   0     0     0      0   0     0     1                 
 271.6  2.0     1.5   1.7e-6   0.0   2    0.0045    1.9e-2  0.9e-2  1   0     0     0      0   0     0     1
 304.6  2.0     0.5   1.1e-3   0.0   0    0.0045    0.10    0.05    1   0     0     0      0   0     0     0                          
 358.6  2.0     1.5   1.5e-4   0.0   2    0.0045    0.19    0.09    1   0     0     0      0   0     0     1
 436.6  3.0     1.5   2.3e-3   0.0   2    0.0045    0.14    0.07    1   0     0     0      0   0     0     1
 446.6  2.0     0.5   0.26     0.0   0    0.0045    0.10    0.05    1   0     0     0      0   0     0     0
 534.6  2.0     1.5   0.032    0.0   2    0.0045    0.33    0.17    1   0     0     0      0   0     0     1
 561.6  2.0     1.5   0.057    0.0   2    0.0045    0.10    0.05    1   0     0     0      0   0     0     0
 639.6  2.0     0.5   18.4     0.0   0    0.0045    0.10    0.05    1   0     0     0      0   0     0     0
 669.6  2.0     1.5   10.3     0.0   1    0.0045    0.10    0.05    1   0     0     0      0   0     0     0
 740.6  5.0     1.5   29.2     0.0   1    0.0045    0.10    0.05    1   0     0     0      0   0     0     0
 788.6  2.0     0.5   152.0    0.0   0    0.0045    0.10    0.05    1   0     0     0      0   0     0     0
 818.6  2.0     1.5   78.0     0.0   1    0.0045    0.10    0.05    1   0     0     0      0   0     0     0
****************************************************************************************************************
Interference between Resonances [numerical integration only]
Note: + for positive, - for negative interference; +- if interference sign is unknown
Ecm	    DEcm	Jr      G1      DG1   L1   PT     G2     DG2    L2  PT     G3     DG3    L3  PT      Exf  
!+- 
0.0     0.0     0.0     0.0     0.0   0    0      0.0    0.0    0   0      0.0    0.0    0   0       0.0  
0.0     0.0     0.0     0.0     0.0   0    0      0.0    0.0    0   0      0.0    0.0    0   0       0.0  
****************************************************************************************************************
Reaction Rate and PDF at NT selected temperatures only
Note: default values are used for reaction rate range if Min=Max=0.0
T9	   Min	Max
0.01   0.0	0.0
0.1	   0.0	0.0
****************************************************************************************************************
Comments:
1. Excitation energies, spin-parities and lifetimes from Endt (1990) and ENSDF, unless noted otherwise.
2. Resonance energies and strengths adopted from Haenninen (1984). In that work, all strengths were normalized
   to Erlab=1394 keV, wg=1.30+-0.25 eV. This strength is in agreement with the earlier measurement of Maripuu 
   (1970). 
3. Direct capture S-factor calcuted using measured spectroscopic factors from (3He,d) work of Knoepfle et al.
   (1974).  
4. For Ercm=165 keV, Gp is calculated from measured C2S=0.25 of Knoepfle et al. (1974); assumption of Gg=0.1
   eV is inconsequential.
5. For resonance upper limit calculation: (i) smalles possible orbital angular momentum ell assumed; (ii) high
   -spin states at 6434, 6475, 7092, 7142 keV have been disregarded since ell>4; (iii) for 6546 kev (T=3/2) 
   level, C2S=0.245 adopted from (3He,d) study of Knoepfle et al. (1974); (iv) Gg=0.1 eV assumption for
   Ercm=15, 29, 77, 84 keV is inconsequential since Gg>>Gp; 
6. Gg=0.1 eV for Ercm=305, 447, 562, 640, 670, 741, 789, 819 keV is a crude guess.
7. For Ercm=120, 147, 272, 359, 437 keV spin-parity is ambiguous; we assumed smallest possible spin for upper
   limit calculation.
   
   
   


   
\end{verbatim}
\normalsize
\vspace{5mm}

\begin{figure}[ht]
\centering
\includegraphics[scale=0.5]{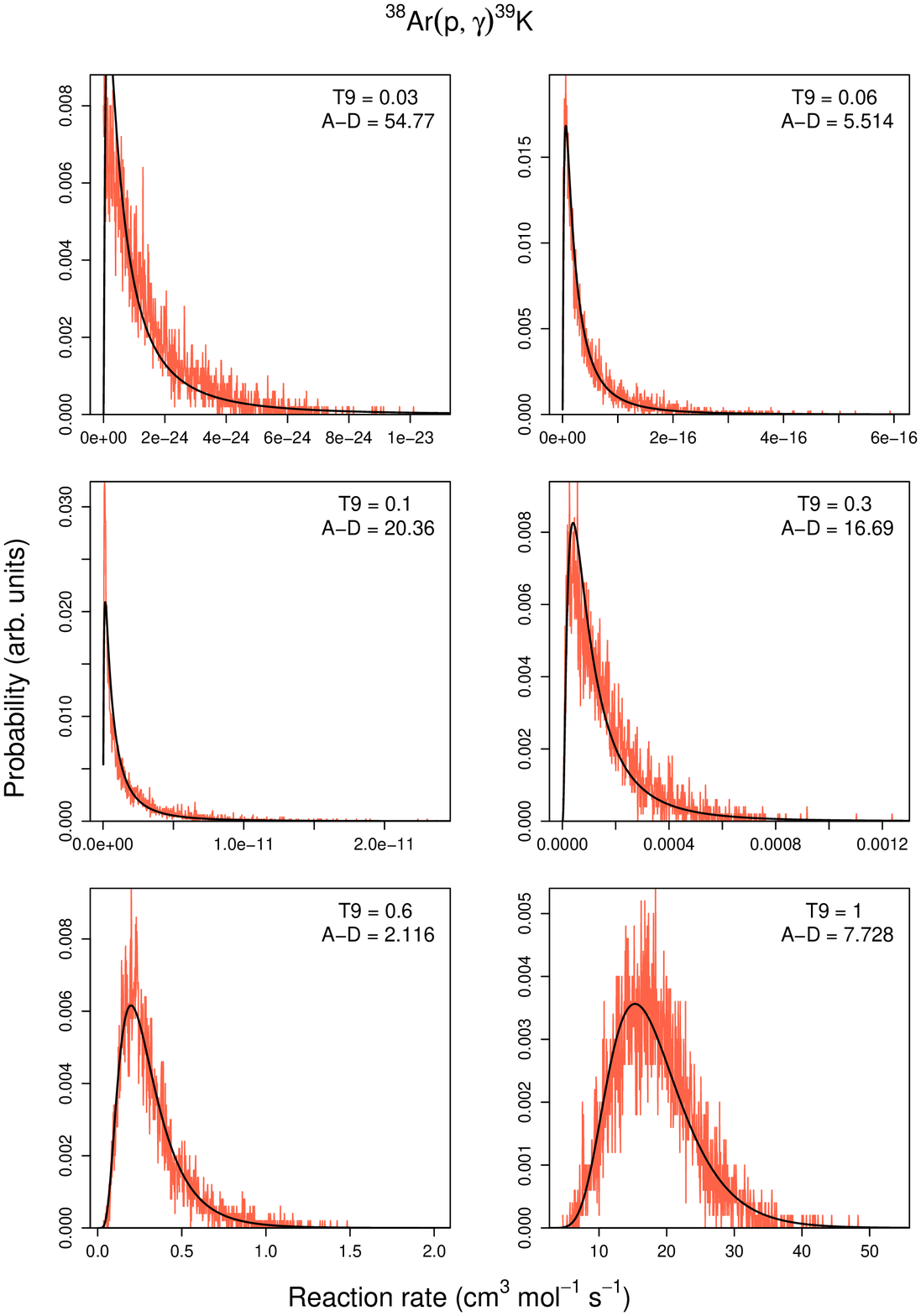}
\label{38ar_ag}
\end{figure}
\begin{figure}[ht]
\centering
\includegraphics[scale=0.5]{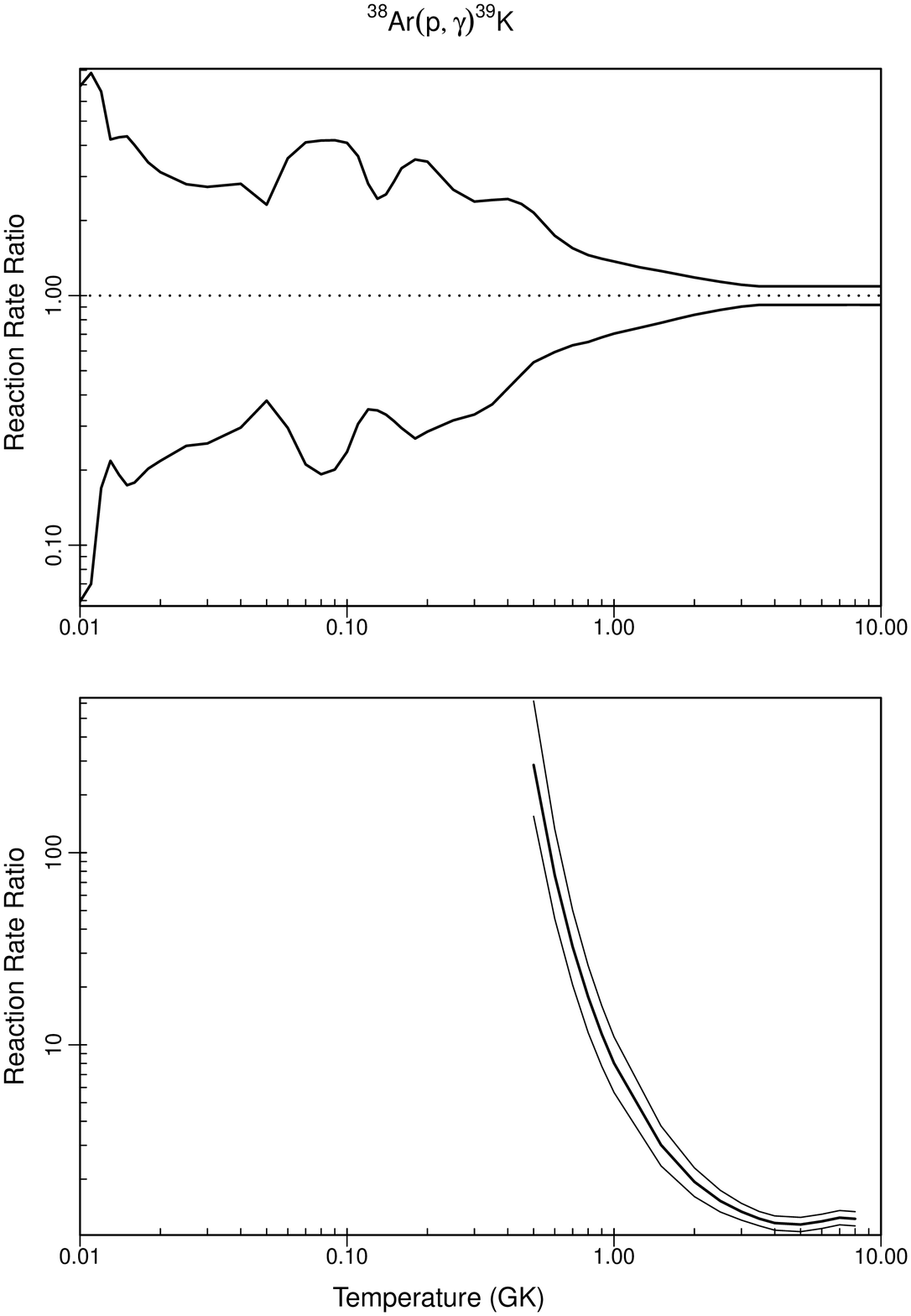}
\label{38ar_ag_1}
\end{figure}

\clearpage

%







%




\begin{thebibliography}{100}
\expandafter\ifx\csname natexlab\endcsname\relax\def\natexlab#1{#1}\fi

\bibitem[{{Adelberger} {et~al.}(1998)}]{adelberger1}
{Adelberger}, E.~G., {et~al.} 1998, Rev. Mod. Phys., 70, 1265

\bibitem[{{Adelberger} {et~al.}(2011)}]{adelberger2}
---. 2011, Rev. Mod. Phys., 83, 195

\bibitem[{{Almanza} {et~al.}(1975){Almanza}, {Murillo}, {Darden}, \&
  {Sen}}]{almanza}
{Almanza}, R., {Murillo}, G., {Darden}, S.~E., \& {Sen}, S. 1975, Nucl. Phys.
  A, 248, 214

\bibitem[{{Angulo} {et~al.}(1999){Angulo}, {Arnould}, {Rayet}, {Descouvemont},
  {Baye}, {Leclercq-Willain}, {Coc}, {Barhoumi}, {Aguer}, {Rolfs}, {Kunz},
  {Hammer}, {Mayer}, {Paradellis}, {Kossionides}, {Chronidou}, {Spyrou},
  {degl'Innocenti}, {Fiorentini}, {Ricci}, {Zavatarelli}, {Providencia},
  {Wolters}, {Soares}, {Grama}, {Rahighi}, {Shotter}, \& {Lamehi
  Rachti}}]{nacre}
{Angulo}, C., {Arnould}, M., {Rayet}, M., {et~al.} 1999, Nucl. Phys. A, 656, 3

\bibitem[{{Arnould} \& {Goriely}(2003)}]{arnould}
{Arnould}, M., \& {Goriely}, S. 2003, \physrep, 384, 1

\bibitem[{{Arnould} {et~al.}(2007){Arnould}, {Goriely}, \&
  {Takahashi}}]{arnould07}
{Arnould}, M., {Goriely}, S., \& {Takahashi}, K. 2007, \physrep, 450, 97

\bibitem[{{Audi} {et~al.}(2003){Audi}, {Wapstra}, \& {Thibault}}]{audi}
{Audi}, G., {Wapstra}, A.~H., \& {Thibault}, C. 2003, Nucl. Phys. A, 729, 337

\bibitem[{{Bao} {et~al.}(2000){Bao}, {Beer}, {K{\"a}ppeler}, {Voss}, {Wisshak},
  \& {Rauscher}}]{Bao00}
{Bao}, Z.~Y., {Beer}, H., {K{\"a}ppeler}, F., {et~al.} 2000, At. Data and Nucl.
  Data Tables, 76, 70

\bibitem[{{Bardayan} {et~al.}(2007){Bardayan}, {Blackmon}, {Fitzgerald}, {Hix},
  {Jones}, {Kozub}, {Liang}, {Livesay}, {Ma}, {Roberts}, {Smith}, {Thomas}, \&
  {Visser}}]{bard}
{Bardayan}, D.~W., {Blackmon}, J.~C., {Fitzgerald}, R.~P., {et~al.} 2007, \prc,
  76, 045803

\bibitem[{{Becker} {et~al.}(1995){Becker}, {Bahr}, {Berheide}, {Borucki},
  {Buschmann}, {Rolfs}, {Roters}, {Schmidt}, {Schulte}, {Mitchell}, \&
  {Schweitzer}}]{becker}
{Becker}, H.~W., {Bahr}, M., {Berheide}, M., {et~al.} 1995, Zeitschrift fur
  Physik A Hadrons and Nuclei, 351, 453

\bibitem[{{Buckner} {et~al.}(2012)}]{bu12}
{Buckner}, M., {et~al.} 2012, submitted to Phys. Rev. C

\bibitem[{{Capote} {et~al.}(2009){Capote}, {Herman}, {Oblo{\v z}insk{\'y}},
  {Young}, {Goriely}, {Belgya}, {Ignatyuk}, {Koning}, {Hilaire}, {Plujko},
  {Avrigeanu}, {Bersillon}, {Chadwick}, {Fukahori}, {Ge}, {Han}, {Kailas},
  {Kopecky}, {Maslov}, {Reffo}, {Sin}, {Soukhovitskii}, \& {Talou}}]{Capote09}
{Capote}, R., {Herman}, M., {Oblo{\v z}insk{\'y}}, P., {et~al.} 2009, Nuclear
  Data Sheets, 110, 3107

\bibitem[{{Caughlan} \& {Fowler}(1988)}]{CF88}
{Caughlan}, G.~R., \& {Fowler}, W.~A. 1988, At. Data and Nucl. Data Tables, 40,
  283

\bibitem[{Coc {et~al.}(1999)Coc, Porquet, \& Nowacki}]{coc}
Coc, A., Porquet, M.-G., \& Nowacki, F. 1999, Phys. Rev. C, 61, 015801

\bibitem[{{Coc} {et~al.}(2002){Coc}, {Vangioni-Flam}, {Cass{\'e}}, \&
  {Rabiet}}]{coc2}
{Coc}, A., {Vangioni-Flam}, E., {Cass{\'e}}, M., \& {Rabiet}, M. 2002, \prd,
  65, 043510

\bibitem[{{Coc} {et~al.}(2004){Coc}, {Vangioni-Flam}, {Descouvemont},
  {Adahchour}, \& {Angulo}}]{coc04}
{Coc}, A., {Vangioni-Flam}, E., {Descouvemont}, P., {Adahchour}, A., \&
  {Angulo}, C. 2004, \apj, 600, 544

\bibitem[{{Crow} \& {Shimizu}(1988)}]{lognormal}
{Crow}, E.~L., \& {Shimizu}, K., eds. 1988, Lognormal Distributions (M. Dekker)

\bibitem[{{Cyburt} {et~al.}(2010)}]{reaclib}
{Cyburt}, R.~H., {et~al.} 2010, Ap. J. Suppl. Ser., 189, 240

\bibitem[{{Descouvemont} {et~al.}(2004){Descouvemont}, {Adahchour}, {Angulo},
  {Coc}, \& {Vangioni-Flam}}]{des}
{Descouvemont}, P., {Adahchour}, A., {Angulo}, C., {Coc}, A., \&
  {Vangioni-Flam}, E. 2004, At. Data and Nucl. Data Tables, 88, 203

\bibitem[{{Endt}(1990)}]{endt90}
{Endt}, P.~M. 1990, Nucl. Phys. A, 521, 1

\bibitem[{{Endt}(1998)}]{endt98}
---. 1998, Nucl. Phys. A, 633, 1

\bibitem[{{Evans} {et~al.}(2000){Evans}, {Hastings}, \& {Peacock}}]{evansbook}
{Evans}, M., {Hastings}, N., \& {Peacock}, B. 2000, Statistical Distributions,
  3rd ed. (Wiley)

\bibitem[{{Fuller} {et~al.}(1982){Fuller}, {Fowler}, \& {Newman}}]{fuller}
{Fuller}, G.~M., {Fowler}, W.~A., \& {Newman}, M.~J. 1982, \apjs, 48, 279

\bibitem[{{Fynbo.} {et~al.}(2000)}]{fyn}
{Fynbo.}, H.~O.~U., {et~al.} 2000, Nucl. Phys. A, 677, 38

\bibitem[{{Giesen} {et~al.}(1993){Giesen}, {Browne}, {G{\"o}rres}, {Graff},
  {Iliadis}, {Trautvetter}, {Wiescher}, {Harms}, {Kratz}, {Pfeiffer}, {Azuma},
  {Buckby}, \& {King}}]{giesen}
{Giesen}, U., {Browne}, C.~P., {G{\"o}rres}, J., {et~al.} 1993, Nucl. Phys. A,
  561, 95

\bibitem[{{Goriely}(1999)}]{goriely99}
{Goriely}, S. 1999, \aap, 342, 881

\bibitem[{{Goriely} {et~al.}(2010){Goriely}, {Chamel}, \&
  {Pearson}}]{Goriely10}
{Goriely}, S., {Chamel}, N., \& {Pearson}, J.~M. 2010, \prc, 82, 035804

\bibitem[{{Goriely} {et~al.}(2008{\natexlab{a}}){Goriely}, {Hilaire}, \&
  {Koning}}]{Goriely08a}
{Goriely}, S., {Hilaire}, S., \& {Koning}, A.~J. 2008{\natexlab{a}}, \aap, 487,
  767

\bibitem[{{Goriely} {et~al.}(2008{\natexlab{b}}){Goriely}, {Hilaire}, \&
  {Koning}}]{Goriely08}
---. 2008{\natexlab{b}}, \prc, 78, 064307

\bibitem[{{Goriely} {et~al.}(2004){Goriely}, {Khan}, \& {Samyn}}]{Goriely04}
{Goriely}, S., {Khan}, E., \& {Samyn}, M. 2004, Nucl. Phys. A, 739, 331

\bibitem[{{H{\"a}nninen}(1984)}]{hanninen}
{H{\"a}nninen}, R. 1984, Nucl. Phys. A, 420, 351

\bibitem[{{Hix} {et~al.}(2003){Hix}, {Smith}, {Starrfield}, {Mezzacappa}, \&
  {Smith}}]{hix}
{Hix}, W.~R., {Smith}, M.~S., {Starrfield}, S., {Mezzacappa}, A., \& {Smith},
  D.~L. 2003, Nucl. Phys. A, 718, 620

\bibitem[{{Iacob} {et~al.}(2006){Iacob}, {Zhai}, {Al-Abdullah}, {Fu}, {Hardy},
  {Nica}, {Park}, {Tabacaru}, {Trache}, \& {Tribble}}]{iacob}
{Iacob}, V.~E., {Zhai}, Y., {Al-Abdullah}, T., {et~al.} 2006, Phys. Rev. C, 74,
  045810

\bibitem[{Iliadis(2007)}]{iliadisbook}
Iliadis, C. 2007, Nuclear Physics of Stars (Weinheim: Wiley-VCH)

\bibitem[{{Iliadis} {et~al.}(2011){Iliadis}, {Champagne}, {Chieffi}, \&
  {Limongi}}]{il_al26}
{Iliadis}, C., {Champagne}, A., {Chieffi}, A., \& {Limongi}, M. 2011, \apjs,
  193, 16

\bibitem[{{Iliadis} {et~al.}(2001){Iliadis}, {D'Auria}, {Starrfield},
  {Thompson}, \& {Wiescher}}]{ilcomp}
{Iliadis}, C., {D'Auria}, J.~M., {Starrfield}, S., {Thompson}, W.~J., \&
  {Wiescher}, M. 2001, \apjs, 134, 151

\bibitem[{{Iliadis} {et~al.}(2010{\natexlab{a}}){Iliadis}, {Longland},
  {Champagne}, \& {Coc}}]{iliadis_4}
{Iliadis}, C., {Longland}, R., {Champagne}, A., \& {Coc}, A.
  2010{\natexlab{a}}, Nucl. Phys. A, 841, 323

\bibitem[{{Iliadis} {et~al.}(2010{\natexlab{b}}){Iliadis}, {Longland},
  {Champagne}, \& {Coc}}]{iliadis_3}
{Iliadis}, C., {Longland}, R., {Champagne}, A.~E., \& {Coc}, A.
  2010{\natexlab{b}}, Nucl. Phys. A, 841, 251

\bibitem[{{Iliadis} \& {Wiescher}(2004)}]{ilw}
{Iliadis}, C., \& {Wiescher}, M. 2004, \prc, 69, 064305

\bibitem[{{Iliadis} {et~al.}(2010{\natexlab{c}})}]{iliadis_2}
{Iliadis}, C., {et~al.} 2010{\natexlab{c}}, Nucl. Phys. A, 841, 31

\bibitem[{{Izzard} {et~al.}(2007){Izzard}, {Lugaro}, {Karakas}, {Iliadis}, \&
  {van Raai}}]{izzard}
{Izzard}, R.~G., {Lugaro}, M., {Karakas}, A.~I., {Iliadis}, C., \& {van Raai},
  M. 2007, \aap, 466, 641

\bibitem[{{Jenkins} {et~al.}(2004)}]{jenkins}
{Jenkins}, D.~G., {et~al.} 2004, Phys. Rev. Lett., 92, 031101

\bibitem[{{Kn{\"o}pfle} {et~al.}(1974){Kn{\"o}pfle}, {Doll}, {Mairle}, \&
  {Wagner}}]{knopfle}
{Kn{\"o}pfle}, K.~T., {Doll}, P., {Mairle}, G., \& {Wagner}, G.~J. 1974, Nucl.
  Phys. A, 233, 317

\bibitem[{{Koning} {et~al.}(2004){Koning}, {Hilaire}, \&
  {Duijvestijn}}]{Koning04}
{Koning}, A.~J., {Hilaire}, S., \& {Duijvestijn}, M. 2004, in TALYS: A nuclear
  reaction program (NRG-report 21297/04.62741/P); also available at {\it
  http//www.talys.eu}.

\bibitem[{{Koning} {et~al.}(2008){Koning}, {Hilaire}, \&
  {Duijvestijn}}]{Koning08a}
---. 2008, \emph{Nuclear Data for Science and Technology} (EDP Sciences; eds O.
  Bersillon et al.), pp. 211.

\bibitem[{{Koning} \& {Rochman}(2009)}]{koningrochman}
{Koning}, A.~J., \& {Rochman}, D. 2009, jEFF-Doc 1310 (2009), also available at
  http://www.talys.eu/tendl-2009

\bibitem[{{Krauss} {et~al.}(1992){Krauss}, {Grun}, T., \& {Oberhummer}}]{tedca}
{Krauss}, H., {Grun}, K., T., R., \& {Oberhummer}, H. 1992, {\it computer code}
  {\tt TEDCA} (TU Wien, Vienna, Austria).

\bibitem[{{Krauss} \& {Chaboyer}(2003)}]{krauss}
{Krauss}, L.~M., \& {Chaboyer}, B. 2003, Science, 299, 65

\bibitem[{{Krauss} \& {Romanelli}(1990)}]{krauss90}
{Krauss}, L.~M., \& {Romanelli}, P. 1990, \apj, 358, 47

\bibitem[{{Kuhlmann} {et~al.}(1973){Kuhlmann}, {Albrecht}, \&
  {Hoffmann}}]{kuhl}
{Kuhlmann}, E., {Albrecht}, W., \& {Hoffmann}, A. 1973, Nucl. Phys. A, 213, 82

\bibitem[{{La Cognata} {et~al.}(2010{\natexlab{a}}){La Cognata}, {Spitaleri},
  \& {Mukhamedzhanov}}]{lacogb}
{La Cognata}, M., {Spitaleri}, C., \& {Mukhamedzhanov}, A.~M.
  2010{\natexlab{a}}, \apj, 723, 1512

\bibitem[{{La Cognata} {et~al.}(2010{\natexlab{b}}){La Cognata}, {Spitaleri},
  {Mukhamedzhanov}, {Banu}, {Cherubini}, {Coc}, {Crucill{\`a}}, {Goldberg},
  {Gulino}, {Irgaziev}, {Kiss}, {Lamia}, {Mrazek}, {Pizzone}, {Puglia},
  {Rapisarda}, {Romano}, {Sergi}, {Tabacaru}, {Trache}, {Tribble}, {Trzaska},
  \& {Tumino}}]{lacoga}
{La Cognata}, M., {Spitaleri}, C., {Mukhamedzhanov}, A., {et~al.}
  2010{\natexlab{b}}, \apj, 708, 796

\bibitem[{{Langanke} \& {Mart{\'{\i}}nez-Pinedo}(2001)}]{karl}
{Langanke}, K., \& {Mart{\'{\i}}nez-Pinedo}, G. 2001, At. Data and Nucl. Data
  Tables, 79, 1

\bibitem[{{Longland}(2012)}]{longlandtobesub}
{Longland}, R. 2012, in press for Astron. Astrophys;
  http://arxiv.org/abs/1210.4495

\bibitem[{{Longland} {et~al.}(2010{\natexlab{a}}){Longland}, {Iliadis},
  {Cesaratto}, {Champagne}, {Daigle}, {Newton}, \&
  {Fitzgerald}}]{2010PhRvC..81e5804L}
{Longland}, R., {Iliadis}, C., {Cesaratto}, J.~M., {et~al.} 2010{\natexlab{a}},
  \prc, 81, 055804

\bibitem[{{Longland} {et~al.}(2012){Longland}, {Iliadis}, \& {Karakas}}]{lo12}
{Longland}, R., {Iliadis}, C., \& {Karakas}, A.~I. 2012, \prc, 85, 065809

\bibitem[{{Longland} {et~al.}(2010{\natexlab{b}})}]{iliadis_1}
{Longland}, R., {et~al.} 2010{\natexlab{b}}, Nucl. Phys. A, 841, 1

\bibitem[{{Lorenz-Wirzba} {et~al.}(1979){Lorenz-Wirzba}, {Schmalbrock},
  {Trautvetter}, {Wiescher}, {Rolfs}, \& {Rodney}}]{LW}
{Lorenz-Wirzba}, H., {Schmalbrock}, P., {Trautvetter}, H.~P., {et~al.} 1979,
  Nucl. Phys. A, 313, 346

\bibitem[{{Mackh} {et~al.}(1973){Mackh}, {Oeschler}, {Wagner}, {Dehnhard}, \&
  {Ohnuma}}]{mackh}
{Mackh}, H., {Oeschler}, H., {Wagner}, G.~J., {Dehnhard}, D., \& {Ohnuma}, H.
  1973, Nucl. Phys. A, 202, 497

\bibitem[{{Mak} {et~al.}(1978){Mak}, {Evans}, {Ewan}, \& {MacArthur}}]{mak}
{Mak}, H.-B., {Evans}, H.~C., {Ewan}, G.~T., \& {MacArthur}, J.~D. 1978, Nucl.
  Phys. A, 304, 210

\bibitem[{{Maripuu}(1970)}]{maripuu}
{Maripuu}, S. 1970, Nucl. Phys. A, 151, 465

\bibitem[{{Murillo} {et~al.}(1979){Murillo}, {Fern{\'a}nde}, {P{\'e}rez},
  {Ram{\'{\i}}rez}, {Darden}, {Cobian-Rozak}, \& {Montestruque}}]{murillo}
{Murillo}, G., {Fern{\'a}nde}, M., {P{\'e}rez}, P., {et~al.} 1979, Nucl. Phys.
  A, 318, 352

\bibitem[{{Newton} {et~al.}(2008){Newton}, {Longland}, \& {Iliadis}}]{newton}
{Newton}, J.~R., {Longland}, R., \& {Iliadis}, C. 2008, \prc, 78, 025805

\bibitem[{{Oda} {et~al.}(1994){Oda}, {Hino}, {Muto}, {Takahara}, \&
  {Sato}}]{oda}
{Oda}, T., {Hino}, M., {Muto}, K., {Takahara}, M., \& {Sato}, K. 1994, At. Data
  and Nucl. Data Tables, 56, 231

\bibitem[{{Orihara} {et~al.}(1973){Orihara}, {Rudolf}, \&
  {Gorodetzky}}]{orihara}
{Orihara}, H., {Rudolf}, G., \& {Gorodetzky}, P. 1973, Nucl. Phys. A, 203, 78

\bibitem[{{Paddock}(1972)}]{paddock}
{Paddock}, R.~A. 1972, \prc, 5, 485

\bibitem[{{Parikh} {et~al.}(2008){Parikh}, {Jos{\'e}}, {Moreno}, \&
  {Iliadis}}]{parikh}
{Parikh}, A., {Jos{\'e}}, J., {Moreno}, F., \& {Iliadis}, C. 2008, \apjs, 178,
  110

\bibitem[{{Per{\"a}j{\"a}rvi} {et~al.}(2000){Per{\"a}j{\"a}rvi}, {Siiskonen},
  {Honkanen}, {Dendooven}, {Jokinen}, {Lipas}, {Oinonen}, {Penttil{\"a}}, \&
  {{\"A}yst{\"o}}}]{per}
{Per{\"a}j{\"a}rvi}, K., {Siiskonen}, T., {Honkanen}, A., {et~al.} 2000, Phys.
  Lett. B, 492, 1

\bibitem[{{Rauscher}(2012{\natexlab{a}})}]{rauscher2}
{Rauscher}, T. 2012{\natexlab{a}}, \apjl, 755, L10

\bibitem[{{Rauscher}(2012{\natexlab{b}})}]{rauscher3}
---. 2012{\natexlab{b}}, \apjs, 201, 26

\bibitem[{{Rauscher} {et~al.}(2011){Rauscher}, {Mohr}, {Dillmann}, \&
  {Plag}}]{rauscher1}
{Rauscher}, T., {Mohr}, P., {Dillmann}, I., \& {Plag}, R. 2011, \apj, 738, 143

\bibitem[{{Rauscher} \& {Thielemann}(2000)}]{partition}
{Rauscher}, T., \& {Thielemann}, F.-K. 2000, At. Data and Nucl. Data Tables,
  75, 1

\bibitem[{REACLIB(2012)}]{reaclibformat}
REACLIB. 2012,
  https://groups.nscl.msu.edu/jina/reaclib/db/docs/reaclibFormat.pdf

\bibitem[{{Roberts} {et~al.}(2006){Roberts}, {Hix}, {Smith}, \&
  {Fisker}}]{nicIX}
{Roberts}, L.~F., {Hix}, W.~R., {Smith}, M.~S., \& {Fisker}, J.~L. 2006, in
  Proc. Science, International Symp. on Nuclear Astrophysics (PoS NIC-IX;
  Geneva: CERN), 202

\bibitem[{Rolfs \& Rodney(1988)}]{RR}
Rolfs, C.~E., \& Rodney, W.~S. 1988, Cauldrons in the Cosmos: Nuclear
  Astrophysics (The University of Chicago Press)

\bibitem[{{Runkle} {et~al.}(2001){Runkle}, {Champagne}, \& {Engel}}]{runkle}
{Runkle}, R.~C., {Champagne}, A.~E., \& {Engel}, J. 2001, \apj, 556, 970

\bibitem[{{Saastamoinen} {et~al.}(2011)}]{2011PhRvC..83d5808S}
{Saastamoinen}, A., {et~al.} 2011, Phys. Rev. C, 83, 045808

\bibitem[{Sallaska(2010)}]{sallaska}
Sallaska, A.~L. 2010, PhD thesis, University of Washington

\bibitem[{{Sallaska} \& {Iliadis}(2012)}]{nicXII}
{Sallaska}, A.~L., \& {Iliadis}, C. 2012, in Proc. Science, International Symp.
  on Nuclear Astrophysics (PoS NIC-XII; Cairns: AUSTRALIA), 90

\bibitem[{{Sallaska} {et~al.}(2012){Sallaska}, {Iliadis}, {Champagne},
  {Timmes}, {Starrfield}, \& S}]{nicXIIa}
{Sallaska}, A.~L., {Iliadis}, C., {Champagne}, A.~E., {et~al.} 2012, in Proc.
  Science, International Symp. on Nuclear Astrophysics (PoS NIC-XII; Cairns:
  AUSTRALIA), 230

\bibitem[{{Sallaska} {et~al.}(2010)}]{2010PhRvL.105o2501S}
{Sallaska}, A.~L., {et~al.} 2010, Phys. Rev. Lett., 105, 152501

\bibitem[{{Sallaska} {et~al.}(2011)}]{2011PhRvC..83c4611S}
---. 2011, Phys. Rev. C, 83, 034611

\bibitem[{{Sellin} {et~al.}(1969){Sellin}, {Newson}, \& {Bilpuch}}]{sellin}
{Sellin}, D.~L., {Newson}, H.~W., \& {Bilpuch}, E.~G. 1969, Ann. of Phys., 51,
  461

\bibitem[{{Setoodehnia} {et~al.}(2010){Setoodehnia}, {Chen}, {Chen}, {Clark},
  {Deibel}, {Geraedts}, {Kahl}, {Parker}, {Seiler}, \& {Wrede}}]{kiana2010}
{Setoodehnia}, K., {Chen}, A.~A., {Chen}, J., {et~al.} 2010, \prc, 82, 022801

\bibitem[{{Setoodehnia} {et~al.}(2011)}]{kiana2011}
{Setoodehnia}, K., {et~al.} 2011, \prc, 83, 018803

\bibitem[{{Seuthe} {et~al.}(1990)}]{seuthe}
{Seuthe}, S., {et~al.} 1990, Nucl. Phys. A, 514, 471

\bibitem[{{Smith} {et~al.}(1993){Smith}, {Kawano}, \& {Malaney}}]{smith93}
{Smith}, M.~S., {Kawano}, L.~H., \& {Malaney}, R.~A. 1993, \apjs, 85, 219

\bibitem[{{Smith} {et~al.}(2002)}]{smithnova}
{Smith}, M.~S., {et~al.} 2002, in Classical Nova Explosions, ed. M.~{Hernanz}
  \& J.~{Jos{\'e}}, 161

\bibitem[{{Stegm{\"u}ller} {et~al.}(1996)}]{stegmuller}
{Stegm{\"u}ller}, F., {et~al.} 1996, Nucl. Phys. A, 601, 168

\bibitem[{{Takahashi} \& {Yokoi}(1987)}]{takahashi}
{Takahashi}, K., \& {Yokoi}, K. 1987, At. Data and Nucl. Data Tables, 36, 375

\bibitem[{{The} {et~al.}(2000){The}, {El Eid}, \& {Meyer}}]{meyer}
{The}, L.-S., {El Eid}, M.~F., \& {Meyer}, B.~S. 2000, \apj, 533, 998

\bibitem[{{Thompson} \& {Iliadis}(1999)}]{ilthomp}
{Thompson}, W.~J., \& {Iliadis}, C. 1999, Nucl. Phys. A, 647, 259

\bibitem[{{Tilley} {et~al.}(1995){Tilley}, {Weller}, {Cheves}, \&
  {Chasteler}}]{tilley}
{Tilley}, D.~R., {Weller}, H.~R., {Cheves}, C.~M., \& {Chasteler}, R.~M. 1995,
  Nucl. Phys. A, 595, 1

\bibitem[{{Ugalde}(2008)}]{claudio}
{Ugalde}, C. 2008, private communication

\bibitem[{{Wang} {et~al.}(2012){Wang}, {Audi}, {Wapstra}, {Kondev},
  {MacCormick}, {Xu}, \& B.}]{ame2012}
{Wang}, M., {Audi}, G., {Wapstra}, A.~H., {et~al.} 2012, Chin. Phys. C, 36,
  1287

\bibitem[{{Ward} \& {Fowler}(1980)}]{ward}
{Ward}, R.~A., \& {Fowler}, W.~A. 1980, \apj, 238, 266

\bibitem[{{Wietfeldt} \& {Greene}(2011)}]{neutron}
{Wietfeldt}, F.~E., \& {Greene}, G.~L. 2011, Rev. Mod. Phys., 83, 1173

\bibitem[{{Xu} {et~al.}(2013){Xu}, {Goriely}, {Jorissen}, {Chen}, \&
  {Arnould}}]{bruslib}
{Xu}, Y., {Goriely}, S., {Jorissen}, A., {Chen}, G., \& {Arnould}, M. 2013,
  Astron. Astrophys., 549, A106, http://www.astro.ulb.ac.be/bruslib

\bibitem[{{Yagi}(1962)}]{yagi}
{Yagi}, K. 1962, J. of the Phys. Soc. of Jpn, 17, 604

\bibitem[{{Yokota} {et~al.}(1982){Yokota}, {Fujioka}, {Ichimaru}, {Mihara}, \&
  {Chiba}}]{yokota}
{Yokota}, H., {Fujioka}, K., {Ichimaru}, K., {Mihara}, Y., \& {Chiba}, R. 1982,
  Nucl. Phys. A, 383, 298

\end{thebibliography}
\end{document}